\documentclass{aa_mod}  

\usepackage{graphicx}
\usepackage{comment}
\usepackage{soul} % to use strike out 

\usepackage{txfonts}
\usepackage[colorlinks=true,
    linkcolor=blue,
    filecolor=magenta,      
    urlcolor=cyan,
    citecolor=blue]{hyperref}

\usepackage{xcolor}
\usepackage[normalem]{ulem} % for \sout to work 
\makeatletter
\renewcommand*\aa@pageof{, page \thepage{} of \pageref*{LastPage}}
\makeatother

\newcommand{\simpropto}{\mathrel{\vcenter{
  \offinterlineskip\halign{\hfil$##$\cr
    \propto\cr\noalign{\kern2pt}\sim\cr\noalign{\kern-2pt}}}}}

\newcommand{\RoNSNS}{\ensuremath{188_{-138}^{+290}\,\mathrm{Gpc^{-3}\,yr^{-1}}}}

\newcommand{\nsnsETTgw}{$1.16^{+1.79}_{-0.85}\times10^4$}
\newcommand{\nsnsETTknJ}{$2.4^{+3.8}_{-1.8}$}
\newcommand{\nsnsETTknz}{$271^{+419}_{-200}$}
\newcommand{\nsnsETTkng}{$2.54^{+3.93}_{-1.88}\times10^3$}
\newcommand{\nsnsETTafterradio}{$15^{+23}_{-11}$}
\newcommand{\nsnsETTafteroptic}{$15^{+23}_{-11}$}
\newcommand{\nsnsETTafterx}{$33^{+51}_{-24}$}
\newcommand{\nsnsETTafterSSC}{$0.03^{+0.04}_{-0.02}$}
\newcommand{\nsnsETTpromptGBM}{$21^{+53}_{-13}$}

\newcommand{\nsnsETTgwsky}{$130^{+202}_{-96}$}
\newcommand{\nsnsETTknJsky}{$1.8^{+2.8}_{-1.3}$}
\newcommand{\nsnsETTknzsky}{$42^{+65}_{-31}$}
\newcommand{\nsnsETTkngsky}{$104^{+160}_{-76}$}
\newcommand{\nsnsETTafterradiosky}{$2.8^{+4.4}_{-2.1}$}
\newcommand{\nsnsETTafteropticsky}{$1.2^{+1.8}_{-0.9}$}
\newcommand{\nsnsETTafterxsky}{$1.7^{+2.7}_{-1.3}$}
\newcommand{\nsnsETTafterSSCsky}{$0.01^{+0.02}_{-0.01}$}
\newcommand{\nsnsETTpromptGBMsky}{$1.3^{+2.1}_{-1.0}$}

\newcommand{\nsnsETTgwskyten}{$4.9^{+7.6}_{-3.6}$}
\newcommand{\nsnsETTknJskyten}{$0.64^{+0.98}_{-0.47}$}
\newcommand{\nsnsETTknzskyten}{$3.9^{+6.0}_{-2.9}$}
\newcommand{\nsnsETTkngskyten}{$4.0^{+6.2}_{-3.0}$}
\newcommand{\nsnsETTafterradioskyten}{$0.30^{+0.47}_{-0.22}$}
\newcommand{\nsnsETTafteropticskyten}{$0.15^{+0.22}_{-0.11}$}
\newcommand{\nsnsETTafterxskyten}{$0.20^{+0.31}_{-0.15}$}
\newcommand{\nsnsETTafterSSCskyten}{$0.003^{+0.005}_{-0.002}$}
\newcommand{\nsnsETTpromptGBMskyten}{$0.15^{+0.23}_{-0.11}$}

\newcommand{\nsnsETLgw}{$2.40^{+3.70}_{-1.77}\times10^4$}
\newcommand{\nsnsETLknJ}{$2.4^{+3.8}_{-1.8}$}
\newcommand{\nsnsETLknz}{$268^{+415}_{-198}$}
\newcommand{\nsnsETLkng}{$2.89^{+4.47}_{-2.13}\times10^3$}
\newcommand{\nsnsETLafterradio}{$15^{+23}_{-11}$}
\newcommand{\nsnsETLafteroptic}{$20^{+31}_{-15}$}
\newcommand{\nsnsETLafterx}{$45^{+69}_{-33}$}
\newcommand{\nsnsETLafterSSC}{$0.03^{+0.04}_{-0.02}$}
\newcommand{\nsnsETLpromptGBM}{$27^{+68}_{-17}$}

\newcommand{\nsnsETLgwsky}{$412^{+636}_{-304}$}
\newcommand{\nsnsETLknJsky}{$2.0^{+3.2}_{-1.5}$}
\newcommand{\nsnsETLknzsky}{$125^{+193}_{-92}$}
\newcommand{\nsnsETLkngsky}{$283^{+437}_{-208}$}
\newcommand{\nsnsETLafterradiosky}{$6.6^{+10.1}_{-4.8}$}
\newcommand{\nsnsETLafteropticsky}{$4.0^{+6.1}_{-2.9}$}
\newcommand{\nsnsETLafterxsky}{$5.7^{+8.8}_{-4.2}$}
\newcommand{\nsnsETLafterSSCsky}{$0.02^{+0.03}_{-0.01}$}
\newcommand{\nsnsETLpromptGBMsky}{$3.9^{+6.1}_{-2.9}$}

\newcommand{\nsnsETLgwskyten}{$17^{+26}_{-13}$}
\newcommand{\nsnsETLknJskyten}{$1.5^{+2.3}_{-1.1}$}
\newcommand{\nsnsETLknzskyten}{$13^{+20}_{-10}$}
\newcommand{\nsnsETLkngskyten}{$14^{+21}_{-10}$}
\newcommand{\nsnsETLafterradioskyten}{$0.88^{+1.35}_{-0.64}$}
\newcommand{\nsnsETLafteropticskyten}{$0.33^{+0.50}_{-0.24}$}
\newcommand{\nsnsETLafterxskyten}{$0.47^{+0.72}_{-0.35}$}
\newcommand{\nsnsETLafterSSCskyten}{$0.007^{+0.010}_{-0.005}$}
\newcommand{\nsnsETLpromptGBMskyten}{$0.36^{+0.56}_{-0.27}$}

\newcommand{\nsnsETTCEgw}{$6.70^{+10.34}_{-4.93}\times10^4$}
\newcommand{\nsnsETTCEknJ}{$2.4^{+3.8}_{-1.8}$}
\newcommand{\nsnsETTCEknz}{$278^{+430}_{-205}$}
\newcommand{\nsnsETTCEkng}{$3.29^{+5.09}_{-2.43}\times10^3$}
\newcommand{\nsnsETTCEafterradio}{$16^{+25}_{-12}$}
\newcommand{\nsnsETTCEafteroptic}{$25^{+38}_{-18}$}
\newcommand{\nsnsETTCEafterx}{$65^{+100}_{-45}$}
\newcommand{\nsnsETTCEafterSSC}{$0.03^{+0.04}_{-0.02}$}
\newcommand{\nsnsETTCEpromptGBM}{$37^{+100}_{-23}$}

\newcommand{\nsnsETTCEgwsky}{$4.87^{+7.53}_{-3.60}\times10^4$}
\newcommand{\nsnsETTCEgwskyten}{$3.68^{+5.68}_{-2.71}\times10^3$}
\newcommand{\nsnsETLCEgw}{$7.81^{+12.06}_{-5.76}\times10^4$}
\newcommand{\nsnsETLCEknJ}{$2.4^{+3.8}_{-1.8}$}
\newcommand{\nsnsETLCEknz}{$278^{+430}_{-205}$}
\newcommand{\nsnsETLCEkng}{$3.31^{+5.12}_{-2.44}\times10^3$}
\newcommand{\nsnsETLCEafterradio}{$16^{+25}_{-12}$}
\newcommand{\nsnsETLCEafteroptic}{$25^{+39}_{-19}$}
\newcommand{\nsnsETLCEafterx}{$68^{+100}_{-50}$}
\newcommand{\nsnsETLCEafterSSC}{$0.03^{+0.04}_{-0.02}$}
\newcommand{\nsnsETLCEpromptGBM}{$38^{+100}_{-23}$}

\newcommand{\nsnsETLCEgwsky}{$6.04^{+9.32}_{-4.45}\times10^4$}
\newcommand{\nsnsETLCEknJsky}{$2.4^{+3.8}_{-1.8}$}
\newcommand{\nsnsETLCEknzsky}{$270^{+417}_{-200}$}
\newcommand{\nsnsETLCEkngsky}{$3.10^{+4.78}_{-2.28}\times10^3$}
\newcommand{\nsnsETLCEafterradiosky}{$16^{+25}_{-12}$}
\newcommand{\nsnsETLCEafteropticsky}{$24^{+38}_{-18}$}
\newcommand{\nsnsETLCEafterxsky}{$63^{+97}_{-46}$}
\newcommand{\nsnsETLCEafterSSCsky}{$0.03^{+0.04}_{-0.02}$}
\newcommand{\nsnsETLCEpromptGBMsky}{$36^{+56}_{-27}$}

\newcommand{\nsnsETLCEgwskyten}{$4.98^{+7.69}_{-3.67}\times10^3$}
\newcommand{\nsnsETLCEknJskyten}{$2.4^{+3.8}_{-1.8}$}
\newcommand{\nsnsETLCEknzskyten}{$251^{+388}_{-185}$}
\newcommand{\nsnsETLCEkngskyten}{$1.98^{+3.05}_{-1.46}\times10^3$}
\newcommand{\nsnsETLCEafterradioskyten}{$13^{+20}_{-10}$}
\newcommand{\nsnsETLCEafteropticskyten}{$11.5^{+17.8}_{-8.5}$}
\newcommand{\nsnsETLCEafterxskyten}{$20^{+31}_{-15}$}
\newcommand{\nsnsETLCEafterSSCskyten}{$0.03^{+0.04}_{-0.02}$}
\newcommand{\nsnsETLCEpromptGBMskyten}{$13^{+20}_{-10}$}

\newcommand{\bhnsETTgw}{$1.39^{+1.41}_{-0.81}\times10^4$}
\newcommand{\bhnsETTknJ}{$0.05^{+0.05}_{-0.03}$}
\newcommand{\bhnsETTknz}{$4.2^{+4.2}_{-2.4}$}
\newcommand{\bhnsETTkng}{$20^{+20}_{-11}$}
\newcommand{\bhnsETTafterradio}{$0.21^{+0.21}_{-0.12}$}
\newcommand{\bhnsETTafteroptic}{$1.1^{+1.1}_{-0.6}$}
\newcommand{\bhnsETTafterx}{$1.4^{+1.4}_{-0.8}$}
\newcommand{\bhnsETTafterSSC}{$<10^{-3}$}
\newcommand{\bhnsETTpromptGBM}{$0.43^{+0.43}_{-0.24}$}

\newcommand{\bhnsETTgwsky}{$94^{+95}_{-55}$}
\newcommand{\bhnsETTknJsky}{$0.04^{+0.04}_{-0.03}$}
\newcommand{\bhnsETTknzsky}{$0.8^{+0.8}_{-0.5}$}
\newcommand{\bhnsETTkngsky}{$2.6^{+2.6}_{-1.5}$}
\newcommand{\bhnsETTafterradiosky}{$0.08^{+0.08}_{-0.05}$}
\newcommand{\bhnsETTafteropticsky}{$0.10^{+0.11}_{-0.06}$}
\newcommand{\bhnsETTafterxsky}{$0.14^{+0.14}_{-0.08}$}

\newcommand{\bhnsETTpromptGBMsky}{$0.06^{+0.06}_{-0.04}$}

\newcommand{\bhnsETTgwskyten}{$5.0^{+5.0}_{-2.9}$}
\newcommand{\bhnsETTknJskyten}{$0.02^{+0.02}_{-0.01}$}
\newcommand{\bhnsETTknzskyten}{$0.12^{+0.12}_{-0.07}$}
\newcommand{\bhnsETTkngskyten}{$0.37^{+0.37}_{-0.21}$}
\newcommand{\bhnsETTafterradioskyten}{$0.014^{+0.014}_{-0.008}$}
\newcommand{\bhnsETTafteropticskyten}{$0.011^{+0.011}_{-0.007}$}
\newcommand{\bhnsETTafterxskyten}{$0.013^{+0.013}_{-0.008}$}

\newcommand{\bhnsETTpromptGBMskyten}{$0.006^{+0.006}_{-0.004}$}

\newcommand{\bhnsETLgw}{$2.51^{+2.54}_{-1.46}\times10^4$}
\newcommand{\bhnsETLknJ}{$0.05^{+0.05}_{-0.03}$}
\newcommand{\bhnsETLknz}{$4.2^{+4.2}_{-2.4}$}
\newcommand{\bhnsETLkng}{$20^{+20}_{-12}$}
\newcommand{\bhnsETLafterradio}{$0.20^{+0.21}_{-0.12}$}
\newcommand{\bhnsETLafteroptic}{$1.6^{+1.6}_{-0.9}$}
\newcommand{\bhnsETLafterx}{$2.1^{+2.1}_{-1.2}$}
\newcommand{\bhnsETLafterSSC}{$<10^{-3}$}
\newcommand{\bhnsETLpromptGBM}{$0.47^{+0.47}_{-0.27}$}

\newcommand{\bhnsETLgwsky}{$317^{+321}_{-184}$}
\newcommand{\bhnsETLknJsky}{$0.05^{+0.05}_{-0.03}$}
\newcommand{\bhnsETLknzsky}{$1.9^{+2.0}_{-1.1}$}
\newcommand{\bhnsETLkngsky}{$6.1^{+6.2}_{-3.6}$}
\newcommand{\bhnsETLafterradiosky}{$0.15^{+0.15}_{-0.09}$}
\newcommand{\bhnsETLafteropticsky}{$0.24^{+0.25}_{-0.14}$}
\newcommand{\bhnsETLafterxsky}{$0.28^{+0.29}_{-0.17}$}

\newcommand{\bhnsETLpromptGBMsky}{$0.14^{+0.14}_{-0.08}$}

\newcommand{\bhnsETLgwskyten}{$13.9^{+14.1}_{-8.0}$}
\newcommand{\bhnsETLknJskyten}{$0.04^{+0.04}_{-0.02}$}
\newcommand{\bhnsETLknzskyten}{$0.38^{+0.38}_{-0.22}$}
\newcommand{\bhnsETLkngskyten}{$0.45^{+0.45}_{-0.26}$}
\newcommand{\bhnsETLafterradioskyten}{$0.04^{+0.04}_{-0.02}$}
\newcommand{\bhnsETLafteropticskyten}{$0.03^{+0.03}_{-0.02}$}
\newcommand{\bhnsETLafterxskyten}{$0.03^{+0.03}_{-0.02}$}

\newcommand{\bhnsETLpromptGBMskyten}{$0.02^{+0.02}_{-0.01}$}

\newcommand{\bhnsETTCEgw}{$6.28^{+6.35}_{-3.64}\times10^4$}
\newcommand{\bhnsETTCEknJ}{$0.06^{+0.05}_{-0.03}$}
\newcommand{\bhnsETTCEknz}{$4.2^{+4.2}_{-2.4}$}
\newcommand{\bhnsETTCEkng}{$20^{+21}_{-12}$}
\newcommand{\bhnsETTCEafterradio}{$0.21^{+0.22}_{-0.12}$}
\newcommand{\bhnsETTCEafteroptic}{$2.8^{+2.9}_{-1.7}$}
\newcommand{\bhnsETTCEafterx}{$4.6^{+4.7}_{-2.7}$}
\newcommand{\bhnsETTCEafterSSC}{$<10^{-3}$}
\newcommand{\bhnsETTCEpromptGBM}{$0.64^{+0.65}_{-0.37}$}

\newcommand{\bhnsETTCEgwsky}{$4.87^{+4.92}_{-2.82}\times10^4$}
\newcommand{\bhnsETTCEknJsky}{$0.05^{+0.05}_{-0.03}$}
\newcommand{\bhnsETTCEknzsky}{$4.2^{+4.2}_{-2.4}$}
\newcommand{\bhnsETTCEkngsky}{$20^{+20}_{-11}$}
\newcommand{\bhnsETTCEafterradiosky}{$0.21^{+0.21}_{-0.12}$}
\newcommand{\bhnsETTCEafteropticsky}{$2.8^{+2.9}_{-1.6}$}
\newcommand{\bhnsETTCEafterxsky}{$4.6^{+4.6}_{-2.6}$}

\newcommand{\bhnsETTCEpromptGBMsky}{$0.64^{+0.64}_{-0.37}$}

\newcommand{\bhnsETTCEgwskyten}{$4.66^{+4.71}_{-2.70}\times10^3$}
\newcommand{\bhnsETTCEknJskyten}{$0.05^{+0.05}_{-0.03}$}
\newcommand{\bhnsETTCEknzskyten}{$4.2^{+4.2}_{-2.4}$}
\newcommand{\bhnsETTCEkngskyten}{$19^{+19}_{-11}$}
\newcommand{\bhnsETTCEafterradioskyten}{$0.20^{+0.20}_{-0.11}$}
\newcommand{\bhnsETTCEafteropticskyten}{$0.86^{+0.87}_{-0.50}$}
\newcommand{\bhnsETTCEafterxskyten}{$1.0^{+1.0}_{-0.6}$}

\newcommand{\bhnsETTCEpromptGBMskyten}{$0.35^{+0.36}_{-0.20}$}

\newcommand{\bhnsETLCEgw}{$7.15^{+7.24}_{-4.15}\times10^4$}
\newcommand{\bhnsETLCEknJ}{$0.05^{+0.05}_{-0.03}$}
\newcommand{\bhnsETLCEknz}{$4.2^{+4.2}_{-2.4}$}
\newcommand{\bhnsETLCEkng}{$20^{+21}_{-12}$}
\newcommand{\bhnsETLCEafterradio}{$0.23^{+0.23}_{-0.13}$}
\newcommand{\bhnsETLCEafteroptic}{$2.9^{+2.9}_{-1.7}$}
\newcommand{\bhnsETLCEafterx}{$4.6^{+4.7}_{-2.7}$}
\newcommand{\bhnsETLCEafterSSC}{$<10^{-3}$}
\newcommand{\bhnsETLCEpromptGBM}{$0.65^{+0.65}_{-0.37}$}

\newcommand{\bhnsETLCEgwsky}{$5.82^{+5.89}_{-3.38}\times10^4$}
\newcommand{\bhnsETLCEknJsky}{$0.05^{+0.05}_{-0.03}$}
\newcommand{\bhnsETLCEknzsky}{$4.2^{+4.2}_{-2.4}$}
\newcommand{\bhnsETLCEkngsky}{$20^{+20}_{-11}$}
\newcommand{\bhnsETLCEafterradiosky}{$0.21^{+0.21}_{-0.12}$}
\newcommand{\bhnsETLCEafteropticsky}{$2.8^{+2.9}_{-1.6}$}
\newcommand{\bhnsETLCEafterxsky}{$4.6^{+4.6}_{-2.6}$}

\newcommand{\bhnsETLCEpromptGBMsky}{$0.64^{+0.64}_{-0.37}$}

\newcommand{\bhnsETLCEgwskyten}{$6.10^{+6.17}_{-3.54}\times10^3$}
\newcommand{\bhnsETLCEknJskyten}{$0.05^{+0.05}_{-0.03}$}
\newcommand{\bhnsETLCEknzskyten}{$4.1^{+4.2}_{-2.4}$}
\newcommand{\bhnsETLCEkngskyten}{$19^{+19}_{-11}$}
\newcommand{\bhnsETLCEafterradioskyten}{$0.20^{+0.20}_{-0.12}$}
\newcommand{\bhnsETLCEafteropticskyten}{$0.88^{+0.89}_{-0.51}$}
\newcommand{\bhnsETLCEafterxskyten}{$1.11^{+1.12}_{-0.65}$}

\newcommand{\bhnsETLCEpromptGBMskyten}{$0.36^{+0.37}_{-0.21}$}

\newcommand{\UCSD}{Department of Astronomy and Astrophysics, University of California, San Diego, La Jolla, CA 92093, USA}
\begin{document}

   \title{Multi-messenger observations in the Einstein Telescope era: binary neutron star and black hole - neutron star mergers\thanks{The data produced in this work are publicly available on Zenodo through the link \url{https://doi.org/10.5281/zenodo.15411012}. The scripts and files to reproduce the main figures in the text are publicly available at \url{https://github.com/acolombo140/ET_MM}.}}
   
   \author{{Alberto} Colombo\inst{1,2}\thanks{\email{alberto.colombo@inaf.it}}, {Om Sharan} Salafia\inst{1,2}, {Giancarlo} Ghirlanda\inst{1,2}, {Francesco} Iacovelli\inst{3,4,5}, {Michele} Mancarella\inst{6}, {Floor~S.} Broekgaarden\inst{7}, {Lara} Nava\inst{1}, {Bruno} Giacomazzo\inst{8,2}, {Monica} Colpi\inst{8,2,1}}

   \institute{
            INAF -- Osservatorio Astronomico di Brera, via Emilio Bianchi 46, I-23807 Merate (LC), Italy
        \and
             INFN -- Sezione di Milano-Bicocca, Piazza della Scienza 3, I-20126 Milano (MI), Italy
        \and
            D\'epartement de Physique Th\'eorique, Universit\'e de Gen\`eve, 24 quai Ernest Ansermet, 1211 Gen\`eve 4, Switzerland
        \and
            Gravitational Wave Science Center (GWSC), Universit\'e de Gen\`eve, 24 quai E. Ansermet, CH-1211 Geneva, Switzerland
        \and
            William H. Miller III Department of Physics and Astronomy, Johns Hopkins University, 3400 North Charles Street, Baltimore, Maryland, 21218, USA
        \and
            Aix-Marseille Universit\'e, Universit\'e de Toulon, CNRS, CPT, Marseille, France
        \and 
            \UCSD
        \and
            Università degli Studi di Milano-Bicocca, Dipartimento di Fisica “G. Occhialini”, Piazza della Scienza 3, I-20126 Milano (MI), Italy
            }

    \authorrunning{A.\ Colombo et al.}
    \titlerunning{Multi-messenger observations in the Einstein Telescope era}
  
   %\date{Received xxx; accepted xxx}

% \abstract{}{}{}{}{} 
% 5 {} token are mandatory
 
 \abstract{The Einstein Telescope (ET), a proposed next-generation gravitational wave (GW) observatory, will expand the reach of GW astronomy of stellar-mass compact object binaries to unprecedented distances, enhancing opportunities for multi-messenger observations. 
 Here we investigate multi-messenger emission properties of binary neutron star (NSNS) and black hole-neutron star (BHNS) mergers detectable by ET, providing projections to optimize observational strategies and maximize scientific insights from these sources. Using a synthetic population of compact binary mergers, we model each source's GW signal-to-noise ratio, sky localization uncertainty, kilonova (KN) light curves in optical and near-infrared bands, fluence of the relativistic jet gamma-ray burst (GRB) prompt emission and afterglow light curves across radio, optical, X-ray and very high energy wavelengths. We analyze multi-messenger detectability prospects for ET as a standalone observatory with two different configurations and within a network of next-generation GW detectors. ET will detect over $10^4$ NSNS mergers annually, enabling  potential observation of tens to hundreds of electromagnetic (EM) counterparts. 
BHNS mergers have more limited multi-messenger prospects, but joint GW-EM  rates will increase by an order of magnitude compared to current-generation instruments. We quantify uncertainties due to the NS equation of state (EoS) and mass distribution of NSNSs, as well as the NS EoS and BH spin for BHNSs. While a single ET will achieve an impressive GW detection rate, the fraction of well-localized events  (< 100 deg$^2$) is orders of magnitude lower than in a network with additional detectors. This significantly limits efficient EM follow-up and science cases requiring well-characterized counterparts or early observations. The challenge is even greater for BHNS mergers due to their low EM rate. Thus, multi-messenger astronomy in the next decade will critically depend on a network of at least two detectors.
}


   \maketitle

   %
%-------------------------------------------------------------------

\section{Introduction}
On August 17, 2017, the first gravitational wave (GW) signal consistent with the coalescence of a binary neutron star (NSNS) system, GW170817, was detected \citep{Abbott2017,Abbott2019_GW170817_properties} by the LIGO \citep{AdvancedLIGO2015}, Virgo \citep{AdvancedVirgo2015}, and KAGRA \citep{KAGRA2013} (LVK) GW detectors. Remarkably, less than two seconds later a short gamma-ray burst (GRB), GRB170817A, was detected by the Fermi and INTEGRAL satellites \citep{Abbott2017_GRB170817A}, marking the beginning of the multi-messenger (MM) era with GWs \citep{Abbott2017_MM}. The presence of Virgo in the GW detector network enabled GW170817 to be localized within an area of 28 square degrees in the sky \citep{Veitch2015,Abbott2019_GW170817_properties}, prompting an extensive observing campaign that spanned the entire electromagnetic (EM) spectrum \citep{Abbott2017_MM}. Approximately 11 hours after the merger, a faint and rapidly evolving optical/near-infrared transient was discovered in the nearby galaxy NGC 4993 \citep{Coulter2017}. This transient was spectroscopically identified as a kilonova \citep[KN, ][]{Pian2017}, characterized by quasi-thermal emission from the merger's expanding ejecta \citep{li1998,metzger2019_kn}. Subsequently, a non-thermal broadband source (radio to X-rays) was detected at the same location and was identified months later as the afterglow of an off-axis relativistic jet, confirmed through very long baseline interferometry observations \citep{Lazzati2018,Mooley2018,ghirlanda2019}.

In 2019, a second NSNS merger, GW190425, was discovered \citep{Abbott2019_GW190425}. However, no EM counterpart was identified, possibly due to the poorly constrained sky localization and the relatively far away distance. No significant NSNS merger candidates have been reported in the first part of the fourth observing run (O4a) of the LVK detector network\footnote{The plan for the observing run can be seen here: \url{https://observing.docs.ligo.org/plan/}}, despite the improved GW detection range with respect to the previous runs. Therefore, although GW170817 showcased the potential of multi-messenger astronomy, it remains a single case so far. More recently, two long GRBs, GRB 211211A \citep{rastinejad2022,troja2022,Mei2022} and GRB 230307A \citep{levan2024,yang2024}, exhibited KN signatures, suggesting an NSNS origin. Although these events were within the nominal detection horizon of GW detectors, the instruments were unfortunately offline for upgrades at the time.

Even black hole-neutron star (BHNS) mergers could be associated with EM counterparts. This is expected to occur when the distance $d_\mathrm{tidal}$ at which the tidal disruption of the NS takes place is larger than the BH innermost stable circular orbit (ISCO): $d_\mathrm{tidal}>R_\mathrm{ISCO}$ \citep{kawagichi2015,foucart2018,foucart2019,barbieri2020}. This requirement depends on several factors, being more favorable in the case of larger NS tidal deformability, higher  BH spins, and lower BH masses. The presence of matter outside the BH can potentially power a KN \citep{li1998,kawagichi2015} and launch a relativistic jet, which may generate prompt and afterglow emission, contributing to a subclass of GRBs \citep{Lattimer1974,Eichler1989,Mochkovitch1993}.

To date, no EM counterpart has been conclusively associated with BHNS mergers, even though these events have been observed through GWs \citep{Abbott2021,lvk2024_bhns}. Notably, during O4a in April 2024, an event named GW230529 was reported \citep{lvk2024_bhns}. The mass of the BH in this event was $3.6^{+0.8}_{-1.2} M_\odot$, placing it within the lower mass gap between NSs and BHs. Although the probability of EM emission from this event was limited (around 10\%), the potential existence of systems with low-mass BHs increases the likelihood of EM counterparts \citep{xing2024}, even in scenarios with non-spinning BHs or soft equations of state (EoS), which appear to be the most favored cases based on LVK constraints \citep{GWTC3}.

The development of next-generation GW observatories, such as the Einstein Telescope (ET) \citep{punturo2010} and Cosmic Explorer (CE) \citep{Abbott2017_CE,Reitze2019}, is anticipated to lead to a monumental advancement in our ability to observe GW sources. These future facilities are set to revolutionize the field in the next decades by offering unprecedented detection capabilities far beyond what is achievable with current instruments. ET is planned to be constructed as an underground observatory for best insulation from seismic and environmental noise. Until recently, the leading design concept was that of a triangular `xylophone', consisting of three pairs of low- and high-frequency interferometers with 10 km arms. A recent cost-benefit comparison of different designs \citep{branchesi2023} considered a different configuration consisting of two L-shaped interferometers with 15 km arms, misaligned by 45 degrees among them, again featuring both a low- and a high-frequency instrument. Instead, the current CE design consists of an L-shaped surface-based interferometer with an arm length of 40 km, located in the United States. A possible second 20 km CE interferometer is being considered \citep{evans2023,gupta2023}.

The current second-generation GW detectors, including LVK and the upcoming LIGO-India, are projected to detect a few to several hundred NSNS mergers annually at design sensitivity, with a maximum observable redshift of approximately $z\sim0.2$ \citep{abbott2020}. 
By contrast, next-generation observatories like ET and CE will significantly extend the observable volume of the Universe \citep{maggiore2020,Kalogera2021}. ET is expected to detect NSNS mergers up to redshift $z\sim 4$ (for optimally located and oriented systems), well beyond the peak of star formation, while CE will push the boundary even further, with a detection capability virtually extending to $z\sim10$  \citep{maggiore2020,branchesi2023,evans2021,evans2023,Iacovelli2022,gupta2023}. This remarkable increase in sensitivity and detection range will enable the observation of compact binary mergers across cosmic history up to the epoch of the first forming structures.

Beyond sheer detection rates, ET and CE will also deliver precise measurements of source parameters. For the most favourable systems, and in presence of more than one detector, these include improved estimates of sky localization, luminosity distance and binary inclination angle, thereby enhancing our ability to study the astrophysical and cosmological properties of GW sources. Such advancements will not only enable more accurate source characterization but also facilitate more effective coordination with EM observatories, bolstering the field of multi-messenger astronomy \citep{maggiore2020,branchesi2023}.

The era of ET will coincide with significant advancements in EM facilities, enabling a deeper exploration of multi-messenger astrophysics. Among the most anticipated developments in the optical and near-infrared is the Vera C. Rubin Observatory \citep{ivezic2019,andreoni2022}, whose exceptional sensitivity and wide field of view make it uniquely suited to detect KN\ae\ associated with GW events. Moreover, the introduction of highly sensitive spectroscopic telescopes, such as the Extremely Large Telescope \citep[ELT,][]{marconi2022} and the Wide-field Spectroscopic Telescope project \citep[WST,][]{mainieri2024}, will be pivotal in studying the EM counterparts of GW events. In the radio domain, the Square Kilometre Array \citep[SKA,][]{braun2019} and the Next Generation Very Large Array \citep[ngVLA,][]{corsi2019} will provide remarkable sensitivity for detecting GRB afterglows. The detection of these counterparts will also benefit from new X-ray facilities such as NewAthena \citep{nandra2013}. For the MeV and GeV bands, there are currently only proposed but not yet confirmed projects, such as the Transient High-Energy Sky and Early Universe Surveyor \citep[THESEUS,][]{amati2021}, aimed at enhancing high-energy transient monitoring. However, in this energy range, the new generation of AstroSats, such as the High-Energy Rapid Modular Ensemble of Satellites \citep[HERMES, ][]{fiore2020,ghirlanda2024} could play a crucial role due to their lower costs and rapid development timelines, making them a valuable asset for future observations. Finally, in the very high energy (VHE, $>100$ GeV) gamma-ray band, the advent of the Cherenkov Telescope Array \citep[CTA,][]{CTA2019} will offer unprecedented opportunities to detect transient emission associated with the most extreme astrophysical events.

To optimize the scientific impact of next-generation GW detectors in a multi-messenger framework, it is crucial to assess the expected joint detections and the properties of the expected EM emission, in order to define the necessary instrument requirements, and determine the most effective observation strategies. In this work, we present our projections for the rates and properties of NSNS and BHNS events that will be detected as multi-messenger sources in the ET era. This study is based on a state-of-the-art population synthesis model, considering the emission from KN, GRB prompt, and GRB afterglow, including the VHE band, extending and improving the methods presented in \cite{colombo2022,colombo2023}. The results are reported considering ET both as a standalone detector with either of the two most likely configurations, and as part of a network with two CE detectors, representing the most optimistic scenario. Additionally, we also investigate the impact of the mass distribution and the NS EoS on the NSNS results, as well as the effects of the NS EoS and the BH spin on the BHNS outcomes.

The structure of the paper is as follows. Section \ref{sec::gw_em_models} outlines the distribution of binary parameters used in our NSNS and BHNS population models, describes the emission models for both GW and EM signals, and details the representative multi-messenger detection limits assumed to compute the detection rates. In Section \ref{sec::mm_prospects}, we present our findings on multi-messenger detection rates, as well as variations in the starting NSNS and BHNS populations. Additionally, we discuss the impact of different detectors network configurations. In Section \ref{sec::EM_properties}, we analyse the expected properties of the various EM counterparts. Section \ref{sec::conclusions} provides a summary discussion of our results, including a comparison with similar studies in the literature, and the conclusions. Throughout this study, we adopt a flat cosmology with parameters from \cite{Planck2020}.

\section{GW and EM population models}\label{sec::gw_em_models}
\subsection{NSNS and BHNS populations}
For the analysis conducted in this work, we considered two populations of merging compact objects, one of NSNS and one of BHNS. 

The NSNS population was the same as assumed in \cite{colombo2022}, in which
the mass distribution was calibrated with current observational data from GW detections and Galactic NSNS systems (see Figure \ref{fig:nsns_mass}). To analyze how the NSNS mass distribution impacts our results, we also considered two variations: a Gaussian and a uniform mass distribution \citep[see also][]{schwab2010,valentim2011,Kiziltan2013,you2024}. The Gaussian has a mean value of 1.33 $M_\odot$ with standard deviation of 0.09 $M_\odot$ and it is based on the masses of galactic NSNS binaries \citep{ozel2012,ozel2016}. The uniform mass distribution ranges between 1 $M_\odot$ and $M_\mathrm{TOV}$, the latter representing the maximum mass for a non-rotating NS, whose value depends on the choice of the NS EoS.
We assumed two different EoS models: the soft SFHo \citep{Hempel2012} and the stiff DD2 EoS \citep{Steiner2013}. The SFHo EoS predicts a maximum non-rotating NS mass of $M_\mathrm{TOV} = 2.06\,\mathrm{M_\odot}$ and a radius of $R_{1.4}=11.30\,\mathrm{km}$ for a 1.4 $\mathrm{M_\odot}$ NS, while the DD2 has $M_\mathrm{TOV} = 2.46\,\mathrm{M_\odot}$ and $R_{1.4}=13.25\,\mathrm{km}$.

Hereafter, we call `fiducial' the NSNS population constructed adopting the mass distribution from \cite{colombo2022} and the SFHo EoS, treating the other combinations of mass distributions and EoS as variations over this fiducial model.

For all variations, we assumed the same cosmic merger rate density as in \cite{colombo2022}, derived by convolving a power law delay time distribution $P(t_\mathrm{d})\propto t_\mathrm{d}^{-1}$ (with a minimal delay of $t_\mathrm{d,min}=50\,\mathrm{Myr}$, \citealt{Mapelli2018,Safarzadeh2019,zevin2020}) with the cosmic star formation rate from \citealt{Madau2014}. This was normalized to align with a local rate density $R_0=\RoNSNS$, obtained ensuring a consistent match with the observed frequency of significant NSNS mergers in LVK observing runs up to O4a. Given the absence of NSNS detections in the latter run, this is equivalent to the NSNS local rate density obtained in \cite{colombo2022} (see Appendix A.1) with a correction factor that stems from the increased time-volume surveyed after O4a\footnote{A detailed derivation of the correction factor for the local merger rate density is provided here: \url{https://dcc.ligo.org/LIGO-P2400022/public}}.

For BHNS we considered the population described in \cite{colombo2023}. Specifically, we relied on the BH and NS mass distributions from the standard parameter set (model A) detailed in \citet{Broekgaarden2021}. We also incorporated the fiducial metallicity-specific star formation rate density from the same study, grounded in the phenomenological model of \citet{Neijssel2019}. As we did for the NSNS population, the rate density  was normalized to be consistent with  the merger rate density
$R_0 = 81^{+82}_{-47} \rm Gpc^{-3} \rm yr^{-1}$ at redshift $z = 0$, correcting the value presented in \cite{colombo2023} to reflect all the BHNS events observed until the end of O4a \citep{lvk2024_bhns}\footnote{We assume four BHNS events with false alarm rate less than 1 in 4 years, as reported in \url{https://emfollow.docs.ligo.org/userguide/capabilities.html}}.

The binary stellar evolution model behind our BHNS population adopts the prescriptions from \citet{fryer2012}'s `delayed' supernova explosion mechanism. This results in a BH mass distribution that extends into the lower mass gap, particularly with BHs having masses below $5$ $M_\odot$. Prior to GW observations, the existence of such low-mass BHs was widely debated, largely due to X-ray binary observations in the Milky Way, which suggested a sharp cutoff at around $5$ $M_\odot$ for BH masses \citep{ozel2010,farr2011}. However, recent GW detections indicate that the lower mass gap may not be as empty as once believed, with observations of a number of systems that likely contain a component within this mass range \citep{abbott2020_190814,zevin2020,lvk2024_bhns,xing2024}.

As in \cite{colombo2023}, we considered two different configurations for the BH spin parameter $\chi_\mathrm{BH}$ prior to the merger: a conservative approach with $\chi_\mathrm{BH}=0$ for all binaries, and a more optimistic scenario with a uniform distribution in the range $\chi_\mathrm{BH} \in [0,0.5]$. These values align with the typical spin values obtained in various simulations, assuming that the helium star progenitors of BHs that form second in the binary can spin-up through tidal interactions \citep{Fuller2019,Belczynski2020,RomanGarza2021,Bavera2020,Bavera2021,Bavera2023}. If the BH forms first, it is believed to have zero spin as a consequence of efficient angular momentum transport \citep{Fragos2015,Qin2018,Fuller2019,Belczynski2020}.

In order to compute the NS compactness, we assumed the SFHo and the DD2 EoS, as we did for the NSNS population. In what follows, we adopt the most conservative setup with non-spinning BHs and the SFHo EoS as our fiducial population.

\subsection{GW signal models and parameters estimation}
\begin{table*}
    \caption{Summary of the interferometer assumptions, including arm length, configuration type, location, duty cycle and the applied $\rm S/N$ threshold.}
\tiny
\centering
    \begin{tabular}{lccccc}
        \hline
        \hline
         & Arm length (km) & Configuration & Location & Duty cycle & $\rm S/N$ threshold \\
        \hline
        ET$\Delta$ & 10 & Triangular & Sardinia & 85\% & 12 \\
        ET2L & 15 & 2L-Shaped (45° misalignment) & Sardinia \& Meuse-Rhine & 85\% & 12 \\
        ET$\Delta$+2CE & 10 (ET), 40 \& 20 (CE) & Triangular + 2 CEs & Sardinia (ET), US (CE) & 85\% & 12 \\
        ET2L+2CE & 15 (ET), 40 \& 20 (CE) & 2L-Shaped + 2 CEs & Sardinia \& Meuse-Rhine (ET), US (CE) & 85\% & 12 \\
        \hline
    \end{tabular}

    \label{tab:interferometers}
\end{table*}

For each merging event within our study, we have computed the optimal matched-filter signal-to-noise ratio (S/N) (see e.g.\ Chap.~7 of \citealt{Maggiore:2007ulw}) 
\begin{equation}
    ({\rm S/N})^2 = \sum\nolimits_{i} ({\rm S/N})^2_i\,, \quad ({\rm S/N})^2_i = 4 \int_{f_{\rm min}}^{f_{\rm cut}} \dfrac{|\tilde{h}_{(i)}(f)|^2}{S_{n,i}(f)} {\rm d}f \,,
\end{equation}
with the index $i$ running over the considered detectors in the case of a network, $f_{\rm min}=2~{\rm Hz}$ for ET and $f_{\rm min}=5~{\rm Hz}$ for CE, $f_{\rm cut}$ being a cutoff frequency depending on the events' parameters, $\tilde{h}_{(i)}(f)$ denoting the Fourier-domain GW strain projected onto the detector $i$ and $S_{n,i}(f)$ the noise power spectral density (PSD) of the $i$\textsuperscript{th} interferometer. Additionally, we calculated the 90\% credible area for sky localization $\Delta \Omega_{90\%}$ for each detected signal.  We assumed two different configurations for ET: a triangular configuration with 10 km arms (ET$\Delta$), locating the detector in Sardinia, and a 2L-shaped interferometer design with 15 km arms misaligned by 45 deg among them (ET2L), locating one detector in Sardinia and the other one in the Meuse-Rhine region. For both the configurations, we also considered the possibility of ET operating in a global network with either one CE with 40 km arms or two CEs one with 40~km and one with 20~km arms (ET$\Delta$+2CE, ET2L+2CE), located in the US. For ET, we adopted the same sensitivity curves as in \citealt{branchesi2023}\footnote{The ET PSDs are publicly available at \url{https://apps.et-gw.eu/tds/?content=3&r=18213}.}, while for CE we used the latest publicly available
official PSDs\footnote{The CE PSDs we used are publicly available at \url{https://dcc.cosmicexplorer.org/CE-T2000017/public}.}.
Moreover, for each detector (and each interferometer in the case of a triangle) we incorporated a 85\% uncorrelated duty cycle, aligning with the standard set in \citealt{branchesi2023}. Given that the considered signals can stay in band as long as ${\cal O}(1~{\rm day})$ at 3G instruments, resulting in improved localization capabilities especially for a single detector, we include the effect of Earth's rotation in the reconstruction as described in \citet{IacovelliGWFAST,Iacovelli2022}. 
The main assumptions about the interferometers are summarized in Table~\ref{tab:interferometers}.

The computations of S/N and sky localization were performed via the public \texttt{GWFAST} package \citep{IacovelliGWFAST,Iacovelli2022}, adopting the \texttt{IMRPhenomD\_NRTidalv2} \citep{dietrich2019} waveform approximant for NSNS, and the \texttt{IMRPhenomNSBH} approximant \citep{pannarale2015,dietrich2019} for BHNS.  These waveform models depend on several parameters, including the detector-frame chirp mass, mass ratio, dimensionless spin parameters of the binary components, luminosity distance, sky position, binary inclination angle, polarization angle, time and phase of coalescence, and the NS tidal deformability \citep{IacovelliGWFAST}. For parameters not explicitly discussed previously, values were drawn from non-informative priors within their physically relevant ranges, as detailed in \citet{IacovelliGWFAST}.
The sky localization areas are computed within the Fisher-information-matrix formalism, valid in the high-S/N limit, in which the GW likelihood is approximated as a multivariate Gaussian near the peak (see e.g.\ \cite{Vallisneri:2007ev} for a comprehensive discussion of the formalism and its limitations). To avoid numerical instabilities in the sky position subspace due to degeneracies between some parameters (in particular distance and inclination for nearly face-on/-off systems, more likely associated to a GRB detection), we resort to a singular value decomposition as in \citet{Dupletsa:2022scg, ronchini2022} and eliminate from the inversion singular values below a threshold of $10^{-10}$.

\subsection{Ejecta properties and EM emission models}

\begin{figure}
    \centering
    \includegraphics[width=\columnwidth]{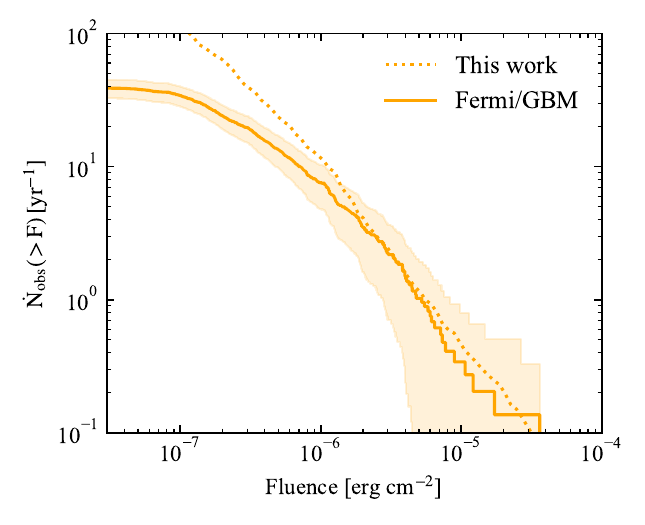}
    \caption{Cumulative distribution of events above a threshold bolometric fluence. The solid line represents the distribution for {\it Fermi}/GBM observed SGRBs (with a duration below the customary 2 s threshold), constructed using the spectral information available in the online catalog. The colored band shows the associated Poisson uncertainty.  The dotted line shows our model distribution for comparison.}
    \label{fig:logn_logs}
\end{figure}

Our framework for modeling ejecta properties and their EM emission is based on the methods presented in \cite{colombo2022, colombo2023}. A detailed description of how ejecta masses and velocities, and accretion disk masses are estimated for NSNS and BHNS systems using analytical formulae informed by numerical relativity is provided in these previous studies. The EM counterparts, such as KN and GRB prompt and afterglow, are modeled through semi-analytical approaches using the ejecta properties as inputs. Specifically, for each binary, we computed the KN light curves between $0.1$ and $30$ days in the $g$ (484 nm), $z$ (900 nm), and $J$ (1250 nm) bands. The GRB afterglow light curves over $0.1$ to $1000$ days were computed in the radio, optical, and X-ray bands. 

To explore the potential synergy between CTA and ET, we also calculated the afterglow light curves at $0.1$ and $1$ TeV  over
0.01 to 1000 days. We assumed VHE emission to be produced by the jet-driven external forward shock as a result of the
synchrotron-self Compton (SSC) radiative process. The method we followed in computing such emission is described in \citet{Salafia2022b}. In brief, this model adopts a simplified description of the emissivity obtained from a delta function approximation of the average single-particle spectrum. Klein-Nishina effects are accounted for roughly by assuming a vanishing cross section for the electron-photon interaction beyond the Klein-Nishina limit. The observed flux is then computed accounting for Doppler beaming and for light-travel-time effects. The latter effects smooth out the spectra, reducing the impact of the employed approximations, producing results that deviate from a more exact treatment \citep[e.g.][]{Miceli2022} only by factors of order one.

In computing the GRB prompt emission, we modified slightly some model parameters with respect to \citet{colombo2022,colombo2023}, for the following reasons. Our methodology, which is similar to that in \citet{salafia2019}, assumes that a constant fraction \(\eta_\gamma\) of the jet energy density, restricted to regions with a bulk Lorentz factor \(\Gamma \geq 10\), is emitted in the form of photons. The spectrum of these photons in the jet comoving frame is assumed to be the same at all angles. The observed time-integrated spectrum is then obtained by transforming the spectrum to the observer frame and integrating the emission over the jet's solid angle. In our previous works, the photon flux spectrum was then calculated by dividing the time-integrated spectrum by a fixed duration for all bursts, regardless of the viewing angle. This means that the photon flux computed from such model represents an average over the GRB duration, rather than a peak photon flux. In previous works we compared the distribution of such photon fluxes with the peak photon flux distribution of \textit{Fermi}/GBM short GRBs based on light curves with a 64~ms binning. To avoid this inconsistency, in this work we compare the distribution of bolometric fluence
$E_\mathrm{iso}(1+z)/4 \pi d_L^2$ from our model population with that obtained from the \textit{Fermi}/GBM sample of short GRBs with spectral information available in the online catalog (see Figure \ref{fig:logn_logs}). To improve the agreement between the observed and model distributions, we decreased the \(\eta_\gamma\) parameter to 0.1 (it was 0.15 in \citealt{colombo2022,colombo2023}). We also decreased the parameter that sets the efficiency of conversion of disk mass into jet energy (indicated with the symbol $\epsilon$ in equation B1 of \citealt{colombo2022}) by a factor 2.4. The value  remains well within the uncertainty limits of this parameter \citep{Salafia2021}. We note that, while these adjustments allow for better consistency between our fiducial population and the observed GRB properties, the impact on the predictions is less than the uncertainties induced by the poorly constrained local rate density and by the large number of model parameters.

 For NSNS systems observed within a viewing angle $\theta_\mathrm{v}\leq 60^\circ$, we also included a cocoon shock breakout component, modeled following the characteristics of GRB 170817A \citep{Abbott2017_GRB170817A}, namely a luminosity $L_\mathrm{SB}=10^{47}\,\mathrm{erg/s}$ and a cut-off power-law spectrum with $\nu F_\nu$ peak photon energy $E_\mathrm{p,SB}=185\,\mathrm{keV}$ and low-energy photon index $\alpha=-0.62$. In BHNS we did not consider this additional emission, because of the lack of observing constraints from this kind of sources.

 In both the populations, we assumed a jet angular structure motivated by the one of GRB 170817A \citep{ghirlanda2019}, characterized by a uniform core with a half-opening angle of $\theta_j = 3.4^\circ$ \citep[see again][for more details]{colombo2022}. However, BHNS and NSNS jets may differ due to the different environments in which they form. BHNS jets may experience less self-collimation because of the lower polar region density compared to NSNS systems \citep{Bromberg2011,Duffell2015,Lazzati2019,Urrutia2021,Salafia2020,Hamidani2021,Gottlieb2022,Salafia2022}. Additionally, NSNS mergers produce more isotropic ejecta and stronger post-merger winds \citep{Foucart2020,kawaguchi2016,Fernandez2013,Just2015}. For this reason, in the BHNS population we also considered a broader jet opening angle of $\theta_j = 15^\circ$,\footnote{The effect of this variation on the rates is included in the error bars in the right panel of Figure \ref{fig:variations}. The effect on the GRB properties is shown as an orange curve in the right panels of Figures \ref{fig:et_after} and \ref{fig:et_prompt}.} while keeping other parameters the same, adjusting only the jet core isotropic-equivalent energy $E_\mathrm{c}=E(0)$ to maintain the total jet energy constant.

\subsection{Multi-messenger detection criteria}
\label{sec:et_limits}
In order to represent the multi-messenger detection of our model sources in a simple and general way, avoiding facility-specific simulations, we opted for representing the detection condition as a threshold on integrated photon flux (for gamma-rays) or flux density (for radio and X-rays) and, equivalently, apparent magnitude for ultraviolet-to-infrared (UVOIR) bands. These thresholds are given below and are summarized in Table \ref{tab:et_det_rates}. For the GW signal detection, we imposed a network S/N threshold of 12. This relatively stringent S/N threshold also enhances the reliability of the parameter estimation forecasts based on the \texttt{GWFAST} Fisher-information-matrix \citep{Iacovelli2022}.

Regarding EM follow-up, we anticipate substantial improvements in radio and optical search depths, owing to new instrumentation. In the radio spectrum, we expect a tenfold sensitivity increase to $0.01\,{\rm mJy}$ \citep{Dobie2021} with respect to the representative limits we assumed for O4 \citep{colombo2023}. Such a depth is achievable by next-generation instruments like the SKA2 \citep{braun2019}, Next-Generation VLA \citep{corsi2019}, or DSA-2000 \citep{hallinan2019}. In optical observations, advancements are anticipated with the coming on line of large FoV instruments like the Vera Rubin Observatory \citep{ivezic2008}. We considered magnitude thresholds of 26 in the $g$ band and 24.4 in the $z$ band, corresponding to expectations for Rubin Observatory's target-of-opportunity program \citep{andreoni2022}. For the X-ray band we assumed a flux density limit of $10^{-13}\,{\rm erg/cm}^2{\rm /s/keV}$ at 1\,keV, achievable by \textit{Swift}/XRT. 

For the GRB prompt emission, given the uncertainty on the future observational landscape, we conservatively assumed a \textit{Fermi}/GBM like instrument. We represent its sensitivity with a threshold on bolometric fluence of $3.09\times10^{-7}\,{\rm erg\, cm^{-2}}$, based on a visual comparison of the fluence distribution predicted by our model with the one observed by \textit{Fermi}/GBM (see Figure \ref{fig:logn_logs}). To determine the final GRB prompt emission detection rates, we accounted for the restricted field of view and duty cycle of {\it Fermi}/GBM  by applying a correction factor of 0.60 \citep{Burns2016}.

Regarding the VHE afterglow band, we used the sensitivity curves of CTA North and South\footnote{The sensitivity curves are reported here: \url{https://www.ctao.org/for-scientists/performance/}} relative to photon energies of 0.1 and 1 TeV, assuming an integration time ranging from the minimum time of the light curve (0.01 days) up to 50 hours. We considered an event as detected if at least one point in its light curve was above this sensitivity threshold. We additionally assumed a 15\% duty cycle to account for weather and moon constraints and a 50\% reduction in sky visibility, considering that the sub-arrays are unable to observe the sky beyond a zenith angle of 60 degrees.

It is crucial to note that our analysis is predicated on the assumption that the GW sky localization areas for NSNS and BHNS mergers will be thoroughly surveyed to the outlined detection thresholds that can be obtained with extensive follow up. A more comprehensive analysis of practical detection rates would require simulations mimicking the search strategies of individual observatories. Nevertheless, as we will show, the possibility to cover the GW sky localization regions with a realistic investment of resources critically depends on the presence of multiple next-generation GW detectors in a global network.

\section{Multi-messenger detection prospects}\label{sec::mm_prospects}
\begin{figure}
    \centering
    \includegraphics[width=\columnwidth]{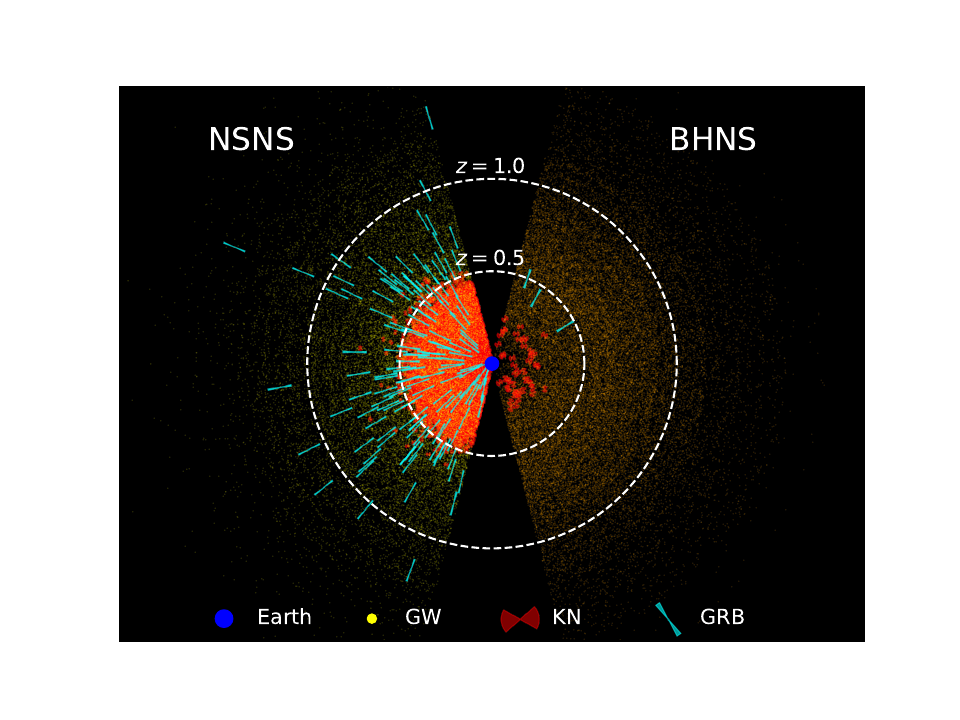}
    \footnotesize
    \caption{Representation of two of our fiducial multi-messenger synthetic populations in a geocentric Universe. Compact binaries that are detected by ET$\Delta$ are represented by yellow (NSNS -- left semicircle) and orange (BHNS -- right semicircle) dots. The distance of each dot from the Earth (blue circle) is proportional to the redshift of the corresponding compact binary. If the simulated merger produces a relativistic jet whose prompt or afterglow emission is detectable, according to the limits reported in Table \ref{tab:et_det_rates}, a cyan jet is plotted centered on the dot, with its axis inclined by the actual viewing angle with respect to the line of sight to the Earth. If a KN is also produced and if it is detectable, then a red butterfly shape is also plotted. The total number of binaries is representative of 5 years of ET operation.}
    
    \label{fig:donut}
\end{figure}

\begin{figure*}
    \centering
    \includegraphics[width=\textwidth]{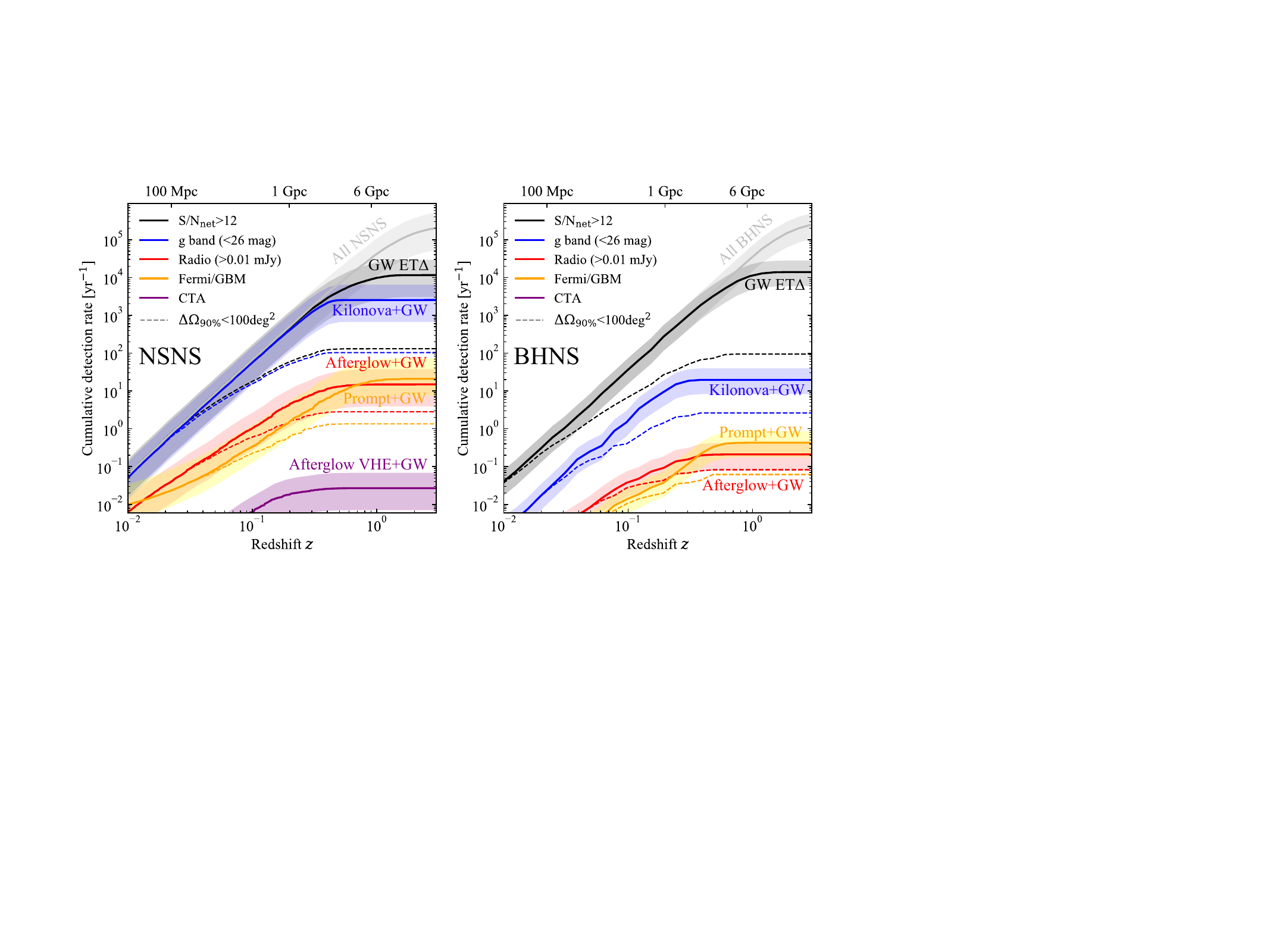}
    \caption{Cumulative multi-messenger detection rates as a function of redshift (luminosity distance) for our fiducial NSNS and BHNS population (SFHo EoS, non-spinning BHs), assuming the ET triangle 10 km configuration.  \textit{Left-hand panel}: NSNS population. The light grey line (“All NSNS”) represents the intrinsic merger rate for the NSNS population, with the grey band showing its uncertainty due to that on the local merger rate. This uncertainty propagates as a constant relative error to all the other rates. The black (“GW ET”) line is the cumulative GW detection rate (events per year with network S/N $\geq 12$). The blue (“Kilonova+GW”), red (“Afterglow+GW”), purple (“Afterglow VHE+GW”) and orange (“Prompt+GW”) lines are the cumulative detection rates for the joint detection of ET GW plus either a KN ($g$ band), a GRB afterglow (radio and VHE bands) or a GRB prompt (the orange and purple lines account, respectively, for the \textit{Fermi}/GBM and CTA duty cycle and field of view). The dashed lines are the cumulative detection rates assuming only the binaries with $\Delta\Omega_{{\rm 90}\%}<100\mathrm{deg}^2$. The assumed thresholds or instruments sensitivity are shown in the legend. \textit{Right-hand panel:} same as the left-hand panel, but for the BHNS population.}
    \label{fig:detection_rates}
\end{figure*}

\begin{figure*}
    \centering
    \includegraphics[width=\textwidth]{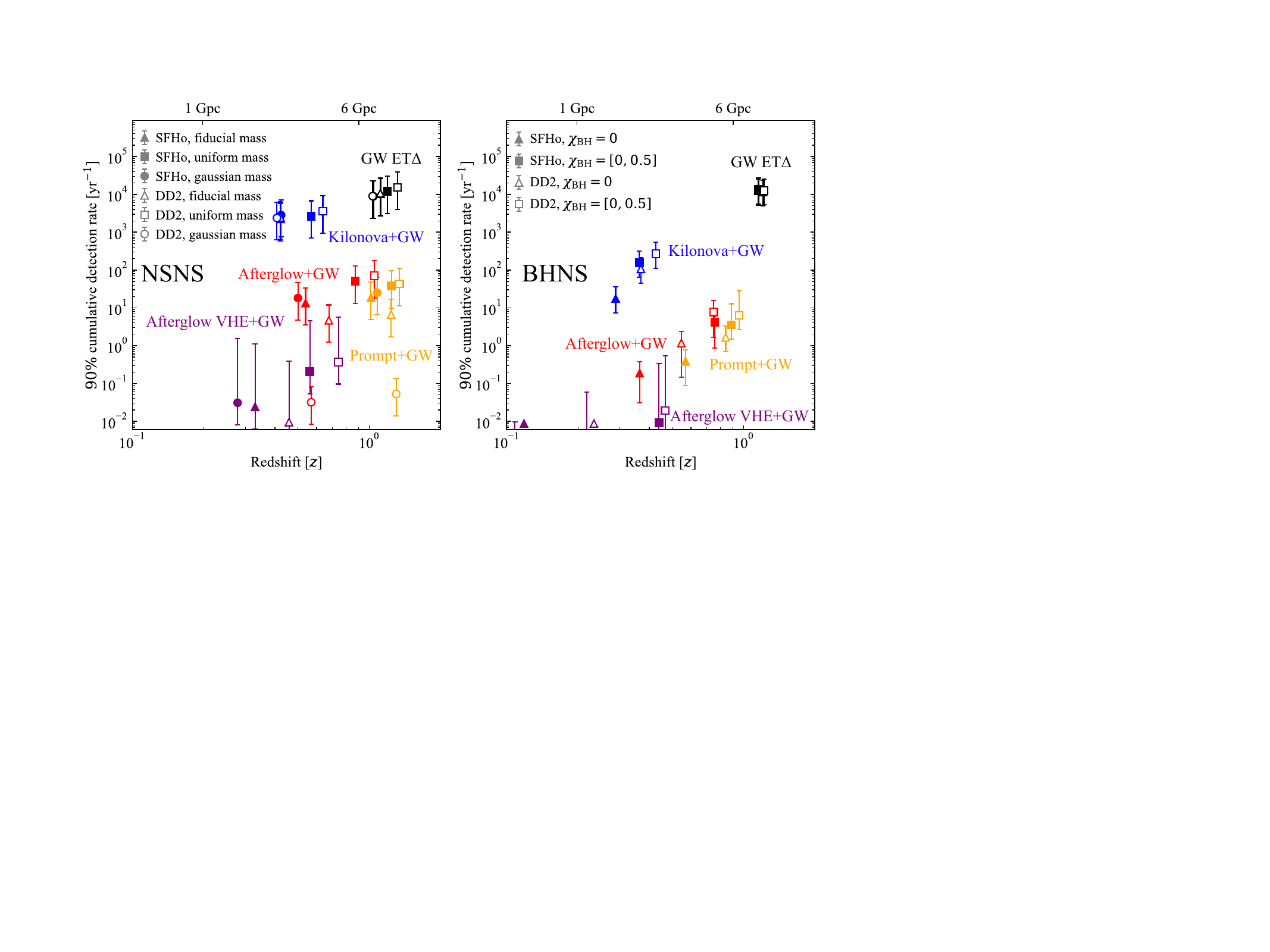}
    \caption{Predicted 90$^\mathrm{th}$ percentile of the cumulative multi-messenger detection rates for our NSNS and BHNS population model variations. \textit{Left-hand panel:} NSNS fiducial population, assuming ET triangle 10 km configuration and EM bands and detection limits reported in Figure \ref{fig:detection_rates}. Different colors refer to different counterparts as in Figure \ref{fig:detection_rates}. Different marker shapes indicate different adopted mass distributions (triangle: fiducial; square: uniform; circle: Gaussian). Filled markers indicate the SFHo EoS, while empty markers are for the DD2 EoS. The error bars indicate the uncertainty on the local merger rate and for the “Afterglow VHE+GW” channel also a variation on the median circum-burst density. \textit{Right-hand panel:} BHNS fiducial population, assuming ET triangle 10 km configuration and EM limits reported in Table \ref{tab:et_det_rates}. Different marker shapes indicate different adopted BH spin distributions (triangle: $\chi_\mathrm{BH} = 0$; square: uniform between 0 and 0.5). Filled markers indicate the SFHo EoS, while empty markers are for the DD2 EoS. The error bars indicate the uncertainty on the local merger rate. For GRB afterglow and prompt they also take into account a variation on the jet core half-opening angle ($\theta_j = 15^{\circ}$) and for the Afterglow VHE also a variation on the median circum-burst density.}
    \label{fig:variations}
\end{figure*}

\subsection{Bird's eye view of the accessible population}

Figure \ref{fig:donut} represents a ``bird's eye view'' of the accessible multi-messenger populations of NSNS and BHNS mergers according to our fiducial models. The blue dot in the center of the figure represents the Earth (not to scale). The distance scale is linear in the redshift $z$ and two circles of constant distance (corresponding to $z=0.5$ and $z=1.0$) are represented with white dashed lines. Yellow (orange) dots in the left (right) half of the plot represent NSNS (BHNS) mergers that produce GW signals that exceed the assumed detection threshold in ET$\Delta$. A red butterfly-shaped symbol is drawn around GW-detectable mergers that produce a KN whose emission exceeds the assumed detection thresholds in at least one of the considered UVOIR bands; a light blue elongated jet-like symbol is drawn around GW-detectable mergers that produce a jet whose emission (prompt or afterglow) exceeds the assumed thresholds in at least one of the considered bands. The inclination of the jet and KN symbol symmetry axes with respect to the direction towards the Earth reflect those in the actual synthetic population. The number of sources represented corresponds to five hypothetical years of ET$\Delta$ operation and EM follow up.

\subsection{Detection rates}

In the left-hand panel of Figure \ref{fig:detection_rates} we present our forecasts for the NSNS EM counterpart detection scenario in conjunction with ET$\Delta$, based on the limits established in the previous section. The light grey line, labeled “All NSNS”, depicts the intrinsic cumulative merger rate within redshift $z$, with the corresponding light grey band illustrating its uncertainty. The latter originates from the assumed uncertainty in the intrinsic local NSNS merger rate density, and it propagates as a constant relative error to all subsequent rate estimates, displayed in the Figure as coloured bands. In this section, no other sources of error related to the initial population properties or the EM model are considered. These will be discussed in detail in the following section.

Each of the other solid lines represents a cumulative detection rate.
The black line, labelled “GW ET$\Delta$”, represents NSNS mergers with detectable GW assuming an ET triangle 10 km configuration. The other colors refer to NSNS mergers with detectable GW and one particular counterpart that exceeds the assumed detection threshold: KN in $g$ band (“Kilonova+GW”, blue); GRB radio afterglow (“Afterglow+GW”, red); GRB prompt emission, assuming \textit{Fermi}/GBM as a representative instrument (“Prompt+GW”, orange); GRB VHE afterglow, with limits representative of CTA (“Afterglows VHE+GW”). Results for other spectral bands are summarized in Table \ref{tab:et_det_rates}. The dashed lines represent the cumulative detection rates considering only binaries with a sky localization $\Delta\Omega_{{\rm 90}\%}<100$ $\mathrm{deg}^2$.

We find that ET will detect \nsnsETTgw\ NSNS mergers per year,  with the 90$^\mathrm{th}$ percentile of the redshift distribution of detectable events at $z_{90\%} \sim 1.0$.
Among these GW detectable binaries, \nsnsETTkng\ have also a KN that exceeds, up to redshift $z_{90\%} \sim 0.4$, the brightness thresholds at its peak. The large number of detectable events and the rapid evolution of these sources poses a significant challenge to the EM follow up infrastructure. The ability of ET to localize the GW source conditions the effectiveness of the EM search for a counterpart, by determining the number of telescope pointings needed to adequately cover the GW localization region \citep{branchesi2023}. For this reason, in Table~\ref{tab:et_det_rates} and Figure \ref{fig:detection_rates} we also report the detections rates for the best localized binaries, having $\Delta\Omega_{{\rm 90}\%}<100\mathrm{deg}^2$ or $\Delta\Omega_{{\rm 90}\%}<10\mathrm{deg}^2$. The rate of GW-detectable NSNS mergers with a potentially detectable KN that are also localized to within such an accuracy decreases to \nsnsETTkngsky\ yr$^{-1}$, which represents a more realistically manageable number.

The high fraction (98\%) of off-axis jets within the population makes the predicted joint GW and GRB rates comparatively lower, with a detection rate of \nsnsETTafterradio\ $\mathrm{yr}^{-1}$ for GRB radio afterglows and \nsnsETTpromptGBM\ $\mathrm{yr}^{-1}$ for GRB prompt. For events localized within 100 deg$^2$, these rates decrease to \nsnsETTafterradiosky\ $\mathrm{yr}^{-1}$ and \nsnsETTpromptGBMsky\  $\mathrm{yr}^{-1}$, respectively. According to this analysis, the majority ($\sim 54\%$) of SGRBs detectable by \textit{Fermi}/GBM will have an associated detectable GW signal, in agreement with the estimate from \cite{ronchini2022}. This makes the GRB prompt emission the most promising probe for multi-messenger astronomy in the distant Universe.  In the cumulative distribution of the GRB prompt emission, an increase in rates can be observed at low redshifts, corresponding to events dominated by the cocoon shock breakout component. This component becomes relevant for events occurring at distances closer than about 100 Mpc.

Regarding the VHE afterglow band, the detection rates are not promising, with values around \nsnsETTafterSSC\ yr$^{-1}$. This low rate is primarily due to the abundance of off-axis events in the population, which produce flux levels that are too low compared to the projected sensitivity of CTA. Of course, this rate is dependent on the microphysical parameters of the afterglow model and the assumptions about the average density of the circum-burst medium. In particular, increasing the circum-burst medium density to $n = 0.1$ cm$^{-3}$ produces detection rates higher by a factor of 10, as discussed in the following section.

In the right-hand panel of Figure \ref{fig:detection_rates}, we report the same cumulative detection rates for the BHNS population. We expect a GW detection rate of \bhnsETTgw\ $\mathrm{yr}^{-1}$, again with $z_{90\%} \sim 1$. In our fiducial population, only about 2\% of the mergers produce some mass remnant  ($m_\mathrm{out}>0$), potentially powering an EM counterpart. This results in significantly lower rates of multi-messenger detectable events compared to the NSNS population. 
In particular, we predict a detectable KN rate of \bhnsETTkng\ $\mathrm{yr}^{-1}$, that decreases to \bhnsETTkngsky\ $\mathrm{yr}^{-1}$ considering only the events with $\Delta\Omega_{{\rm 90}\%}<100\mathrm{deg}^2$. The KN horizon is the same as the NSNS case, $z_{90\%} \sim 0.4$, determined by the assumed sensitivity of EM facilities.

GW-detectable systems with an observable radio afterglow and GRB prompt are forecasted to achieve a total rate of \bhnsETTafterradio\ ${\rm yr}^{-1}$ and \bhnsETTpromptGBM\ ${\rm yr}^{-1}$ (\bhnsETTafterradiosky\ ${\rm yr}^{-1}$ and \bhnsETTpromptGBMsky\ ${\rm yr}^{-1}$ for the events with $\Delta\Omega_{{\rm 90}\%}<100\mathrm{deg}^2$). The GRB prompt horizon is smaller than the GW one: this occurs because only events with low BH masses, and hence an intrinsically fainter GW signal, lead to a NS disruption outside of the BH ISCO. Consequently, the combined GW+GRB detection horizon is defined by the GW detection of events involving BHs below a specific mass threshold. For radio afterglows, on the other hand, the value at which the curve saturates is still set by the assumed EM detection limit.

Although the detection rates for the BHNS population are significantly lower compared to NSNS cases, it is important to stress here that we are considering the most conservative population scenario, with non-spinning BHs and a soft EoS for NSs. In the next section we discuss some variations in the progenitor BHNS population to explore also more optimistic scenarios. Nevertheless, the detection rates predicted in our fiducial set up increase by more than a factor of 10 with respect to those of the O5 run (see \citealt{colombo2023}). This suggests that we may have to wait for next-generation GW detectors for the first identification of EM counterparts from these types of sources: thus, ET could represent a pivotal advancement for multi-messenger astronomy in the context of BHNS observations.

\subsection{Variations in the NSNS and BHNS populations}

The results shown in Figure \ref{fig:detection_rates} refer to the NSNS and BHNS populations that we have defined as fiducial. 
In Figure \ref{fig:variations} we examine the impact on the detection rates of changing some initial population assumptions. For clarity, we only show the 90$^\mathrm{th}$ percentile of the redshift distributions of multi-messenger detectable events, assuming the same EM bands and detection limits as in the previous section, without constraints on the sky localization. The error bars always include the uncertainty due to the merger rate density. In some cases, described below, they also include some additional sources of uncertainty.

In the left-hand panel of Figure \ref{fig:variations}, we show the variations in the NSNS population, assuming the ET triangular configuration with a 10 km arm length. The different markers represent different mass distribution choices: triangles denote the fiducial distribution used in Figure \ref{fig:detection_rates}, while squares and circles represent a Gaussian distribution centered at 1.33 with a standard deviation of 0.9, and a uniform distribution between 1 and $M_\mathrm{TOV}$, respectively. A filled marker represents a result obtained assuming the SFHo EoS, while an empty marker indicates the DD2 EoS has been assumed. 

Regarding GW detections, represented by black markers, we observe an increase in detection rates as we move from the Gaussian distribution to the fiducial and uniform distributions, which correspond to more massive NSs on average. The variation in EoS results in only marginal changes, as expected for the inspiral signal.

For KN+GW detections (blue symbols), the only noticeable effect is an increase in the horizon, from $z_{90\%} \sim 0.4$ to $z_{90\%} \sim 0.6$, when assuming the uniform mass distribution, due to an increased fraction of events with a massive accretion disc, whose winds produce a strong KN emission.

In the Afterglow+GW and Prompt+GW channels (red and orange symbols, respectively) we observe differences of up to a few orders of magnitude in the detection rates depending on the population assumptions. For the SFHo EoS (filled symbols), the rates do not differ appreciably between the Gaussian and fiducial mass distributions, but they increase when assuming the uniform distribution, due to the increased fraction of events with massive accretion discs (see Figure \ref{fig:nsns_mass} for a comparison between mass distribution and ejecta and accretion disc masses). On the other hand, for the DD2 EoS (empty symbols), the rates vary significantly with different mass distributions. This is because our jet-launching conditions require both a non-zero accretion disk and the formation of a BH within a timescale much shorter than the accretion timescale. This latter condition is enucleated in the requirement that $M_\mathrm{rem}>1.2M_\mathrm{TOV}$. In the case of DD2 we have $M_\mathrm{TOV}=2.46M_\odot$, while for SFHo it is only $M_\mathrm{TOV}=2.06M_\odot$. Therefore, when assuming DD2, the fiducial and Gaussian distributions mostly produce supramassive or stable NS remnants, leading to a reduced jet production rate (see Figure \ref{fig:nsns_mass}, where the HMNS formation condition is indicated by a pink line). 

A similar trend is visible in the VHE afterglow band (purple markers), with significant spread for the same reasons. In this case, the error bars also include the effects of a variation in the assumed median circum-burst density, from the fiducial value $n=5\times10^{-3}$ to a higher $n=0.1$ $\mathrm{cm^{-3}}$. The latter more optimistic assumption leads to an increase in detection rates by an order of magnitude. 

Here it is important to note that our assumed jet structure and the model for the GRB prompt emission were calibrated to reproduce the bolometric fluence distribution observed by \textit{Fermi}/GBM. The variations in the mass distribution and EoS naturally alter the fluence distribution, preventing a match with SGRB observations unless the jet and prompt emission models are adjusted. Therefore, the changes in the detection rates for jet-related emission should be understood as an exploration aimed at clarifying the model's dependencies and how these factors influence the results.

In the right-hand panel of Figure \ref{fig:variations}, we show the variations in the BHNS mergers population. As in the previous case, filled and empty markers represent the SFHo and DD2 EoS, respectively, while triangles and circles indicate the assumed BH spin distribution, either $\chi_\mathrm{BH}=0$ or a uniform distribution between 0 and 0.5.
In this case, we did not consider variations in the BH mass distribution. While it is clear that different assumptions on the mass distribution can affect the multi-messenger detection rates, we focused on the uncertainties associated with the BH spin, the NS EoS, and the overall merger rate. These factors, as well the systematics in the modeling, tend to dominate the impact on the fraction of events producing EM emission, as shown in \cite[][]{Broekgaarden2021, RomanGarza:2021ApJ...912L..23R, Biscoveanu:2023MNRAS.518.5298B, Chen:2024, Xing:2024A&A...683A.144X}.
We note, however, that the role of BH spin must be interpreted carefully. While our models include a wide range of spin assumptions to bracket potential outcomes, realistic spin distributions from isolated binary evolution are expected to favor low BH spins, with only a small fraction of systems potentially reaching high spins due to formation order reversal \citep[e.g.][]{Fuller2019, Xing:2024A&A...683A.144X}. Therefore, the impact of BH spin on EM detectability may be somewhat overestimated under more optimistic spin assumptions. 
Additionally, we acknowledge that recent observational results challenge some long-standing assumptions about the BH mass distribution. In particular, the detection of GW230529 \citep{lvk2024_bhns} not only questions the existence of a lower-mass gap—reproduced by the ‘rapid’ SN model of \citet{fryer2012}—but also suggests that the typical BH mass in BHNS systems may be lower than previously assumed. While this potential shift in the BH mass distribution is not explicitly modeled in our current work, it is an important direction for future studies, since it could further enhance EM detectability in the BHNS population.

The GW detection rate is essentially insensitive to the variations considered in the EoS and $\chi_\mathrm{BH}$ distribution. On the other hand, the impact of varying population assumptions on EM+GW detection rates is larger compared to the NSNS case. For KN+GW detections, we observe an increase in the rate by more than an order of magnitude, which also extends the redshift range at which these events can be detected. This is due both to the change in the EoS --where DD2 leads to more deformable NSs and, consequently, larger ejecta masses -- and to the inclusion of the more optimistic spin assumption, which increases the fraction of events capable of producing an EM counterpart (see Figure \ref{fig:bhns_mass} for a comparison of the mass distribution with the ejecta and accretion disc masses). The error bars for these channels, in addition to the uncertainty related to the local merger rate, include the uncertainty due to variations in the jet opening angle, ranging from 3.4 to 15 degrees. In the case of the VHE afterglow band, we also include the effect of the variation in the median circum-burst density: similarly to the NSNS case, this allows for an increase in the detection rate by roughly a factor of 10.

A similar trend can be observed in the jet-related emission, where, unlike the NSNS population, there is no longer a dependence on the conditions for launching a jet, since all events result in the formation of a black hole. The strong dependence of multi-messenger detection rates on the initial population properties demonstrates how future multi-messenger observations could provide important constraints on the population and, consequently, on its formation channels.

\def\arraystretch{1.3}% 
\begin{table*}
\tiny
\centering
\caption{Detection limits and predicted detection rates for NSNS and BHNS, assuming ET triangle 10 km. We report in parenthesis the detection rates assuming $\Delta\Omega_{{\rm 90}\%}<100\mathrm{deg}^2$ and $\Delta\Omega_{{\rm 90}\%}<10\mathrm{deg}^2$. The GW detection limits refer to the $\mathrm{S/N_{net}}$ threshold. Near infrared and optical limiting magnitudes are in the AB system; radio limiting flux densities are in mJy @ 1.4 GHz; X-ray limiting flux densities are in erg cm$^{-2}$ s$^{-1}$ keV$^{-1}$ @ 1 keV; gamma-ray limiting fluence is in erg cm$^{-2}$ (\textit{Fermi}/GBM). Detection rates are in $\mathrm{yr}^{-1}$. The reported errors, given at the 90\% credible level, stem from the uncertainty on the overall merger rate, while systematic errors are not included.}
\makebox[\textwidth]{
\begin{tabular}{lccccccccc}
\hline
\hline
\multicolumn{1}{c}{} & \multicolumn{1}{c}{GW ET} & \multicolumn{3}{c}{KN+GW} & \multicolumn{4}{c}{GRB Afterglow+GW} & \multicolumn{1}{c}{GRB Prompt+GW} \\
 & ~ & \textit{J} & \textit{z} & \textit{g} & Radio & Optical & X-rays & VHE & \textit{Fermi} \\ \hline
%\multicolumn{11}{c}{}\\
%\textbf{Search} & \multicolumn{10}{c}{}\\\hline
\textbf{NSNS} & ~ & ~ & ~ & ~ & ~ &  &  &  & \\
Limit & 12 & 21 & 24.4 & 26 & 0.01 & 26 & $10^{-13}$ & CTA  & $3.09\times10^{-7}$  \vspace{0.15cm}\\  
ET$\Delta$ & ~ & ~ & ~ & ~ & ~ &  &  &  & \\
Rate  & \nsnsETTgw & \nsnsETTknJ & \nsnsETTknz & \nsnsETTkng & \nsnsETTafterradio & \nsnsETTafteroptic & \nsnsETTafterx & \nsnsETTafterSSC & \nsnsETTpromptGBM \\ 
($\Delta\Omega_{{\rm 90}\%}<100\mathrm{deg}^2$) & (\nsnsETTgwsky) & (\nsnsETTknJsky) & (\nsnsETTknzsky) & (\nsnsETTkngsky) & (\nsnsETTafterradiosky) & (\nsnsETTafteropticsky) & (\nsnsETTafterxsky) & (\nsnsETTafterSSCsky) & (\nsnsETTpromptGBMsky) \vspace{0.15cm}\\ 
($\Delta\Omega_{{\rm 90}\%}<10\mathrm{deg}^2$) & (\nsnsETTgwskyten) & (\nsnsETTknJskyten) & (\nsnsETTknzskyten) & (\nsnsETTkngskyten) & (\nsnsETTafterradioskyten) & (\nsnsETTafteropticskyten) & (\nsnsETTafterxskyten) & (\nsnsETTafterSSCskyten) & (\nsnsETTpromptGBMskyten) \vspace{0.15cm}\\ 
ET2L & ~ & ~ & ~ & ~ & ~ &  &  &  & \\
Rate  & \nsnsETLgw & \nsnsETLknJ & \nsnsETLknz & \nsnsETLkng & \nsnsETLafterradio & \nsnsETLafteroptic & \nsnsETLafterx & \nsnsETLafterSSC & \nsnsETLpromptGBM \\ 
($\Delta\Omega_{{\rm 90}\%}<100\mathrm{deg}^2$) & (\nsnsETLgwsky) & (\nsnsETLknJsky) & (\nsnsETLknzsky) & (\nsnsETLkngsky) & (\nsnsETLafterradiosky) & (\nsnsETLafteropticsky) & (\nsnsETLafterxsky) & (\nsnsETLafterSSCsky) & (\nsnsETLpromptGBMsky) \vspace{0.15cm}\\ 
($\Delta\Omega_{{\rm 90}\%}<10\mathrm{deg}^2$) & (\nsnsETLgwskyten) & (\nsnsETLknJskyten) & (\nsnsETLknzskyten) & (\nsnsETLkngskyten) & (\nsnsETLafterradioskyten) & (\nsnsETLafteropticskyten) & (\nsnsETLafterxskyten) & (\nsnsETLafterSSCskyten) & (\nsnsETLpromptGBMskyten) \vspace{0.15cm}\\ 
ET$\Delta$ + 2CE & ~ & ~ & ~ & ~ & ~ &  &  &  & \\
Rate  & \nsnsETTCEgw & \nsnsETTCEknJ & \nsnsETTCEknz & \nsnsETTCEkng & \nsnsETTCEafterradio & \nsnsETTCEafteroptic & \nsnsETTCEafterx & \nsnsETTCEafterSSC & \nsnsETTCEpromptGBM \\ 
($\Delta\Omega_{{\rm 90}\%}<100\mathrm{deg}^2$) & (\nsnsETTCEgwsky) & (\nsnsETLCEknJsky) & (\nsnsETLCEknzsky) & (\nsnsETLCEkngsky) & (\nsnsETLCEafterradiosky) & (\nsnsETLCEafteropticsky) & (\nsnsETLCEafterxsky) & (\nsnsETLCEafterSSCsky) & (\nsnsETLCEpromptGBMsky) \vspace{0.15cm}\\ 
($\Delta\Omega_{{\rm 90}\%}<10\mathrm{deg}^2$) & (\nsnsETTCEgwskyten) & (\nsnsETLCEknJskyten) & (\nsnsETLCEknzskyten) & (\nsnsETLCEkngskyten) & (\nsnsETLCEafterradioskyten) & (\nsnsETLCEafteropticskyten) & (\nsnsETLCEafterxskyten) & (\nsnsETLCEafterSSCskyten) & (\nsnsETLCEpromptGBMskyten) \vspace{0.15cm}\\ 
ET2L + 2CE & ~ & ~ & ~ & ~ & ~ &  &  &  & \\
Rate  & \nsnsETLCEgw & \nsnsETLCEknJ & \nsnsETLCEknz & \nsnsETLCEkng & \nsnsETLCEafterradio & \nsnsETLCEafteroptic & \nsnsETLCEafterx & \nsnsETLCEafterSSC & \nsnsETLCEpromptGBM \\ 
($\Delta\Omega_{{\rm 90}\%}<100\mathrm{deg}^2$) & (\nsnsETLCEgwsky) & (\nsnsETLCEknJsky) & (\nsnsETLCEknzsky) & (\nsnsETLCEkngsky) & (\nsnsETLCEafterradiosky) & (\nsnsETLCEafteropticsky) & (\nsnsETLCEafterxsky) & (\nsnsETLCEafterSSCsky) & (\nsnsETLCEpromptGBMsky) \vspace{0.15cm}\\
($\Delta\Omega_{{\rm 90}\%}<10\mathrm{deg}^2$) & (\nsnsETLCEgwskyten) & (\nsnsETLCEknJskyten) & (\nsnsETLCEknzskyten) & (\nsnsETLCEkngskyten) & (\nsnsETLCEafterradioskyten) & (\nsnsETLCEafteropticskyten) & (\nsnsETLCEafterxskyten) & (\nsnsETLCEafterSSCskyten) & (\nsnsETLCEpromptGBMskyten) \vspace{0.15cm}\\

\hline
 %&  &  &  &  &  &  &  &  &  &\\
%\textbf{Monitoring} & \multicolumn{10}{c}{} \\\hline
\textbf{BHNS} & ~ & ~ & ~ & ~ & ~ &  &  &  & \\
Limit  & 12 & 21 & 24.4 & 26 & 0.01 & 26 & $10^{-13}$ & CTA  & $3.09\times10^{-7}$  \vspace{0.15cm}\\ 
ET$\Delta$ & ~ & ~ & ~ & ~ & ~ &  &  &  & \\ 
Rate  & \bhnsETTgw & \bhnsETTknJ & \bhnsETTknz & \bhnsETTkng & \bhnsETTafterradio & \bhnsETTafteroptic & \bhnsETTafterx & \bhnsETTafterSSC & \bhnsETTpromptGBM \\ 
($\Delta\Omega_{{\rm 90}\%}<100\mathrm{deg}^2$) & (\bhnsETTgwsky) & (\bhnsETTknJsky) & (\bhnsETTknzsky) & (\bhnsETTkngsky) & (\bhnsETTafterradiosky) & (\bhnsETTafteropticsky) & (\bhnsETTafterxsky) &  & (\bhnsETTpromptGBMsky) \vspace{0.15cm}\\
($\Delta\Omega_{{\rm 90}\%}<10\mathrm{deg}^2$) & (\bhnsETTgwskyten) & (\bhnsETTknJskyten) & (\bhnsETTknzskyten) & (\bhnsETTkngskyten) & (\bhnsETTafterradioskyten) & (\bhnsETTafteropticskyten) & (\bhnsETTafterxskyten) &  & (\bhnsETTpromptGBMskyten) \vspace{0.15cm}\\
ET2L & ~ & ~ & ~ & ~ & ~ &  &  &  & \\ 
Rate  & \bhnsETLgw & \bhnsETLknJ & \bhnsETLknz & \bhnsETLkng & \bhnsETLafterradio & \bhnsETLafteroptic & \bhnsETLafterx & \bhnsETLafterSSC & \bhnsETLpromptGBM \\ 
($\Delta\Omega_{{\rm 90}\%}<100\mathrm{deg}^2$) & (\bhnsETLgwsky) & (\bhnsETLknJsky) & (\bhnsETLknzsky) & (\bhnsETLkngsky) & (\bhnsETLafterradiosky) & (\bhnsETLafteropticsky) & (\bhnsETLafterxsky) &  & (\bhnsETLpromptGBMsky) \vspace{0.15cm}\\  
($\Delta\Omega_{{\rm 90}\%}<10\mathrm{deg}^2$) & (\bhnsETLgwskyten) & (\bhnsETLknJskyten) & (\bhnsETLknzskyten) & (\bhnsETLkngskyten) & (\bhnsETLafterradioskyten) & (\bhnsETLafteropticskyten) & (\bhnsETLafterxskyten) &  & (\bhnsETLpromptGBMskyten) \vspace{0.15cm}\\  
ET$\Delta$ + 2CE & ~ & ~ & ~ & ~ & ~ &  &  &  & \\ 
Rate  & \bhnsETTCEgw & \bhnsETTCEknJ & \bhnsETTCEknz & \bhnsETTCEkng & \bhnsETTCEafterradio & \bhnsETTCEafteroptic & \bhnsETTCEafterx & \bhnsETTCEafterSSC & \bhnsETTCEpromptGBM \\ 
($\Delta\Omega_{{\rm 90}\%}<100\mathrm{deg}^2$) & (\bhnsETTCEgwsky) & (\bhnsETTCEknJsky) & (\bhnsETTCEknzsky) & (\bhnsETTCEkngsky) & (\bhnsETTCEafterradiosky) & (\bhnsETTCEafteropticsky) & (\bhnsETTCEafterxsky) &  & (\bhnsETTCEpromptGBMsky) \vspace{0.15cm}\\
($\Delta\Omega_{{\rm 90}\%}<10\mathrm{deg}^2$) & (\bhnsETTCEgwskyten) & (\bhnsETTCEknJskyten) & (\bhnsETTCEknzskyten) & (\bhnsETTCEkngskyten) & (\bhnsETTCEafterradioskyten) & (\bhnsETTCEafteropticskyten) & (\bhnsETTCEafterxskyten) &  & (\bhnsETTCEpromptGBMskyten) \vspace{0.15cm}\\
ET2L + 2CE & ~ & ~ & ~ & ~ & ~ &  &  &  & \\ 
Rate  & \bhnsETLCEgw & \bhnsETLCEknJ & \bhnsETLCEknz & \bhnsETLCEkng & \bhnsETLCEafterradio & \bhnsETLCEafteroptic & \bhnsETLCEafterx & \bhnsETLCEafterSSC & \bhnsETLCEpromptGBM \\ 
($\Delta\Omega_{{\rm 90}\%}<100\mathrm{deg}^2$) & (\bhnsETLCEgwsky) & (\bhnsETLCEknJsky) & (\bhnsETLCEknzsky) & (\bhnsETLCEkngsky) & (\bhnsETLCEafterradiosky) & (\bhnsETLCEafteropticsky) & (\bhnsETLCEafterxsky) &  & (\bhnsETLCEpromptGBMsky) \\
($\Delta\Omega_{{\rm 90}\%}<10\mathrm{deg}^2$) & (\bhnsETLCEgwskyten) & (\bhnsETLCEknJskyten) & (\bhnsETLCEknzskyten) & (\bhnsETLCEkngskyten) & (\bhnsETLCEafterradioskyten) & (\bhnsETLCEafteropticskyten) & (\bhnsETLCEafterxskyten) &  & (\bhnsETLCEpromptGBMskyten) \\

\end{tabular}
}
\label{tab:et_det_rates}
\end{table*}

\subsection{GW sky localization}
\begin{figure*}
    \centering
    \includegraphics[width=\textwidth]{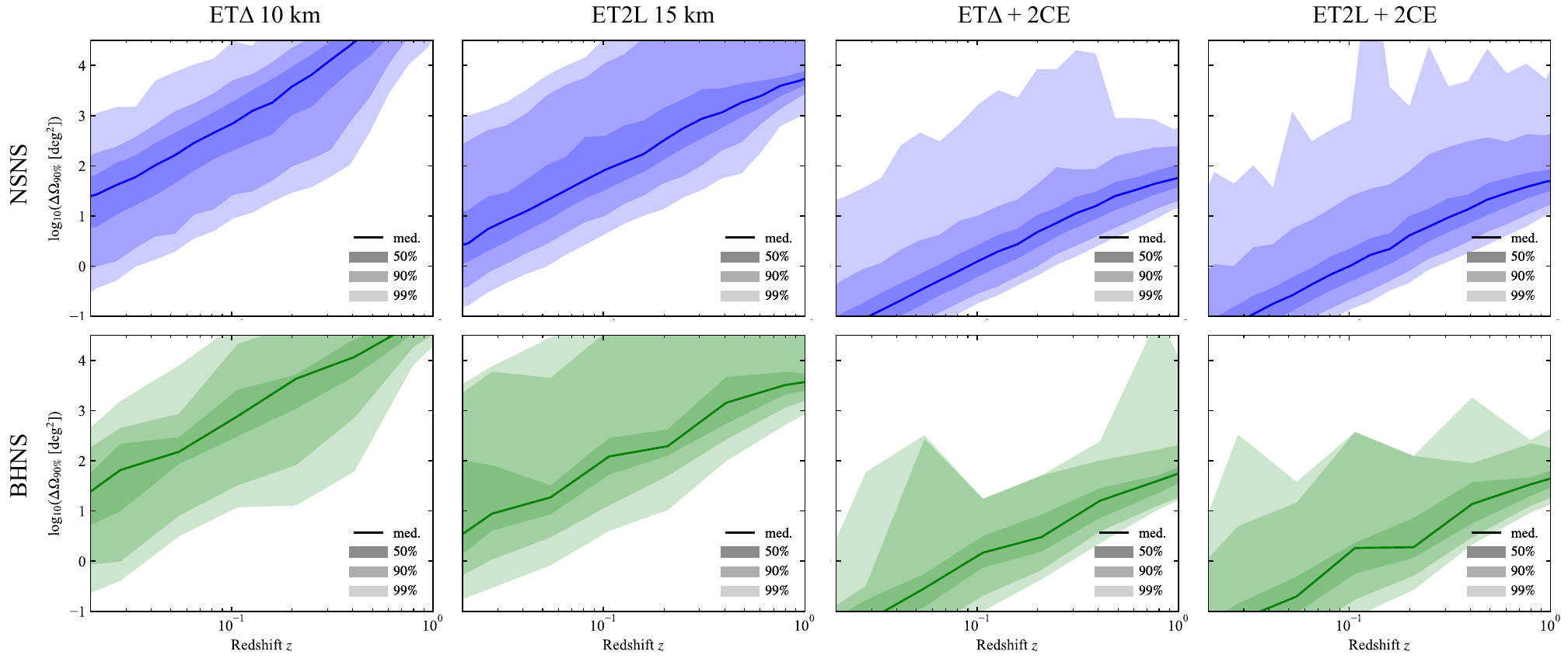}
    \caption{Distribution of the extent of the 90\% sky localization region of GW-detected NSNS events (upper row) and BHNS events (lower row) that are capable of producing an EM counterpart, as a function of redshift. Different columns refer to ET$\Delta$ (first column), ET2L (second column), ET$\Delta$ + 2CE (third column) and ET2L + 2CE (four column). The solid line represents the median, while the coloured bands encompass the 50$\%$, 90$\%$, and 99$\%$ credible interval at each fixed redshift.}
    \label{fig:skyloc_redshift}
\end{figure*}

Fig.~\ref{fig:skyloc_redshift} displays the $\Delta\Omega_{{\rm 90}\%}$ sky localization distribution as a function of redshift for the NSNS population (top panel, blue) and the BHNS population (bottom panel, green), considering only events capable of producing an EM counterpart. This condition requires the presence of either non-zero ejecta or an accretion disk, or both. The shaded regions show the extent of the 50\%, 90\%, and 99\% confidence intervals at each redshift, with the median indicated by a solid line. Panels in different columns refer to different configurations: ET triangle configuration ($\mathrm{ET}\Delta$,  first column), ET 2L configuration ($\mathrm{ET2L}$, second column), $\mathrm{ET}\Delta$ combined with two CE detectors ($\mathrm{ET}\Delta$+2CE, third column), and $\mathrm{ET2L}$ combined with two CE detectors (ETL+2CE, fourth column).

The figure shows how sky localization deteriorates as redshift increases. For the triangle configuration, at redshifts below approximately $z \lesssim 0.04$, more than 50\% of events are localized within 100 deg$^2$. However, beyond $z \gtrsim 0.2$, more than 50\% of events exhibit sky localization areas exceeding 1000 deg$^2$. In contrast, the 2L configuration generally shows improved localization capabilities, with median values reduced by nearly an order of magnitude. Specifically, at $z \lesssim 0.04$, the majority of events are localized to within 1 to 10 deg$^2$, while at redshifts greater than $z \gtrsim 0.5$, 50\% of events have localization areas larger than 1000 deg$^2$. The broader distribution of sky localization areas in the 2L configuration, compared to the triangle configuration, is due to instances where only a single detector is active, as dictated by the assumed duty cycle.

Considering ET as part of a network with two CE detectors, the sky localization improves even more significantly, as expected. Specifically, for both the NSNS and BHNS populations, 50\% of the events within a redshift of $z=1$ achieve a sky localization better than 100 square degrees. In this case, the differences between the two ET configurations become negligible.

\subsection{ET configurations and detectors network} \label{sec:ET_config}
\begin{figure*}
    \centering
    \includegraphics[width=\textwidth]{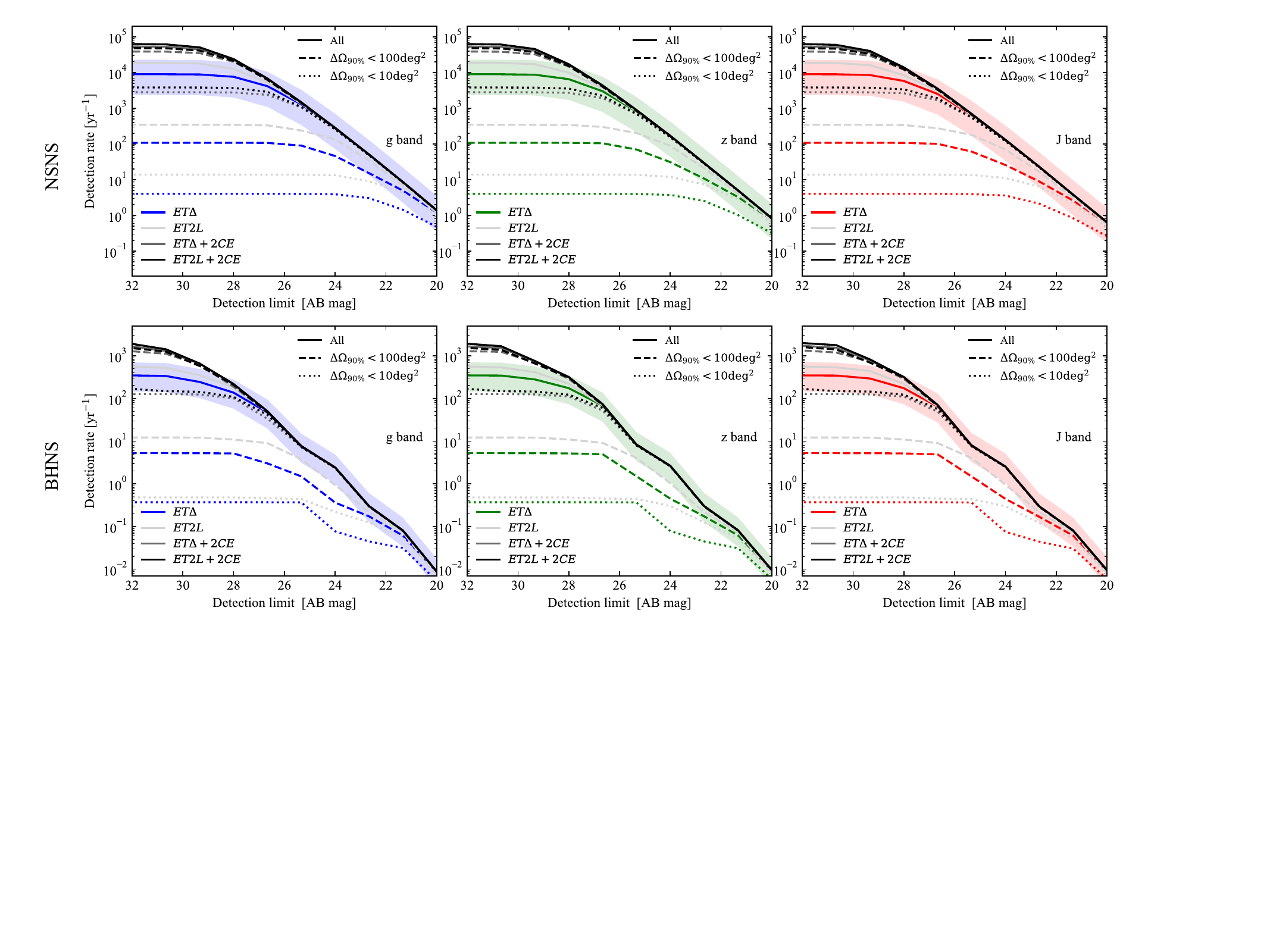}
    \caption{Kilonova detection rate as a function of the EM detection limit threshold for our fiducial NSNS  (upper panels) and BHNS (lower panels) populations. The blue, green and red colors indicate the $g$, $z$ and $J$ bands, respectively, assuming the ET$\Delta$ configuration. In each panel we also report in gray and black the ET2L and ET$\Delta$+2CE configurations. The solid line indicates all the detectable binaries, the dashed and dotted lines the detectable binaries with $\Delta\Omega_{{\rm 90}\%}<100\mathrm{deg}^2$ and the ones with $\Delta\Omega_{{\rm 90}\%}<10\mathrm{deg}^2$, respectively.}
    \label{fig:kn_rate_lim}
\end{figure*}

In order to show how our predicted detection rates depend on the chosen EM detection thresholds, as well as different ET configurations and detectors networks, we provide in Fig. \ref{fig:kn_rate_lim}, the detection rates as functions of EM detection limits for the joint KN+GW channels analyzed in this study, for ET$\Delta$, ET2L, ET$\Delta$+2CE, and ET2L+2CE, corresponding respectively to the colored, light gray, dark gray, and black lines. Panels in the top row refer to the NSNS population, while the bottom row refers to the BHNS population. The blue, green, and red colors correspond to the $g$, $z$, and $J$ bands, respectively. The solid line considers all GW detected events, the dashed line represents events with a sky localization $\Delta\Omega_{{\rm 90}\%}<100\mathrm{deg}^2$, and the dotted line corresponds to events with $\Delta\Omega_{{\rm 90}\%}<10\mathrm{deg}^2$.

For a fixed detector network and band, one can observe an increase in the detection rate as the considered magnitude limit increases, until reaching saturation, which corresponds to the detection of all KN\ae\ associated with those GW events. The plot also allows us to compare the impact of different ET configurations and detectors networks on rates and sky localizations. Assuming a magnitude limit of 26 in the $g$ band for the NSNS population, the rate of events with $\Delta\Omega_{{\rm 90}\%}<10 (100)\mathrm{deg}^2$ increases by approximately a factor of 3.5 (3) in the ET2L configuration. Assuming instead ET$\Delta$ in a network with two CEs, the rate increases by a factor of 400 (30). In particular, the differences between the two ET configurations become negligible in a network with two CEs, as the source distance is limited by the considered EM detection limit.
For higher magnitude limits, the differences become more significant, as they correspond to larger EM horizons.

In Appendix \ref{App:det_lim} we report the same figures for the joint GRB afterglow+GW and GRB prompt+GW channels, for which similar considerations apply.

\section{EM properties}\label{sec::EM_properties}
\subsection{Kilonova}
\begin{figure*}
    \centering
    \includegraphics[width=\textwidth]{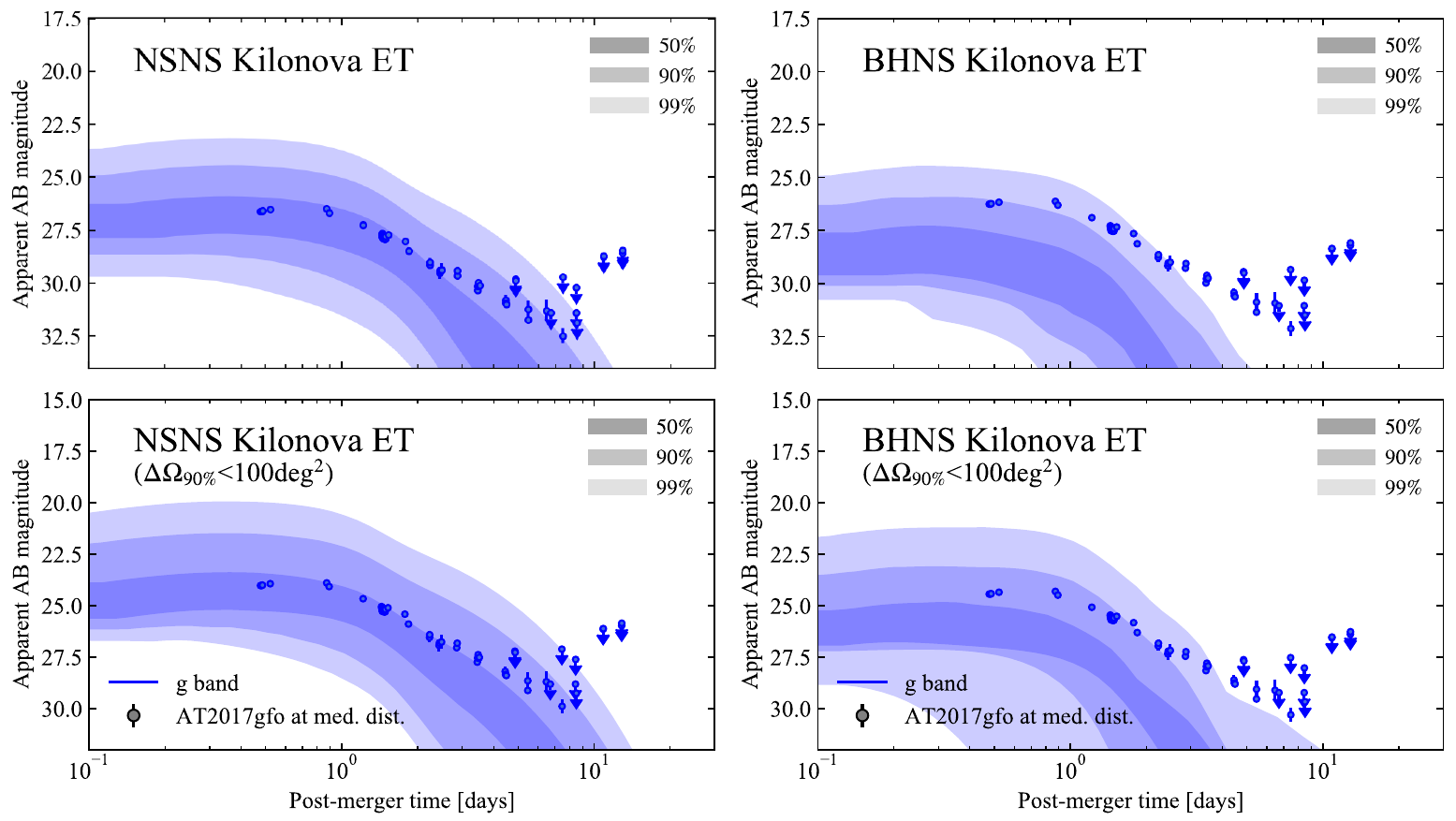}
    \caption{Distribution of optical KN brightness as a function of time for ET-detectable (assuming ET$\Delta$) events in our NSNS (left-hand column) and BHNS (right-hand column) populations. The shaded regions contain $50\%$, $90\%$ and $99\%$ of the KN light curves in the $g$ (484 nm) band. Coloured circles show extinction-corrected AT2017gfo data rescaled to the median distance of the considered populations (upper limits are marked with a downward arrow, data from \citealt{Villar2017}). The bottom row shows the result restricted to binaries with $\Delta\Omega_{{\rm 90}\%}<100\mathrm{deg}^2$.}.
    \label{fig:et_kn}
\end{figure*}

In Figure \ref{fig:et_kn}, we display the distribution of KN apparent AB magnitude as a function of time in the $g$ band for ET-detectable (assuming ET$\Delta$) binary systems in our NSNS (left column) and BHNS (right column) populations. Filled regions show the ranges encompassing 50\%, 90\%, and 99\% of the brightness at each fixed time, considering all ET-detectable binaries (top row) or only the binaries with $\Delta\Omega_{{\rm 90}\%}<100\mathrm{deg}^2$ (bottom row). For comparison, we also report the observed data of AT2017gfo \citep{Villar2017} at the median distance of the GW-detectable events: $3.5$ Gpc for NSNS ($1.0$ Gpc for the well-localized events) and $2.4$ Gpc for BHNS ($1.3$ Gpc for the well-localized events).

For the NSNS population we find that, considering all the ET-detectable binaries, the apparent AB magnitude of the KN at peak spans from 23 down to 29, with 50\% being concentrated in the 24.5-27.5 interval. When applying the constraint on the sky localization, the majority of the peaks are found in the 23.5-25.5 interval, making almost all the KN\ae\ accessible to the Rubin Observatory. 

Considering all simulated BHNS binary mergers detectable by ET which do not produce a direct plunge of the NS, the KN peak apparent AB magnitudes span the range 24.5 to 32, with 50\% clustered between 27 and 29. When focusing on those with $\Delta\Omega_{{\rm 90}\%}<100\mathrm{deg}^2$, most of the peaks fall within the 25-27.5 range. Hence, we predict fainter KN\ae\ on average with respect to NSNS. This is also made apparent by the comparison to GW170817 observational data in Figure \ref{fig:et_kn}.

The plot also shows the rapid KN brightness decline, particularly for the BHNS population. A possible strategy to counteract the challenge of rapidly dimming KN\ae\ involves directly seeking out non-thermal counterparts, like radio afterglows, using tiling instruments, as discussed in \cite{colombo2023}. This approach is well-suited for upcoming radio surveys, enabling another possible search for these transient events.

\subsection{GRB Afterglow}
\begin{figure*}
    \centering
    \includegraphics[width=\textwidth]{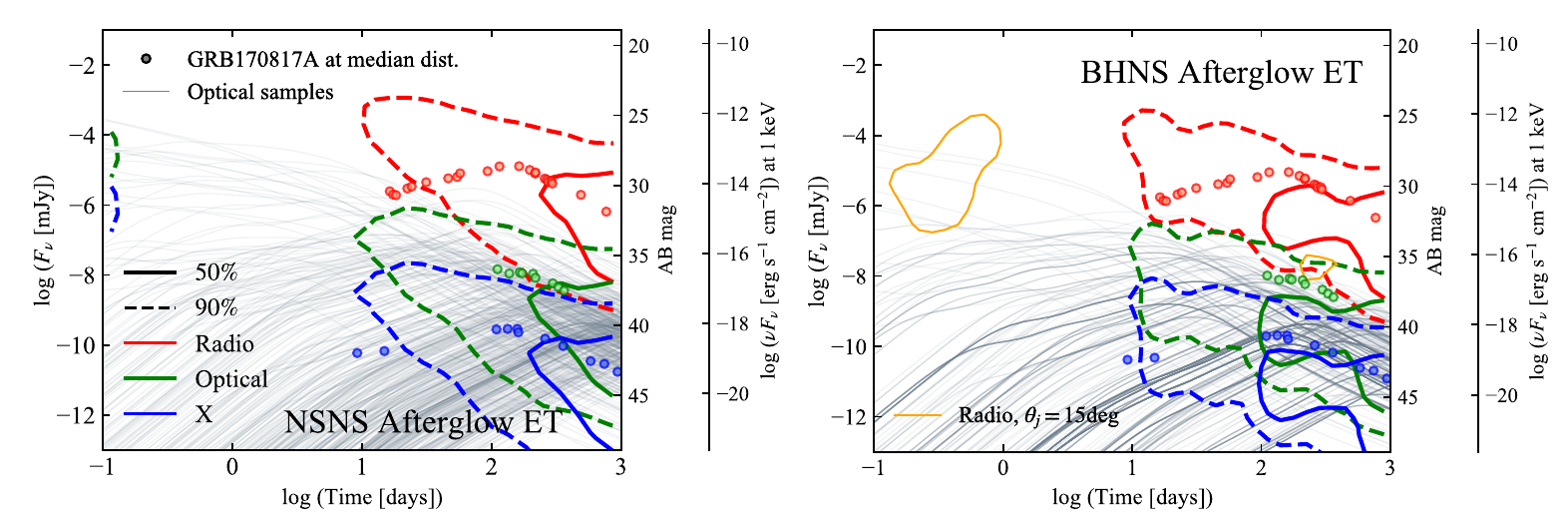}
    \caption{Distribution of brightness versus time for GRB afterglows associated to ET-detectable NSNS (left panel) and BHNS (right panel) mergers. Solid and dashed contours contain $50\%$ and $90\%$ of the peaks, respectively. Red, green and blue colors refer to the radio ($F_\nu$ at $1.4 \times 10^9$Hz), optical (AB magnitude at $4.8 \times 10^{14}$Hz), and X-rays ($\nu F_\nu$ at 1 keV), respectively. The colored circles are the observed data of GRB170817A \citep{Makhathini2021} at the median distance of our population. The grey lines in the background are a random sample of optical light curves from the underlying population. We assume for both the populations the same jet half-opening angle, $\theta_\mathrm{j} = 3.4$ deg. In the BHNS panel, we also show with orange solid lines the 50\% contour for radio peaks when assuming $\theta_\mathrm{j}=15$ deg.}
    \label{fig:et_after}
\end{figure*}

In Figure \ref{fig:et_after}, we illustrate the properties of GRB afterglows associated to ET-detectable NSNS (left panel) and BHNS (right panel) binaries. The figure presents contours encompassing 50\% (solid lines) and 90\% (dashed lines) of the peaks of GRB afterglow light curves in the radio (1.4 GHz, red), optical ($g$ band, green) and X-rays (1 keV, blue). 
Most of the peaks occur beyond $10^2$ days. We note that light curve calculations were restricted between $10^{-1}$ and $10^3$ rest-frame days. 
To enhance visualization of the underlying light curve behavior, we included a sample of randomly chosen optical light curves (thin grey lines) in the background. For context, we also include GRB170817A data \citep[][small circles]{Makhathini2021} rescaled to the median distance of the simulated populations.

These observations are largely influenced by the strong dependency of GRB afterglow light curves on the viewing angle, combined with the viewing angle distribution skewed by GW detection (which favours somewhat smaller angles compared to an isotropic distribution, with a peak at about $30^\circ$ -- \citealt{schutz2011}). As a result, the majority of peaks occur several months to years post-GW, with a minority peaking earlier (around hours) in the optical and X-rays, exhibiting bright emission due to closer viewing angle.

In the right-hand panel, referring to the BHNS population, we also include for comparison the region containing 50\% of the afterglow peaks in the radio band assuming a jet opening angle of $\theta_j = 15$ deg, represented by the solid orange line. In this case, more events are expected to be observed within the jet core or close to its border. As a result, the 50\% region now comprises light curve peaks at $<1$ day. The impact on the optical and X-ray bands (not shown here solely to avoid clutter) is qualitatively similar.

\subsection{GRB Prompt}
\begin{figure*}
    \centering
    \includegraphics[width=\textwidth]{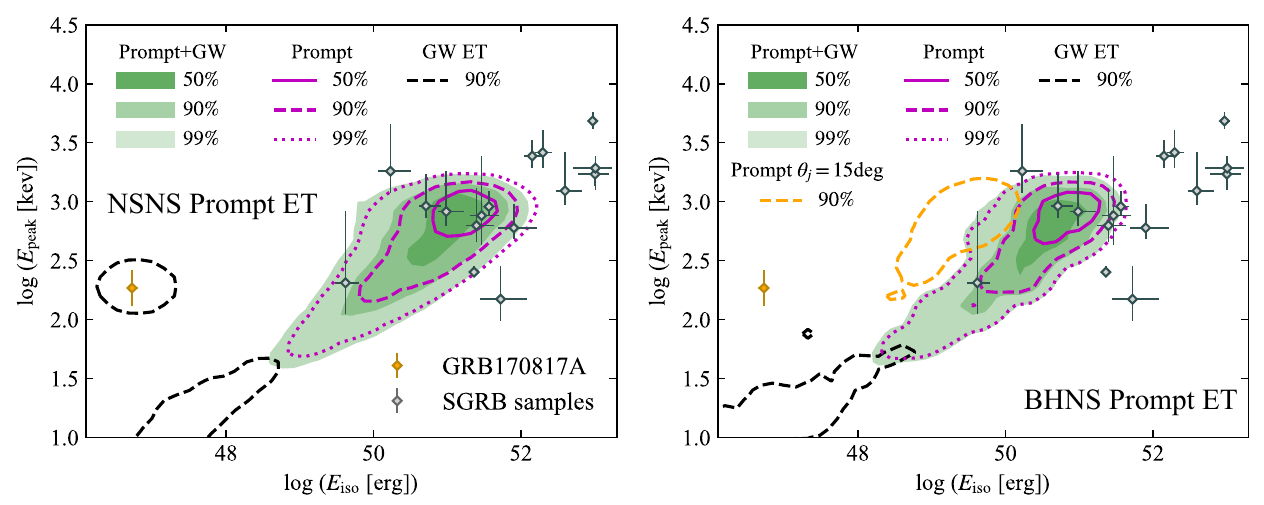}
    \caption{Rest-frame SED peak photon energy $E_\mathrm{peak}$ versus the isotropic-equivalent energy $E_\mathrm{iso}$ for our NSNS (left panel) and BHNS (right panel) populations. The filled green colored regions contain $50\%$, $90\%$ and $99\%$ of the binaries both GRB Prompt- and ET-detectable. The magenta lines contain $50\%$, $90\%$ and $99\%$ (solid, dashed and dotted, respectively) of the GRB Prompt-detectable binaries. The black dashed line contains $90\%$ of the ET-detectable binaries. The black dots with error bars represent a sGRB sample for comparison \citep{Salafia2023}. The orange dot is GRB170817A. We assume for both the populations the same jet half-opening angle $\theta_\mathrm{j} = 3.4$ deg. In the BHNS panel, we also show with orange solid lines the 50\% contour for radio peaks when assuming $\theta_\mathrm{j}=15$ deg.}
    \label{fig:et_prompt}
\end{figure*}

In Figure \ref{fig:et_prompt}, we present the distribution of rest-frame spectral energy distribution (SED) peak energy $E_\mathrm{peak}$ versus isotropic-equivalent energy $E_\mathrm{iso}$ for NSNS (left-hand panel) and BHNS (right-hand panel) events that meet our detection criteria for both the GW signal and the GRB prompt emission.

The green shaded areas encompass 50\%, 90\%, and 99\% of the binary systems detectable by both \textit{Fermi}/GBM and the ET. The magenta lines, varying in style from solid to dashed to dotted, represent the 50\%, 90\%, and 99\% containment levels, respectively, of binary systems that are detectable through GRB prompt emission without the requirement of ET detection.

To contextualize \textit{Fermi}/GBM-detected GRB prompt events within the broader cosmological population, we incorporate a sample of SGRBs with known redshift, symbolized by gray diamonds \citep{Salafia2023}. The specific case of GRB 170817A is marked as an orange diamond on the plot, serving as a reference point.

A dashed black line is included to denote the 90\% confidence region for binaries detectable by ET, independent of the GRB prompt emission detectability constraint. This curve displays the most significant differences between the two populations. For NSNS, the curve shows a bimodal distribution, as we also consider a cocoon shock breakout component for events with a viewing angle less than 60 degrees, replicating the properties of GRB170817A. This component is significant for particularly close events (within 100 Mpc), thus it is not as relevant in comparison to the cosmological population of SGRBs. Since our BHNS population lacks this cocoon shock breakout component, the black curve does not intersect with the cocoon shock breakout event cluster and extends into lower $E_\mathrm{iso}$ values. 

For the BHNS population, we also report a dashed orange line corresponding to 90\% of binary systems that are detectable through GRB prompt emission, assuming a jet half-opening angle $\theta_\mathrm{j}=15$ deg. Varying the jet core half-opening angle while keeping the total jet energy fixed primarily results in a shift of the $E_\mathrm{iso}$ distribution: wider jets correspond to lower on-axis $E_\mathrm{iso}$ values, as this quantity scales approximately as $\theta_\mathrm{j}^{-2}$. As a result, populations with systematically larger opening angles would trace a distinct locus in the $E_\mathrm{peak}$–$E_\mathrm{iso}$ plane compared to BNS-associated GRBs, assuming similar intrinsic emission properties.

\section{Summary and conclusions}\label{sec::conclusions}
In this study, we explored the multi-messenger detection prospects for NSNS and BHNS mergers in the era of ET. By leveraging state-of-the-art population synthesis models and incorporating both GW and EM emission, we have provided detailed forecasts of detection rates and outlined the observable characteristics of the EM counterparts, including KN, GRB prompt and GRB afterglow. Our analysis considered ET as a standalone detector with different configurations, as well as its potential integration within a global network that includes two CE observatories.

Our findings indicate that ET will significantly enhance our ability to observe NSNS mergers. We project that ET could detect more than $10^{4}$ NSNS mergers per year, with thousands of these events also producing detectable KN emission. 

The rates of events with detectable jet-related emission are on the order of a few tens, primarily due to the high number of off-axis events. Specifically, we find that most SGRBs currently observed by \textit{Fermi}/GBM are likely to have a detectable GW counterpart. On the other hand, the prospects for VHE afterglow detection are not promising, with a rate of less than $10^{-1}$ events per year with our fiducial assumptions and with a CTA-like sensitivity.

For BHNS mergers, our results are less optimistic: in fact, only a small fraction of these events are expected to generate EM emission, primarily because tidal disruption of the NS occurs outside of the BH ISCO only under favorable conditions, such as low black hole mass, high BH spin, and large NS deformability. Nevertheless, we anticipate a tenfold increase in the detection rate of BHNS mergers compared to current expectations for the O5 run, with a few KN detections per year and one GRB detection every few years, providing a unique opportunity to study these peculiar systems.

With an ET$\Delta$ detector alone, the above rates are reduced by a factor of $10^2$ when considering only events with a sky localization uncertainty $\Delta\Omega_{{\rm 90}\%}<100\mathrm{deg}^2$. The fraction of reasonably well localized events increases by a factor of around four with a network of two L-shaped ET detectors, while a much more drastic improvement comes when considering ET in tandem with two CE detectors, bringing the fraction of well localized events (for which a multi-wavelength follow-up can be realized with conceivable resources) up to more than 70\% of the detected sources.

We examined the impact of variations in the NS EoS, BH spin distribution, and mass distributions on our detection rates. For NSNS systems, the primary source of error in the detection rates is the uncertainty in the local merger rate, while variations in the mass distribution and EoS generally result in smaller deviations, except in certain cases involving GRB emission, due to the specific conditions assumed for the launch of relativistic jets. For BHNS systems, variations in the EoS and the BH spin distribution can lead to an increase in rates by more than an order of magnitude, surpassing the uncertainty associated with the local merger rate. These dependencies demonstrate the potential of multi-messenger observations to constrain the astrophysical properties of compact objects and refine our understanding of their formation channels.

Prospects for the observation of EM counterparts in the era of ET have already been studied in the literature using different approaches, considering optical \citep{chen2021,branchesi2023,loffredo2024}, high-energy \citep{ronchini2022,hendriks2023}, and very-high-energy \citep{banerjee2023} observations. In general, these studies focus on a specific band and the corresponding observational facilities, considering only counterparts from NSNS mergers. For BHNS, studies have considered observations of KN and afterglow \citep{zhu2021,boersma2022}.

In this work, for the first time, we consider a population of both NSNS and BHNS with a consistent approach, simultaneously analyzing the joint detection of KN, GRB afterglow, and GRB prompt emission across multiple EM bands. Additionally, we provide prospects for the VHE band of the afterglow assuming CTA, an estimate not yet present in the literature.

In conclusion, ET, particularly when operating as part of a network with other third-generation GW observatories, will open new frontiers in multi-messenger astronomy. It will allow for a comprehensive investigation of the origin and properties of compact binary mergers, advancing our knowledge of dense matter physics, jet dynamics, and heavy element nucleosynthesis. The next decade promises transformative discoveries that will bridge GW and EM observations, providing a holistic view of some of the Universe’s most powerful events.

\begin{acknowledgements}
We acknowledge useful discussion with E. Loffredo, S. Ronchini, T. Fragos, P. Schmidt, F. Gabrielli.
M.M. is supported by the French government under the France 2030 investment plan, as part of the Initiative d'Excellence d'Aix-Marseille Universit\'e -- A*MIDEX AMX-22-CEI-02. The work of F.~I. is supported by  the  Swiss National Science Foundation, grant 200020$\_$191957, by the SwissMap National Center for Competence in Research, and by a Miller Postdoctoral Fellowship. Part of the numerical computations made use of the Baobab cluster at the University of Geneva.
\end{acknowledgements}

\bibliographystyle{aa}
\footnotesize
\bibliography{main}

\begin{thebibliography}{127}
\expandafter\ifx\csname natexlab\endcsname\relax\def\natexlab#1{#1}\fi

\bibitem[{{Aasi} {et~al.}(2015){Aasi}, {Abbott}, {Abbott}, {the LIGO Scientific Collaboration}, {et~al.}}]{AdvancedLIGO2015}
{Aasi}, J., {Abbott}, B.~P., {Abbott}, R., {the LIGO Scientific Collaboration}, {et~al.} 2015, Classical and Quantum Gravity, 32, 074001

\bibitem[{{Abac} {et~al.}(2024){Abac}, {Abbott}, {Abouelfettouh}, {Acernese}, {Ackley}, {Adhicary}, {Adhikari}, {Adhikari}, {Adkins}, {Agarwal}, {Agathos}, {Abchouyeh}, {Aguiar}, {Aguilar}, {Aiello}, {Ain}, {Ajith}, {Ak{\c{c}}ay}, {Akutsu}, {Albanesi}, {Alfaidi}, {Al-Jodah}, {All{\'e}n{\'e}}, {Allocca}, {Al-Shammari}, {Altin}, {Alvarez-Lopez}, {Amato}, {Amez-Droz}, {Amorosi}, {Amra}, {Ananyeva}, {Anderson}, {Anderson}, {Andia}, {Ando}, {Andrade}, {Andres}, {Andr{\'e}s-Carcasona}, {Andri{\'c}}, {Anglin}, {Ansoldi}, {Antelis}, {Antier}, {Aoumi}, {Appavuravther}, {Appert}, {Apple}, {Arai}, {Araya}, {Araya}, {Areeda}, {Argianas}, {Aritomi}, {Armato}, {Arnaud}, {Arogeti}, {Aronson}, {Arun}, {Ashton}, {Aso}, {Assiduo}, {de Souza Melo}, {Aston}, {Astone}, {Attadio}, {Aubin}, {Aultoneal}, {Avallone}, {Azrad}, {Babak}, {Badaracco}, {Badger}, {Bae}, {Bagnasco}, {Bagui}, {Baier}, {Baiotti}, {Bajpai}, {Baka}, {Ball}, {Ballardin}, {Ballmer}, {Banagiri}, {Banerjee}, {Bankar}, {Baral}, {Barayoga}, {Barish}, {Barker},
  {Barneo}, {Barone}, {Barr}, {Barsotti}, {Barsuglia}, {Barta}, {Bartoletti}, {Barton}, {Bartos}, {Basak}, {Basalaev}, {Bassiri}, {Basti}, {Bates}, {Bawaj}, {Baxi}, {Bayley}, {Baylor}, {Baynard}, {Bazzan}, {Bedakihale}, {Beirnaert}, {Bejger}, {Belardinelli}, {Bell}, {Benedetto}, {Benoit}, {Bentara}, {Bentley}, {Ben Yaala}, {Bera}, {Berbel}, {Bergamin}, {Berger}, {Bernuzzi}, {Beroiz}, {Berry}, {Bersanetti}, {Bertolini}, {Betzwieser}, {Beveridge}, {Bevins}, {Bhandare}, {Bhardwaj}, {Bhatt}, {Bhattacharjee}, {Bhaumik}, {Bhowmick}, {Bianchi}, {Bilenko}, {Billingsley}, {Binetti}, {Bini}, {Birnholtz}, {Biscoveanu}, {Bisht}, {Bitossi}, {Bizouard}, {Blackburn}, {Blagg}, {Blair}, {Blair}, {Bobba}, {Bode}, {Boileau}, {Boldrini}, {Bolingbroke}, {Bolliand}, {Bonavena}, {Bondarescu}, {Bondu}, {Bonilla}, {Bonilla}, {Bonino}, {Bonnand}, {Booker}, {Borchers}, {Boschi}, {Bose}, {Bossilkov}, {Boudart}, {Boudon}, {Bozzi}, {Bradaschia}, {Brady}, {Braglia}, {Branch}, {Branchesi}, {Brandt}, {Braun}, {Breschi}, {Briant}, {Brillet},
  {Brinkmann}, {Brockill}, {Brockmueller}, {Brooks}, {Brown}, {Brown}, {Brozzetti}, {Brunett}, {Bruno}, {Bruntz}, {Bryant}, {Bucci}, {Buchanan}, {Bulashenko}, {Bulik}, {Bulten}, {Buonanno}, {Burtnyk}, {Buscicchio}, {Buskulic}, {Buy}, {Byer}, {Cabourn Davies}, {Cabras}, {Cabrita}, {C{\'a}ceres-Barbosa}, {Cadonati}, {Cagnoli}, {Cahillane}, {Bustillo}, {Callister}, {Calloni}, {Camp}, {Canepa}, {Caneva Santoro}, {Cannon}, {Cao}, {Capistran}, {Capocasa}, {Capote}, {Carapella}, {Carbognani}, {Carlassara}, {Carlin}, {Carpinelli}, {Carrillo}, {Carter}, {Carullo}, {Casanueva Diaz}, {Casentini}, {Castro-Lucas}, {Caudill}, {Cavagli{\`a}}, {Cavalieri}, {Cella}, {Cerd{\'a}-Dur{\'a}n}, {Cesarini}, {Chaibi}, {Chakraborty}, {Subrahmanya}, {Chan}, {Chan}, {Chandra}, {Chang}, {Chao}, {Char}, {Charlton}, {Charlton}, {Chassande-Mottin}, {Chatterjee}, {Chatterjee}, {Chatterjee}, {Chattopadhyay}, {Chaturvedi}, {Chaty}, {Chatziioannou}, {Chen}, {Chen}, {Chen}, {Chen}, {Chen}, {Chen}, {Chen}, {Chen}, {Chen}, {Chen}, {Cheng},
  {Chessa}, {Cheung}, {Cheung}, {Chiadini}, {Chiarini}, {Chierici}, {Chincarini}, {Chiofalo}, {Chiummo}, {Chou}, {Choudhary}, {Christensen}, {Chua}, {Chugh}, {Ciani}, {Ciecielag}, {Cie{\'s}lar}, {Cifaldi}, {Ciolfi}, {Clara}, {Clark}, {Clarke}, {Clarke}, {Clearwater}, {Clesse}, {Coccia}, {Codazzo}, {Cohadon}, {Colace}, {Colleoni}, {Collette}, {Collins}, {Colloms}, {Colombo}, {Colpi}, {Compton}, {Connolly}, {Conti}, {Corbitt}, {Cordero-Carri{\'o}n}, {Corezzi}, {Cornish}, {Corsi}, {Cortese}, {Costa}, {Cottingham}, {Coughlin}, {Couineaux}, {Coulon}, {Countryman}, {Coupechoux}, {Couvares}, {Coward}, {Cowart}, {Coyne}, {Craig}, {Creed}, {Creighton}, {Creighton}, {Cremonese}, {Criswell}, {Crockett-Gray}, {Crook}, {Crouch}, {Csizmazia}, {Cudell}, {Cullen}, {Cumming}, {Cuoco}, {Cusinato}, {Dabadie}, {Dal Canton}, {Dall'Osso}, {Dal Pra}, {D{\'a}lya}, {D'Angelo}, {Danilishin}, {D'Antonio}, {Danzmann}, {Darroch}, {Dartez}, {Dasgupta}, {Datta}, {Dattilo}, {Daumas}, {Davari}, {Dave}, {Davenport}, {Davier}, {Davies},
  {Davis}, {Davis}, {Davis}, {Davis}, {Dax}, {de Bolle}, {Deenadayalan}, {Degallaix}, {de Laurentis}, {Del{\'e}glise}, {de Lillo}, {Dell'Aquila}, {Del Pozzo}, {De Marco}, {de Matteis}, {D'Emilio}, {Demos}, {Dent}, {Depasse}, {Depergola}, {de Pietri}, {De Rosa}, {de Rossi}, {Desalvo}, {de Simone}, {Dhani}, {Diab}, {D{\'\i}az}, {di Cesare}, {Dideron}, {Didio}, {Dietrich}, {di Fiore}, {di Fronzo}, {di Giovanni}, {di Girolamo}, {Diksha}, {di Michele}, {Ding}, {di Pace}, {di Palma}, {di Renzo}, {Divyajyoti}, {Dmitriev}, {Doctor}, {Dohmen}, {Doleva}, {Dominguez}, {D'Onofrio}, {Donovan}, {Dooley}, {Dooney}, {Doravari}, {Dorosh}, {Drago}, {Driggers}, {Ducoin}, {Dunn}, {Dupletsa}, {D'Urso}, {Duval}, {Duverne}, {Dwyer}, {Eassa}, {Ebersold}, {Eckhardt}, {Eddolls}, {Edelman}, {Edo}, {Edy}, {Effler}, {Eichholz}, {Einsle}, {Eisenmann}, {Eisenstein}, {Ejlli}, {Eleveld}, {Emma}, {Endo}, {Engl}, {Enloe}, {Errico}, {Essick}, {Estell{\'e}s}, {Estevez}, {Etzel}, {Evans}, {Evstafyeva}, {Ewing}, {Ezquiaga}, {Fabrizi}, {Faedi},
  {Fafone}, {Fairhurst}, {Farah}, {Farr}, {Farr}, {Favaro}, {Favata}, {Fays}, {Fazio}, {Feicht}, {Fejer}, {Felicetti}, {Fenyvesi}, {Ferguson}, {Ferraiuolo}, {Ferrante}, {Ferreira}, {Fidecaro}, {Figura}, {Fiori}, {Fiori}, {Fishbach}, {Fisher}, {Fittipaldi}, {Fiumara}, {Flaminio}, {Fleischer}, {Fleming}, {Floden}, {Foley}, {Fong}, {Font}, {Fornal}, {Forsyth}, {Franceschetti}, {Franchini}, {Frasca}, {Frasconi}, {Mascioli}, {Frei}, {Freise}, {Freitas}, {Frey}, {Frischhertz}, {Fritschel}, {Frolov}, {Fronz{\'e}}, {Fuentes-Garcia}, {Fujii}, {Fujimori}, {Fulda}, {Fyffe}, {Gadre}, {Gair}, {Galaudage}, {Galdi}, {Gallagher}, {Gallardo}, {Gallego}, {Gamba}, {Gamboa}, {Ganapathy}, {Ganguly}, {Garaventa}, {Garc{\'\i}a-Bellido}, {Garc{\'\i}a N{\'u}{\~n}ez}, {Garc{\'\i}a-Quir{\'o}s}, {Gardner}, {Gardner}, {Gargiulo}, {Garron}, {Garufi}, {Gasbarra}, {Gateley}, {Gayathri}, {Gemme}, {Gennai}, {Gennari}, {George}, {George}, {Gerberding}, {Gergely}, {Ghonge}, {Ghosh}, {Ghosh}, {Ghosh}, {Ghosh}, {Ghosh}, {Ghosh}, {Giacoppo},
  {Giaime}, {Giardina}, {Gibson}, {Gibson}, {Gier}, {Giri}, {Gissi}, {Gkaitatzis}, {Glanzer}, {Glotin}, {Godfrey}, {Godwin}, {Goebbels}, {Goetz}, {Golomb}, {Gomez Lopez}, {Goncharov}, {Gong}, {Gonz{\'a}lez}, {Goodarzi}, {Goode}, {Goodwin-Jones}, {Gosselin}, {G{\"o}ttel}, {Gouaty}, {Gould}, {Govorkova}, {Goyal}, {Grace}, {Grado}, {Graham}, {Granados}, {Granata}, {Granata}, {Gras}, {Grassia}, {Gray}, {Gray}, {Gray}, {Greco}, {Green}, {Green}, {Green}, {Gretarsson}, {Gretarsson}, {Griffith}, {Griffiths}, {Griggs}, {Grignani}, {Grimaldi}, {Grimaud}, {Grote}, {Guerra}, {Guetta}, {Guidi}, {Guimaraes}, {Gulati}, {Gulminelli}, {Gunny}, {Guo}, {Guo}, {Guo}, {Gupta}, {Gupta}, {Gupta}, {Gupta}, {Gupta}, {Gupta}, {Gupta}, {Gupte}, {Gurs}, {Gutierrez}, {Guzman}, {H}, {Haba}, {Haberland}, {Haino}, {Hall}, {Hamilton}, {Hammond}, {Han}, {Haney}, {Hanks}, {Hanna}, {Hannam}, {Hannuksela}, {Hanselman}, {Hansen}, {Hanson}, {Harada}, {Hardison}, {Haris}, {Harmark}, {Harms}, {Harry}, {Harry}, {Hart}, {Haskell}, {Haster},
  {Hathaway}, {Haughian}, {Hayakawa}, {Hayama}, {Hayes}, {Heffernan}, {Heidmann}, {Heintze}, {Heinze}, {Heinzel}, {Heitmann}, {Hellman}, {Hello}, {Helmling-Cornell}, {Hemming}, {Henderson-Sapir}, {Hendry}, {Heng}, {Hennes}, {Henshaw}, {Hertog}, {Heurs}, {Hewitt}, {Heyns}, {Higginbotham}, {Hild}, {Hill}, {Himemoto}, {Hirata}, {Hirose}, {Hoang}, {Hochheim}, {Hofman}, {Holland}, {Holley-Bockelmann}, {Holmes}, {Holz}, {Honet}, {Hong}, {Hornung}, {Hoshino}, {Hough}, {Hourihane}, {Howell}, {Hoy}, {Hrishikesh}, {Hsieh}, {Hsiung}, {Hsu}, {Hsu}, {Hu}, {Hu}, {Huang}, {Huang}, {Huddart}, {Hughey}, {Hui}, {Hui}, {Husa}, {Huxford}, {Huynh-Dinh}, {Iampieri}, {Iandolo}, {Ianni}, {Iess}, {Imafuku}, {Inayoshi}, {Inoue}, {Iorio}, {Iqbal}, {Irwin}, {Ishikawa}, {Isi}, {Ismail}, {Itoh}, {Iwanaga}, {Iwaya}, {Iyer}, {Jaberianhamedan}, {Jacquet}, {Jacquet}, {Jadhav}, {Jadhav}, {Jain}, {James}, {James}, {Jamshidi}, {Janquart}, {Janssens}, {Janthalur}, {Jaraba}, {Jaranowski}, {Jaume}, {Javed}, {Jennings}, {Jia}, {Jiang}, {Kubisz},
  {Johanson}, {Johns}, {Johnson}, {Johnson-McDaniel}, {Johnston}, {Johnston}, {Johny}, {Jones}, {Jones}, {Jones}, {Jose}, {Joshi}, {Ju}, {Jung}, {Junker}, {Juste}, {Kajita}, {Kaku}, {Kalaghatgi}, {Kalogera}, {Kamiizumi}, {Kanda}, {Kandhasamy}, {Kang}, {Kanner}, {Kapadia}, {Kapasi}, {Karat}, {Karathanasis}, {Kashyap}, {Kasprzack}, {Kastaun}, {Kato}, {Katsavounidis}, {Katzman}, {Kaushik}, {Kawabe}, {Kawamoto}, {Kazemi}, {Kedia}, {Keitel}, {Kelley-Derzon}, {Kennington}, {Kesharwani}, {Key}, {Khadela}, {Khadka}, {Khalili}, {Khan}, {Khan}, {Khanam}, {Khursheed}, {Khusid}, {Kiendrebeogo}, {Kijbunchoo}, {Kim}, {Kim}, {Kim}, {Kim}, {Kim}, {Kim}, {Kimball}, {Kinley-Hanlon}, {Kinnear}, {Kissel}, {Klimenko}, {Knee}, {Knust}, {Kobayashi}, {Koch}, {Koehlenbeck}, {Koekoek}, {Kohri}, {Kokeyama}, {Koley}, {Kolitsidou}, {Kolstein}, {Komori}, {Kong}, {Kontos}, {Korobko}, {Kossak}, {Kou}, {Koushik}, {Kouvatsos}, {Kovalam}, {Kozak}, {Kranzhoff}, {Kringel}, {Krishnendu}, {Kr{\'o}lak}, {Kruska}, {Kuehn}, {Kuijer}, {Kulkarni},
  {Ramamohan}, {Kumar}, {Kumar}, {Kumar}, {Kumar}, {Kumar}, {Kume}, {Kuns}, {Kuntimaddi}, {Kuroyanagi}, {Kurth}, {Kuwahara}, {Kwak}, {Kwan}, {Kwok}, {Lacaille}, {Lagabbe}, {Laghi}, {Lai}, {Laity}, {Lakkis}, {Lalande}, {Lalleman}, {Lalremruati}, {Landry}, {Landry}, {Lane}, {Lang}, {Lange}, {Lantz}, {La Rana}, {La Rosa}, {Lartaux-Vollard}, {Lasky}, {Lawrence}, {Lawrence}, {Laxen}, {Lazzarini}, {Lazzaro}, {Leaci}, {Lecoeuche}, {Lee}, {Lee}, {Lee}, {Lee}, {Lee}, {Lee}, {Lee}, {Legred}, {Lehmann}, {Lehner}, {Le Jean}, {Lema{\^\i}tre}, {Lenti}, {Leonardi}, {Lequime}, {Leroy}, {Lesovsky}, {Letendre}, {Lethuillier}, {Levin}, {Levin}, {Leyde}, {Li}, {Li}, {Li}, {Li}, {Li}, {Lihos}, {Lin}, {Lin}, {Lin}, {Lin}, {Lin}, {Lin}, {Lin}, {Linde}, {Linker}, {Littenberg}, {Liu}, {Liu}, {Liu}, {Villarreal}, {Llobera-Querol}, {Lo}, {Locquet}, {London}, {Longo}, {Lopez}, {Lopez Portilla}, {Lorenzini}, {Lorenzo-Medina}, {Loriette}, {Lormand}, {Losurdo}, {}, {Lough}, {Loughlin}, {Lousto}, {Lowry}, {Lu}, {L{\"u}ck}, {Lumaca},
  {Lundgren}, {Lussier}, {Ma}, {Ma}, {Ma'Arif}, {Macas}, {Macedo}, {Macinnis}, {Maciy}, {MacLeod}, {MacMillan}, {Macquet}, {Macri}, {Maeda}, {Maenaut}, {Hernandez}, {Magare}, {Magazz{\`u}}, {Magee}, {Maggio}, {Maggiore}, {Magnozzi}, {Mahesh}, {Mahesh}, {Maini}, {Majhi}, {Majorana}, {Makarem}, {Makelele}, {Malaquias-Reis}, {Mali}, {Maliakal}, {Malik}, {Man}, {Mandic}, {Mangano}, {Mannix}, {Mansell}, {Mansingh}, {Manske}, {Mantovani}, {Mapelli}, {Marchesoni}, {Mar{\'\i}n Pina}, {Marion}, {M{\'a}rka}, {M{\'a}rka}, {Markosyan}, {Markowitz}, {Maros}, {Marsat}, {Martelli}, {Martin}, {Martin}, {Martinez}, {Martinez}, {Martinez}, {Martini}, {Martinovic}, {Martins}, {Martynov}, {Marx}, {Massaro}, {Masserot}, {Masso-Reid}, {Mastrodicasa}, {Mastrogiovanni}, {Matcovich}, {Matiushechkina}, {Matsuyama}, {Mavalvala}, {Maxwell}, {McCarrol}, {McCarthy}, {McClelland}, {McCormick}, {McCuller}, {McEachin}, {McElhenny}, {McGhee}, {McGinn}, {McGowan}, {McIver}, {McLeod}, {McRae}, {Meacher}, {Meijer}, {Melatos}, {Mellaerts},
  {Menendez-Vazquez}, {Menoni}, {Mera}, {Mercer}, {Mereni}, {Merfeld}, {Merilh}, {M{\'e}rou}, {Merritt}, {Merzougui}, {Messenger}, {Messick}, {Meyer-Conde}, {Meylahn}, {Mhaske}, {Miani}, {Miao}, {Michaloliakos}, {Michel}, {Michimura}, {Middleton}, {Miller}, {Miller}, {Millhouse}, {Milotti}, {Milotti}, {Minenkov}, {Mio}, {Mir}, {Mirasola}, {Miravet-Ten{\'e}s}, {Miritescu}, {Mishra}, {Mishra}, {Mishra}, {Mishra}, {Mitchell}, {Mitchell}, {Mitra}, {Mitrofanov}, {Mittleman}, {Miyakawa}, {Miyamoto}, {Miyoki}, {Mo}, {Mobilia}, {Mohapatra}, {Mohite}, {Molina-Ruiz}, {Mondal}, {Mondin}, {Montani}, {Moore}, {Moraru}, {More}, {More}, {Moreno}, {Morgan}, {Morisaki}, {Moriwaki}, {Morras}, {Moscatello}, {Mourier}, {Mours}, {Mow-Lowry}, {Muciaccia}, {Mukherjee}, {Mukherjee}, {Mukherjee}, {Mukherjee}, {Mukherjee}, {Mukherjee}, {Mukund}, {Mullavey}, {Munch}, {Mundi}, {Mungioli}, {Oberg}, {Murakami}, {Murakoshi}, {Murray}, {Muusse}, {Nabari}, {Nadji}, {Nagar}, {Nagarajan}, {Nagler}, {Nakagaki}, {Nakamura}, {Nakano}, {Nakano},
  {Nandi}, {Napolano}, {Narayan}, {Nardecchia}, {Narikawa}, {Narola}, {Naticchioni}, {Nayak}, {Neilson}, {Nelson}, {Nelson}, {Nery}, {Neunzert}, {Ng}, {Quynh}, {Nichols}, {Nielsen}, {Nieradka}, {Niko}, {Nishino}, {Nishizawa}, {Nissanke}, {Nitoglia}, {Niu}, {Nocera}, {Norman}, {North}, {Novak}, {Nu{\~n}o Siles}, {Nuttall}, {Obayashi}, {Oberling}, {O'Dell}, {Oertel}, {Offermans}, {Oganesyan}, {Oh}, {Oh}, {O'Hanlon}, {Ohashi}, {Ohkawa}, {Ohme}, {Oliveira}, {Oliveri}, {O'Neal}, {Oohara}, {O'Reilly}, {Ormsby}, {Orselli}, {O'Shaughnessy}, {O'Shea}, {Oshima}, {Oshino}, {Ossokine}, {Osthelder}, {Ota}, {Ottaway}, {Ouzriat}, {Overmier}, {Owen}, {Pace}, {Pagano}, {Page}, {Pai}, {Pal}, {Pal}, {Palaia}, {P{\'a}lfi}, {Palma}, {Palomba}, {Palud}, {Pan}, {Pan}, {Pan}, {Panai}, {Panda}, {Pandey}, {Panebianco}, {Pang}, {Pannarale}, {Pannone}, {Pant}, {Panther}, {Paoletti}, {Paolone}, {Papalexakis}, {Papalini}, {Papigkiotis}, {Paquis}, {Parisi}, {Park}, {Park}, {Parker}, {Pascale}, {Pascucci}, {Pasqualetti}, {Passaquieti},
  {Passenger}, {Passuello}, {Patane}, {Pathak}, {Pathak}, {Patra}, {Patricelli}, {Patron}, {Paul}, {Paul}, {Payne}, {Pearce}, {Pedraza}, {Pegna}, {Pele}, {Arellano}, {Penn}, {Penuliar}, {Perego}, {Pereira}, {Perez}, {P{\'e}rigois}, {Perna}, {Perreca}, {Perret}, {Perri{\`e}s}, {Perry}, {Pesios}, {Petracca}, {Petrillo}, {Pfeiffer}, {Pham}, {Pham}, {Phukon}, {Phurailatpam}, {Piarulli}, {Piccari}, {Piccinni}, {Pichot}, {Piendibene}, {Piergiovanni}, {Pierini}, {Pierra}, {Pierro}, {Pietrzak}, {Pillas}, {Pilo}, {Pinard}, {Pinto}, {Pinto}, {Piotrzkowski}, {Pirello}, {Pitkin}, {Placidi}, {Placidi}, {Planas}, {Plastino}, {Poggiani}, {Polini}, {Pompili}, {Poon}, {Porcelli}, {Porter}, {Posnansky}, {Poulton}, {Powell}, {Pracchia}, {Pradhan}, {Pradier}, {Prajapati}, {Prasai}, {Prasanna}, {Prasia}, {Pratten}, {Principe}, {Principe}, {Prodi}, {Prokhorov}, {Prosposito}, {Puecher}, {Pullin}, {Punturo}, {Puppo}, {P{\"u}rrer}, {Qi}, {Qin}, {Qu{\'e}m{\'e}ner}, {Quetschke}, {Quigley}, {Quinonez}, {Raab}, {Raabith}, {Raaijmakers},
  {Raja}, {Rajan}, {Rajbhandari}, {Ramirez}, {Vidal}, {Ramos-Buades}, {Rana}, {Ranjan}, {Ransom}, {Rapagnani}, {Ratto}, {Rawat}, {Ray}, {Raymond}, {Razzano}, {Read}, {Payo}, {Regimbau}, {Rei}, {Reid}, {Reitze}, {Relton}, {Renzini}, {Rettegno}, {Revenu}, {Reyes}, {Rezaei}, {Ricci}, {Ricci}, {Ricciardone}, {Richardson}, {Richardson}, {Rijal}, {Riles}, {Riley}, {Rinaldi}, {Rittmeyer}, {Robertson}, {Robinet}, {Robinson}, {Rocchi}, {Rolland}, {Rollins}, {Romano}, {Romano}, {Romero}, {Romero-Shaw}, {Romie}, {Ronchini}, {Roocke}, {Rosa}, {Rosauer}, {Rose}, {Rosi{\'n}ska}, {Ross}, {Rossello}, {Rowan}, {Roy}, {Roy}, {Rozza}, {Ruggi}, {Ruhama}, {Morales}, {Ruiz-Rocha}, {Sachdev}, {Sadecki}, {Sadiq}, {Saffarieh}, {Sah}, {Saha}, {Saha}, {Sainrat}, {Menon}, {Sakai}, {Sakellariadou}, {Sakon}, {Salafia}, {Salces-Carcoba}, {Salconi}, {Saleem}, {Salemi}, {Sall{\'e}}, {Salvador}, {Sanchez}, {Sanchez}, {Sanchez}, {Sanchez}, {Sanchis-Gual}, {Sanders}, {S{\"a}nger}, {Santoliquido}, {Saravanan}, {Sarin}, {Sasaoka}, {Sasli},
  {Sassi}, {Sassolas}, {Satari}, {Sathyaprakash}, {Sato}, {Sato}, {Sauter}, {Savage}, {Sawada}, {Sawant}, {Sayah}, {Scacco}, {Schaetzl}, {Scheel}, {Schiebelbein}, {Schiworski}, {Schmidt}, {Schmidt}, {Schnabel}, {Schneewind}, {Schofield}, {Schouteden}, {Schulte}, {Schutz}, {Schwartz}, {Scialpi}, {Scott}, {Scott}, {Seetharamu}, {Seglar-Arroyo}, {Sekiguchi}, {Sellers}, {Sengupta}, {Sentenac}, {Seo}, {Seo}, {Sequino}, {Serra}, {Servignat}, {Sevrin}, {Shaffer}, {Shah}, {Shaikh}, {Shao}, {Sharma}, {Sharma}, {Sharma-Chaudhary}, {Shaw}, {Shawhan}, {Shcheblanov}, {Sheridan}, {Shikano}, {Shikauchi}, {Shimode}, {Shinkai}, {Shiota}, {Shoemaker}, {Shoemaker}, {Short}, {Shyamsundar}, {Sider}, {Siegel}, {Sieniawska}, {Sigg}, {Silenzi}, {Simmonds}, {Singer}, {Singh}, {Singh}, {Singh}, {Singh}, {Singha}, {Sintes}, {Sipala}, {Skliris}, {Slagmolen}, {Slaven-Blair}, {Smetana}, {Smith}, {Smith}, {Smith}, {Smith}, {Soldateschi}, {Somiya}, {Song}, {Soni}, {Soni}, {Sordini}, {Sorrentino}, {Sorrentino}, {Sotani}, {Soulard},
  {Southgate}, {Spagnuolo}, {Spencer}, {Spera}, {Spinicelli}, {Spoon}, {Sprague}, {Srivastava}, {Stachurski}, {Steer}, {Steinlechner}, {Steinlechner}, {Stergioulas}, {Stevens}, {Stevenson}, {Stpierre}, {Stratta}, {Strong}, {Strunk}, {Sturani}, {Stuver}, {Suchenek}, {Sudhagar}, {Sueltmann}, {Suleiman}, {Sullivan}, {Sun}, {Sunil}, {Suresh}, {Sutton}, {Suzuki}, {Suzuki}, {Swinkels}, {Syx}, {Szczepa{\'n}czyk}, {Szewczyk}, {Tacca}, {Tagoshi}, {Tait}, {Takahashi}, {Takahashi}, {Takamori}, {Takase}, {Takatani}, {Takeda}, {Takeshita}, {Talbot}, {Tamaki}, {Tamanini}, {Tanabe}, {Tanaka}, {Tanaka}, {Tanaka}, {Tang}, {Tanioka}, {Tanner}, {Tao}, {Tapia}, {San Mart{\'\i}n}, {Tarafder}, {Taranto}, {Taruya}, {Tasson}, {Teloi}, {Tenorio}, {Themann}, {Theodoropoulos}, {Thirugnanasambandam}, {Thomas}, {Thomas}, {Thomas}, {Thompson}, {Thondapu}, {Thorne}, {Thrane}, {Tissino}, {Tiwari}, {Tiwari}, {Tiwari}, {Tiwari}, {Todd}, {Toivonen}, {Toland}, {Tolley}, {Tomaru}, {Tomita}, {Tomura}, {Tong}, {Tong-Yu}, {Toriyama}, {Toropov},
  {Torres-Forn{\'e}}, {Torrie}, {Toscani}, {E Melo}, {Tournefier}, {Trapananti}, {Travasso}, {Traylor}, {Trevor}, {Tringali}, {Tripathee}, {Troian}, {Troiano}, {Trovato}, {Trozzo}, {Trudeau}, {Tsang}, {Tso}, {Tsuchida}, {Tsukada}, {Tsutsui}, {Turbang}, {Turconi}, {Turski}, {Ubach}, {Uchikata}, {Uchiyama}, {Udall}, {Uehara}, {Uematsu}, {Ueno}, {Ueno}, {Undheim}, {Ushiba}, {Vacatello}, {Vahlbruch}, {Vaidya}, {Vajente}, {Vajpeyi}, {Valdes}, {Valencia}, {Valentini}, {Vallejo-Pe{\~n}a}, {Vallero}, {Valsan}, {van Bakel}, {van Beuzekom}, {van Dael}, {van den Brand}, {Broeck}, {Vander-Hyde}, {van der Sluys}, {van de Walle}, {van Dongen}, {Vandra}, {van Haevermaet}, {van Heijningen}, {van Hove}, {Vankeuren}, {Vanosky}, {van Putten}, {van Ranst}, {van Remortel}, {Vardaro}, {Vargas}, {Varghese}, {Varma}, {Vas{\'u}th}, {Vecchio}, {Vedovato}, {Veitch}, {Veitch}, {Venikoudis}, {Venneberg}, {Verdier}, {Verkindt}, {Verma}, {Verma}, {Verma}, {Vermeulen}, {Vetrano}, {Veutro}, {Vibhute}, {Vicer{\'e}}, {Vidyant}, {Viets},
  {Vijaykumar}, {Vilkha}, {Villa-Ortega}, {Vincent}, {Vinet}, {Viret}, {Virtuoso}, {Vitale}, {Vives}, {Vocca}, {Voigt}, {von Reis}, {von Wrangel}, {Vyatchanin}, {Wade}, {Wade}, {Wagner}, {Wajid}, {Walker}, {Wallace}, {Wallace}, {Wang}, {Wang}, {Wang}, {Wang}, {Waratkar}, {Warner}, {Was}, {Washimi}, {Washington}, {Watarai}, {Wayt}, {Weaver}, {Weaver}, {Weaving}, {Webster}, {Weinert}, {Weinstein}, {Weiss}, {Wellmann}, {Wen}, {We{\ss}els}, {Wette}, {Whelan}, {Whiting}, {Whittle}, {Wildberger}, {Wilk}, {Wilken}, {Wilkin}, {Willadsen}, {Willetts}, {Williams}, {Williams}, {Williams}, {Willis}, {Willke}, {Wils}, {Winterflood}, {Wipf}, {Woan}, {Woehler}, {Wofford}, {Wolfe}, {Wong}, {Wong}, {Wong}, {Wright}, {Wright}, {Wu}, {Wu}, {Wu}, {Wuchner}, {Wysocki}, {Xu}, {Xu}, {Yadav}, {Yamamoto}, {Yamamoto}, {Yamamoto}, {Yamamoto}, {Yamamura}, {Yamazaki}, {Yan}, {Yan}, {Yang}, {Yang}, {Yang}, {Yang}, {Yarbrough}, {Yasui}, {Yeh}, {Yelikar}, {Yin}, {Yokoyama}, {Yokozawa}, {Yoo}, {Yu}, {Yuan}, {Yuzurihara}, {Zadro{\.z}ny},
  {Zanolin}, {Zeeshan}, {Zelenova}, {Zendri}, {Zeoli}, {Zerrad}, {Zevin}, {Zhang}, {Zhang}, {Zhang}, {Zhang}, {Zhang}, {Zhao}, {Zhao}, {Zhao}, {Zheng}, {Zhong}, {Zhou}, {Zhu}, {Zhu}, {Zimmerman}, {Zucker}, {Zweizig}, {Ligo Scientific Collaboration}, {VIRGO Collaboration}, \& {Kagra Collaboration}}]{lvk2024_bhns}
{Abac}, A.~G., {Abbott}, R., {Abouelfettouh}, I., {et~al.} 2024, \apjl, 970, L34

\bibitem[{{Abbott} {et~al.}(2017{\natexlab{a}}){Abbott}, {Abbott}, {Abbott}, {Abernathy}, {Ackley}, {Adams}, {Addesso}, {Adhikari}, {Adya}, {Affeldt}, {Aggarwal}, {Aguiar}, {Ain}, {Ajith}, {Allen}, {Altin}, {Anderson}, {Anderson}, {Arai}, {Araya}, {Arceneaux}, {Areeda}, {Arun}, {Ashton}, {Ast}, {Aston}, {Aufmuth}, {Aulbert}, {Babak}, {Baker}, {Ballmer}, {Barayoga}, {Barclay}, {Barish}, {Barker}, {Barr}, {Barsotti}, {Bartlett}, {Bartos}, {Bassiri}, {Batch}, {Baune}, {Bell}, {Berger}, {Bergmann}, {Berry}, {Betzwieser}, {Bhagwat}, {Bhandare}, {Bilenko}, {Billingsley}, {Birch}, {Birney}, {Biscans}, {Bisht}, {Biwer}, {Blackburn}, {Blair}, {Blair}, {Blair}, {Bock}, {Bogan}, {Bohe}, {Bond}, {Bork}, {Bose}, {Brady}, {Braginsky}, {Brau}, {Brinkmann}, {Brockill}, {Broida}, {Brooks}, {Brown}, {Brown}, {Brown}, {Brunett}, {Buchanan}, {Buikema}, {Buonanno}, {Byer}, {Cabero}, {Cadonati}, {Cahillane}, {Calder{\'o}n Bustillo}, {Callister}, {Camp}, {Cannon}, {Cao}, {Capano}, {Caride}, {Caudill}, {Cavagli{\`a}}, {Cepeda},
  {Chamberlin}, {Chan}, {Chao}, {Charlton}, {Cheeseboro}, {Chen}, {Chen}, {Cheng}, {Cho}, {Cho}, {Chow}, {Christensen}, {Chu}, {Chung}, {Ciani}, {Clara}, {Clark}, {Collette}, {Cominsky}, {Constancio}, {Cook}, {Corbitt}, {Cornish}, {Corsi}, {Costa}, {Coughlin}, {Coughlin}, {Countryman}, {Couvares}, {Cowan}, {Coward}, {Cowart}, {Coyne}, {Coyne}, {Craig}, {Creighton}, {Cripe}, {Crowder}, {Cumming}, {Cunningham}, {Dal Canton}, {Danilishin}, {Danzmann}, {Darman}, {Dasgupta}, {Da Silva Costa}, {Dave}, {Davies}, {Daw}, {De}, {DeBra}, {Del Pozzo}, {Denker}, {Dent}, {Dergachev}, {DeRosa}, {DeSalvo}, {Devine}, {Dhurandhar}, {D{\'\i}az}, {Di Palma}, {Donovan}, {Dooley}, {Doravari}, {Douglas}, {Downes}, {Drago}, {Drever}, {Driggers}, {Dwyer}, {Edo}, {Edwards}, {Effler}, {Eggenstein}, {Ehrens}, {Eichholz}, {Eikenberry}, {Engels}, {Essick}, {Etzel}, {Evans}, {Evans}, {Everett}, {Factourovich}, {Fair}, {Fairhurst}, {Fan}, {Fang}, {Farr}, {Farr}, {Favata}, {Fays}, {Fehrmann}, {Fejer}, {Fenyvesi}, {Ferreira}, {Fisher},
  {Fletcher}, {Frei}, {Freise}, {Frey}, {Fritschel}, {Frolov}, {Fulda}, {Fyffe}, {Gabbard}, {Gair}, {Gaonkar}, {Gaur}, {Gehrels}, {Geng}, {George}, {Gergely}, {Ghosh}, {Ghosh}, {Giaime}, {Giardina}, {Gill}, {Glaefke}, {Goetz}, {Goetz}, {Gondan}, {Gonz{\'a}lez}, {Gopakumar}, {Gordon}, {Gorodetsky}, {Gossan}, {Graef}, {Graff}, {Grant}, {Gras}, {Gray}, {Green}, {Grote}, {Grunewald}, {Guo}, {Gupta}, {Gupta}, {Gushwa}, {Gustafson}, {Gustafson}, {Hacker}, {Hall}, {Hall}, {Hammond}, {Haney}, {Hanke}, {Hanks}, {Hanna}, {Hannam}, {Hanson}, {Hardwick}, {Harry}, {Harry}, {Hart}, {Hartman}, {Haster}, {Haughian}, {Heintze}, {Hendry}, {Heng}, {Hennig}, {Henry}, {Heptonstall}, {Heurs}, {Hild}, {Hoak}, {Holt}, {Holz}, {Hopkins}, {Hough}, {Houston}, {Howell}, {Hu}, {Huang}, {Huerta}, {Hughey}, {Husa}, {Huttner}, {Huynh-Dinh}, {Indik}, {Ingram}, {Inta}, {Isa}, {Isi}, {Isogai}, {Iyer}, {Izumi}, {Jang}, {Jani}, {Jawahar}, {Jian}, {Jim{\'e}nez-Forteza}, {Johnson}, {Jones}, {Jones}, {Ju}, {Haris}, {Kalaghatgi}, {Kalogera},
  {Kandhasamy}, {Kang}, {Kanner}, {Kapadia}, {Karki}, {Karvinen}, {Kasprzack}, {Katsavounidis}, {Katzman}, {Kaufer}, {Kaur}, {Kawabe}, {Kehl}, {Keitel}, {Kelley}, {Kells}, {Kennedy}, {Key}, {Khalili}, {Khan}, {Khan}, {Khazanov}, {Kijbunchoo}, {Kim}, {Kim}, {Kim}, {Kim}, {Kim}, {Kim}, {Kim}, {Kimbrell}, {King}, {King}, {Kissel}, {Klein}, {Kleybolte}, {Klimenko}, {Koehlenbeck}, {Kondrashov}, {Kontos}, {Korobko}, {Korth}, {Kozak}, {Kringel}, {Krueger}, {Kuehn}, {Kumar}, {Kumar}, {Kuo}, {Lackey}, {Landry}, {Lange}, {Lantz}, {Lasky}, {Laxen}, {Lazzarini}, {Leavey}, {Lebigot}, {Lee}, {Lee}, {Lee}, {Lee}, {Lenon}, {Leong}, {Levin}, {Lewis}, {Li}, {Libson}, {Littenberg}, {Lockerbie}, {Lombardi}, {London}, {Lord}, {Lormand}, {Lough}, {L{\"u}ck}, {Lundgren}, {Lynch}, {Ma}, {Machenschalk}, {MacInnis}, {Macleod}, {Maga{\~n}a-Sandoval}, {Maga{\~n}a Zertuche}, {Magee}, {Mandic}, {Mangano}, {Mansell}, {Manske}, {M{\'a}rka}, {M{\'a}rka}, {Markosyan}, {Maros}, {Martin}, {Martynov}, {Mason}, {Massinger}, {Masso-Reid},
  {Matichard}, {Matone}, {Mavalvala}, {Mazumder}, {McCarthy}, {McClelland}, {McCormick}, {McGuire}, {McIntyre}, {McIver}, {McManus}, {McRae}, {McWilliams}, {Meacher}, {Meadors}, {Melatos}, {Mendell}, {Mercer}, {Merilh}, {Meshkov}, {Messenger}, {Messick}, {Meyers}, {Miao}, {Middleton}, {Mikhailov}, {Miller}, {Miller}, {Miller}, {Miller}, {Millhouse}, {Ming}, {Mirshekari}, {Mishra}, {Mitra}, {Mitrofanov}, {Mitselmakher}, {Mittleman}, {Mohapatra}, {Moore}, {Moore}, {Moraru}, {Moreno}, {Morriss}, {Mossavi}, {Mow-Lowry}, {Mueller}, {Muir}, {Mukherjee}, {Mukherjee}, {Mukherjee}, {Mukund}, {Mullavey}, {Munch}, {Murphy}, {Murray}, {Mytidis}, {Nayak}, {Nedkova}, {Nelson}, {Neunzert}, {Newton}, {Nguyen}, {Nielsen}, {Nitz}, {Nolting}, {Normandin}, {Nuttall}, {Oberling}, {Ochsner}, {O'Dell}, {Oelker}, {Ogin}, {Oh}, {Oh}, {Ohme}, {Oliver}, {Oppermann}, {Oram}, {O'Reilly}, {O'Shaughnessy}, {Ottaway}, {Overmier}, {Owen}, {Pai}, {Pai}, {Palamos}, {Palashov}, {Pal-Singh}, {Pan}, {Pankow}, {Pannarale}, {Pant}, {Papa}, {Paris},
  {Parker}, {Pascucci}, {Patrick}, {Pearlstone}, {Pedraza}, {Pekowsky}, {Pele}, {Penn}, {Perreca}, {Perri}, {Phelps}, {Pierro}, {Pinto}, {Pitkin}, {Poe}, {Post}, {Powell}, {Prasad}, {Predoi}, {Prestegard}, {Price}, {Prijatelj}, {Principe}, {Privitera}, {Prokhorov}, {Puncken}, {P{\"u}rrer}, {Qi}, {Qin}, {Qiu}, {Quetschke}, {Quintero}, {Quitzow-James}, {Raab}, {Rabeling}, {Radkins}, {Raffai}, {Raja}, {Rajan}, {Rakhmanov}, {Raymond}, {Read}, {Reed}, {Reid}, {Reitze}, {Rew}, {Reyes}, {Riles}, {Rizzo}, {Robertson}, {Robie}, {Rollins}, {Roma}, {Romanov}, {Romie}, {Rowan}, {R{\"u}diger}, {Ryan}, {Sachdev}, {Sadecki}, {Sadeghian}, {Sakellariadou}, {Saleem}, {Salemi}, {Samajdar}, {Sammut}, {Sanchez}, {Sandberg}, {Sandeen}, {Sanders}, {Sathyaprakash}, {Saulson}, {Sauter}, {Savage}, {Sawadsky}, {Schale}, {Schilling}, {Schmidt}, {Schmidt}, {Schnabel}, {Schofield}, {Sch{\"o}nbeck}, {Schreiber}, {Schuette}, {Schutz}, {Scott}, {Scott}, {Sellers}, {Sengupta}, {Sergeev}, {Shaddock}, {Shaffer}, {Shahriar}, {Shaltev},
  {Shapiro}, {Shawhan}, {Sheperd}, {Shoemaker}, {Shoemaker}, {Siellez}, {Siemens}, {Sigg}, {Silva}, {Singer}, {Singer}, {Singh}, {Singh}, {Sintes}, {Slagmolen}, {Smith}, {Smith}, {Smith}, {Son}, {Sorazu}, {Souradeep}, {Srivastava}, {Staley}, {Steinke}, {Steinlechner}, {Steinlechner}, {Steinmeyer}, {Stephens}, {Stone}, {Strain}, {Strauss}, {Strigin}, {Sturani}, {Stuver}, {Summerscales}, {Sun}, {Sunil}, {Sutton}, {Szczepa{\'n}czyk}, {Talukder}, {Tanner}, {T{\'a}pai}, {Tarabrin}, {Taracchini}, {Taylor}, {Theeg}, {Thirugnanasambandam}, {Thomas}, {Thomas}, {Thomas}, {Thorne}, {Thrane}, {Tiwari}, {Tokmakov}, {Toland}, {Tomlinson}, {Tornasi}, {Torres}, {Torrie}, {T{\"o}yr{\"a}}, {Traylor}, {Trifir{\`o}}, {Tse}, {Tuyenbayev}, {Ugolini}, {Unnikrishnan}, {Urban}, {Usman}, {Vahlbruch}, {Vajente}, {Valdes}, {Vander-Hyde}, {van Veggel}, {Vass}, {Vaulin}, {Vecchio}, {Veitch}, {Veitch}, {Venkateswara}, {Vinciguerra}, {Vine}, {Vitale}, {Vo}, {Vorvick}, {Voss}, {Vousden}, {Vyatchanin}, {Wade}, {Wade}, {Wade}, {Walker},
  {Wallace}, {Walsh}, {Wang}, {Wang}, {Wang}, {Wang}, {Ward}, {Warner}, {Weaver}, {Weinert}, {Weinstein}, {Weiss}, {Wen}, {We{\ss}els}, {Westphal}, {Wette}, {Whelan}, {Whiting}, {Williams}, {Williamson}, {Willis}, {Willke}, {Wimmer}, {Winkler}, {Wipf}, {Wittel}, {Woan}, {Woehler}, {Worden}, {Wright}, {Wu}, {Wu}, {Yablon}, {Yam}, {Yamamoto}, {Yancey}, {Yu}, {Zanolin}, {Zevin}, {Zhang}, {Zhang}, {Zhang}, {Zhao}, {Zhou}, {Zhou}, {Zhu}, {Zucker}, {Zuraw}, {Zweizig}, {(LIGO Scientific Collaboration}, \& {Harms}}]{Abbott2017_CE}
{Abbott}, B.~P., {Abbott}, R., {Abbott}, T.~D., {et~al.} 2017{\natexlab{a}}, Classical and Quantum Gravity, 34, 044001

\bibitem[{{Abbott} {et~al.}(2020{\natexlab{a}}){Abbott}, {Abbott}, {Abbott}, {Abraham}, {Acernese}, {Ackley}, {Adams}, {Adya}, {Affeldt}, {Agathos}, {Agatsuma}, {Aggarwal}, {Aguiar}, {Aiello}, {Ain}, {Ajith}, {Akutsu}, {Allen}, {Allocca}, {Aloy}, {Altin}, {Amato}, {Ananyeva}, {Anderson}, {Anderson}, {Ando}, {Angelova}, {Antier}, {Appert}, {Arai}, {Arai}, {Arai}, {Araki}, {Araya}, {Araya}, {Areeda}, {Ar{\`e}ne}, {Aritomi}, {Arnaud}, {Arun}, {Ascenzi}, {Ashton}, {Aso}, {Aston}, {Astone}, {Aubin}, {Aufmuth}, {Aultoneal}, {Austin}, {Avendano}, {Avila-Alvarez}, {Babak}, {Bacon}, {Badaracco}, {Bader}, {Bae}, {Bae}, {Baiotti}, {Bajpai}, {Baker}, {Baldaccini}, {Ballardin}, {Ballmer}, {Banagiri}, {Barayoga}, {Barclay}, {Barish}, {Barker}, {Barkett}, {Barnum}, {Barone}, {Barr}, {Barsotti}, {Barsuglia}, {Barta}, {Bartlett}, {Barton}, {Bartos}, {Bassiri}, {Basti}, {Bawaj}, {Bayley}, {Bazzan}, {B{\'e}csy}, {Bejger}, {Belahcene}, {Bell}, {Beniwal}, {Berger}, {Bergmann}, {Bernuzzi}, {Bero}, {Berry}, {Bersanetti},
  {Bertolini}, {Betzwieser}, {Bhandare}, {Bidler}, {Bilenko}, {Bilgili}, {Billingsley}, {Birch}, {Birney}, {Birnholtz}, {Biscans}, {Biscoveanu}, {Bisht}, {Bitossi}, {Bizouard}, {Blackburn}, {Blair}, {Blair}, {Blair}, {Bloemen}, {Bode}, {Boer}, {Boetzel}, {Bogaert}, {Bondu}, {Bonilla}, {Bonnand}, {Booker}, {Boom}, {Booth}, {Bork}, {Boschi}, {Bose}, {Bossie}, {Bossilkov}, {Bosveld}, {Bouffanais}, {Bozzi}, {Bradaschia}, {Brady}, {Bramley}, {Branchesi}, {Brau}, {Briant}, {Briggs}, {Brighenti}, {Brillet}, {Brinkmann}, {Brisson}, {Brockill}, {Brooks}, {Brown}, {Brown}, {Brunett}, {Buikema}, {Bulik}, {Bulten}, {Buonanno}, {Buskulic}, {Buy}, {Byer}, {Cabero}, {Cadonati}, {Cagnoli}, {Cahillane}, {Bustillo}, {Callister}, {Calloni}, {Camp}, {Campbell}, {Canepa}, {Cannon}, {Cannon}, {Cao}, {Cao}, {Capocasa}, {Carbognani}, {Caride}, {Carney}, {Carullo}, {Diaz}, {Casentini}, {Caudill}, {Cavagli{\`a}}, {Cavalier}, {Cavalieri}, {Cella}, {Cerd{\'a}-Dur{\'a}n}, {Cerretani}, {Cesarini}, {Chaibi}, {Chakravarti}, {Chamberlin},
  {Chan}, {Chan}, {Chao}, {Charlton}, {Chase}, {Chassande-Mottin}, {Chatterjee}, {Chaturvedi}, {Chatziioannou}, {Cheeseboro}, {Chen}, {Chen}, {Chen}, {Chen}, {Chen}, {Chen}, {Cheng}, {Cheong}, {Chia}, {Chincarini}, {Chiummo}, {Cho}, {Cho}, {Cho}, {Christensen}, {Chu}, {Chu}, {Chu}, {Chua}, {Chung}, {Chung}, {Ciani}, {Ciobanu}, {Ciolfi}, {Cipriano}, {Cirone}, {Clara}, {Clark}, {Clearwater}, {Cleva}, {Cocchieri}, {Coccia}, {Cohadon}, {Cohen}, {Colgan}, {Colleoni}, {Collette}, {Collins}, {Cominsky}, {Constancio}, {Conti}, {Cooper}, {Corban}, {Corbitt}, {Cordero-Carri{\'o}n}, {Corley}, {Cornish}, {Corsi}, {Cortese}, {Costa}, {Cotesta}, {Coughlin}, {Coughlin}, {Coulon}, {Countryman}, {Couvares}, {Covas}, {Cowan}, {Coward}, {Cowart}, {Coyne}, {Coyne}, {Creighton}, {Creighton}, {Cripe}, {Croquette}, {Crowder}, {Cullen}, {Cumming}, {Cunningham}, {Cuoco}, {Canton}, {D{\'a}lya}, {Danilishin}, {D'Antonio}, {Danzmann}, {Dasgupta}, {da Silva Costa}, {Datrier}, {Dattilo}, {Dave}, {Davier}, {Davis}, {Daw}, {Debra},
  {Deenadayalan}, {Degallaix}, {de Laurentis}, {Del{\'e}glise}, {Pozzo}, {Demarchi}, {Demos}, {Dent}, {de Pietri}, {Derby}, {De Rosa}, {de Rossi}, {Desalvo}, {de Varona}, {Dhurandhar}, {D{\'\i}az}, {Dietrich}, {Fiore}, {Giovanni}, {Girolamo}, {Lieto}, {Ding}, {Pace}, {Palma}, {Renzo}, {Dmitriev}, {Doctor}, {Doi}, {Donovan}, {Dooley}, {Doravari}, {Dorrington}, {Downes}, {Drago}, {Driggers}, {Du}, {Ducoin}, {Dupej}, {Dwyer}, {Easter}, {Edo}, {Edwards}, {Effler}, {Eguchi}, {Ehrens}, {Eichholz}, {Eikenberry}, {Eisenmann}, {Eisenstein}, {Enomoto}, {Essick}, {Estelles}, {Estevez}, {Etienne}, {Etzel}, {Evans}, {Evans}, {Fafone}, {Fair}, {Fairhurst}, {Fan}, {Farinon}, {Farr}, {Farr}, {Fauchon-Jones}, {Favata}, {Fays}, {Fazio}, {Fee}, {Feicht}, {Fejer}, {Feng}, {Fernandez-Galiana}, {Ferrante}, {Ferreira}, {Ferreira}, {Ferrini}, {Fidecaro}, {Fiori}, {Fiorucci}, {Fishbach}, {Fisher}, {Fishner}, {Fitz-Axen}, {Flaminio}, {Fletcher}, {Flynn}, {Fong}, {Font}, {Forsyth}, {Fournier}, {Frasca}, {Frasconi}, {Frei}, {Freise},
  {Frey}, {Frey}, {Fritschel}, {Frolov}, {Fujii}, {Fukunaga}, {Fukushima}, {Fulda}, {Fyffe}, {Gabbard}, {Gadre}, {Gaebel}, {Gair}, {Gammaitoni}, {Ganija}, {Gaonkar}, {Garcia}, {Garc{\'\i}a-Quir{\'o}s}, {Garufi}, {Gateley}, {Gaudio}, {Gaur}, {Gayathri}, {Ge}, {Gemme}, {Genin}, {Gennai}, {George}, {George}, {Gergely}, {Germain}, {Ghonge}, {Ghosh}, {Ghosh}, {Ghosh}, {Giacomazzo}, {Giaime}, {Giardina}, {Giazotto}, {Gill}, {Giordano}, {Glover}, {Godwin}, {Goetz}, {Goetz}, {Goncharov}, {Gonz{\'a}lez}, {Castro}, {Gopakumar}, {Gorodetsky}, {Gossan}, {Gosselin}, {Gouaty}, {Grado}, {Graef}, {Granata}, {Grant}, {Gras}, {Grassia}, {Gray}, {Gray}, {Greco}, {Green}, {Green}, {Gretarsson}, {Groot}, {Grote}, {Grunewald}, {Gruning}, {Guidi}, {Gulati}, {Guo}, {Gupta}, {Gupta}, {Gustafson}, {Gustafson}, {Haegel}, {Hagiwara}, {Haino}, {Halim}, {Hall}, {Hall}, {Hamilton}, {Hammond}, {Haney}, {Hanke}, {Hanks}, {Hanna}, {Hannam}, {Hannuksela}, {Hanson}, {Hardwick}, {Haris}, {Harms}, {Harry}, {Harry}, {Hasegawa}, {Haster},
  {Haughian}, {Hayakawa}, {Hayama}, {Hayes}, {Healy}, {Heidmann}, {Heintze}, {Heitmann}, {Hello}, {Hemming}, {Hendry}, {Heng}, {Hennig}, {Heptonstall}, {Heurs}, {Hild}, {Himemoto}, {Hinderer}, {Hiranuma}, {Hirata}, {Hirose}, {Hoak}, {Hochheim}, {Hofman}, {Holgado}, {Holland}, {Holt}, {Holz}, {Hong}, {Hopkins}, {Horst}, {Hough}, {Howell}, {Hoy}, {Hreibi}, {Hsieh}, {Huang}, {Huang}, {Huang}, {Huerta}, {Huet}, {Hughey}, {Hulko}, {Husa}, {Huttner}, {Huynh-Dinh}, {Idzkowski}, {Iess}, {Ikenoue}, {Imam}, {Inayoshi}, {Ingram}, {Inoue}, {Inta}, {Intini}, {Ioka}, {Irwin}, {Isa}, {Isac}, {Isi}, {Itoh}, {Iyer}, {Izumi}, {Jacqmin}, {Jadhav}, {Jani}, {Janthalur}, {Jaranowski}, {Jenkins}, {Jiang}, {Johnson}, {Jones}, {Jones}, {Jones}, {Jonker}, {Ju}, {Jung}, {Jung}, {Junker}, {Kajita}, {Kalaghatgi}, {Kalogera}, {Kamai}, {Kamiizumi}, {Kanda}, {Kandhasamy}, {Kang}, {Kanner}, {Kapadia}, {Karki}, {Karvinen}, {Kashyap}, {Kasprzack}, {Katsanevas}, {Katsavounidis}, {Katzman}, {Kaufer}, {Kawabe}, {Kawaguchi}, {Kawai}, {Kawasaki},
  {Keerthana}, {K{\'e}f{\'e}lian}, {Keitel}, {Kennedy}, {Key}, {Khalili}, {Khan}, {Khan}, {Khan}, {Khan}, {Khazanov}, {Khursheed}, {Kijbunchoo}, {Kim}, {Kim}, {Kim}, {Kim}, {Kim}, {Kim}, {Kim}, {Kim}, {Kimball}, {Kimura}, {King}, {King}, {Kinley-Hanlon}, {Kirchhoff}, {Kissel}, {Kita}, {Kitazawa}, {Kleybolte}, {Klika}, {Klimenko}, {Knowles}, {Knyazev}, {Koch}, {Koehlenbeck}, {Koekoek}, {Kojima}, {Kokeyama}, {Koley}, {Komori}, {Kondrashov}, {Kong}, {Kontos}, {Koper}, {Korobko}, {Korth}, {Kotake}, {Kowalska}, {Kozak}, {Kozakai}, {Kozu}, {Kringel}, {Krishnendu}, {Kr{\'o}lak}, {Kuehn}, {Kumar}, {Kumar}, {Kumar}, {Kumar}, {Kumar}, {Kume}, {Kuo}, {Kuo}, {Kuo}, {Kuroyanagi}, {Kusayanagi}, {Kutynia}, {Kwak}, {Kwang}, {Lackey}, {Lai}, {Lam}, {Landry}, {Lane}, {Lang}, {Lange}, {Lantz}, {Lanza}, {Lartaux-Vollard}, {Lasky}, {Laxen}, {Lazzarini}, {Lazzaro}, {Leaci}, {Leavey}, {Lecoeuche}, {Lee}, {Lee}, {Lee}, {Lee}, {Lee}, {Lee}, {Lee}, {Lehmann}, {Lenon}, {Leonardi}, {Leroy}, {Letendre}, {Levin}, {Li}, {Li}, {Li}, {Li},
  {Lin}, {Lin}, {Lin}, {Lin}, {Linde}, {Linker}, {Littenberg}, {Liu}, {Liu}, {Liu}, {Lo}, {Lockerbie}, {London}, {Longo}, {Lorenzini}, {Loriette}, {Lormand}, {Losurdo}, {Lough}, {Lousto}, {Lovelace}, {Lower}, {L{\"u}ck}, {Lumaca}, {Lundgren}, {Luo}, {Lynch}, {Ma}, {Macas}, {Macfoy}, {Macinnis}, {MacLeod}, {Macquet}, {Maga{\~n}a-Sandoval}, {Zertuche}, {Magee}, {Majorana}, {Maksimovic}, {Malik}, {Man}, {Mandic}, {Mangano}, {Mansell}, {Manske}, {Mantovani}, {Marchesoni}, {Marchio}, {Marion}, {M{\'a}rka}, {M{\'a}rka}, {Markakis}, {Markosyan}, {Markowitz}, {Maros}, {Marquina}, {Marsat}, {Martelli}, {Martin}, {Martin}, {Martynov}, {Mason}, {Massera}, {Masserot}, {Massinger}, {Masso-Reid}, {Mastrogiovanni}, {Matas}, {Matichard}, {Matone}, {Mavalvala}, {Mazumder}, {McCann}, {McCarthy}, {McClelland}, {McCormick}, {McCuller}, {McGuire}, {McIver}, {McManus}, {McRae}, {McWilliams}, {Meacher}, {Meadors}, {Mehmet}, {Mehta}, {Meidam}, {Melatos}, {Mendell}, {Mercer}, {Mereni}, {Merilh}, {Merzougui}, {Meshkov}, {Messenger},
  {Messick}, {Metzdorff}, {Meyers}, {Miao}, {Michel}, {Michimura}, {Middleton}, {Mikhailov}, {Milano}, {Miller}, {Miller}, {Millhouse}, {Mills}, {Milovich-Goff}, {Minazzoli}, {Minenkov}, {Mio}, {Mishkin}, {Mishra}, {Mistry}, {Mitra}, {Mitrofanov}, {Mitselmakher}, {Mittleman}, {Miyakawa}, {Miyamoto}, {Miyazaki}, {Miyo}, {Miyoki}, {Mo}, {Moffa}, {Mogushi}, {Mohapatra}, {Montani}, {Moore}, {Moraru}, {Moreno}, {Morisaki}, {Moriwaki}, {Mours}, {Mow-Lowry}, {Mukherjee}, {Mukherjee}, {Mukherjee}, {Mukund}, {Mullavey}, {Munch}, {Mu{\~n}iz}, {Muratore}, {Murray}, {Nagano}, {Nagano}, {Nagar}, {Nakamura}, {Nakano}, {Nakano}, {Nakashima}, {Nardecchia}, {Narikawa}, {Naticchioni}, {Nayak}, {Negishi}, {Neilson}, {Nelemans}, {Nelson}, {Nery}, {Neunzert}, {Ng}, {Ng}, {Nguyen}, {Ni}, {Nichols}, {Nishizawa}, {Nissanke}, {Nocera}, {North}, {Nuttall}, {Obergaulinger}, {Oberling}, {O'Brien}, {Obuchi}, {O'Dea}, {Ogaki}, {Ogin}, {Oh}, {Oh}, {Ohashi}, {Ohishi}, {Ohkawa}, {Ohme}, {Ohta}, {Okada}, {Okutomi}, {Oliver}, {Oohara}, {Ooi},
  {Oppermann}, {Oram}, {O'Reilly}, {Ormiston}, {Ortega}, {O'Shaughnessy}, {Oshino}, {Ossokine}, {Ottaway}, {Overmier}, {Owen}, {Pace}, {Pagano}, {Page}, {Pai}, {Pai}, {Palamos}, {Palashov}, {Palomba}, {Pal-Singh}, {Pan}, {Pan}, {Pang}, {Pang}, {Pang}, {Pankow}, {Pannarale}, {Pant}, {Paoletti}, {Paoli}, {Papa}, {Parida}, {Park}, {Parker}, {Pascucci}, {Pasqualetti}, {Passaquieti}, {Passuello}, {Patil}, {Patricelli}, {Pearlstone}, {Pedersen}, {Pedraza}, {Pedurand}, {Pele}, {Arellano}, {Penn}, {Perez}, {Perreca}, {Pfeiffer}, {Phelps}, {Phukon}, {Piccinni}, {Pichot}, {Piergiovanni}, {Pillant}, {Pinard}, {Pinto}, {Pirello}, {Pitkin}, {Poggiani}, {Pong}, {Ponrathnam}, {Popolizio}, {Porter}, {Powell}, {Prajapati}, {Prasad}, {Prasai}, {Prasanna}, {Pratten}, {Prestegard}, {Privitera}, {Prodi}, {Prokhorov}, {Puncken}, {Punturo}, {Puppo}, {P{\"u}rrer}, {Qi}, {Quetschke}, {Quinonez}, {Quintero}, {Quitzow-James}, {Raab}, {Radkins}, {Radulescu}, {Raffai}, {Raja}, {Rajan}, {Rajbhandari}, {Rakhmanov}, {Ramirez},
  {Ramos-Buades}, {Rana}, {Rao}, {Rapagnani}, {Raymond}, {Razzano}, {Read}, {Regimbau}, {Rei}, {Reid}, {Reitze}, {Ren}, {Ricci}, {Richardson}, {Richardson}, {Ricker}, {Riles}, {Rizzo}, {Robertson}, {Robie}, {Robinet}, {Rocchi}, {Rolland}, {Rollins}, {Roma}, {Romanelli}, {Romano}, {Romel}, {Romie}, {Rose}, {Rosi{\'n}ska}, {Rosofsky}, {Ross}, {Rowan}, {R{\"u}diger}, {Ruggi}, {Rutins}, {Ryan}, {Sachdev}, {Sadecki}, {Sago}, {Saito}, {Saito}, {Sakai}, {Sakai}, {Sakamoto}, {Sakellariadou}, {Sakuno}, {Salconi}, {Saleem}, {Samajdar}, {Sammut}, {Sanchez}, {Sanchez}, {Sanchis-Gual}, {Sandberg}, {Sanders}, {Santiago}, {Sarin}, {Sassolas}, {Sathyaprakash}, {Sato}, {Sato}, {Sauter}, {Savage}, {Sawada}, {Schale}, {Scheel}, {Scheuer}, {Schmidt}, {Schnabel}, {Schofield}, {Sch{\"o}nbeck}, {Schreiber}, {Schulte}, {Schutz}, {Schwalbe}, {Scott}, {Scott}, {Seidel}, {Sekiguchi}, {Sekiguchi}, {Sellers}, {Sengupta}, {Sennett}, {Sentenac}, {Sequino}, {Sergeev}, {Setyawati}, {Shaddock}, {Shaffer}, {Shahriar}, {Shaner}, {Shao},
  {Sharma}, {Shawhan}, {Shen}, {Shibagaki}, {Shimizu}, {Shimoda}, {Shimode}, {Shink}, {Shinkai}, {Shishido}, {Shoda}, {Shoemaker}, {Shoemaker}, {Shyamsundar}, {Siellez}, {Sieniawska}, {Sigg}, {Silva}, {Singer}, {Singh}, {Singhal}, {Sintes}, {Sitmukhambetov}, {Skliris}, {Slagmolen}, {Slaven-Blair}, {Smith}, {Smith}, {Somala}, {Somiya}, {Son}, {Sorazu}, {Sorrentino}, {Sotani}, {Souradeep}, {Sowell}, {Spencer}, {Srivastava}, {Srivastava}, {Staats}, {Stachie}, {Standke}, {Steer}, {Steinke}, {Steinlechner}, {Steinlechner}, {Steinmeyer}, {Stevenson}, {Stocks}, {Stone}, {Stops}, {Strain}, {Stratta}, {Strigin}, {Strunk}, {Sturani}, {Stuver}, {Sudhir}, {Sugimoto}, {Summerscales}, {Sun}, {Sunil}, {Suresh}, {Sutton}, {Suzuki}, {Suzuki}, {Swinkels}, {Szczepa{\'n}czyk}, {Tacca}, {Tagoshi}, {Tait}, {Takahashi}, {Takahashi}, {Takamori}, {Takano}, {Takeda}, {Takeda}, {Talbot}, {Talukder}, {Tanaka}, {Tanaka}, {Tanaka}, {Tanaka}, {Tanaka}, {Tanioka}, {Tanner}, {T{\'a}pai}, {Martin}, {Taracchini}, {Tasson}, {Taylor}, {Telada},
  {Thies}, {Thomas}, {Thomas}, {Thondapu}, {Thorne}, {Thrane}, {Tiwari}, {Tiwari}, {Tiwari}, {Toland}, {Tomaru}, {Tomigami}, {Tomura}, {Tonelli}, {Tornasi}, {Torres-Forn{\'e}}, {Torrie}, {T{\"o}yr{\"a}}, {Travasso}, {Traylor}, {Tringali}, {Trovato}, {Trozzo}, {Trudeau}, {Tsang}, {Tsang}, {Tse}, {Tso}, {Tsubono}, {Tsuchida}, {Tsukada}, {Tsuna}, {Tsuzuki}, {Tuyenbayev}, {Uchikata}, {Uchiyama}, {Ueda}, {Uehara}, {Ueno}, {Ueshima}, {Ugolini}, {Unnikrishnan}, {Uraguchi}, {Urban}, {Ushiba}, {Usman}, {Vahlbruch}, {Vajente}, {Valdes}, {Bakel}, {Beuzekom}, {Brand}, {Broeck}, {Vander-Hyde}, {Schaaf}, {Heijningen}, {Putten}, {Veggel}, {Vardaro}, {Varma}, {Vass}, {Vas{\'u}th}, {Vecchio}, {Vedovato}, {Veitch}, {Veitch}, {Venkateswara}, {Venugopalan}, {Verkindt}, {Vetrano}, {Vicer{\'e}}, {Viets}, {Vine}, {Vinet}, {Vitale}, {Vivanco}, {Vo}, {Vocca}, {Vorvick}, {Vyatchanin}, {Wade}, {Wade}, {Wade}, {Walet}, {Walker}, {Wallace}, {Walsh}, {Wang}, {Wang}, {Wang}, {Wang}, {Wang}, {Wang}, {Ward}, {Warden}, {Warner}, {Was},
  {Watchi}, {Weaver}, {Wei}, {Weinert}, {Weinstein}, {Weiss}, {Wellmann}, {Wen}, {Wessel}, {We{\ss}els}, {Westhouse}, {Wette}, {Whelan}, {Whiting}, {Whittle}, {Wilken}, {Williams}, {Williamson}, {Willis}, {Willke}, {Wimmer}, {Winkler}, {Wipf}, {Wittel}, {Woan}, {Woehler}, {Wofford}, {Worden}, {Wright}, {Wu}, {Wu}, {Wu}, {Wu}, {Wysocki}, {Xiao}, {Xu}, {Yamada}, {Yamamoto}, {Yamamoto}, {Yamamoto}, {Yamamoto}, {Yancey}, {Yang}, {Yap}, {Yazback}, {Yeeles}, {Yokogawa}, {Yokoyama}, {Yokozawa}, {Yoshioka}, {Yu}, {Yu}, {Yuen}, {Yuzurihara}, {Yvert}, {Zadro{\.z}ny}, {Zanolin}, {Zeidler}, {Zelenova}, {Zendri}, {Zevin}, {Zhang}, {Zhang}, {Zhang}, {Zhao}, {Zhao}, {Zhou}, {Zhou}, {Zhu}, {Zhu}, {Zimmerman}, {Zucker}, {Zweizig}, {Kagra Collaboration}, \& {VIRGO Collaboration}}]{abbott2020}
{Abbott}, B.~P., {Abbott}, R., {Abbott}, T.~D., {et~al.} 2020{\natexlab{a}}, Living Reviews in Relativity, 23, 3

\bibitem[{{Abbott} {et~al.}(2017{\natexlab{b}}){Abbott}, {Abbott}, {Abbott}, {Acernese}, {Ackley}, {Adams}, {Adams}, {Addesso}, {Adhikari}, {Adya}, {Affeldt}, {Afrough}, {Agarwal}, {Agathos}, {Agatsuma}, {Aggarwal}, {Aguiar}, {Aiello}, {Ain}, {Ajith}, {Allen}, {Allen}, {Allocca}, {Altin}, {Amato}, {Ananyeva}, {Anderson}, {Anderson}, {Angelova}, {Antier}, {Appert}, {Arai}, {Araya}, {Areeda}, {Arnaud}, {Arun}, {Ascenzi}, {Ashton}, {Ast}, {Aston}, {Astone}, {Atallah}, {Aufmuth}, {Aulbert}, {AultONeal}, {Austin}, {Avila-Alvarez}, {Babak}, {Bacon}, {Bader}, {Bae}, {Bailes}, {Baker}, {Baldaccini}, {Ballardin}, {Ballmer}, {Banagiri}, {Barayoga}, {Barclay}, {Barish}, {Barker}, {Barkett}, {Barone}, {Barr}, {Barsotti}, {Barsuglia}, {Barta}, {Barthelmy}, {Bartlett}, {Bartos}, {Bassiri}, {Basti}, {Batch}, {Bawaj}, {Bayley}, {Bazzan}, {B{\'e}csy}, {Beer}, {Bejger}, {Belahcene}, {Bell}, {Berger}, {Bergmann}, {Bernuzzi}, {Bero}, {Berry}, {Bersanetti}, {Bertolini}, {Betzwieser}, {Bhagwat}, {Bhandare}, {Bilenko},
  {Billingsley}, {Billman}, {Birch}, {Birney}, {Birnholtz}, {Biscans}, {Biscoveanu}, {Bisht}, {Bitossi}, {Biwer}, {Bizouard}, {Blackburn}, {Blackman}, {Blair}, {Blair}, {Blair}, {Bloemen}, {Bock}, {Bode}, {Boer}, {Bogaert}, {Bohe}, {Bondu}, {Bonilla}, {Bonnand}, {Boom}, {Bork}, {Boschi}, {Bose}, {Bossie}, {Bouffanais}, {Bozzi}, {Bradaschia}, {Brady}, {Branchesi}, {Brau}, {Briant}, {Brillet}, {Brinkmann}, {Brisson}, {Brockill}, {Broida}, {Brooks}, {Brown}, {Brown}, {Brunett}, {Buchanan}, {Buikema}, {Bulik}, {Bulten}, {Buonanno}, {Buskulic}, {Buy}, {Byer}, {Cabero}, {Cadonati}, {Cagnoli}, {Cahillane}, {Calder{\'o}n Bustillo}, {Callister}, {Calloni}, {Camp}, {Canepa}, {Canizares}, {Cannon}, {Cao}, {Cao}, {Capano}, {Capocasa}, {Carbognani}, {Caride}, {Carney}, {Carullo}, {Casanueva Diaz}, {Casentini}, {Caudill}, {Cavagli{\`a}}, {Cavalier}, {Cavalieri}, {Cella}, {Cepeda}, {Cerd{\'a}-Dur{\'a}n}, {Cerretani}, {Cesarini}, {Chamberlin}, {Chan}, {Chao}, {Charlton}, {Chase}, {Chassande-Mottin}, {Chatterjee},
  {Chatziioannou}, {Cheeseboro}, {Chen}, {Chen}, {Chen}, {Cheng}, {Chia}, {Chincarini}, {Chiummo}, {Chmiel}, {Cho}, {Cho}, {Chow}, {Christensen}, {Chu}, {Chua}, {Chua}, {Chung}, {Chung}, {Ciani}, {Ciolfi}, {Cirelli}, {Cirone}, {Clara}, {Clark}, {Clearwater}, {Cleva}, {Cocchieri}, {Coccia}, {Cohadon}, {Cohen}, {Colla}, {Collette}, {Cominsky}, {Constancio}, {Conti}, {Cooper}, {Corban}, {Corbitt}, {Cordero-Carri{\'o}n}, {Corley}, {Cornish}, {Corsi}, {Cortese}, {Costa}, {Coughlin}, {Coughlin}, {Coulon}, {Countryman}, {Couvares}, {Covas}, {Cowan}, {Coward}, {Cowart}, {Coyne}, {Coyne}, {Creighton}, {Creighton}, {Cripe}, {Crowder}, {Cullen}, {Cumming}, {Cunningham}, {Cuoco}, {Dal Canton}, {D{\'a}lya}, {Danilishin}, {D'Antonio}, {Danzmann}, {Dasgupta}, {Da Silva Costa}, {Dattilo}, {Dave}, {Davier}, {Davis}, {Daw}, {Day}, {De}, {DeBra}, {Degallaix}, {De Laurentis}, {Del{\'e}glise}, {Del Pozzo}, {Demos}, {Denker}, {Dent}, {De Pietri}, {Dergachev}, {De Rosa}, {DeRosa}, {De Rossi}, {DeSalvo}, {de Varona}, {Devenson},
  {Dhurandhar}, {D{\'\i}az}, {Dietrich}, {Di Fiore}, {Di Giovanni}, {Di Girolamo}, {Di Lieto}, {Di Pace}, {Di Palma}, {Di Renzo}, {Doctor}, {Dolique}, {Donovan}, {Dooley}, {Doravari}, {Dorrington}, {Douglas}, {Dovale {\'A}lvarez}, {Downes}, {Drago}, {Dreissigacker}, {Driggers}, {Du}, {Ducrot}, {Dudi}, {Dupej}, {Dwyer}, {Edo}, {Edwards}, {Effler}, {Eggenstein}, {Ehrens}, {Eichholz}, {Eikenberry}, {Eisenstein}, {Essick}, {Estevez}, {Etienne}, {Etzel}, {Evans}, {Evans}, {Factourovich}, {Fafone}, {Fair}, {Fairhurst}, {Fan}, {Farinon}, {Farr}, {Farr}, {Fauchon-Jones}, {Favata}, {Fays}, {Fee}, {Fehrmann}, {Feicht}, {Fejer}, {Fernandez-Galiana}, {Ferrante}, {Ferreira}, {Ferrini}, {Fidecaro}, {Finstad}, {Fiori}, {Fiorucci}, {Fishbach}, {Fisher}, {Fitz-Axen}, {Flaminio}, {Fletcher}, {Fong}, {Font}, {Forsyth}, {Forsyth}, {Fournier}, {Frasca}, {Frasconi}, {Frei}, {Freise}, {Frey}, {Frey}, {Fries}, {Fritschel}, {Frolov}, {Fulda}, {Fyffe}, {Gabbard}, {Gadre}, {Gaebel}, {Gair}, {Gammaitoni}, {Ganija}, {Gaonkar},
  {Garcia-Quiros}, {Garufi}, {Gateley}, {Gaudio}, {Gaur}, {Gayathri}, {Gehrels}, {Gemme}, {Genin}, {Gennai}, {George}, {George}, {Gergely}, {Germain}, {Ghonge}, {Ghosh}, {Ghosh}, {Ghosh}, {Giaime}, {Giardina}, {Giazotto}, {Gill}, {Glover}, {Goetz}, {Goetz}, {Gomes}, {Goncharov}, {Gonz{\'a}lez}, {Gonzalez Castro}, {Gopakumar}, {Gorodetsky}, {Gossan}, {Gosselin}, {Gouaty}, {Grado}, {Graef}, {Granata}, {Grant}, {Gras}, {Gray}, {Greco}, {Green}, {Gretarsson}, {Groot}, {Grote}, {Grunewald}, {Gruning}, {Guidi}, {Guo}, {Gupta}, {Gupta}, {Gushwa}, {Gustafson}, {Gustafson}, {Halim}, {Hall}, {Hall}, {Hamilton}, {Hammond}, {Haney}, {Hanke}, {Hanks}, {Hanna}, {Hannam}, {Hannuksela}, {Hanson}, {Hardwick}, {Harms}, {Harry}, {Harry}, {Hart}, {Haster}, {Haughian}, {Healy}, {Heidmann}, {Heintze}, {Heitmann}, {Hello}, {Hemming}, {Hendry}, {Heng}, {Hennig}, {Heptonstall}, {Heurs}, {Hild}, {Hinderer}, {Ho}, {Hoak}, {Hofman}, {Holt}, {Holz}, {Hopkins}, {Horst}, {Hough}, {Houston}, {Howell}, {Hreibi}, {Hu}, {Huerta}, {Huet},
  {Hughey}, {Husa}, {Huttner}, {Huynh-Dinh}, {Indik}, {Inta}, {Intini}, {Isa}, {Isac}, {Isi}, {Iyer}, {Izumi}, {Jacqmin}, {Jani}, {Jaranowski}, {Jawahar}, {Jim{\'e}nez-Forteza}, {Johnson}, {Johnson-McDaniel}, {Jones}, {Jones}, {Jonker}, {Ju}, {Junker}, {Kalaghatgi}, {Kalogera}, {Kamai}, {Kandhasamy}, {Kang}, {Kanner}, {Kapadia}, {Karki}, {Karvinen}, {Kasprzack}, {Kastaun}, {Katolik}, {Katsavounidis}, {Katzman}, {Kaufer}, {Kawabe}, {K{\'e}f{\'e}lian}, {Keitel}, {Kemball}, {Kennedy}, {Kent}, {Key}, {Khalili}, {Khan}, {Khan}, {Khan}, {Khazanov}, {Kijbunchoo}, {Kim}, {Kim}, {Kim}, {Kim}, {Kim}, {Kim}, {Kimbrell}, {King}, {King}, {Kinley-Hanlon}, {Kirchhoff}, {Kissel}, {Kleybolte}, {Klimenko}, {Knowles}, {Koch}, {Koehlenbeck}, {Koley}, {Kondrashov}, {Kontos}, {Korobko}, {Korth}, {Kowalska}, {Kozak}, {Kr{\"a}mer}, {Kringel}, {Krishnan}, {Kr{\'o}lak}, {Kuehn}, {Kumar}, {Kumar}, {Kumar}, {Kuo}, {Kutynia}, {Kwang}, {Lackey}, {Lai}, {Landry}, {Lang}, {Lange}, {Lantz}, {Lanza}, {Larson}, {Lartaux-Vollard}, {Lasky},
  {Laxen}, {Lazzarini}, {Lazzaro}, {Leaci}, {Leavey}, {Lee}, {Lee}, {Lee}, {Lee}, {Lee}, {Lehmann}, {Lenon}, {Leon}, {Leonardi}, {Leroy}, {Letendre}, {Levin}, {Li}, {Linker}, {Littenberg}, {Liu}, {Liu}, {Lo}, {Lockerbie}, {London}, {Lord}, {Lorenzini}, {Loriette}, {Lormand}, {Losurdo}, {Lough}, {Lousto}, {Lovelace}, {L{\"u}ck}, {Lumaca}, {Lundgren}, {Lynch}, {Ma}, {Macas}, {Macfoy}, {Machenschalk}, {MacInnis}, {Macleod}, {Maga{\~n}a Hernandez}, {Maga{\~n}a-Sandoval}, {Maga{\~n}a Zertuche}, {Magee}, {Majorana}, {Maksimovic}, {Man}, {Mandic}, {Mangano}, {Mansell}, {Manske}, {Mantovani}, {Marchesoni}, {Marion}, {M{\'a}rka}, {M{\'a}rka}, {Markakis}, {Markosyan}, {Markowitz}, {Maros}, {Marquina}, {Marsh}, {Martelli}, {Martellini}, {Martin}, {Martin}, {Martynov}, {Marx}, {Mason}, {Massera}, {Masserot}, {Massinger}, {Masso-Reid}, {Mastrogiovanni}, {Matas}, {Matichard}, {Matone}, {Mavalvala}, {Mazumder}, {McCarthy}, {McClelland}, {McCormick}, {McCuller}, {McGuire}, {McIntyre}, {McIver}, {McManus}, {McNeill}, {McRae},
  {McWilliams}, {Meacher}, {Meadors}, {Mehmet}, {Meidam}, {Mejuto-Villa}, {Melatos}, {Mendell}, {Mercer}, {Merilh}, {Merzougui}, {Meshkov}, {Messenger}, {Messick}, {Metzdorff}, {Meyers}, {Miao}, {Michel}, {Middleton}, {Mikhailov}, {Milano}, {Miller}, {Miller}, {Miller}, {Millhouse}, {Milovich-Goff}, {Minazzoli}, {Minenkov}, {Ming}, {Mishra}, {Mitra}, {Mitrofanov}, {Mitselmakher}, {Mittleman}, {Moffa}, {Moggi}, {Mogushi}, {Mohan}, {Mohapatra}, {Molina}, {Montani}, {Moore}, {Moraru}, {Moreno}, {Morisaki}, {Morriss}, {Mours}, {Mow-Lowry}, {Mueller}, {Muir}, {Mukherjee}, {Mukherjee}, {Mukherjee}, {Mukund}, {Mullavey}, {Munch}, {Mu{\~n}iz}, {Muratore}, {Murray}, {Nagar}, {Napier}, {Nardecchia}, {Naticchioni}, {Nayak}, {Neilson}, {Nelemans}, {Nelson}, {Nery}, {Neunzert}, {Nevin}, {Newport}, {Newton}, {Ng}, {Nguyen}, {Nguyen}, {Nichols}, {Nielsen}, {Nissanke}, {Nitz}, {Noack}, {Nocera}, {Nolting}, {North}, {Nuttall}, {Oberling}, {O'Dea}, {Ogin}, {Oh}, {Oh}, {Ohme}, {Okada}, {Oliver}, {Oppermann}, {Oram}, {O'Reilly},
  {Ormiston}, {Ortega}, {O'Shaughnessy}, {Ossokine}, {Ottaway}, {Overmier}, {Owen}, {Pace}, {Page}, {Page}, {Pai}, {Pai}, {Palamos}, {Palashov}, {Palomba}, {Pal-Singh}, {Pan}, {Pan}, {Pang}, {Pang}, {Pankow}, {Pannarale}, {Pant}, {Paoletti}, {Paoli}, {Papa}, {Parida}, {Parker}, {Pascucci}, {Pasqualetti}, {Passaquieti}, {Passuello}, {Patil}, {Patricelli}, {Pearlstone}, {Pedraza}, {Pedurand}, {Pekowsky}, {Pele}, {Penn}, {Perez}, {Perreca}, {Perri}, {Pfeiffer}, {Phelps}, {Piccinni}, {Pichot}, {Piergiovanni}, {Pierro}, {Pillant}, {Pinard}, {Pinto}, {Pirello}, {Pitkin}, {Poe}, {Poggiani}, {Popolizio}, {Porter}, {Post}, {Powell}, {Prasad}, {Pratt}, {Pratten}, {Predoi}, {Prestegard}, {Prijatelj}, {Principe}, {Privitera}, {Prix}, {Prodi}, {Prokhorov}, {Puncken}, {Punturo}, {Puppo}, {P{\"u}rrer}, {Qi}, {Quetschke}, {Quintero}, {Quitzow-James}, {Raab}, {Rabeling}, {Radkins}, {Raffai}, {Raja}, {Rajan}, {Rajbhandari}, {Rakhmanov}, {Ramirez}, {Ramos-Buades}, {Rapagnani}, {Raymond}, {Razzano}, {Read}, {Regimbau}, {Rei},
  {Reid}, {Reitze}, {Ren}, {Reyes}, {Ricci}, {Ricker}, {Rieger}, {Riles}, {Rizzo}, {Robertson}, {Robie}, {Robinet}, {Rocchi}, {Rolland}, {Rollins}, {Roma}, {Romano}, {Romano}, {Romel}, {Romie}, {Rosi{\'n}ska}, {Ross}, {Rowan}, {R{\"u}diger}, {Ruggi}, {Rutins}, {Ryan}, {Sachdev}, {Sadecki}, {Sadeghian}, {Sakellariadou}, {Salconi}, {Saleem}, {Salemi}, {Samajdar}, {Sammut}, {Sampson}, {Sanchez}, {Sanchez}, {Sanchis-Gual}, {Sandberg}, {Sanders}, {Sassolas}, {Sathyaprakash}, {Saulson}, {Sauter}, {Savage}, {Sawadsky}, {Schale}, {Scheel}, {Scheuer}, {Schmidt}, {Schmidt}, {Schnabel}, {Schofield}, {Sch{\"o}nbeck}, {Schreiber}, {Schuette}, {Schulte}, {Schutz}, {Schwalbe}, {Scott}, {Scott}, {Seidel}, {Sellers}, {Sengupta}, {Sentenac}, {Sequino}, {Sergeev}, {Shaddock}, {Shaffer}, {Shah}, {Shahriar}, {Shaner}, {Shao}, {Shapiro}, {Shawhan}, {Sheperd}, {Shoemaker}, {Shoemaker}, {Siellez}, {Siemens}, {Sieniawska}, {Sigg}, {Silva}, {Singer}, {Singh}, {Singhal}, {Sintes}, {Slagmolen}, {Smith}, {Smith}, {Smith}, {Somala},
  {Son}, {Sonnenberg}, {Sorazu}, {Sorrentino}, {Souradeep}, {Spencer}, {Srivastava}, {Staats}, {Staley}, {Steinke}, {Steinlechner}, {Steinlechner}, {Steinmeyer}, {Stevenson}, {Stone}, {Stops}, {Strain}, {Stratta}, {Strigin}, {Strunk}, {Sturani}, {Stuver}, {Summerscales}, {Sun}, {Sunil}, {Suresh}, {Sutton}, {Swinkels}, {Szczepa{\'n}czyk}, {Tacca}, {Tait}, {Talbot}, {Talukder}, {Tanner}, {T{\'a}pai}, {Taracchini}, {Tasson}, {Taylor}, {Taylor}, {Tewari}, {Theeg}, {Thies}, {Thomas}, {Thomas}, {Thomas}, {Thorne}, {Thorne}, {Thrane}, {Tiwari}, {Tiwari}, {Tokmakov}, {Toland}, {Tonelli}, {Tornasi}, {Torres-Forn{\'e}}, {Torrie}, {T{\"o}yr{\"a}}, {Travasso}, {Traylor}, {Trinastic}, {Tringali}, {Trozzo}, {Tsang}, {Tse}, {Tso}, {Tsukada}, {Tsuna}, {Tuyenbayev}, {Ueno}, {Ugolini}, {Unnikrishnan}, {Urban}, {Usman}, {Vahlbruch}, {Vajente}, {Valdes}, {Vallisneri}, {van Bakel}, {van Beuzekom}, {van den Brand}, {Van Den Broeck}, {Vander-Hyde}, {van der Schaaf}, {van Heijningen}, {van Veggel}, {Vardaro}, {Varma}, {Vass},
  {Vas{\'u}th}, {Vecchio}, {Vedovato}, {Veitch}, {Veitch}, {Venkateswara}, {Venugopalan}, {Verkindt}, {Vetrano}, {Vicer{\'e}}, {Viets}, {Vinciguerra}, {Vine}, {Vinet}, {Vitale}, {Vo}, {Vocca}, {Vorvick}, {Vyatchanin}, {Wade}, {Wade}, {Wade}, {Walet}, {Walker}, {Wallace}, {Walsh}, {Wang}, {Wang}, {Wang}, {Wang}, {Wang}, {Ward}, {Warner}, {Was}, {Watchi}, {Weaver}, {Wei}, {Weinert}, {Weinstein}, {Weiss}, {Wen}, {Wessel}, {We{\ss}els}, {Westerweck}, {Westphal}, {Wette}, {Whelan}, {Whitcomb}, {Whiting}, {Whittle}, {Wilken}, {Williams}, {Williams}, {Williamson}, {Willis}, {Willke}, {Wimmer}, {Winkler}, {Wipf}, {Wittel}, {Woan}, {Woehler}, {Wofford}, {Wong}, {Worden}, {Wright}, {Wu}, {Wysocki}, {Xiao}, {Yamamoto}, {Yancey}, {Yang}, {Yap}, {Yazback}, {Yu}, {Yu}, {Yvert}, {Zadro{\.z}ny}, {Zanolin}, {Zelenova}, {Zendri}, {Zevin}, {Zhang}, {Zhang}, {Zhang}, {Zhang}, {Zhao}, {Zhou}, {Zhou}, {Zhu}, {Zhu}, {Zimmerman}, {Zucker}, {Zweizig}, {LIGO Scientific Collaboration}, \& {Virgo Collaboration}}]{Abbott2017}
{Abbott}, B.~P., {Abbott}, R., {Abbott}, T.~D., {et~al.} 2017{\natexlab{b}}, \prl, 119, 161101

\bibitem[{{Abbott} {et~al.}(2017{\natexlab{c}}){Abbott}, {Abbott}, {Abbott}, {Acernese}, {Ackley}, {Adams}, {Adams}, {Addesso}, {Adhikari}, {Adya}, {Affeldt}, {Afrough}, {Agarwal}, {Agathos}, {Agatsuma}, {Aggarwal}, {Aguiar}, {Aiello}, {Ain}, {Ajith}, {Allen}, {Allen}, {Allocca}, {Altin}, {Amato}, {Ananyeva}, {Anderson}, {Anderson}, {Angelova}, {Antier}, {Appert}, {Arai}, {Araya}, {Areeda}, {Arnaud}, {Arun}, {Ascenzi}, {Ashton}, {Ast}, {Aston}, {Astone}, {Atallah}, {Aufmuth}, {Aulbert}, {AultONeal}, {Austin}, {Avila-Alvarez}, {Babak}, {Bacon}, {Bader}, {Bae}, {Baker}, {Baldaccini}, {Ballardin}, {Ballmer}, {Banagiri}, {Barayoga}, {Barclay}, {Barish}, {Barker}, {Barkett}, {Barone}, {Barr}, {Barsotti}, {Barsuglia}, {Barta}, {Barthelmy}, {Bartlett}, {Bartos}, {Bassiri}, {Basti}, {Batch}, {Bawaj}, {Bayley}, {Bazzan}, {B{\'e}csy}, {Beer}, {Bejger}, {Belahcene}, {Bell}, {Berger}, {Bergmann}, {Bero}, {Berry}, {Bersanetti}, {Bertolini}, {Betzwieser}, {Bhagwat}, {Bhandare}, {Bilenko}, {Billingsley}, {Billman}, {Birch},
  {Birney}, {Birnholtz}, {Biscans}, {Biscoveanu}, {Bisht}, {Bitossi}, {Biwer}, {Bizouard}, {Blackburn}, {Blackman}, {Blair}, {Blair}, {Blair}, {Bloemen}, {Bock}, {Bode}, {Boer}, {Bogaert}, {Bohe}, {Bondu}, {Bonilla}, {Bonnand}, {Boom}, {Bork}, {Boschi}, {Bose}, {Bossie}, {Bouffanais}, {Bozzi}, {Bradaschia}, {Brady}, {Branchesi}, {Brau}, {Briant}, {Brillet}, {Brinkmann}, {Brisson}, {Brockill}, {Broida}, {Brooks}, {Brown}, {Brown}, {Brunett}, {Buchanan}, {Buikema}, {Bulik}, {Bulten}, {Buonanno}, {Buskulic}, {Buy}, {Byer}, {Cabero}, {Cadonati}, {Cagnoli}, {Cahillane}, {Calder{\'o}n Bustillo}, {Callister}, {Calloni}, {Camp}, {Canepa}, {Canizares}, {Cannon}, {Cao}, {Cao}, {Capano}, {Capocasa}, {Carbognani}, {Caride}, {Carney}, {Casanueva Diaz}, {Casentini}, {Caudill}, {Cavagli{\`a}}, {Cavalier}, {Cavalieri}, {Cella}, {Cepeda}, {Cerd{\'a}-Dur{\'a}n}, {Cerretani}, {Cesarini}, {Chamberlin}, {Chan}, {Chao}, {Charlton}, {Chase}, {Chassande-Mottin}, {Chatterjee}, {Chatziioannou}, {Cheeseboro}, {Chen}, {Chen}, {Chen},
  {Cheng}, {Chia}, {Chincarini}, {Chiummo}, {Chmiel}, {Cho}, {Cho}, {Chow}, {Christensen}, {Chu}, {Chua}, {Chua}, {Chung}, {Chung}, {Ciani}, {Ciolfi}, {Cirelli}, {Cirone}, {Clara}, {Clark}, {Clearwater}, {Cleva}, {Cocchieri}, {Coccia}, {Cohadon}, {Cohen}, {Colla}, {Collette}, {Cominsky}, {Constancio}, {Conti}, {Cooper}, {Corban}, {Corbitt}, {Cordero-Carri{\'o}n}, {Corley}, {Cornish}, {Corsi}, {Cortese}, {Costa}, {Coughlin}, {Coughlin}, {Coulon}, {Countryman}, {Couvares}, {Covas}, {Cowan}, {Coward}, {Cowart}, {Coyne}, {Coyne}, {Creighton}, {Creighton}, {Cripe}, {Crowder}, {Cullen}, {Cumming}, {Cunningham}, {Cuoco}, {Dal Canton}, {D{\'a}lya}, {Danilishin}, {D'Antonio}, {Danzmann}, {Dasgupta}, {Da Silva Costa}, {Dattilo}, {Dave}, {Davier}, {Davis}, {Daw}, {Day}, {De}, {DeBra}, {Degallaix}, {De Laurentis}, {Del{\'e}glise}, {Del Pozzo}, {Demos}, {Denker}, {Dent}, {De Pietri}, {Dergachev}, {De Rosa}, {DeRosa}, {De Rossi}, {DeSalvo}, {de Varona}, {Devenson}, {Dhurandhar}, {D{\'\i}az}, {Di Fiore}, {Di Giovanni}, {Di
  Girolamo}, {Di Lieto}, {Di Pace}, {Di Palma}, {Di Renzo}, {Doctor}, {Dolique}, {Donovan}, {Dooley}, {Doravari}, {Dorrington}, {Douglas}, {Dovale {\'A}lvarez}, {Downes}, {Drago}, {Dreissigacker}, {Driggers}, {Du}, {Ducrot}, {Dupej}, {Dwyer}, {Edo}, {Edwards}, {Effler}, {Ehrens}, {Eichholz}, {Eikenberry}, {Eisenstein}, {Essick}, {Estevez}, {Etienne}, {Etzel}, {Evans}, {Evans}, {Factourovich}, {Fafone}, {Fair}, {Fairhurst}, {Fan}, {Farinon}, {Farr}, {Farr}, {Fauchon-Jones}, {Favata}, {Fays}, {Fee}, {Fehrmann}, {Feicht}, {Fejer}, {Fernandez-Galiana}, {Ferrante}, {Ferreira}, {Ferrini}, {Fidecaro}, {Finstad}, {Fiori}, {Fiorucci}, {Fishbach}, {Fisher}, {Fitz-Axen}, {Flaminio}, {Fletcher}, {Fong}, {Font}, {Forsyth}, {Forsyth}, {Fournier}, {Frasca}, {Frasconi}, {Frei}, {Freise}, {Frey}, {Frey}, {Fries}, {Fritschel}, {Frolov}, {Fulda}, {Fyffe}, {Gabbard}, {Gadre}, {Gaebel}, {Gair}, {Gammaitoni}, {Ganija}, {Gaonkar}, {Garcia-Quiros}, {Garufi}, {Gateley}, {Gaudio}, {Gaur}, {Gayathri}, {Gehrels}, {Gemme}, {Genin},
  {Gennai}, {George}, {George}, {Gergely}, {Germain}, {Ghonge}, {Ghosh}, {Ghosh}, {Ghosh}, {Giaime}, {Giardina}, {Giazotto}, {Gill}, {Glover}, {Goetz}, {Goetz}, {Gomes}, {Goncharov}, {Gonz{\'a}lez}, {Gonzalez Castro}, {Gopakumar}, {Gorodetsky}, {Gossan}, {Gosselin}, {Gouaty}, {Grado}, {Graef}, {Granata}, {Grant}, {Gras}, {Gray}, {Greco}, {Green}, {Gretarsson}, {Griswold}, {Groot}, {Grote}, {Grunewald}, {Gruning}, {Guidi}, {Guo}, {Gupta}, {Gupta}, {Gushwa}, {Gustafson}, {Gustafson}, {Halim}, {Hall}, {Hall}, {Hamilton}, {Hammond}, {Haney}, {Hanke}, {Hanks}, {Hanna}, {Hannam}, {Hannuksela}, {Hanson}, {Hardwick}, {Harms}, {Harry}, {Harry}, {Hart}, {Haster}, {Haughian}, {Healy}, {Heidmann}, {Heintze}, {Heitmann}, {Hello}, {Hemming}, {Hendry}, {Heng}, {Hennig}, {Heptonstall}, {Heurs}, {Hild}, {Hinderer}, {Hoak}, {Hofman}, {Holt}, {Holz}, {Hopkins}, {Horst}, {Hough}, {Houston}, {Howell}, {Hreibi}, {Hu}, {Huerta}, {Huet}, {Hughey}, {Husa}, {Huttner}, {Huynh-Dinh}, {Indik}, {Inta}, {Intini}, {Isa}, {Isac}, {Isi},
  {Iyer}, {Izumi}, {Jacqmin}, {Jani}, {Jaranowski}, {Jawahar}, {Jim{\'e}nez-Forteza}, {Johnson}, {Jones}, {Jones}, {Jonker}, {Ju}, {Junker}, {Kalaghatgi}, {Kalogera}, {Kamai}, {Kandhasamy}, {Kang}, {Kanner}, {Kapadia}, {Karki}, {Karvinen}, {Kasprzack}, {Katolik}, {Katsavounidis}, {Katzman}, {Kaufer}, {Kawabe}, {K{\'e}f{\'e}lian}, {Keitel}, {Kemball}, {Kennedy}, {Kent}, {Key}, {Khalili}, {Khan}, {Khan}, {Khan}, {Khazanov}, {Kijbunchoo}, {Kim}, {Kim}, {Kim}, {Kim}, {Kim}, {Kim}, {Kimbrell}, {King}, {King}, {Kinley-Hanlon}, {Kirchhoff}, {Kissel}, {Kleybolte}, {Klimenko}, {Knowles}, {Koch}, {Koehlenbeck}, {Koley}, {Kondrashov}, {Kontos}, {Korobko}, {Korth}, {Kowalska}, {Kozak}, {Kr{\"a}mer}, {Kringel}, {Krishnan}, {Kr{\'o}lak}, {Kuehn}, {Kumar}, {Kumar}, {Kumar}, {Kuo}, {Kutynia}, {Kwang}, {Lackey}, {Lai}, {Landry}, {Lang}, {Lange}, {Lantz}, {Lanza}, {Larson}, {Lartaux-Vollard}, {Lasky}, {Laxen}, {Lazzarini}, {Lazzaro}, {Leaci}, {Leavey}, {Lee}, {Lee}, {Lee}, {Lee}, {Lee}, {Lehmann}, {Lenon}, {Leonardi}, {Leroy},
  {Letendre}, {Levin}, {Li}, {Linker}, {Littenberg}, {Liu}, {Lo}, {Lockerbie}, {London}, {Lord}, {Lorenzini}, {Loriette}, {Lormand}, {Losurdo}, {Lough}, {Lousto}, {Lovelace}, {L{\"u}ck}, {Lumaca}, {Lundgren}, {Lynch}, {Ma}, {Macas}, {Macfoy}, {Machenschalk}, {MacInnis}, {Macleod}, {Maga{\~n}a Hernandez}, {Maga{\~n}a-Sandoval}, {Maga{\~n}a Zertuche}, {Magee}, {Majorana}, {Maksimovic}, {Man}, {Mandic}, {Mangano}, {Mansell}, {Manske}, {Mantovani}, {Marchesoni}, {Marion}, {M{\'a}rka}, {M{\'a}rka}, {Markakis}, {Markosyan}, {Markowitz}, {Maros}, {Marquina}, {Marsh}, {Martelli}, {Martellini}, {Martin}, {Martin}, {Martynov}, {Mason}, {Massera}, {Masserot}, {Massinger}, {Masso-Reid}, {Mastrogiovanni}, {Matas}, {Matichard}, {Matone}, {Mavalvala}, {Mazumder}, {McCarthy}, {McClelland}, {McCormick}, {McCuller}, {McGuire}, {McIntyre}, {McIver}, {McManus}, {McNeill}, {McRae}, {McWilliams}, {Meacher}, {Meadors}, {Mehmet}, {Meidam}, {Mejuto-Villa}, {Melatos}, {Mendell}, {Mercer}, {Merilh}, {Merzougui}, {Meshkov}, {Messenger},
  {Messick}, {Metzdorff}, {Meyers}, {Miao}, {Michel}, {Middleton}, {Mikhailov}, {Milano}, {Miller}, {Miller}, {Miller}, {Millhouse}, {Milovich-Goff}, {Minazzoli}, {Minenkov}, {Ming}, {Mishra}, {Mitra}, {Mitrofanov}, {Mitselmakher}, {Mittleman}, {Moffa}, {Moggi}, {Mogushi}, {Mohan}, {Mohapatra}, {Montani}, {Moore}, {Moraru}, {Moreno}, {Morriss}, {Mours}, {Mow-Lowry}, {Mueller}, {Muir}, {Mukherjee}, {Mukherjee}, {Mukherjee}, {Mukund}, {Mullavey}, {Munch}, {Mu{\~n}iz}, {Muratore}, {Murray}, {Napier}, {Nardecchia}, {Naticchioni}, {Nayak}, {Neilson}, {Nelemans}, {Nelson}, {Nery}, {Neunzert}, {Nevin}, {Newport}, {Newton}, {Ng}, {Nguyen}, {Nguyen}, {Nichols}, {Nielsen}, {Nissanke}, {Nitz}, {Noack}, {Nocera}, {Nolting}, {North}, {Nuttall}, {Oberling}, {O'Dea}, {Ogin}, {Oh}, {Oh}, {Ohme}, {Okada}, {Oliver}, {Oppermann}, {Oram}, {O'Reilly}, {Ormiston}, {Ortega}, {O'Shaughnessy}, {Ossokine}, {Ottaway}, {Overmier}, {Owen}, {Pace}, {Page}, {Page}, {Pai}, {Pai}, {Palamos}, {Palashov}, {Palomba}, {Pal-Singh}, {Pan}, {Pan},
  {Pang}, {Pang}, {Pankow}, {Pannarale}, {Pant}, {Paoletti}, {Paoli}, {Papa}, {Parida}, {Parker}, {Pascucci}, {Pasqualetti}, {Passaquieti}, {Passuello}, {Patil}, {Patricelli}, {Pearlstone}, {Pedraza}, {Pedurand}, {Pekowsky}, {Pele}, {Penn}, {Perez}, {Perreca}, {Perri}, {Pfeiffer}, {Phelps}, {Piccinni}, {Pichot}, {Piergiovanni}, {Pierro}, {Pillant}, {Pinard}, {Pinto}, {Pirello}, {Pitkin}, {Poe}, {Poggiani}, {Popolizio}, {Porter}, {Post}, {Powell}, {Prasad}, {Pratt}, {Pratten}, {Predoi}, {Prestegard}, {Price}, {Prijatelj}, {Principe}, {Privitera}, {Prodi}, {Prokhorov}, {Puncken}, {Punturo}, {Puppo}, {P{\"u}rrer}, {Qi}, {Quetschke}, {Quintero}, {Quitzow-James}, {Raab}, {Rabeling}, {Radkins}, {Raffai}, {Raja}, {Rajan}, {Rajbhandari}, {Rakhmanov}, {Ramirez}, {Ramos-Buades}, {Rapagnani}, {Raymond}, {Razzano}, {Read}, {Regimbau}, {Rei}, {Reid}, {Reitze}, {Ren}, {Reyes}, {Ricci}, {Ricker}, {Rieger}, {Riles}, {Rizzo}, {Robertson}, {Robie}, {Robinet}, {Rocchi}, {Rolland}, {Rollins}, {Roma}, {Romano}, {Romel}, {Romie},
  {Rosi{\'n}ska}, {Ross}, {Rowan}, {R{\"u}diger}, {Ruggi}, {Rutins}, {Ryan}, {Sachdev}, {Sadecki}, {Sadeghian}, {Sakellariadou}, {Salconi}, {Saleem}, {Salemi}, {Samajdar}, {Sammut}, {Sampson}, {Sanchez}, {Sanchez}, {Sanchis-Gual}, {Sandberg}, {Sanders}, {Sassolas}, {Sathyaprakash}, {Saulson}, {Sauter}, {Savage}, {Sawadsky}, {Schale}, {Scheel}, {Scheuer}, {Schmidt}, {Schmidt}, {Schnabel}, {Schofield}, {Sch{\"o}nbeck}, {Schreiber}, {Schuette}, {Schulte}, {Schutz}, {Schwalbe}, {Scott}, {Scott}, {Seidel}, {Sellers}, {Sengupta}, {Sentenac}, {Sequino}, {Sergeev}, {Shaddock}, {Shaffer}, {Shah}, {Shahriar}, {Shaner}, {Shao}, {Shapiro}, {Shawhan}, {Sheperd}, {Shoemaker}, {Shoemaker}, {Siellez}, {Siemens}, {Sieniawska}, {Sigg}, {Silva}, {Singer}, {Singh}, {Singhal}, {Sintes}, {Slagmolen}, {Smith}, {Smith}, {Smith}, {Somala}, {Son}, {Sonnenberg}, {Sorazu}, {Sorrentino}, {Souradeep}, {Spencer}, {Srivastava}, {Staats}, {Staley}, {Steinke}, {Steinlechner}, {Steinlechner}, {Steinmeyer}, {Stevenson}, {Stone}, {Stops},
  {Strain}, {Stratta}, {Strigin}, {Strunk}, {Sturani}, {Stuver}, {Summerscales}, {Sun}, {Sunil}, {Suresh}, {Sutton}, {Swinkels}, {Szczepa{\'n}czyk}, {Tacca}, {Tait}, {Talbot}, {Talukder}, {Tanner}, {T{\'a}pai}, {Taracchini}, {Tasson}, {Taylor}, {Taylor}, {Tewari}, {Theeg}, {Thies}, {Thomas}, {Thomas}, {Thomas}, {Thorne}, {Thorne}, {Thrane}, {Tiwari}, {Tiwari}, {Tokmakov}, {Toland}, {Tonelli}, {Tornasi}, {Torres-Forn{\'e}}, {Torrie}, {T{\"o}yr{\"a}}, {Travasso}, {Traylor}, {Trinastic}, {Tringali}, {Trozzo}, {Tsang}, {Tse}, {Tso}, {Tsukada}, {Tsuna}, {Tuyenbayev}, {Ueno}, {Ugolini}, {Unnikrishnan}, {Urban}, {Usman}, {Vahlbruch}, {Vajente}, {Valdes}, {van Bakel}, {van Beuzekom}, {van den Brand}, {Van Den Broeck}, {Vander-Hyde}, {van der Schaaf}, {van Heijningen}, {van Veggel}, {Vardaro}, {Varma}, {Vass}, {Vas{\'u}th}, {Vecchio}, {Vedovato}, {Veitch}, {Veitch}, {Venkateswara}, {Venugopalan}, {Verkindt}, {Vetrano}, {Vicer{\'e}}, {Viets}, {Vinciguerra}, {Vine}, {Vinet}, {Vitale}, {Vo}, {Vocca}, {Vorvick},
  {Vyatchanin}, {Wade}, {Wade}, {Wade}, {Walet}, {Walker}, {Wallace}, {Walsh}, {Wang}, {Wang}, {Wang}, {Wang}, {Wang}, {Ward}, {Warner}, {Was}, {Watchi}, {Weaver}, {Wei}, {Weinert}, {Weinstein}, {Weiss}, {Wen}, {Wessel}, {Wessels}, {Westerweck}, {Westphal}, {Wette}, {Whelan}, {Whitcomb}, {Whiting}, {Whittle}, {Wilken}, {Williams}, {Williams}, {Williamson}, {Willis}, {Willke}, {Wimmer}, {Winkler}, {Wipf}, {Wittel}, {Woan}, {Woehler}, {Wofford}, {Wong}, {Worden}, {Wright}, {Wu}, {Wysocki}, {Xiao}, {Yamamoto}, {Yancey}, {Yang}, {Yap}, {Yazback}, {Yu}, {Yu}, {Yvert}, {Zadro{\.z}ny}, {Zanolin}, {Zelenova}, {Zendri}, {Zevin}, {Zhang}, {Zhang}, {Zhang}, {Zhang}, {Zhao}, {Zhou}, {Zhou}, {Zhu}, {Zhu}, {Zimmerman}, {Zucker}, {Zweizig}, {LIGO Scientific Collaboration}, {Virgo Collaboration}, {Wilson-Hodge}, {Bissaldi}, {Blackburn}, {Briggs}, {Burns}, {Cleveland}, {Connaughton}, {Gibby}, {Giles}, {Goldstein}, {Hamburg}, {Jenke}, {Hui}, {Kippen}, {Kocevski}, {McBreen}, {Meegan}, {Paciesas}, {Poolakkil}, {Preece},
  {Racusin}, {Roberts}, {Stanbro}, {Veres}, {von Kienlin}, {GBM}, {Savchenko}, {Ferrigno}, {Kuulkers}, {Bazzano}, {Bozzo}, {Brandt}, {Chenevez}, {Courvoisier}, {Diehl}, {Domingo}, {Hanlon}, {Jourdain}, {Laurent}, {Lebrun}, {Lutovinov}, {Martin-Carrillo}, {Mereghetti}, {Natalucci}, {Rodi}, {Roques}, {Sunyaev}, {Ubertini}, {INTEGRAL}, {Aartsen}, {Ackermann}, {Adams}, {Aguilar}, {Ahlers}, {Ahrens}, {Samarai}, {Altmann}, {Andeen}, {Anderson}, {Ansseau}, {Anton}, {Arg{\"u}elles}, {Auffenberg}, {Axani}, {Bagherpour}, {Bai}, {Barron}, {Barwick}, {Baum}, {Bay}, {Beatty}, {Becker Tjus}, {Bernardini}, {Besson}, {Binder}, {Bindig}, {Blaufuss}, {Blot}, {Bohm}, {B{\"o}rner}, {Bos}, {Bose}, {B{\"o}ser}, {Botner}, {Bourbeau}, {Bourbeau}, {Bradascio}, {Braun}, {Brayeur}, {Brenzke}, {Bretz}, {Bron}, {Brostean-Kaiser}, {Burgman}, {Carver}, {Casey}, {Casier}, {Cheung}, {Chirkin}, {Christov}, {Clark}, {Classen}, {Coenders}, {Collin}, {Conrad}, {Cowen}, {Cross}, {Day}, {de Andr{\'e}}, {De Clercq}, {DeLaunay}, {Dembinski}, {De
  Ridder}, {Desiati}, {de Vries}, {de Wasseige}, {de With}, {DeYoung}, {D{\'\i}az-V{\'e}lez}, {di Lorenzo}, {Dujmovic}, {Dumm}, {Dunkman}, {Dvorak}, {Eberhardt}, {Ehrhardt}, {Eichmann}, {Eller}, {Evenson}, {Fahey}, {Fazely}, {Felde}, {Filimonov}, {Finley}, {Flis}, {Franckowiak}, {Friedman}, {Fuchs}, {Gaisser}, {Gallagher}, {Gerhardt}, {Ghorbani}, {Giang}, {Glauch}, {Gl{\"u}senkamp}, {Goldschmidt}, {Gonzalez}, {Grant}, {Griffith}, {Haack}, {Hallgren}, {Halzen}, {Hanson}, {Hebecker}, {Heereman}, {Helbing}, {Hellauer}, {Hickford}, {Hignight}, {Hill}, {Hoffman}, {Hoffmann}, {Hokanson-Fasig}, {Hoshina}, {Huang}, {Huber}, {Hultqvist}, {H{\"u}nnefeld}, {In}, {Ishihara}, {Jacobi}, {Japaridze}, {Jeong}, {Jero}, {Jones}, {Kalaczynski}, {Kang}, {Kappes}, {Karg}, {Karle}, {Kauer}, {Keivani}, {Kelley}, {Kheirandish}, {Kim}, {Kim}, {Kintscher}, {Kiryluk}, {Kittler}, {Klein}, {Kohnen}, {Koirala}, {Kolanoski}, {K{\"o}pke}, {Kopper}, {Kopper}, {Koschinsky}, {Koskinen}, {Kowalski}, {Krings}, {Kroll}, {Kr{\"u}ckl}, {Kunnen},
  {Kunwar}, {Kurahashi}, {Kuwabara}, {Kyriacou}, {Labare}, {Lanfranchi}, {Larson}, {Lauber}, {Lesiak-Bzdak}, {Leuermann}, {Liu}, {Lu}, {L{\"u}nemann}, {Luszczak}, {Madsen}, {Maggi}, {Mahn}, {Mancina}, {Maruyama}, {Mase}, {Maunu}, {McNally}, {Meagher}, {Medici}, {Meier}, {Menne}, {Merino}, {Meures}, {Miarecki}, {Micallef}, {Moment{\'e}}, {Montaruli}, {Moore}, {Moulai}, {Nahnhauer}, {Nakarmi}, {Naumann}, {Neer}, {Niederhausen}, {Nowicki}, {Nygren}, {Obertacke Pollmann}, {Olivas}, {O'Murchadha}, {Palczewski}, {Pandya}, {Pankova}, {Peiffer}, {Pepper}, {P{\'e}rez de los Heros}, {Pieloth}, {Pinat}, {Price}, {Przybylski}, {Raab}, {R{\"a}del}, {Rameez}, {Rawlins}, {Rea}, {Reimann}, {Relethford}, {Relich}, {Resconi}, {Rhode}, {Richman}, {Robertson}, {Rongen}, {Rott}, {Ruhe}, {Ryckbosch}, {Rysewyk}, {S{\"a}lzer}, {Sanchez Herrera}, {Sandrock}, {Sandroos}, {Santander}, {Sarkar}, {Sarkar}, {Satalecka}, {Schlunder}, {Schmidt}, {Schneider}, {Schoenen}, {Sch{\"o}neberg}, {Schumacher}, {Seckel}, {Seunarine}, {Soedingrekso},
  {Soldin}, {Song}, {Spiczak}, {Spiering}, {Stachurska}, {Stamatikos}, {Stanev}, {Stasik}, {Stettner}, {Steuer}, {Stezelberger}, {Stokstad}, {St{\"o}ssl}, {Strotjohann}, {Stuttard}, {Sullivan}, {Sutherland}, {Taboada}, {Tatar}, {Tenholt}, {Ter-Antonyan}, {Terliuk}, {Te{\v{s}}i{\'c}}, {Tilav}, {Toale}, {Tobin}, {Toscano}, {Tosi}, {Tselengidou}, {Tung}, {Turcati}, {Turley}, {Ty}, {Unger}, {Usner}, {Vandenbroucke}, {Van Driessche}, {van Eijndhoven}, {Vanheule}, {van Santen}, {Vehring}, {Vogel}, {Vraeghe}, {Walck}, {Wallace}, {Wallraff}, {Wandler}, {Wandkowsky}, {Waza}, {Weaver}, {Weiss}, {Wendt}, {Werthebach}, {Whelan}, {Wiebe}, {Wiebusch}, {Wille}, {Williams}, {Wills}, {Wolf}, {Wood}, {Woolsey}, {Woschnagg}, {Xu}, {Xu}, {Xu}, {Yanez}, {Yodh}, {Yoshida}, {Yuan}, {Zoll}, {IceCube Collaboration}, {Balasubramanian}, {Mate}, {Bhalerao}, {Bhattacharya}, {Vibhute}, {Dewangan}, {Rao}, {Vadawale}, {AstroSat Cadmium Zinc Telluride Imager Team}, {Svinkin}, {Hurley}, {Aptekar}, {Frederiks}, {Golenetskii}, {Kozlova},
  {Lysenko}, {Oleynik}, {Tsvetkova}, {Ulanov}, {Cline}, {IPN Collaboration}, {Li}, {Xiong}, {Zhang}, {Lu}, {Song}, {Cao}, {Chang}, {Chen}, {Chen}, {Chen}, {Chen}, {Chen}, {Chen}, {Cui}, {Cui}, {Deng}, {Dong}, {Du}, {Fu}, {Gao}, {Gao}, {Gao}, {Ge}, {Gu}, {Guan}, {Guo}, {Han}, {Hu}, {Huang}, {Huo}, {Jia}, {Jiang}, {Jiang}, {Jin}, {Jin}, {Li}, {Li}, {Li}, {Li}, {Li}, {Li}, {Li}, {Li}, {Li}, {Li}, {Li}, {Liang}, {Liao}, {Liu}, {Liu}, {Liu}, {Liu}, {Liu}, {Liu}, {Liu}, {Lu}, {Lu}, {Luo}, {Ma}, {Meng}, {Nang}, {Nie}, {Ou}, {Qu}, {Sai}, {Sun}, {Tan}, {Tao}, {Tao}, {Tuo}, {Wang}, {Wang}, {Wang}, {Wang}, {Wang}, {Wen}, {Wu}, {Wu}, {Xiao}, {Xu}, {Xu}, {Yan}, {Yang}, {Yang}, {Yang}, {Zhang}, {Zhang}, {Zhang}, {Zhang}, {Zhang}, {Zhang}, {Zhang}, {Zhang}, {Zhang}, {Zhang}, {Zhang}, {Zhang}, {Zhang}, {Zhang}, {Zhang}, {Zhang}, {Zhang}, {Zhang}, {Zhao}, {Zhao}, {Zhao}, {Zheng}, {Zhu}, {Zhu}, {Zou}, {Insight-HXMT Collaboration}, {Albert}, {Andr{\'e}}, {Anghinolfi}, {Ardid}, {Aubert}, {Aublin}, {Avgitas}, {Baret},
  {Barrios-Mart{\'\i}}, {Basa}, {Belhorma}, {Bertin}, {Biagi}, {Bormuth}, {Bourret}, {Bouwhuis}, {Br{\^a}nza{\c{s}}}, {Bruijn}, {Brunner}, {Busto}, {Capone}, {Caramete}, {Carr}, {Celli}, {Cherkaoui El Moursli}, {Chiarusi}, {Circella}, {Coelho}, {Coleiro}, {Coniglione}, {Costantini}, {Coyle}, {Creusot}, {D{\'\i}az}, {Deschamps}, {De Bonis}, {Distefano}, {Di Palma}, {Domi}, {Donzaud}, {Dornic}, {Drouhin}, {Eberl}, {El Bojaddaini}, {El Khayati}, {Els{\"a}sser}, {Enzenh{\"o}fer}, {Ettahiri}, {Fassi}, {Felis}, {Fusco}, {Gay}, {Giordano}, {Glotin}, {Gr{\'e}goire}, {Ruiz}, {Graf}, {Hallmann}, {van Haren}, {Heijboer}, {Hello}, {Hern{\'a}ndez-Rey}, {H{\"o}ssl}, {Hofest{\"a}dt}, {Hugon}, {Illuminati}, {James}, {de Jong}, {Jongen}, {Kadler}, {Kalekin}, {Katz}, {Kiessling}, {Kouchner}, {Kreter}, {Kreykenbohm}, {Kulikovskiy}, {Lachaud}, {Lahmann}, {Lef{\`e}vre}, {Leonora}, {Lotze}, {Loucatos}, {Marcelin}, {Margiotta}, {Marinelli}, {Mart{\'\i}nez-Mora}, {Mele}, {Melis}, {Michael}, {Migliozzi}, {Moussa}, {Navas}, {Nezri},
  {Organokov}, {P{\u{a}}v{\u{a}}la{\c{s}}}, {Pellegrino}, {Perrina}, {Piattelli}, {Popa}, {Pradier}, {Quinn}, {Racca}, {Riccobene}, {S{\'a}nchez-Losa}, {Salda{\~n}a}, {Salvadori}, {Samtleben}, {Sanguineti}, {Sapienza}, {Sieger}, {Spurio}, {Stolarczyk}, {Taiuti}, {Tayalati}, {Trovato}, {Turpin}, {T{\"o}nnis}, {Vallage}, {Van Elewyck}, {Versari}, {Vivolo}, {Vizzoca}, {Wilms}, {Zornoza}, {Z{\'u}{\~n}iga}, {ANTARES Collaboration}, {Beardmore}, {Breeveld}, {Burrows}, {Cenko}, {Cusumano}, {D'A{\`\i}}, {de Pasquale}, {Emery}, {Evans}, {Giommi}, {Gronwall}, {Kennea}, {Krimm}, {Kuin}, {Lien}, {Marshall}, {Melandri}, {Nousek}, {Oates}, {Osborne}, {Pagani}, {Page}, {Palmer}, {Perri}, {Siegel}, {Sbarufatti}, {Tagliaferri}, {Tohuvavohu}, {Swift Collaboration}, {Tavani}, {Verrecchia}, {Bulgarelli}, {Evangelista}, {Pacciani}, {Feroci}, {Pittori}, {Giuliani}, {Del Monte}, {Donnarumma}, {Argan}, {Trois}, {Ursi}, {Cardillo}, {Piano}, {Longo}, {Lucarelli}, {Munar-Adrover}, {Fuschino}, {Labanti}, {Marisaldi}, {Minervini},
  {Fioretti}, {Parmiggiani}, {Gianotti}, {Trifoglio}, {Di Persio}, {Antonelli}, {Barbiellini}, {Caraveo}, {Cattaneo}, {Costa}, {Colafrancesco}, {D'Amico}, {Ferrari}, {Morselli}, {Paoletti}, {Picozza}, {Pilia}, {Rappoldi}, {Soffitta}, {Vercellone}, {AGILE Team}, {Foley}, {Coulter}, {Kilpatrick}, {Drout}, {Piro}, {Shappee}, {Siebert}, {Simon}, {Ulloa}, {Kasen}, {Madore}, {Murguia-Berthier}, {Pan}, {Prochaska}, {Ramirez-Ruiz}, {Rest}, {Rojas-Bravo}, {1M2H Team}, {Berger}, {Soares-Santos}, {Annis}, {Alexander}, {Allam}, {Balbinot}, {Blanchard}, {Brout}, {Butler}, {Chornock}, {Cook}, {Cowperthwaite}, {Diehl}, {Drlica-Wagner}, {Drout}, {Durret}, {Eftekhari}, {Finley}, {Fong}, {Frieman}, {Fryer}, {Garc{\'\i}a-Bellido}, {Gruendl}, {Hartley}, {Herner}, {Kessler}, {Lin}, {Lopes}, {Louren{\c{c}}o}, {Margutti}, {Marshall}, {Matheson}, {Medina}, {Metzger}, {Mu{\~n}oz}, {Muir}, {Nicholl}, {Nugent}, {Palmese}, {Paz-Chinch{\'o}n}, {Quataert}, {Sako}, {Sauseda}, {Schlegel}, {Scolnic}, {Secco}, {Smith}, {Sobreira}, {Villar},
  {Vivas}, {Wester}, {Williams}, {Yanny}, {Zenteno}, {Zhang}, {Abbott}, {Banerji}, {Bechtol}, {Benoit-L{\'e}vy}, {Bertin}, {Brooks}, {Buckley-Geer}, {Burke}, {Capozzi}, {Carnero Rosell}, {Carrasco Kind}, {Castander}, {Crocce}, {Cunha}, {D'Andrea}, {da Costa}, {Davis}, {DePoy}, {Desai}, {Dietrich}, {Eifler}, {Fernandez}, {Flaugher}, {Fosalba}, {Gaztanaga}, {Gerdes}, {Giannantonio}, {Goldstein}, {Gruen}, {Gschwend}, {Gutierrez}, {Honscheid}, {James}, {Jeltema}, {Johnson}, {Johnson}, {Kent}, {Krause}, {Kron}, {Kuehn}, {Lahav}, {Lima}, {Maia}, {March}, {Martini}, {McMahon}, {Menanteau}, {Miller}, {Miquel}, {Mohr}, {Nichol}, {Ogando}, {Plazas}, {Romer}, {Roodman}, {Rykoff}, {Sanchez}, {Scarpine}, {Schindler}, {Schubnell}, {Sevilla-Noarbe}, {Sheldon}, {Smith}, {Smith}, {Stebbins}, {Suchyta}, {Swanson}, {Tarle}, {Thomas}, {Troxel}, {Tucker}, {Vikram}, {Walker}, {Wechsler}, {Weller}, {Carlin}, {Gill}, {Li}, {Marriner}, {Neilsen}, {Dark Energy Camera GW-EM Collaboration}, {DES Collaboration}, {Haislip}, {Kouprianov},
  {Reichart}, {Sand}, {Tartaglia}, {Valenti}, {Yang}, {DLT40 Collaboration}, {Benetti}, {Brocato}, {Campana}, {Cappellaro}, {Covino}, {D'Avanzo}, {D'Elia}, {Getman}, {Ghirlanda}, {Ghisellini}, {Limatola}, {Nicastro}, {Palazzi}, {Pian}, {Piranomonte}, {Possenti}, {Rossi}, {Salafia}, {Tomasella}, {Amati}, {Antonelli}, {Bernardini}, {Bufano}, {Capaccioli}, {Casella}, {Dadina}, {De Cesare}, {Di Paola}, {Giuffrida}, {Giunta}, {Israel}, {Lisi}, {Maiorano}, {Mapelli}, {Masetti}, {Pescalli}, {Pulone}, {Salvaterra}, {Schipani}, {Spera}, {Stamerra}, {Stella}, {Testa}, {Turatto}, {Vergani}, {Aresu}, {Bachetti}, {Buffa}, {Burgay}, {Buttu}, {Caria}, {Carretti}, {Casasola}, {Castangia}, {Carboni}, {Casu}, {Concu}, {Corongiu}, {Deiana}, {Egron}, {Fara}, {Gaudiomonte}, {Gusai}, {Ladu}, {Loru}, {Leurini}, {Marongiu}, {Melis}, {Melis}, {Migoni}, {Milia}, {Navarrini}, {Orlati}, {Ortu}, {Palmas}, {Pellizzoni}, {Perrodin}, {Pisanu}, {Poppi}, {Righini}, {Saba}, {Serra}, {Serrau}, {Stagni}, {Surcis}, {Vacca}, {Vargiu}, {Hunt},
  {Jin}, {Klose}, {Kouveliotou}, {Mazzali}, {M{\o}ller}, {Nava}, {Piran}, {Selsing}, {Vergani}, {Wiersema}, {Toma}, {Higgins}, {Mundell}, {di Serego Alighieri}, {G{\'o}tz}, {Gao}, {Gomboc}, {Kaper}, {Kobayashi}, {Kopac}, {Mao}, {Starling}, {Steele}, {van der Horst}, {GRAWITA: GRAvitational Wave Inaf TeAm}, {Acero}, {Atwood}, {Baldini}, {Barbiellini}, {Bastieri}, {Berenji}, {Bellazzini}, {Bissaldi}, {Blandford}, {Bloom}, {Bonino}, {Bottacini}, {Bregeon}, {Buehler}, {Buson}, {Cameron}, {Caputo}, {Caraveo}, {Cavazzuti}, {Chekhtman}, {Cheung}, {Chiang}, {Ciprini}, {Cohen-Tanugi}, {Cominsky}, {Costantin}, {Cuoco}, {D'Ammando}, {de Palma}, {Digel}, {Di Lalla}, {Di Mauro}, {Di Venere}, {Dubois}, {Fegan}, {Focke}, {Franckowiak}, {Fukazawa}, {Funk}, {Fusco}, {Gargano}, {Gasparrini}, {Giglietto}, {Giordano}, {Giroletti}, {Glanzman}, {Green}, {Grondin}, {Guillemot}, {Guiriec}, {Harding}, {Horan}, {J{\'o}hannesson}, {Kamae}, {Kensei}, {Kuss}, {La Mura}, {Latronico}, {Lemoine-Goumard}, {Longo}, {Loparco}, {Lovellette},
  {Lubrano}, {Magill}, {Maldera}, {Manfreda}, {Mazziotta}, {McEnery}, {Meyer}, {Michelson}, {Mirabal}, {Monzani}, {Moretti}, {Morselli}, {Moskalenko}, {Negro}, {Nuss}, {Ojha}, {Omodei}, {Orienti}, {Orlando}, {Palatiello}, {Paliya}, {Paneque}, {Pesce-Rollins}, {Piron}, {Porter}, {Principe}, {Rain{\`o}}, {Rando}, {Razzano}, {Razzaque}, {Reimer}, {Reimer}, {Reposeur}, {Rochester}, {Saz Parkinson}, {Sgr{\`o}}, {Siskind}, {Spada}, {Spandre}, {Suson}, {Takahashi}, {Tanaka}, {Thayer}, {Thayer}, {Thompson}, {Tibaldo}, {Torres}, {Torresi}, {Troja}, {Venters}, {Vianello}, {Zaharijas}, {Fermi Large Area Telescope Collaboration}, {Allison}, {Bannister}, {Dobie}, {Kaplan}, {Lenc}, {Lynch}, {Murphy}, {Sadler}, {Australia Telescope Compact Array}, {Hotan}, {James}, {Oslowski}, {Raja}, {Shannon}, {Whiting}, {Australian SKA Pathfinder}, {Arcavi}, {Howell}, {McCully}, {Hosseinzadeh}, {Hiramatsu}, {Poznanski}, {Barnes}, {Zaltzman}, {Vasylyev}, {Maoz}, {Las Cumbres Observatory Group}, {Cooke}, {Bailes}, {Wolf}, {Deller},
  {Lidman}, {Wang}, {Gendre}, {Andreoni}, {Ackley}, {Pritchard}, {Bessell}, {Chang}, {M{\"o}ller}, {Onken}, {Scalzo}, {Ridden-Harper}, {Sharp}, {Tucker}, {Farrell}, {Elmer}, {Johnston}, {Venkatraman Krishnan}, {Keane}, {Green}, {Jameson}, {Hu}, {Ma}, {Sun}, {Wu}, {Wang}, {Shang}, {Hu}, {Ashley}, {Yuan}, {Li}, {Tao}, {Zhu}, {Zhang}, {Suntzeff}, {Zhou}, {Yang}, {Orange}, {Morris}, {Cucchiara}, {Giblin}, {Klotz}, {Staff}, {Thierry}, {Schmidt}, {OzGrav}, {(Deeper}, {Wider}, {program}, {AST3}, {CAASTRO Collaborations}, {Tanvir}, {Levan}, {Cano}, {de Ugarte-Postigo}, {Gonz{\'a}lez-Fern{\'a}ndez}, {Greiner}, {Hjorth}, {Irwin}, {Kr{\"u}hler}, {Mandel}, {Milvang-Jensen}, {O'Brien}, {Rol}, {Rosetti}, {Rosswog}, {Rowlinson}, {Steeghs}, {Th{\"o}ne}, {Ulaczyk}, {Watson}, {Bruun}, {Cutter}, {Figuera Jaimes}, {Fujii}, {Fruchter}, {Gompertz}, {Jakobsson}, {Hodosan}, {J{\`e}rgensen}, {Kangas}, {Kann}, {Rabus}, {Schr{\o}der}, {Stanway}, {Wijers}, {VINROUGE Collaboration}, {Lipunov}, {Gorbovskoy}, {Kornilov}, {Tyurina},
  {Balanutsa}, {Kuznetsov}, {Vlasenko}, {Podesta}, {Lopez}, {Podesta}, {Levato}, {Saffe}, {Mallamaci}, {Budnev}, {Gress}, {Kuvshinov}, {Gorbunov}, {Vladimirov}, {Zimnukhov}, {Gabovich}, {Yurkov}, {Sergienko}, {Rebolo}, {Serra-Ricart}, {Tlatov}, {Ishmuhametova}, {MASTER Collaboration}, {Abe}, {Aoki}, {Aoki}, {Asakura}, {Baar}, {Barway}, {Bond}, {Doi}, {Finet}, {Fujiyoshi}, {Furusawa}, {Honda}, {Itoh}, {Kanda}, {Kawabata}, {Kawabata}, {Kim}, {Koshida}, {Kuroda}, {Lee}, {Liu}, {Matsubayashi}, {Miyazaki}, {Morihana}, {Morokuma}, {Motohara}, {Murata}, {Nagai}, {Nagashima}, {Nagayama}, {Nakaoka}, {Nakata}, {Ohsawa}, {Ohshima}, {Ohta}, {Okita}, {Saito}, {Saito}, {Sako}, {Sekiguchi}, {Sumi}, {Tajitsu}, {Takahashi}, {Takayama}, {Tamura}, {Tanaka}, {Tanaka}, {Terai}, {Tominaga}, {Tristram}, {Uemura}, {Utsumi}, {Yamaguchi}, {Yasuda}, {Yoshida}, {Zenko}, {J-GEM}, {Adams}, {Anupama}, {Bally}, {Barway}, {Bellm}, {Blagorodnova}, {Cannella}, {Chandra}, {Chatterjee}, {Clarke}, {Cobb}, {Cook}, {Copperwheat}, {De}, {Emery},
  {Feindt}, {Foster}, {Fox}, {Frail}, {Fremling}, {Frohmaier}, {Garcia}, {Ghosh}, {Giacintucci}, {Goobar}, {Gottlieb}, {Grefenstette}, {Hallinan}, {Harrison}, {Heida}, {Helou}, {Ho}, {Horesh}, {Hotokezaka}, {Ip}, {Itoh}, {Jacobs}, {Jencson}, {Kasen}, {Kasliwal}, {Kassim}, {Kim}, {Kiran}, {Kuin}, {Kulkarni}, {Kupfer}, {Lau}, {Madsen}, {Mazzali}, {Miller}, {Miyasaka}, {Mooley}, {Myers}, {Nakar}, {Ngeow}, {Nugent}, {Ofek}, {Palliyaguru}, {Pavana}, {Perley}, {Peters}, {Pike}, {Piran}, {Qi}, {Quimby}, {Rana}, {Rosswog}, {Rusu}, {Sadler}, {Van Sistine}, {Sollerman}, {Xu}, {Yan}, {Yatsu}, {Yu}, {Zhang}, {Zhao}, {GROWTH}, {JAGWAR}, {Caltech-NRAO}, {TTU-NRAO}, {NuSTAR Collaborations}, {Chambers}, {Huber}, {Schultz}, {Bulger}, {Flewelling}, {Magnier}, {Lowe}, {Wainscoat}, {Waters}, {Willman}, {Pan-STARRS}, {Ebisawa}, {Hanyu}, {Harita}, {Hashimoto}, {Hidaka}, {Hori}, {Ishikawa}, {Isobe}, {Iwakiri}, {Kawai}, {Kawai}, {Kawamuro}, {Kawase}, {Kitaoka}, {Makishima}, {Matsuoka}, {Mihara}, {Morita}, {Morita}, {Nakahira},
  {Nakajima}, {Nakamura}, {Negoro}, {Oda}, {Sakamaki}, {Sasaki}, {Serino}, {Shidatsu}, {Shimomukai}, {Sugawara}, {Sugita}, {Sugizaki}, {Tachibana}, {Takao}, {Tanimoto}, {Tomida}, {Tsuboi}, {Tsunemi}, {Ueda}, {Ueno}, {Yamada}, {Yamaoka}, {Yamauchi}, {Yatabe}, {Yoneyama}, {Yoshii}, {MAXI Team}, {Coward}, {Crisp}, {Macpherson}, {Andreoni}, {Laugier}, {Noysena}, {Klotz}, {Gendre}, {Thierry}, {Turpin}, {Consortium}, {Im}, {Choi}, {Kim}, {Yoon}, {Lim}, {Lee}, {Lee}, {Kim}, {Ko}, {Joe}, {Kwon}, {Kim}, {Lim}, {Choi}, {KU Collaboration}, {Fynbo}, {Malesani}, {Xu}, {Optical Telescope}, {Smartt}, {Jerkstrand}, {Kankare}, {Sim}, {Fraser}, {Inserra}, {Maguire}, {Leloudas}, {Magee}, {Shingles}, {Smith}, {Young}, {Kotak}, {Gal-Yam}, {Lyman}, {Homan}, {Agliozzo}, {Anderson}, {Angus}, {Ashall}, {Barbarino}, {Bauer}, {Berton}, {Botticella}, {Bulla}, {Cannizzaro}, {Cartier}, {Cikota}, {Clark}, {De Cia}, {Della Valle}, {Dennefeld}, {Dessart}, {Dimitriadis}, {Elias-Rosa}, {Firth}, {Fl{\"o}rs}, {Frohmaier}, {Galbany},
  {Gonz{\'a}lez-Gait{\'a}n}, {Gromadzki}, {Guti{\'e}rrez}, {Hamanowicz}, {Harmanen}, {Heintz}, {Hernandez}, {Hodgkin}, {Hook}, {Izzo}, {James}, {Jonker}, {Kerzendorf}, {Kostrzewa-Rutkowska}, {Kromer}, {Kuncarayakti}, {Lawrence}, {Manulis}, {Mattila}, {McBrien}, {M{\"u}ller}, {Nordin}, {O'Neill}, {Onori}, {Palmerio}, {Pastorello}, {Patat}, {Pignata}, {Podsiadlowski}, {Razza}, {Reynolds}, {Roy}, {Ruiter}, {Rybicki}, {Salmon}, {Pumo}, {Prentice}, {Seitenzahl}, {Smith}, {Sollerman}, {Sullivan}, {Szegedi}, {Taddia}, {Taubenberger}, {Terreran}, {Van Soelen}, {Vos}, {Walton}, {Wright}, {Wyrzykowski}, {Yaron}, {pre=''(''>ePESSTO}, {Chen}, {Kr{\"u}hler}, {Schady}, {Wiseman}, {Greiner}, {Rau}, {Schweyer}, {Klose}, {Nicuesa Guelbenzu}, {GROND}, {Palliyaguru}, {Tech University}, {Shara}, {Williams}, {Vaisanen}, {Potter}, {Romero Colmenero}, {Crawford}, {Buckley}, {Mao}, {SALT Group}, {D{\'\i}az}, {Macri}, {Garc{\'\i}a Lambas}, {Mendes de Oliveira}, {Nilo Castell{\'o}n}, {Ribeiro}, {S{\'a}nchez}, {Schoenell}, {Abramo},
  {Akras}, {Alcaniz}, {Artola}, {Beroiz}, {Bonoli}, {Cabral}, {Camuccio}, {Chavushyan}, {Coelho}, {Colazo}, {Costa-Duarte}, {Cuevas Larenas}, {Dom{\'\i}nguez Romero}, {Dultzin}, {Fern{\'a}ndez}, {Garc{\'\i}a}, {Girardini}, {Gon{\c{c}}alves}, {Gon{\c{c}}alves}, {Gurovich}, {Jim{\'e}nez-Teja}, {Kanaan}, {Lares}, {Lopes de Oliveira}, {L{\'o}pez-Cruz}, {Melia}, {Molino}, {Padilla}, {Pe{\~n}uela}, {Placco}, {Qui{\~n}ones}, {Ram{\'\i}rez Rivera}, {Renzi}, {Riguccini}, {R{\'\i}os-L{\'o}pez}, {Rodriguez}, {Sampedro}, {Schneiter}, {Sodr{\'e}}, {Starck}, {Torres-Flores}, {Tornatore}, {Zadro{\.z}ny}, {Castillo}, {TOROS: Transient Robotic Observatory of South Collaboration}, {Castro-Tirado}, {Tello}, {Hu}, {Zhang}, {Cunniffe}, {Castell{\'o}n}, {Hiriart}, {Caballero-Garc{\'\i}a}, {Jel{\'\i}nek}, {Kub{\'a}nek}, {P{\'e}rez del Pulgar}, {Park}, {Jeong}, {Castro Cer{\'o}n}, {Pandey}, {Yock}, {Querel}, {Fan}, {Wang}, {BOOTES Collaboration}, {Beardsley}, {Brown}, {Crosse}, {Emrich}, {Franzen}, {Gaensler}, {Horsley},
  {Johnston-Hollitt}, {Kenney}, {Morales}, {Pallot}, {Sokolowski}, {Steele}, {Tingay}, {Trott}, {Walker}, {Wayth}, {Williams}, {Wu}, {Murchison Widefield Array}, {Yoshida}, {Sakamoto}, {Kawakubo}, {Yamaoka}, {Takahashi}, {Asaoka}, {Ozawa}, {Torii}, {Shimizu}, {Tamura}, {Ishizaki}, {Cherry}, {Ricciarini}, {Penacchioni}, {Marrocchesi}, {CALET Collaboration}, {Pozanenko}, {Volnova}, {Mazaeva}, {Minaev}, {Krugov}, {Kusakin}, {Reva}, {Moskvitin}, {Rumyantsev}, {Inasaridze}, {Klunko}, {Tungalag}, {Schmalz}, {Burhonov}, {IKI-GW Follow-up Collaboration}, {Abdalla}, {Abramowski}, {Aharonian}, {Ait Benkhali}, {Ang{\"u}ner}, {Arakawa}, {Arrieta}, {Aubert}, {Backes}, {Balzer}, {Barnard}, {Becherini}, {Becker Tjus}, {Berge}, {Bernhard}, {Bernl{\"o}hr}, {Blackwell}, {B{\"o}ttcher}, {Boisson}, {Bolmont}, {Bonnefoy}, {Bordas}, {Bregeon}, {Brun}, {Brun}, {Bryan}, {B{\"u}chele}, {Bulik}, {Capasso}, {Caroff}, {Carosi}, {Casanova}, {Cerruti}, {Chakraborty}, {Chaves}, {Chen}, {Chevalier}, {Colafrancesco}, {Condon}, {Conrad},
  {Davids}, {Decock}, {Deil}, {Devin}, {deWilt}, {Dirson}, {Djannati-Ata{\"\i}}, {Donath}, {O'C. Drury}, {Dutson}, {Dyks}, {Edwards}, {Egberts}, {Emery}, {Ernenwein}, {Eschbach}, {Farnier}, {Fegan}, {Fernandes}, {Fiasson}, {Fontaine}, {Funk}, {F{\"u}ssling}, {Gabici}, {Gallant}, {Garrigoux}, {Gat{\'e}}, {Giavitto}, {Giebels}, {Glawion}, {Glicenstein}, {Gottschall}, {Grondin}, {Hahn}, {Haupt}, {Hawkes}, {Heinzelmann}, {Henri}, {Hermann}, {Hinton}, {Hofmann}, {Hoischen}, {Holch}, {Holler}, {Horns}, {Ivascenko}, {Iwasaki}, {Jacholkowska}, {Jamrozy}, {Jankowsky}, {Jankowsky}, {Jingo}, {Jouvin}, {Jung-Richardt}, {Kastendieck}, {Katarzy{\'n}ski}, {Katsuragawa}, {Kerszberg}, {Khangulyan}, {Kh{\'e}lifi}, {King}, {Klepser}, {Klochkov}, {Klu{\'z}niak}, {Komin}, {Kosack}, {Krakau}, {Kraus}, {Kr{\"u}ger}, {Laffon}, {Lamanna}, {Lau}, {Lees}, {Lefaucheur}, {Lemi{\`e}re}, {Lemoine-Goumard}, {Lenain}, {Leser}, {Lohse}, {Lorentz}, {Liu}, {Lypova}, {Malyshev}, {Marandon}, {Marcowith}, {Mariaud}, {Marx}, {Maurin}, {Maxted},
  {Mayer}, {Meintjes}, {Meyer}, {Mitchell}, {Moderski}, {Mohamed}, {Mohrmann}, {Mor{\r{a}}}, {Moulin}, {Murach}, {Nakashima}, {de Naurois}, {Ndiyavala}, {Niederwanger}, {Niemiec}, {Oakes}, {O'Brien}, {Odaka}, {Ohm}, {Ostrowski}, {Oya}, {Padovani}, {Panter}, {Parsons}, {Pekeur}, {Pelletier}, {Perennes}, {Petrucci}, {Peyaud}, {Piel}, {Pita}, {Poireau}, {Poon}, {Prokhorov}, {Prokoph}, {P{\"u}hlhofer}, {Punch}, {Quirrenbach}, {Raab}, {Rauth}, {Reimer}, {Reimer}, {Renaud}, {de los Reyes}, {Rieger}, {Rinchiuso}, {Romoli}, {Rowell}, {Rudak}, {Rulten}, {Sahakian}, {Saito}, {Sanchez}, {Santangelo}, {Sasaki}, {Schlickeiser}, {Sch{\"u}ssler}, {Schulz}, {Schwanke}, {Schwemmer}, {Seglar-Arroyo}, {Settimo}, {Seyffert}, {Shafi}, {Shilon}, {Shiningayamwe}, {Simoni}, {Sol}, {Spanier}, {Spir-Jacob}, {Stawarz}, {Steenkamp}, {Stegmann}, {Steppa}, {Sushch}, {Takahashi}, {Tavernet}, {Tavernier}, {Taylor}, {Terrier}, {Tibaldo}, {Tiziani}, {Tluczykont}, {Trichard}, {Tsirou}, {Tsuji}, {Tuffs}, {Uchiyama}, {van der Walt}, {van Eldik},
  {van Rensburg}, {van Soelen}, {Vasileiadis}, {Veh}, {Venter}, {Viana}, {Vincent}, {Vink}, {Voisin}, {V{\"o}lk}, {Vuillaume}, {Wadiasingh}, {Wagner}, {Wagner}, {Wagner}, {White}, {Wierzcholska}, {Willmann}, {W{\"o}rnlein}, {Wouters}, {Yang}, {Zaborov}, {Zacharias}, {Zanin}, {Zdziarski}, {Zech}, {Zefi}, {Ziegler}, {Zorn}, {{\.Z}ywucka}, {H.~E.~S.~S. Collaboration}, {Fender}, {Broderick}, {Rowlinson}, {Wijers}, {Stewart}, {ter Veen}, {Shulevski}, {LOFAR Collaboration}, {Kavic}, {Simonetti}, {League}, {Tsai}, {Obenberger}, {Nathaniel}, {Taylor}, {Dowell}, {Liebling}, {Estes}, {Lippert}, {Sharma}, {Vincent}, {Farella}, {Wavelength Array}, {Abeysekara}, {Albert}, {Alfaro}, {Alvarez}, {Arceo}, {Arteaga-Vel{\'a}zquez}, {Avila Rojas}, {Ayala Solares}, {Barber}, {Becerra Gonzalez}, {Becerril}, {Belmont-Moreno}, {BenZvi}, {Berley}, {Bernal}, {Braun}, {Brisbois}, {Caballero-Mora}, {Capistr{\'a}n}, {Carrami{\~n}ana}, {Casanova}, {Castillo}, {Cotti}, {Cotzomi}, {Couti{\~n}o de Le{\'o}n}, {De Le{\'o}n}, {De la Fuente},
  {Diaz Hernandez}, {Dichiara}, {Dingus}, {DuVernois}, {D{\'\i}az-V{\'e}lez}, {Ellsworth}, {Engel}, {Enr{\'\i}quez-Rivera}, {Fiorino}, {Fleischhack}, {Fraija}, {Garc{\'\i}a-Gonz{\'a}lez}, {Garfias}, {Gerhardt}, {Gonz{\~o}lez Mu{\~n}oz}, {Gonz{\'a}lez}, {Goodman}, {Hampel-Arias}, {Harding}, {Hernandez}, {Hernandez-Almada}, {Hona}, {H{\"u}ntemeyer}, {Iriarte}, {Jardin-Blicq}, {Joshi}, {Kaufmann}, {Kieda}, {Lara}, {Lauer}, {Lennarz}, {Le{\'o}n Vargas}, {Linnemann}, {Longinotti}, {Raya}, {Luna-Garc{\'\i}a}, {L{\'o}pez-Coto}, {Malone}, {Marinelli}, {Martinez}, {Martinez-Castellanos}, {Mart{\'\i}nez-Castro}, {Mart{\'\i}nez-Huerta}, {Matthews}, {Miranda-Romagnoli}, {Moreno}, {Mostaf{\'a}}, {Nellen}, {Newbold}, {Nisa}, {Noriega-Papaqui}, {Pelayo}, {Pretz}, {P{\'e}rez-P{\'e}rez}, {Ren}, {Rho}, {Rivi{\`e}re}, {Rosa-Gonz{\'a}lez}, {Rosenberg}, {Ruiz-Velasco}, {Salazar}, {Salesa Greus}, {Sandoval}, {Schneider}, {Schoorlemmer}, {Sinnis}, {Smith}, {Springer}, {Surajbali}, {Tibolla}, {Tollefson}, {Torres}, {Ukwatta},
  {Weisgarber}, {Westerhoff}, {Wisher}, {Wood}, {Yapici}, {Yodh}, {Younk}, {Zhou}, {{\'A}lvarez}, {HAWC Collaboration}, {Aab}, {Abreu}, {Aglietta}, {Albuquerque}, {Albury}, {Allekotte}, {Almela}, {Alvarez Castillo}, {Alvarez-Mu{\~n}iz}, {Anastasi}, {Anchordoqui}, {Andrada}, {Andringa}, {Aramo}, {Arsene}, {Asorey}, {Assis}, {Avila}, {Badescu}, {Balaceanu}, {Barbato}, {Barreira Luz}, {Becker}, {Bellido}, {Berat}, {Bertaina}, {Bertou}, {Biermann}, {Biteau}, {Blaess}, {Blanco}, {Blazek}, {Bleve}, {Boh{\'a}{\v{c}}ov{\'a}}, {Bonifazi}, {Borodai}, {Botti}, {Brack}, {Brancus}, {Bretz}, {Bridgeman}, {Briechle}, {Buchholz}, {Bueno}, {Buitink}, {Buscemi}, {Caballero-Mora}, {Caccianiga}, {Cancio}, {Canfora}, {Caruso}, {Castellina}, {Catalani}, {Cataldi}, {Cazon}, {Chavez}, {Chinellato}, {Chudoba}, {Clay}, {Cobos Cerutti}, {Colalillo}, {Coleman}, {Collica}, {Coluccia}, {Concei{\c{c}}{\~a}o}, {Consolati}, {Contreras}, {Cooper}, {Coutu}, {Covault}, {Cronin}, {D'Amico}, {Daniel}, {Dasso}, {Daumiller}, {Dawson}, {Day}, {de
  Almeida}, {de Jong}, {De Mauro}, {de Mello Neto}, {De Mitri}, {de Oliveira}, {de Souza}, {Debatin}, {Deligny}, {D{\'\i}az Castro}, {Diogo}, {Dobrigkeit}, {D'Olivo}, {Dorosti}, {Dos Anjos}, {Dova}, {Dundovic}, {Ebr}, {Engel}, {Erdmann}, {Erfani}, {Escobar}, {Espadanal}, {Etchegoyen}, {Falcke}, {Farmer}, {Farrar}, {Fauth}, {Fazzini}, {Feldbusch}, {Fenu}, {Fick}, {Figueira}, {Filip{\v{c}}i{\v{c}}}, {Freire}, {Fujii}, {Fuster}, {Ga{\"\i}or}, {Garc{\'\i}a}, {Gat{\'e}}, {Gemmeke}, {Gherghel-Lascu}, {Ghia}, {Giaccari}, {Giammarchi}, {Giller}, {G{\l}as}, {Glaser}, {Golup}, {G{\'o}mez Berisso}, {G{\'o}mez Vitale}, {Gonz{\'a}lez}, {Gorgi}, {Gottowik}, {Grillo}, {Grubb}, {Guarino}, {Guedes}, {Halliday}, {Hampel}, {Hansen}, {Harari}, {Harrison}, {Harvey}, {Haungs}, {Hebbeker}, {Heck}, {Heimann}, {Herve}, {Hill}, {Hojvat}, {Holt}, {Homola}, {H{\"o}randel}, {Horvath}, {Hrabovsk{\'y}}, {Huege}, {Hulsman}, {Insolia}, {Isar}, {Jandt}, {Johnsen}, {Josebachuili}, {Jurysek}, {K{\"a}{\"a}p{\"a}}, {Kampert}, {Keilhauer},
  {Kemmerich}, {Kemp}, {Kieckhafer}, {Klages}, {Kleifges}, {Kleinfeller}, {Krause}, {Krohm}, {Kuempel}, {Kukec Mezek}, {Kunka}, {Kuotb Awad}, {Lago}, {LaHurd}, {Lang}, {Lauscher}, {Legumina}, {Leigui de Oliveira}, {Letessier-Selvon}, {Lhenry-Yvon}, {Link}, {Lo Presti}, {Lopes}, {L{\'o}pez}, {L{\'o}pez Casado}, {Lorek}, {Luce}, {Lucero}, {Malacari}, {Mallamaci}, {Mandat}, {Mantsch}, {Mariazzi}, {Maris}, {Marsella}, {Martello}, {Martinez}, {Mart{\'\i}nez Bravo}, {Mas{\'\i}as Meza}, {Mathes}, {Mathys}, {Matthews}, {Matthiae}, {Mayotte}, {Mazur}, {Medina}, {Medina-Tanco}, {Melo}, {Menshikov}, {Merenda}, {Michal}, {Micheletti}, {Middendorf}, {Miramonti}, {Mitrica}, {Mockler}, {Mollerach}, {Montanet}, {Morello}, {Morlino}, {M{\"u}ller}, {M{\"u}ller}, {Muller}, {M{\"u}ller}, {Mussa}, {Naranjo}, {Nguyen}, {Niculescu-Oglinzanu}, {Niechciol}, {Niemietz}, {Niggemann}, {Nitz}, {Nosek}, {Novotny}, {No{\v{z}}ka}, {N{\'u}{\~n}ez}, {Oikonomou}, {Olinto}, {Palatka}, {Pallotta}, {Papenbreer}, {Parente}, {Parra}, {Paul},
  {Pech}, {Pedreira}, {P{\c{e}}kala}, {Pe{\~n}a-Rodriguez}, {Pereira}, {Perlin}, {Perrone}, {Peters}, {Petrera}, {Phuntsok}, {Pierog}, {Pimenta}, {Pirronello}, {Platino}, {Plum}, {Poh}, {Porowski}, {Prado}, {Privitera}, {Prouza}, {Quel}, {Querchfeld}, {Quinn}, {Ramos-Pollan}, {Rautenberg}, {Ravignani}, {Ridky}, {Riehn}, {Risse}, {Ristori}, {Rizi}, {Rodrigues de Carvalho}, {Rodriguez Fernandez}, {Rodriguez Rojo}, {Roncoroni}, {Roth}, {Roulet}, {Rovero}, {Ruehl}, {Saffi}, {Saftoiu}, {Salamida}, {Salazar}, {Saleh}, {Salina}, {S{\'a}nchez}, {Sanchez-Lucas}, {Santos}, {Santos}, {Sarazin}, {Sarmento}, {Sarmiento-Cano}, {Sato}, {Schauer}, {Scherini}, {Schieler}, {Schimp}, {Schmidt}, {Scholten}, {Schov{\'a}nek}, {Schr{\"o}der}, {Schr{\"o}der}, {Schulz}, {Schumacher}, {Sciutto}, {Segreto}, {Shadkam}, {Shellard}, {Sigl}, {Silli}, {{\v{S}}m{\'\i}da}, {Snow}, {Sommers}, {Sonntag}, {Soriano}, {Squartini}, {Stanca}, {Stani{\v{c}}}, {Stasielak}, {Stassi}, {Stolpovskiy}, {Strafella}, {Streich}, {Suarez}, {Suarez-Dur{\'a}n},
  {Sudholz}, {Suomij{\"a}rvi}, {Supanitsky}, {{\v{S}}up{\'\i}k}, {Swain}, {Szadkowski}, {Taboada}, {Taborda}, {Timmermans}, {Todero Peixoto}, {Tomankova}, {Tom{\'e}}, {Torralba Elipe}, {Travnicek}, {Trini}, {Tueros}, {Ulrich}, {Unger}, {Urban}, {Vald{\'e}s Galicia}, {Vali{\~n}o}, {Valore}, {van Aar}, {van Bodegom}, {van den Berg}, {van Vliet}, {Varela}, {Vargas C{\'a}rdenas}, {V{\'a}zquez}, {Veberi{\v{c}}}, {Ventura}, {Vergara Quispe}, {Verzi}, {Vicha}, {Villase{\~n}or}, {Vorobiov}, {Wahlberg}, {Wainberg}, {Walz}, {Watson}, {Weber}, {Weindl}, {Wiede{\'n}ski}, {Wiencke}, {Wilczy{\'n}ski}, {Wirtz}, {Wittkowski}, {Wundheiler}, {Yang}, {Yushkov}, {Zas}, {Zavrtanik}, {Zavrtanik}, {Zepeda}, {Zimmermann}, {Ziolkowski}, {Zong}, {Zuccarello}, {Pierre Auger Collaboration}, {Kim}, {Schulze}, {Bauer}, {Corral-Santana}, {de Gregorio-Monsalvo}, {Gonz{\'a}lez-L{\'o}pez}, {Hartmann}, {Ishwara-Chandra}, {Mart{\'\i}n}, {Mehner}, {Misra}, {Micha{\l}owski}, {Resmi}, {ALMA Collaboration}, {Paragi}, {Agudo}, {An}, {Beswick},
  {Casadio}, {Frey}, {Jonker}, {Kettenis}, {Marcote}, {Moldon}, {Szomoru}, {van Langevelde}, {Yang}, {Euro VLBI Team}, {Cwiek}, {Cwiok}, {Czyrkowski}, {Dabrowski}, {Kasprowicz}, {Mankiewicz}, {Nawrocki}, {Opiela}, {Piotrowski}, {Wrochna}, {Zaremba}, {{\.Z}arnecki}, {Pi of Sky Collaboration}, {Haggard}, {Nynka}, {Ruan}, {Chandra Team at McGill University}, {Bland}, {Booler}, {Devillepoix}, {de Gois}, {Hancock}, {Howie}, {Paxman}, {Sansom}, {Towner}, {Desert Fireball Network}, {Tonry}, {Coughlin}, {Stubbs}, {Denneau}, {Heinze}, {Stalder}, {Weiland}, {ATLAS}, {Eatough}, {Kramer}, {Kraus}, {Time Resolution Universe Survey}, {Troja}, {Piro}, {Becerra Gonz{\'a}lez}, {Butler}, {Fox}, {Khandrika}, {Kutyrev}, {Lee}, {Ricci}, {Ryan}, {S{\'a}nchez-Ram{\'\i}rez}, {Veilleux}, {Watson}, {Wieringa}, {Burgess}, {van Eerten}, {Fontes}, {Fryer}, {Korobkin}, {Wollaeger}, {RIMAS}, {RATIR}, {Camilo}, {Foley}, {Goedhart}, {Makhathini}, {Oozeer}, {Smirnov}, {Fender}, {Woudt}, \& {South Africa/MeerKAT}}]{Abbott2017_MM}
{Abbott}, B.~P., {Abbott}, R., {Abbott}, T.~D., {et~al.} 2017{\natexlab{c}}, \apjl, 848, L12

\bibitem[{{Abbott} {et~al.}(2017{\natexlab{d}}){Abbott}, {Abbott}, {Abbott}, {Acernese}, {et~al.}}]{Abbott2017_GRB170817A}
{Abbott}, B.~P., {Abbott}, R., {Abbott}, T.~D., {Acernese}, F., {et~al.} 2017{\natexlab{d}}, \apjl, 848, L13

\bibitem[{{Abbott} {et~al.}(2019){Abbott}, {LIGO Scientific Collaboration}, \& {Virgo Collaboration}}]{Abbott2019_GW170817_properties}
{Abbott}, B.~P., {LIGO Scientific Collaboration}, \& {Virgo Collaboration}. 2019, Physical Review X, 9, 011001

\bibitem[{{Abbott} {et~al.}(2020{\natexlab{b}}){Abbott}, {LIGO Scientific Collaboration}, \& {Virgo Collaboration}}]{Abbott2019_GW190425}
{Abbott}, B.~P., {LIGO Scientific Collaboration}, \& {Virgo Collaboration}. 2020{\natexlab{b}}, arXiv e-prints, arXiv:2001.01761

\bibitem[{{Abbott} {et~al.}(2021{\natexlab{a}}){Abbott}, {Abbott}, {Abraham}, {Acernese}, {Ackley}, {Adams}, {Adams}, {Adhikari}, {Adya}, {Affeldt}, {Agarwal}, {Agathos}, {Agatsuma}, {Aggarwal}, {Aguiar}, {Aiello}, {Ain}, {Ajith}, {Akutsu}, {Aleman}, {Allen}, {Allocca}, {Altin}, {Amato}, {Anand}, {Ananyeva}, {Anderson}, {Anderson}, {Ando}, {Angelova}, {Ansoldi}, {Antelis}, {Antier}, {Appert}, {Arai}, {Arai}, {Arai}, {Araki}, {Araya}, {Araya}, {Areeda}, {Ar{\`e}ne}, {Aritomi}, {Arnaud}, {Aronson}, {Arun}, {Asada}, {Asali}, {Ashton}, {Aso}, {Aston}, {Astone}, {Aubin}, {Aufmuth}, {Aultoneal}, {Austin}, {Babak}, {Badaracco}, {Bader}, {Bae}, {Bae}, {Baer}, {Bagnasco}, {Bai}, {Baiotti}, {Baird}, {Bajpai}, {Ball}, {Ballardin}, {Ballmer}, {Bals}, {Balsamo}, {Baltus}, {Banagiri}, {Bankar}, {Bankar}, {Barayoga}, {Barbieri}, {Barish}, {Barker}, {Barneo}, {Barone}, {Barr}, {Barsotti}, {Barsuglia}, {Barta}, {Bartlett}, {Barton}, {Bartos}, {Bassiri}, {Basti}, {Bawaj}, {Bayley}, {Baylor}, {Bazzan}, {B{\'e}csy},
  {Bedakihale}, {Bejger}, {Belahcene}, {Benedetto}, {Beniwal}, {Benjamin}, {Benkel}, {Bennett}, {Bentley}, {Benyaala}, {Bergamin}, {Berger}, {Bernuzzi}, {Berry}, {Bersanetti}, {Bertolini}, {Betzwieser}, {Bhandare}, {Bhandari}, {Bhattacharjee}, {Bhaumik}, {Bidler}, {Bilenko}, {Billingsley}, {Birney}, {Birnholtz}, {Biscans}, {Bischi}, {Biscoveanu}, {Bisht}, {Biswas}, {Bitossi}, {Bizouard}, {Blackburn}, {Blackman}, {Blair}, {Blair}, {Blair}, {Bobba}, {Bode}, {Boer}, {Bogaert}, {Boldrini}, {Bondu}, {Bonilla}, {Bonnand}, {Booker}, {Boom}, {Bork}, {Boschi}, {Bose}, {Bose}, {Bossilkov}, {Boudart}, {Bouffanais}, {Bozzi}, {Bradaschia}, {Brady}, {Bramley}, {Branch}, {Branchesi}, {Brau}, {Breschi}, {Briant}, {Briggs}, {Brillet}, {Brinkmann}, {Brockill}, {Brooks}, {Brooks}, {Brown}, {Brunett}, {Bruno}, {Bruntz}, {Bryant}, {Buikema}, {Bulik}, {Bulten}, {Buonanno}, {Buscicchio}, {Buskulic}, {Byer}, {Cadonati}, {Caesar}, {Cagnoli}, {Cahillane}, {Cain}, {Calder{\'o}n Bustillo}, {Callaghan}, {Callister}, {Calloni}, {Camp},
  {Canepa}, {Cannavacciuolo}, {Cannon}, {Cao}, {Cao}, {Cao}, {Capocasa}, {Capote}, {Carapella}, {Carbognani}, {Carlin}, {Carney}, {Carpinelli}, {Carullo}, {Carver}, {Casanueva Diaz}, {Casentini}, {Castaldi}, {Caudill}, {Cavagli{\`a}}, {Cavalier}, {Cavalieri}, {Cella}, {Cerd{\'a}-Dur{\'a}n}, {Cesarini}, {Chaibi}, {Chakravarti}, {Champion}, {Chan}, {Chan}, {Chan}, {Chan}, {Chandra}, {Chanial}, {Chao}, {Charlton}, {Chase}, {Chassande-Mottin}, {Chatterjee}, {Chaturvedi}, {Chatziioannou}, {Chen}, {Chen}, {Chen}, {Chen}, {Chen}, {Chen}, {Chen}, {Chen}, {Chen}, {Cheng}, {Cheong}, {Cheung}, {Chia}, {Chiadini}, {Chiang}, {Chierici}, {Chincarini}, {Chiofalo}, {Chiummo}, {Cho}, {Cho}, {Choate}, {Choudhary}, {Choudhary}, {Christensen}, {Chu}, {Chu}, {Chu}, {Chua}, {Chung}, {Ciani}, {Ciecielag}, {Cie{\'s}lar}, {Cifaldi}, {Ciobanu}, {Ciolfi}, {Cipriano}, {Cirone}, {Clara}, {Clark}, {Clark}, {Clarke}, {Clearwater}, {Clesse}, {Cleva}, {Coccia}, {Cohadon}, {Cohen}, {Cohen}, {Colleoni}, {Collette}, {Colpi}, {Compton},
  {Constancio}, {Conti}, {Cooper}, {Corban}, {Corbitt}, {Cordero-Carri{\'o}n}, {Corezzi}, {Corley}, {Cornish}, {Corre}, {Corsi}, {Cortese}, {Costa}, {Cotesta}, {Coughlin}, {Coughlin}, {Coulon}, {Countryman}, {Cousins}, {Couvares}, {Covas}, {Coward}, {Cowart}, {Coyne}, {Coyne}, {Creighton}, {Creighton}, {Criswell}, {Croquette}, {Crowder}, {Cudell}, {Cullen}, {Cumming}, {Cummings}, {Cuoco}, {Cury{\l}o}, {Dal Canton}, {D{\'a}lya}, {Dana}, {Daneshgaranbajastani}, {D'Angelo}, {Danilishin}, {D'Antonio}, {Danzmann}, {Darsow-Fromm}, {Dasgupta}, {Datrier}, {Dattilo}, {Dave}, {Davier}, {Davies}, {Davis}, {Daw}, {Dean}, {Debra}, {Deenadayalan}, {Degallaix}, {de Laurentis}, {Del{\'e}glise}, {Del Favero}, {de Lillo}, {de Lillo}, {Del Pozzo}, {Demarchi}, {de Matteis}, {D'Emilio}, {Demos}, {Dent}, {Depasse}, {de Pietri}, {De Rosa}, {de Rossi}, {Desalvo}, {de Simone}, {Dhurandhar}, {D{\'\i}az}, {Diaz-Ortiz}, {Didio}, {Dietrich}, {di Fiore}, {di Fronzo}, {di Giorgio}, {di Giovanni}, {di Girolamo}, {di Lieto}, {Ding}, {di
  Pace}, {di Palma}, {di Renzo}, {Divakarla}, {Dmitriev}, {Doctor}, {D'Onofrio}, {Donovan}, {Dooley}, {Doravari}, {Dorrington}, {Drago}, {Driggers}, {Drori}, {Du}, {Ducoin}, {Dupej}, {Durante}, {D'Urso}, {Duverne}, {Dwyer}, {Easter}, {Ebersold}, {Eddolls}, {Edelman}, {Edo}, {Edy}, {Effler}, {Eguchi}, {Eichholz}, {Eikenberry}, {Eisenmann}, {Eisenstein}, {Ejlli}, {Enomoto}, {Errico}, {Essick}, {Estell{\'e}s}, {Estevez}, {Etienne}, {Etzel}, {Evans}, {Evans}, {Ewing}, {Fafone}, {Fair}, {Fairhurst}, {Fan}, {Farah}, {Farinon}, {Farr}, {Farr}, {Farrow}, {Fauchon-Jones}, {Favata}, {Fays}, {Fazio}, {Feicht}, {Fejer}, {Feng}, {Fenyvesi}, {Ferguson}, {Fernandez-Galiana}, {Ferrante}, {Ferreira}, {Fidecaro}, {Figura}, {Fiori}, {Fishbach}, {Fisher}, {Fittipaldi}, {Fiumara}, {Flaminio}, {Floden}, {Flynn}, {Fong}, {Font}, {Fornal}, {Forsyth}, {Franke}, {Frasca}, {Frasconi}, {Frederick}, {Frei}, {Freise}, {Frey}, {Fritschel}, {Frolov}, {Fronz{\'e}}, {Fujii}, {Fujikawa}, {Fukunaga}, {Fukushima}, {Fulda}, {Fyffe}, {Gabbard},
  {Gadre}, {Gaebel}, {Gair}, {Gais}, {Galaudage}, {Gamba}, {Ganapathy}, {Ganguly}, {Gao}, {Gaonkar}, {Garaventa}, {Garc{\'\i}a-N{\'u}{\~n}ez}, {Garc{\'\i}a-Quir{\'o}s}, {Garufi}, {Gateley}, {Gaudio}, {Gayathri}, {Ge}, {Gemme}, {Gennai}, {George}, {Gergely}, {Gewecke}, {Ghonge}, {Ghosh}, {Ghosh}, {Ghosh}, {Ghosh}, {Ghosh}, {Giacomazzo}, {Giacoppo}, {Giaime}, {Giardina}, {Gibson}, {Gier}, {Giesler}, {Giri}, {Gissi}, {Glanzer}, {Gleckl}, {Godwin}, {Goetz}, {Goetz}, {Gohlke}, {Goncharov}, {Gonz{\'a}lez}, {Gopakumar}, {Gosselin}, {Gouaty}, {Grace}, {Grado}, {Granata}, {Granata}, {Grant}, {Gras}, {Grassia}, {Gray}, {Gray}, {Greco}, {Green}, {Green}, {Gretarsson}, {Gretarsson}, {Griffith}, {Griffiths}, {Griggs}, {Grignani}, {Grimaldi}, {Grimes}, {Grimm}, {Grote}, {Grunewald}, {Gruning}, {Guerrero}, {Guidi}, {Guimaraes}, {Guix{\'e}}, {Gulati}, {Guo}, {Guo}, {Gupta}, {Gupta}, {Gupta}, {Gustafson}, {Gustafson}, {Guzman}, {Ha}, {Haegel}, {Hagiwara}, {Haino}, {Halim}, {Hall}, {Hamilton}, {Hammond}, {Han}, {Haney},
  {Hanks}, {Hanna}, {Hannam}, {Hannuksela}, {Hansen}, {Hansen}, {Hanson}, {Harder}, {Hardwick}, {Haris}, {Harms}, {Harry}, {Harry}, {Hartwig}, {Hasegawa}, {Haskell}, {Hasskew}, {Haster}, {Hattori}, {Haughian}, {Hayakawa}, {Hayama}, {Hayes}, {Healy}, {Heidmann}, {Heintze}, {Heinze}, {Heinzel}, {Heitmann}, {Hellman}, {Hello}, {Helmling-Cornell}, {Hemming}, {Hendry}, {Heng}, {Hennes}, {Hennig}, {Hennig}, {Hernandez Vivanco}, {Heurs}, {Hild}, {Hill}, {Himemoto}, {Hinderer}, {Hines}, {Hiranuma}, {Hirata}, {Hirose}, {Ho}, {Hochheim}, {Hofman}, {Hohmann}, {Holgado}, {Holland}, {Hollows}, {Holmes}, {Holt}, {Holz}, {Hong}, {Hopkins}, {Hough}, {Howell}, {Hoy}, {Hoyland}, {Hreibi}, {Hsieh}, {Hsu}, {Huang}, {Huang}, {Huang}, {Huang}, {Huang}, {Huang}, {H{\"u}bner}, {Huddart}, {Huerta}, {Hughey}, {Hui}, {Hui}, {Husa}, {Huttner}, {Huxford}, {Huynh-Dinh}, {Ide}, {Idzkowski}, {Iess}, {Ikenoue}, {Imam}, {Inayoshi}, {Inchauspe}, {Ingram}, {Inoue}, {Intini}, {Ioka}, {Isi}, {Isleif}, {Ito}, {Itoh}, {Iyer}, {Izumi},
  {Jaberianhamedan}, {Jacqmin}, {Jadhav}, {Jadhav}, {James}, {Jan}, {Jani}, {Janssens}, {Janthalur}, {Jaranowski}, {Jariwala}, {Jaume}, {Jenkins}, {Jeon}, {Jeunon}, {Jia}, {Jiang}, {Jin}, {Johns}, {Jones}, {Jones}, {Jones}, {Jones}, {Jones}, {Jonker}, {Ju}, {Jung}, {Jung}, {Junker}, {Kaihotsu}, {Kajita}, {Kakizaki}, {Kalaghatgi}, {Kalogera}, {Kamai}, {Kamiizumi}, {Kanda}, {Kandhasamy}, {Kang}, {Kanner}, {Kao}, {Kapadia}, {Kapasi}, {Karat}, {Karathanasis}, {Karki}, {Kashyap}, {Kasprzack}, {Kastaun}, {Katsanevas}, {Katsavounidis}, {Katzman}, {Kaur}, {Kawabe}, {Kawaguchi}, {Kawai}, {Kawasaki}, {K{\'e}f{\'e}lian}, {Keitel}, {Key}, {Khadka}, {Khalili}, {Khan}, {Khan}, {Khazanov}, {Khetan}, {Khursheed}, {Kijbunchoo}, {Kim}, {Kim}, {Kim}, {Kim}, {Kim}, {Kim}, {Kimball}, {Kimura}, {King}, {Kinley-Hanlon}, {Kirchhoff}, {Kissel}, {Kita}, {Kitazawa}, {Kleybolte}, {Klimenko}, {Knee}, {Knowles}, {Knyazev}, {Koch}, {Koekoek}, {Kojima}, {Kokeyama}, {Koley}, {Kolitsidou}, {Kolstein}, {Komori}, {Kondrashov}, {Kong}, {Kontos},
  {Koper}, {Korobko}, {Kotake}, {Kovalam}, {Kozak}, {Kozakai}, {Kozu}, {Kringel}, {Krishnendu}, {Kr{\'o}lak}, {Kuehn}, {Kuei}, {Kumar}, {Kumar}, {Kumar}, {Kumar}, {Kume}, {Kuns}, {Kuo}, {Kuo}, {Kuromiya}, {Kuroyanagi}, {Kusayanagi}, {Kwak}, {Kwang}, {Laghi}, {Lalande}, {Lam}, {Lamberts}, {Landry}, {Landry}, {Lane}, {Lang}, {Lange}, {Lantz}, {La Rosa}, {Lartaux-Vollard}, {Lasky}, {Laxen}, {Lazzarini}, {Lazzaro}, {Leaci}, {Leavey}, {Lecoeuche}, {Lee}, {Lee}, {Lee}, {Lee}, {Lee}, {Lee}, {Lehmann}, {Lema{\^\i}tre}, {Leon}, {Leonardi}, {Leroy}, {Letendre}, {Levin}, {Leviton}, {Li}, {Li}, {Li}, {Li}, {Li}, {Li}, {Lin}, {Lin}, {Lin}, {Lin}, {Lin}, {Linde}, {Linker}, {Linley}, {Littenberg}, {Liu}, {Liu}, {Liu}, {Liu}, {Llorens-Monteagudo}, {Lo}, {Lockwood}, {Lollie}, {London}, {Longo}, {Lopez}, {Lorenzini}, {Loriette}, {Lormand}, {Losurdo}, {Lough}, {Lousto}, {Lovelace}, {L{\"u}ck}, {Lumaca}, {Lundgren}, {Luo}, {Macas}, {Macinnis}, {MacLeod}, {MacMillan}, {Macquet}, {Maga{\~n}a Hernandez}, {Maga{\~n}a-Sandoval},
  {Magazz{\`u}}, {Magee}, {Maggiore}, {Majorana}, {Makarem}, {Maksimovic}, {Maliakal}, {Malik}, {Man}, {Mandic}, {Mangano}, {Mango}, {Mansell}, {Manske}, {Mantovani}, {Mapelli}, {Marchesoni}, {Marchio}, {Marion}, {Mark}, {M{\'a}rka}, {M{\'a}rka}, {Markakis}, {Markosyan}, {Markowitz}, {Maros}, {Marquina}, {Marsat}, {Martelli}, {Martin}, {Martin}, {Martinez}, {Martinez}, {Martinovic}, {Martynov}, {Marx}, {Masalehdan}, {Mason}, {Massera}, {Masserot}, {Massinger}, {Masso-Reid}, {Mastrogiovanni}, {Matas}, {Mateu-Lucena}, {Matichard}, {Matiushechkina}, {Mavalvala}, {McCann}, {McCarthy}, {McClelland}, {McClincy}, {McCormick}, {McCuller}, {McGhee}, {McGuire}, {McIsaac}, {McIver}, {McManus}, {McRae}, {McWilliams}, {Meacher}, {Mehmet}, {Mehta}, {Melatos}, {Melchor}, {Mendell}, {Menendez-Vazquez}, {Menoni}, {Mercer}, {Mereni}, {Merfeld}, {Merilh}, {Merritt}, {Merzougui}, {Meshkov}, {Messenger}, {Messick}, {Meyers}, {Meylahn}, {Mhaske}, {Miani}, {Miao}, {Michaloliakos}, {Michel}, {Michimura}, {Middleton}, {Milano},
  {Miller}, {Millhouse}, {Mills}, {Milotti}, {Milovich-Goff}, {Minazzoli}, {Minenkov}, {Mio}, {Mir}, {Mishkin}, {Mishra}, {Mishra}, {Mistry}, {Mitra}, {Mitrofanov}, {Mitselmakher}, {Mittleman}, {Miyakawa}, {Miyamoto}, {Miyazaki}, {Miyo}, {Miyoki}, {Mo}, {Mogushi}, {Mohapatra}, {Mohite}, {Molina}, {Molina-Ruiz}, {Mondin}, {Montani}, {Moore}, {Moraru}, {Morawski}, {More}, {Moreno}, {Moreno}, {Mori}, {Morisaki}, {Moriwaki}, {Mours}, {Mow-Lowry}, {Mozzon}, {Muciaccia}, {Mukherjee}, {Mukherjee}, {Mukherjee}, {Mukherjee}, {Mukund}, {Mullavey}, {Munch}, {Mu{\~n}iz}, {Murray}, {Musenich}, {Nadji}, {Nagano}, {Nagano}, {Nagar}, {Nakamura}, {Nakano}, {Nakano}, {Nakashima}, {Nakayama}, {Nardecchia}, {Narikawa}, {Naticchioni}, {Nayak}, {Nayak}, {Negishi}, {Neil}, {Neilson}, {Nelemans}, {Nelson}, {Nery}, {Neunzert}, {Ng}, {Ng}, {Nguyen}, {Nguyen}, {Nguyen}, {Nguyen Quynh}, {Ni}, {Nichols}, {Nishizawa}, {Nissanke}, {Nocera}, {Noh}, {Norman}, {North}, {Nozaki}, {Nuttall}, {Oberling}, {O'Brien}, {Obuchi}, {O'Dell}, {Ogaki},
  {Oganesyan}, {Oh}, {Oh}, {Oh}, {Ohashi}, {Ohishi}, {Ohkawa}, {Ohme}, {Ohta}, {Okada}, {Okutani}, {Okutomi}, {Olivetto}, {Oohara}, {Ooi}, {Oram}, {O'Reilly}, {Ormiston}, {Ormsby}, {Ortega}, {O'Shaughnessy}, {O'Shea}, {Oshino}, {Ossokine}, {Osthelder}, {Otabe}, {Ottaway}, {Overmier}, {Pace}, {Pagano}, {Page}, {Pagliaroli}, {Pai}, {Pai}, {Palamos}, {Palashov}, {Palomba}, {Pan}, {Panda}, {Pang}, {Pang}, {Pankow}, {Pannarale}, {Pant}, {Paoletti}, {Paoli}, {Paolone}, {Parisi}, {Park}, {Parker}, {Pascucci}, {Pasqualetti}, {Passaquieti}, {Passuello}, {Patel}, {Patricelli}, {Payne}, {Pechsiri}, {Pedraza}, {Pegoraro}, {Pele}, {Pe{\~n}a Arellano}, {Penn}, {Perego}, {Pereira}, {Pereira}, {Perez}, {P{\'e}rigois}, {Perreca}, {Perri{\`e}s}, {Petermann}, {Petterson}, {Pfeiffer}, {Pham}, {Phukon}, {Piccinni}, {Pichot}, {Piendibene}, {Piergiovanni}, {Pierini}, {Pierro}, {Pillant}, {Pilo}, {Pinard}, {Pinto}, {Piotrzkowski}, {Piotrzkowski}, {Pirello}, {Pitkin}, {Placidi}, {Plastino}, {Pluchar}, {Poggiani}, {Polini}, {Pong},
  {Ponrathnam}, {Popolizio}, {Porter}, {Powell}, {Pracchia}, {Pradier}, {Prajapati}, {Prasai}, {Prasanna}, {Pratten}, {Prestegard}, {Principe}, {Prodi}, {Prokhorov}, {Prosposito}, {Prudenzi}, {Puecher}, {Punturo}, {Puosi}, {Puppo}, {P{\"u}rrer}, {Qi}, {Quetschke}, {Quinonez}, {Quitzow-James}, {Raab}, {Raaijmakers}, {Radkins}, {Radulesco}, {Raffai}, {Rail}, {Raja}, {Rajan}, {Ramirez}, {Ramirez}, {Ramos-Buades}, {Rana}, {Rapagnani}, {Rapol}, {Ratto}, {Ray}, {Raymond}, {Raza}, {Razzano}, {Read}, {Rees}, {Regimbau}, {Rei}, {Reid}, {Reitze}, {Relton}, {Rettegno}, {Ricci}, {Richardson}, {Richardson}, {Richardson}, {Ricker}, {Riemenschneider}, {Riles}, {Rizzo}, {Robertson}, {Robie}, {Robinet}, {Rocchi}, {Rocha}, {Rodriguez}, {Rodriguez-Soto}, {Rolland}, {Rollins}, {Roma}, {Romanelli}, {Romano}, {Romel}, {Romero}, {Romero-Shaw}, {Romie}, {Rose}, {Rosi{\'n}ska}, {Rosofsky}, {Ross}, {Rowan}, {Rowlinson}, {Roy}, {Roy}, {Rozza}, {Ruggi}, {Ryan}, {Sachdev}, {Sadecki}, {Sadiq}, {Sago}, {Saito}, {Saito}, {Sakai}, {Sakai},
  {Sakellariadou}, {Sakuno}, {Salafia}, {Salconi}, {Saleem}, {Salemi}, {Samajdar}, {Sanchez}, {Sanchez}, {Sanchez}, {Sanchis-Gual}, {Sanders}, {Sanuy}, {Saravanan}, {Sarin}, {Sassolas}, {Satari}, {Sathyaprakash}, {Sato}, {Sato}, {Sauter}, {Savage}, {Savant}, {Sawada}, {Sawant}, {Sawant}, {Sayah}, {Schaetzl}, {Scheel}, {Scheuer}, {Schindler-Tyka}, {Schmidt}, {Schnabel}, {Schneewind}, {Schofield}, {Sch{\"o}nbeck}, {Schulte}, {Schutz}, {Schwartz}, {Scott}, {Scott}, {Seglar-Arroyo}, {Seidel}, {Sekiguchi}, {Sekiguchi}, {Sellers}, {Sengupta}, {Sennett}, {Sentenac}, {Seo}, {Sequino}, {Sergeev}, {Setyawati}, {Shaffer}, {Shahriar}, {Shams}, {Shao}, {Sharifi}, {Sharma}, {Sharma}, {Shawhan}, {Shcheblanov}, {Shen}, {Shibagaki}, {Shikauchi}, {Shimizu}, {Shimoda}, {Shimode}, {Shink}, {Shinkai}, {Shishido}, {Shoda}, {Shoemaker}, {Shoemaker}, {Shukla}, {Shyamsundar}, {Sieniawska}, {Sigg}, {Singer}, {Singh}, {Singh}, {Singha}, {Sintes}, {Sipala}, {Skliris}, {Slagmolen}, {Slaven-Blair}, {Smetana}, {Smith}, {Smith}, {Somala},
  {Somiya}, {Son}, {Soni}, {Soni}, {Sorazu}, {Sordini}, {Sorrentino}, {Sorrentino}, {Sotani}, {Soulard}, {Souradeep}, {Sowell}, {Spagnuolo}, {Spencer}, {Spera}, {Srivastava}, {Srivastava}, {Staats}, {Stachie}, {Steer}, {Steinlechner}, {Steinlechner}, {Stops}, {Stevenson}, {Stover}, {Strain}, {Strang}, {Stratta}, {Strunk}, {Sturani}, {Stuver}, {S{\"u}dbeck}, {Sudhagar}, {Sudhir}, {Sugimoto}, {Suh}, {Summerscales}, {Sun}, {Sun}, {Sunil}, {Sur}, {Suresh}, {Sutton}, {Suzuki}, {Suzuki}, {Swinkels}, {Szczepa{\'n}czyk}, {Szewczyk}, {Tacca}, {Tagoshi}, {Tait}, {Takahashi}, {Takahashi}, {Takamori}, {Takano}, {Takeda}, {Takeda}, {Talbot}, {Tanaka}, {Tanaka}, {Tanaka}, {Tanaka}, {Tanaka}, {Tanasijczuk}, {Tanioka}, {Tanner}, {Tao}, {Tapia}, {Tapia San Martin}, {Tasson}, {Telada}, {Tenorio}, {Terkowski}, {Test}, {Thirugnanasambandam}, {Thomas}, {Thomas}, {Thompson}, {Thondapu}, {Thorne}, {Thrane}, {Tiwari}, {Tiwari}, {Tiwari}, {Toland}, {Tolley}, {Tomaru}, {Tomigami}, {Tomura}, {Tonelli}, {Torres-Forn{\'e}}, {Torrie},
  {Tosta E Melo}, {T{\"o}yr{\"a}}, {Trapananti}, {Travasso}, {Traylor}, {Tringali}, {Tripathee}, {Troiano}, {Trovato}, {Trozzo}, {Trudeau}, {Tsai}, {Tsai}, {Tsang}, {Tsang}, {Tsao}, {Tse}, {Tso}, {Tsubono}, {Tsuchida}, {Tsukada}, {Tsuna}, {Tsutsui}, {Tsuzuki}, {Turconi}, {Tuyenbayev}, {Ubhi}, {Uchikata}, {Uchiyama}, {Udall}, {Ueda}, {Uehara}, {Ueno}, {Ueshima}, {Ugolini}, {Unnikrishnan}, {Uraguchi}, {Urban}, {Ushiba}, {Usman}, {Utina}, {Vahlbruch}, {Vajente}, {Vajpeyi}, {Valdes}, {Valentini}, {Valsan}, {van Bakel}, {van Beuzekom}, {van den Brand}, {van den Broeck}, {Vander-Hyde}, {van der Schaaf}, {van Heijningen}, {Vanosky}, {van Putten}, {Vardaro}, {Vargas}, {Varma}, {Vas{\'u}th}, {Vecchio}, {Vedovato}, {Veitch}, {Veitch}, {Venkateswara}, {Venneberg}, {Venugopalan}, {Verkindt}, {Verma}, {Veske}, {Vetrano}, {Vicer{\'e}}, {Viets}, {Villa-Ortega}, {Vinet}, {Vitale}, {Vo}, {Vocca}, {von Reis}, {von Wrangel}, {Vorvick}, {Vyatchanin}, {Wade}, {Wade}, {Wagner}, {Walet}, {Walker}, {Wallace}, {Wallace}, {Walsh},
  {Wang}, {Wang}, {Wang}, {Ward}, {Warner}, {Was}, {Washimi}, {Washington}, {Watchi}, {Weaver}, {Wei}, {Weinert}, {Weinstein}, {Weiss}, {Weller}, {Wellmann}, {Wen}, {We{\ss}els}, {Westhouse}, {Wette}, {Whelan}, {White}, {Whiting}, {Whittle}, {Wilken}, {Williams}, {Williams}, {Williamson}, {Willis}, {Willke}, {Wilson}, {Winkler}, {Wipf}, {Wlodarczyk}, {Woan}, {Woehler}, {Wofford}, {Wong}, {Wu}, {Wu}, {Wu}, {Wu}, {Wysocki}, {Xiao}, {Xu}, {Yamada}, {Yamamoto}, {Yamamoto}, {Yamamoto}, {Yamamoto}, {Yamashita}, {Yamazaki}, {Yang}, {Yang}, {Yang}, {Yang}, {Yang}, {Yap}, {Yeeles}, {Yelikar}, {Ying}, {Yokogawa}, {Yokoyama}, {Yokozawa}, {Yoon}, {Yoshioka}, {Yu}, {Yu}, {Yuzurihara}, {Zadro{\.z}ny}, {Zanolin}, {Zappa}, {Zeidler}, {Zelenova}, {Zendri}, {Zevin}, {Zhan}, {Zhang}, {Zhang}, {Zhang}, {Zhang}, {Zhang}, {Zhao}, {Zhao}, {Zhao}, {Zhao}, {Zhou}, {Zhu}, {Zhu}, {Zimmerman}, {Zlochower}, {Zucker}, {Zweizig}, {Ligo Scientific Collaboration}, {VIRGO Collaboration}, \& {KAGRA Collaboration}}]{Abbott2021}
{Abbott}, R., {Abbott}, T.~D., {Abraham}, S., {et~al.} 2021{\natexlab{a}}, \apjl, 915, L5

\bibitem[{{Abbott} {et~al.}(2020{\natexlab{c}}){Abbott}, {Abbott}, {Abraham}, {Acernese}, {Ackley}, {Adams}, {Adhikari}, {Adya}, {Affeldt}, {Agathos}, {Agatsuma}, {Aggarwal}, {Aguiar}, {Aich}, {Aiello}, {Ain}, {Ajith}, {Akcay}, {Allen}, {Allocca}, \& et~al.}]{abbott2020_190814}
{Abbott}, R., {Abbott}, T.~D., {Abraham}, S., {et~al.} 2020{\natexlab{c}}, \apjl, 896, L44

\bibitem[{{Abbott} {et~al.}(2021{\natexlab{b}}){Abbott}, {LIGO Scientific Collaboration}, \& {Virgo Collaboration}}]{GWTC3}
{Abbott}, R., {LIGO Scientific Collaboration}, \& {Virgo Collaboration}. 2021{\natexlab{b}}, arXiv e-prints, arXiv:2111.03606

\bibitem[{{Acernese} {et~al.}(2015){Acernese}, {Agathos}, {Agatsuma}, {Aisa}, {et~al.}}]{AdvancedVirgo2015}
{Acernese}, F., {Agathos}, M., {Agatsuma}, K., {Aisa}, D., {et~al.} 2015, Classical and Quantum Gravity, 32, 024001

\bibitem[{{Amati} {et~al.}(2021){Amati}, {O'Brien}, {G{\"o}tz}, {Bozzo}, {Santangelo}, {Tanvir}, {Frontera}, {Mereghetti}, {Osborne}, {Blain}, {Basa}, {Branchesi}, {Burderi}, {Caballero-Garc{\'\i}a}, {Castro-Tirado}, {Christensen}, {Ciolfi}, {De Rosa}, {Doroshenko}, {Ferrara}, {Ghirlanda}, {Hanlon}, {Heddermann}, {Hutchinson}, {Labanti}, {Le Floch}, {Lerman}, {Paltani}, {Reglero}, {Rezzolla}, {Rosati}, {Salvaterra}, {Stratta}, {Tenzer}, \& {Theseus Consortium}}]{amati2021}
{Amati}, L., {O'Brien}, P.~T., {G{\"o}tz}, D., {et~al.} 2021, Experimental Astronomy, 52, 183

\bibitem[{{Andreoni} {et~al.}(2022){Andreoni}, {Margutti}, {Salafia}, {Parazin}, {Villar}, {Coughlin}, {Yoachim}, {Mortensen}, {Brethauer}, {Smartt}, {Kasliwal}, {Alexander}, {Anand}, {Berger}, {Bernardini}, {Bianco}, {Blanchard}, {Bloom}, {Brocato}, {Bulla}, {Cartier}, {Cenko}, {Chornock}, {Copperwheat}, {Corsi}, {D'Ammando}, {D'Avanzo}, {H{\'e}l{\`e}ne Datrier}, {Foley}, {Ghirlanda}, {Goobar}, {Grindlay}, {Hajela}, {Holz}, {Karambelkar}, {Kool}, {Lamb}, {Laskar}, {Levan}, {Maguire}, {May}, {Melandri}, {Milisavljevic}, {Miller}, {Nicholl}, {Nissanke}, {Palmese}, {Piranomonte}, {Rest}, {Sagu{\'e}s-Carracedo}, {Siellez}, {Singer}, {Smith}, {Steeghs}, \& {Tanvir}}]{andreoni2022}
{Andreoni}, I., {Margutti}, R., {Salafia}, O.~S., {et~al.} 2022, \apjs, 260, 18

\bibitem[{{Aso} {et~al.}(2013){Aso}, {Michimura}, {Somiya}, {Ando}, {Miyakawa}, {Sekiguchi}, {Tatsumi}, \& {Yamamoto}}]{KAGRA2013}
{Aso}, Y., {Michimura}, Y., {Somiya}, K., {et~al.} 2013, \prd, 88, 043007

\bibitem[{{Banerjee} {et~al.}(2023){Banerjee}, {Oganesyan}, {Branchesi}, {Dupletsa}, {Aharonian}, {Brighenti}, {Goncharov}, {Harms}, {Mapelli}, {Ronchini}, \& {Santoliquido}}]{banerjee2023}
{Banerjee}, B., {Oganesyan}, G., {Branchesi}, M., {et~al.} 2023, \aap, 678, A126

\bibitem[{{Barbieri} {et~al.}(2021){Barbieri}, {Salafia}, {Colpi}, {Ghirlanda}, \& {Perego}}]{barbieri2021}
{Barbieri}, C., {Salafia}, O.~S., {Colpi}, M., {Ghirlanda}, G., \& {Perego}, A. 2021, \aap, 654, A12

\bibitem[{{Barbieri} {et~al.}(2020){Barbieri}, {Salafia}, {Perego}, {Colpi}, \& {Ghirlanda}}]{barbieri2020}
{Barbieri}, C., {Salafia}, O.~S., {Perego}, A., {Colpi}, M., \& {Ghirlanda}, G. 2020, European Physical Journal A, 56, 8

\bibitem[{{Bavera} {et~al.}(2020){Bavera}, {Fragos}, {Qin}, {Zapartas}, {Neijssel}, {Mandel}, {Batta}, {Gaebel}, {Kimball}, \& {Stevenson}}]{Bavera2020}
{Bavera}, S.~S., {Fragos}, T., {Qin}, Y., {et~al.} 2020, \aap, 635, A97

\bibitem[{{Bavera} {et~al.}(2023){Bavera}, {Fragos}, {Zapartas}, {Andrews}, {Kalogera}, {Berry}, {Kruckow}, {Dotter}, {Kovlakas}, {Misra}, {Rocha}, {Srivastava}, {Sun}, \& {Xing}}]{Bavera2023}
{Bavera}, S.~S., {Fragos}, T., {Zapartas}, E., {et~al.} 2023, Nature Astronomy [\eprint[arXiv]{2212.10924}]

\bibitem[{{Bavera} {et~al.}(2021){Bavera}, {Zevin}, \& {Fragos}}]{Bavera2021}
{Bavera}, S.~S., {Zevin}, M., \& {Fragos}, T. 2021, Research Notes of the American Astronomical Society, 5, 127

\bibitem[{{Belczynski} {et~al.}(2020){Belczynski}, {Klencki}, {Fields}, {Olejak}, {Berti}, {Meynet}, {Fryer}, {Holz}, {O'Shaughnessy}, {Brown}, {Bulik}, {Leung}, {Nomoto}, {Madau}, {Hirschi}, {Kaiser}, {Jones}, {Mondal}, {Chruslinska}, {Drozda}, {Gerosa}, {Doctor}, {Giersz}, {Ekstrom}, {Georgy}, {Askar}, {Baibhav}, {Wysocki}, {Natan}, {Farr}, {Wiktorowicz}, {Coleman Miller}, {Farr}, \& {Lasota}}]{Belczynski2020}
{Belczynski}, K., {Klencki}, J., {Fields}, C.~E., {et~al.} 2020, \aap, 636, A104

\bibitem[{{Biscoveanu} {et~al.}(2023){Biscoveanu}, {Landry}, \& {Vitale}}]{Biscoveanu:2023MNRAS.518.5298B}
{Biscoveanu}, S., {Landry}, P., \& {Vitale}, S. 2023, \mnras, 518, 5298

\bibitem[{{Boersma} \& {van Leeuwen}(2022)}]{boersma2022}
{Boersma}, O.~M. \& {van Leeuwen}, J. 2022, \aap, 664, A160

\bibitem[{{Branchesi} {et~al.}(2023){Branchesi}, {Maggiore}, {Alonso}, {Badger}, {Banerjee}, {Beirnaert}, {Belgacem}, {Bhagwat}, {Boileau}, {Borhanian}, {Brown}, {Leong Chan}, {Cusin}, {Danilishin}, {Degallaix}, {De Luca}, {Dhani}, {Dietrich}, {Dupletsa}, {Foffa}, {Franciolini}, {Freise}, {Gemme}, {Goncharov}, {Ghosh}, {Gulminelli}, {Gupta}, {Kumar Gupta}, {Harms}, {Hazra}, {Hild}, {Hinderer}, {Siong Heng}, {Iacovelli}, {Janquart}, {Janssens}, {Jenkins}, {Kalaghatgi}, {Koroveshi}, {Li}, {Li}, {Loffredo}, {Maggio}, {Mancarella}, {Mapelli}, {Martinovic}, {Maselli}, {Meyers}, {Miller}, {Mondal}, {Muttoni}, {Narola}, {Oertel}, {Oganesyan}, {Pacilio}, {Palomba}, {Pani}, {Pasqualetti}, {Perego}, {P{\'e}rigois}, {Pieroni}, {Piccinni}, {Puecher}, {Puppo}, {Ricciardone}, {Riotto}, {Ronchini}, {Sakellariadou}, {Samajdar}, {Santoliquido}, {Sathyaprakash}, {Steinlechner}, {Steinlechner}, {Utina}, {Van Den Broeck}, \& {Zhang}}]{branchesi2023}
{Branchesi}, M., {Maggiore}, M., {Alonso}, D., {et~al.} 2023, \jcap, 2023, 068

\bibitem[{{Braun} {et~al.}(2019){Braun}, {Bonaldi}, {Bourke}, {Keane}, \& {Wagg}}]{braun2019}
{Braun}, R., {Bonaldi}, A., {Bourke}, T., {Keane}, E., \& {Wagg}, J. 2019, arXiv e-prints, arXiv:1912.12699

\bibitem[{{Broekgaarden} {et~al.}(2021){Broekgaarden}, {Berger}, {Neijssel}, {Vigna-G{\'o}mez}, {Chattopadhyay}, {Stevenson}, {Chruslinska}, {Justham}, {de Mink}, \& {Mandel}}]{Broekgaarden2021}
{Broekgaarden}, F.~S., {Berger}, E., {Neijssel}, C.~J., {et~al.} 2021, \mnras, 508, 5028

\bibitem[{{Bromberg} {et~al.}(2011){Bromberg}, {Nakar}, {Piran}, \& {Sari}}]{Bromberg2011}
{Bromberg}, O., {Nakar}, E., {Piran}, T., \& {Sari}, R. 2011, \apj, 740, 100

\bibitem[{{Burns} {et~al.}(2016){Burns}, {Connaughton}, {Zhang}, {Lien}, {Briggs}, {Goldstein}, {Pelassa}, \& {Troja}}]{Burns2016}
{Burns}, E., {Connaughton}, V., {Zhang}, B.-B., {et~al.} 2016, \apj, 818, 110

\bibitem[{{Chen} {et~al.}(2021){Chen}, {Cowperthwaite}, {Metzger}, \& {Berger}}]{chen2021}
{Chen}, H.-Y., {Cowperthwaite}, P.~S., {Metzger}, B.~D., \& {Berger}, E. 2021, \apjl, 908, L4

\bibitem[{{Chen} {et~al.}(2024){Chen}, {Wang}, {Hayashi}, {Kawaguchi}, {Kiuchi}, \& {Shibata}}]{Chen:2024}
{Chen}, S., {Wang}, L., {Hayashi}, K., {et~al.} 2024, \prd, 110, 063016

\bibitem[{{Cherenkov Telescope Array Consortium} {et~al.}(2019){Cherenkov Telescope Array Consortium}, {Acharya}, {Agudo}, {Al Samarai}, {Alfaro}, {Alfaro}, {Alispach}, {Alves Batista}, {Amans}, {Amato}, {Ambrosi}, {Antolini}, {Antonelli}, {Aramo}, {Araya}, {Armstrong}, {Arqueros}, {Arrabito}, {Asano}, {Ashley}, {Backes}, {Balazs}, {Balbo}, {Ballester}, {Ballet}, {Bamba}, {Barkov}, {Barres de Almeida}, {Barrio}, {Bastieri}, {Becherini}, {Belfiore}, {Benbow}, {Berge}, {Bernardini}, {Bernardini}, {Bernardos}, {Bernl{\"o}hr}, {Bertucci}, {Biasuzzi}, {Bigongiari}, {Biland}, {Bissaldi}, {Biteau}, {Blanch}, {Blazek}, {Boisson}, {Bolmont}, {Bonanno}, {Bonardi}, {Bonavolont{\`a}}, {Bonnoli}, {Bosnjak}, {B{\"o}ttcher}, {Braiding}, {Bregeon}, {Brill}, {Brown}, {Brun}, {Brunetti}, {Buanes}, {Buckley}, {Bugaev}, {B{\"u}hler}, {Bulgarelli}, {Bulik}, {Burton}, {Burtovoi}, {Busetto}, {Canestrari}, {Capalbi}, {Capitanio}, {Caproni}, {Caraveo}, {C{\'a}rdenas}, {Carlile}, {Carosi}, {Carqu{\'\i}n}, {Carr}, {Casanova},
  {Cascone}, {Catalani}, {Catalano}, {Cauz}, {Cerruti}, {Chadwick}, {Chaty}, {Chaves}, {Chen}, {Chen}, {Chernyakova}, {Chikawa}, {Christov}, {Chudoba}, {Cie{\'s}lar}, {Coco}, {Colafrancesco}, {Colin}, {Conforti}, {Connaughton}, {Conrad}, {Contreras}, {Cortina}, {Costa}, {Costantini}, {Cotter}, {Covino}, {Crocker}, {Cuadra}, {Cuevas}, {Cumani}, {D'A{\`\i}}, {D'Ammando}, {D'Avanzo}, {D'Urso}, {Daniel}, {Davids}, {Dawson}, {Dazzi}, {De Angelis}, {de C{\'a}ssia dos Anjos}, {De Cesare}, {De Franco}, {de Gouveia Dal Pino}, {de la Calle}, {de los Reyes Lopez}, {De Lotto}, {De Luca}, {De Lucia}, {de Naurois}, {de O{\~n}a Wilhelmi}, {De Palma}, {De Persio}, {de Souza}, {Deil}, {Del Santo}, {Delgado}, {della Volpe}, {Di Girolamo}, {Di Pierro}, {Di Venere}, {D{\'\i}az}, {Dib}, {Diebold}, {Djannati-Ata{\"\i}}, {Dom{\'\i}nguez}, {Dominis Prester}, {Dorner}, {Doro}, {Drass}, {Dravins}, {Dubus}, {Dwarkadas}, {Ebr}, {Eckner}, {Egberts}, {Einecke}, {Ekoume}, {Els{\"a}sser}, {Ernenwein}, {Espinoza}, {Evoli}, {Fairbairn},
  {Falceta-Goncalves}, {Falcone}, {Farnier}, {Fasola}, {Fedorova}, {Fegan}, {Fernandez-Alonso}, {Fern{\'a}ndez-Barral}, {Ferrand}, {Fesquet}, {Filipovic}, {Fioretti}, {Fontaine}, {Fornasa}, {Fortson}, {Freixas Coromina}, {Fruck}, {Fujita}, {Fukazawa}, {Funk}, {F{\"u}{\ss}ling}, {Gabici}, {Gadola}, {Gallant}, {Garcia}, {Garcia L{\'o}pez}, {Garczarczyk}, {Gaskins}, {Gasparetto}, {Gaug}, {Gerard}, {Giavitto}, {Giglietto}, {Giommi}, {Giordano}, {Giro}, \& {Giroletti}}]{CTA2019}
{Cherenkov Telescope Array Consortium}, {Acharya}, B.~S., {Agudo}, I., {et~al.} 2019, {Science with the Cherenkov Telescope Array} (WORLD SCIENTIFIC)

\bibitem[{{Colombo} {et~al.}(2024){Colombo}, {Duqu{\'e}}, {Salafia}, {Broekgaarden}, {Iacovelli}, {Mancarella}, {Andreoni}, {Gabrielli}, {Ragosta}, {Ghirlanda}, {Fragos}, {Levan}, {Piranomonte}, {Melandri}, {Giacomazzo}, \& {Colpi}}]{colombo2023}
{Colombo}, A., {Duqu{\'e}}, R., {Salafia}, O.~S., {et~al.} 2024, \aap, 686, A265

\bibitem[{{Colombo} {et~al.}(2022){Colombo}, {Salafia}, {Gabrielli}, {Ghirlanda}, {Giacomazzo}, {Perego}, \& {Colpi}}]{colombo2022}
{Colombo}, A., {Salafia}, O.~S., {Gabrielli}, F., {et~al.} 2022, \apj, 937, 79

\bibitem[{{Corsi} {et~al.}(2019){Corsi}, {Lloyd-Ronning}, {Carbone}, {Frail}, {Lazzati}, {Murphy}, {O'Shaughnessy}, {Owen}, {Sand}, {Fong}, {Spekkens}, \& {Seymour}}]{corsi2019}
{Corsi}, A., {Lloyd-Ronning}, N.~M., {Carbone}, D., {et~al.} 2019, arXiv e-prints, arXiv:1903.10589

\bibitem[{{Coulter} {et~al.}(2017){Coulter}, {Foley}, {Kilpatrick}, {Drout}, {Piro}, {Shappee}, {Siebert}, {Simon}, {Ulloa}, {Kasen}, {Madore}, {Murguia-Berthier}, {Pan}, {Prochaska}, {Ramirez-Ruiz}, {Rest}, \& {Rojas-Bravo}}]{Coulter2017}
{Coulter}, D.~A., {Foley}, R.~J., {Kilpatrick}, C.~D., {et~al.} 2017, Science, 358, 1556

\bibitem[{{Dietrich} {et~al.}(2019){Dietrich}, {Samajdar}, {Khan}, {Johnson-McDaniel}, {Dudi}, \& {Tichy}}]{dietrich2019}
{Dietrich}, T., {Samajdar}, A., {Khan}, S., {et~al.} 2019, \prd, 100, 044003

\bibitem[{{Dobie} {et~al.}(2021){Dobie}, {Murphy}, {Kaplan}, {Hotokezaka}, {Bonilla Ataides}, {Mahony}, \& {Sadler}}]{Dobie2021}
{Dobie}, D., {Murphy}, T., {Kaplan}, D.~L., {et~al.} 2021, \mnras, 505, 2647

\bibitem[{{Duffell} {et~al.}(2015){Duffell}, {Quataert}, \& {MacFadyen}}]{Duffell2015}
{Duffell}, P.~C., {Quataert}, E., \& {MacFadyen}, A.~I. 2015, \apj, 813, 64

\bibitem[{Dupletsa {et~al.}(2023)Dupletsa, Harms, Banerjee, Branchesi, Goncharov, Maselli, Oliveira, Ronchini, \& Tissino}]{Dupletsa:2022scg}
Dupletsa, U., Harms, J., Banerjee, B., {et~al.} 2023, Astron. Comput., 42, 100671

\bibitem[{{Eichler} {et~al.}(1989){Eichler}, {Livio}, {Piran}, \& {Schramm}}]{Eichler1989}
{Eichler}, D., {Livio}, M., {Piran}, T., \& {Schramm}, D.~N. 1989, \nat, 340, 126

\bibitem[{{Evans} {et~al.}(2021){Evans}, {Adhikari}, {Afle}, {Ballmer}, {Biscoveanu}, {Borhanian}, {Brown}, {Chen}, {Eisenstein}, {Gruson}, {Gupta}, {Hall}, {Huxford}, {Kamai}, {Kashyap}, {Kissel}, {Kuns}, {Landry}, {Lenon}, {Lovelace}, {McCuller}, {Ng}, {Nitz}, {Read}, {Sathyaprakash}, {Shoemaker}, {Slagmolen}, {Smith}, {Srivastava}, {Sun}, {Vitale}, \& {Weiss}}]{evans2021}
{Evans}, M., {Adhikari}, R.~X., {Afle}, C., {et~al.} 2021, arXiv e-prints, arXiv:2109.09882

\bibitem[{{Evans} {et~al.}(2023){Evans}, {Corsi}, {Afle}, {Ananyeva}, {Arun}, {Ballmer}, {Bandopadhyay}, {Barsotti}, {Baryakhtar}, {Berger}, {Berti}, {Biscoveanu}, {Borhanian}, {Broekgaarden}, {Brown}, {Cahillane}, {Campbell}, {Chen}, {Daniel}, {Dhani}, {Driggers}, {Effler}, {Eisenstein}, {Fairhurst}, {Feicht}, {Fritschel}, {Fulda}, {Gupta}, {Hall}, {Hammond}, {Hannuksela}, {Hansen}, {Haster}, {Kacanja}, {Kamai}, {Kashyap}, {Shapiro Key}, {Khadkikar}, {Kontos}, {Kuns}, {Landry}, {Landry}, {Lantz}, {Li}, {Lovelace}, {Mandic}, {Mansell}, {Martynov}, {McCuller}, {Miller}, {Nitz}, {Owen}, {Palomba}, {Read}, {Phurailatpam}, {Reddy}, {Richardson}, {Rollins}, {Romano}, {Sathyaprakash}, {Schofield}, {Shoemaker}, {Sigg}, {Singh}, {Slagmolen}, {Sledge}, {Smith}, {Soares-Santos}, {Strunk}, {Sun}, {Tanner}, {van Son}, {Vitale}, {Willke}, {Yamamoto}, \& {Zucker}}]{evans2023}
{Evans}, M., {Corsi}, A., {Afle}, C., {et~al.} 2023, arXiv e-prints, arXiv:2306.13745

\bibitem[{{Farr} {et~al.}(2011){Farr}, {Sravan}, {Cantrell}, {Kreidberg}, {Bailyn}, {Mandel}, \& {Kalogera}}]{farr2011}
{Farr}, W.~M., {Sravan}, N., {Cantrell}, A., {et~al.} 2011, \apj, 741, 103

\bibitem[{{Farrow} {et~al.}(2019){Farrow}, {Zhu}, \& {Thrane}}]{farrow2019}
{Farrow}, N., {Zhu}, X.-J., \& {Thrane}, E. 2019, \apj, 876, 18

\bibitem[{{Fern{\'a}ndez} \& {Metzger}(2013)}]{Fernandez2013}
{Fern{\'a}ndez}, R. \& {Metzger}, B.~D. 2013, \mnras, 435, 502

\bibitem[{{Fiore} {et~al.}(2020){Fiore}, {Burderi}, {Lavagna}, {Bertacin}, {Evangelista}, {Campana}, {Fuschino}, {Lunghi}, {Monge}, {Negri}, {Pirrotta}, {Puccetti}, {Sanna}, {Amarilli}, {Ambrosino}, {Amelino-Camelia}, {Anitra}, {Auricchio}, {Barbera}, {Bechini}, {Bellutti}, {Bertuccio}, {Cao}, {Ceraudo}, {Chen}, {Cinelli}, {Citossi}, {Clerici}, {Colagrossi}, {Curzel}, {Della Casa}, {Demenev}, {Del Santo}, {Dilillo}, {Di Salvo}, {Efremov}, {Feroci}, {Feruglio}, {Ferrandi}, {Fiorini}, {Fiorito}, {Frontera}, {Gacnik}, {Galgoczi}, {Gao}, {Gambino}, {Gandola}, {Ghirlanda}, {Gomboc}, {Grassi}, {Guzman}, {Karlica}, {Kostic}, {Labanti}, {La Rosa}, {Lo Cicero}, {Lopez-Fernandez}, {Malcovati}, {Maselli}, {Manca}, {Mele}, {Milankovich}, {Morgante}, {Nava}, {Nogara}, {Ohno}, {Ottolina}, {Pasquale}, {Pal}, {Perri}, {Piazzolla}, {Piccinin}, {Pliego-Caballero}, {Prinetto}, {Pucacco}, {Rashevsky}, {Rashevskaya}, {Riggio}, {Ripa}, {Russo}, {Papitto}, {Piranomonte}, {Santangelo}, {Scala}, {Sciarrone}, {Selcan}, {Silvestrini},
  {Sottile}, {Rotovnik}, {Tenzer}, {Troisi}, {Vacchi}, {Virgilli}, {Werner}, {Wang}, {Xu}, {Zampa}, {Zampa}, \& {Zanotti}}]{fiore2020}
{Fiore}, F., {Burderi}, L., {Lavagna}, M., {et~al.} 2020, in Society of Photo-Optical Instrumentation Engineers (SPIE) Conference Series, Vol. 11444, Space Telescopes and Instrumentation 2020: Ultraviolet to Gamma Ray, ed. J.-W.~A. {den Herder}, S.~{Nikzad}, \& K.~{Nakazawa}, 114441R

\bibitem[{{Foucart}(2020)}]{Foucart2020}
{Foucart}, F. 2020, Frontiers in Astronomy and Space Sciences, 7, 46

\bibitem[{{Foucart} {et~al.}(2019){Foucart}, {Duez}, {Kidder}, {Nissanke}, {Pfeiffer}, \& {Scheel}}]{foucart2019}
{Foucart}, F., {Duez}, M.~D., {Kidder}, L.~E., {et~al.} 2019, \prd, 99, 103025

\bibitem[{{Foucart} {et~al.}(2018){Foucart}, {Hinderer}, \& {Nissanke}}]{foucart2018}
{Foucart}, F., {Hinderer}, T., \& {Nissanke}, S. 2018, \prd, 98, 081501

\bibitem[{{Fragos} \& {McClintock}(2015)}]{Fragos2015}
{Fragos}, T. \& {McClintock}, J.~E. 2015, \apj, 800, 17

\bibitem[{{Fryer} {et~al.}(2012){Fryer}, {Belczynski}, {Wiktorowicz}, {Dominik}, {Kalogera}, \& {Holz}}]{fryer2012}
{Fryer}, C.~L., {Belczynski}, K., {Wiktorowicz}, G., {et~al.} 2012, \apj, 749, 91

\bibitem[{{Fuller} \& {Ma}(2019)}]{Fuller2019}
{Fuller}, J. \& {Ma}, L. 2019, \apjl, 881, L1

\bibitem[{{Ghirlanda} {et~al.}(2024){Ghirlanda}, {Nava}, {Salafia}, {Fiore}, {Campana}, {Salvaterra}, {Sanna}, {Leone}, {Evangelista}, {Dilillo}, {Puccetti}, {Santangelo}, {Trenti}, {Guzm{\'a}n}, {Hedderman}, {Amelino-Camelia}, {Barbera}, {Baroni}, {Bechini}, {Bellutti}, {Bertuccio}, {Borghi}, {Brandonisio}, {Burderi}, {Cabras}, {Chen}, {Citossi}, {Colagrossi}, {Crupi}, {De Cecio}, {Dedolli}, {Del Santo}, {Demenev}, {Di Salvo}, {Ficorella}, {Ga{\v{c}}nik}, {Gandola}, {Gao}, {Gomboc}, {Grassi}, {Iaria}, {La Rosa}, {Lo Cicero}, {Malcovati}, {Manca}, {Marchesini}, {Maselli}, {Mele}, {Nogara}, {Pepponi}, {Perri}, {Picciotto}, {Pirrotta}, {Prinetto}, {Quirino}, {Riggio}, {{\v{R}}{\'\i}pa}, {Russo}, {Sel{\v{c}}an}, {Silvestrini}, {Sottile}, {Thomas}, {Tiberia}, {Trevisan}, {Troisi}, {Tsvetkova}, {Vacchi}, {Werner}, {Zanotti}, \& {Zorzi}}]{ghirlanda2024}
{Ghirlanda}, G., {Nava}, L., {Salafia}, O., {et~al.} 2024, \aap, 689, A175

\bibitem[{{Ghirlanda} {et~al.}(2019){Ghirlanda}, {Salafia}, {Paragi}, {Giroletti}, {Yang}, {Marcote}, {Blanchard}, {Agudo}, {An}, {Bernardini}, {Beswick}, {Branchesi}, {Campana}, {Casadio}, {Chassande-Mottin}, {Colpi}, {Covino}, {D'Avanzo}, {D'Elia}, {Frey}, {Gawronski}, {Ghisellini}, {Gurvits}, {Jonker}, {van Langevelde}, {Melandri}, {Moldon}, {Nava}, {Perego}, {Perez-Torres}, {Reynolds}, {Salvaterra}, {Tagliaferri}, {Venturi}, {Vergani}, \& {Zhang}}]{ghirlanda2019}
{Ghirlanda}, G., {Salafia}, O.~S., {Paragi}, Z., {et~al.} 2019, Science, 363, 968

\bibitem[{{Gottlieb} \& {Nakar}(2022)}]{Gottlieb2022}
{Gottlieb}, O. \& {Nakar}, E. 2022, \mnras, 517, 1640

\bibitem[{{Gupta} {et~al.}(2023){Gupta}, {Afle}, {Arun}, {Bandopadhyay}, {Baryakhtar}, {Biscoveanu}, {Borhanian}, {Broekgaarden}, {Corsi}, {Dhani}, {Evans}, {Hall}, {Hannuksela}, {Kacanja}, {Kashyap}, {Khadkikar}, {Kuns}, {Li}, {Miller}, {Nitz}, {Owen}, {Palomba}, {Pearce}, {Phurailatpam}, {Rajbhandari}, {Read}, {Romano}, {Sathyaprakash}, {Shoemaker}, {Singh}, {Vitale}, {Barsotti}, {Berti}, {Cahillane}, {Chen}, {Fritschel}, {Haster}, {Landry}, {Lovelace}, {McClelland}, {Slagmolen}, {Smith}, {Soares-Santos}, {Sun}, {Tanner}, {Yamamoto}, \& {Zucker}}]{gupta2023}
{Gupta}, I., {Afle}, C., {Arun}, K.~G., {et~al.} 2023, arXiv e-prints, arXiv:2307.10421

\bibitem[{{Hallinan} {et~al.}(2019){Hallinan}, {Ravi}, {Weinreb}, {Kocz}, {Huang}, {Woody}, {Lamb}, {D'Addario}, {Catha}, {Law}, {Kulkarni}, {Phinney}, {Eastwood}, {Bouman}, {McLaughlin}, {Ransom}, {Siemens}, {Cordes}, {Lynch}, {Kaplan}, {Brazier}, {Bhatnagar}, {Myers}, {Walter}, \& {Gaensler}}]{hallinan2019}
{Hallinan}, G., {Ravi}, V., {Weinreb}, S., {et~al.} 2019, in Bulletin of the American Astronomical Society, Vol.~51, 255

\bibitem[{{Hamidani} \& {Ioka}(2021)}]{Hamidani2021}
{Hamidani}, H. \& {Ioka}, K. 2021, \mnras, 500, 627

\bibitem[{{Hempel} {et~al.}(2012){Hempel}, {Fischer}, {Schaffner-Bielich}, \& {Liebend{\"o}rfer}}]{Hempel2012}
{Hempel}, M., {Fischer}, T., {Schaffner-Bielich}, J., \& {Liebend{\"o}rfer}, M. 2012, \apj, 748, 70

\bibitem[{{Hendriks} {et~al.}(2023){Hendriks}, {Yi}, \& {Nelemans}}]{hendriks2023}
{Hendriks}, K., {Yi}, S.-X., \& {Nelemans}, G. 2023, \aap, 672, A74

\bibitem[{{Iacovelli} {et~al.}(2022{\natexlab{a}}){Iacovelli}, {Mancarella}, {Foffa}, \& {Maggiore}}]{Iacovelli2022}
{Iacovelli}, F., {Mancarella}, M., {Foffa}, S., \& {Maggiore}, M. 2022{\natexlab{a}}, \apj, 941, 208

\bibitem[{{Iacovelli} {et~al.}(2022{\natexlab{b}}){Iacovelli}, {Mancarella}, {Foffa}, \& {Maggiore}}]{IacovelliGWFAST}
{Iacovelli}, F., {Mancarella}, M., {Foffa}, S., \& {Maggiore}, M. 2022{\natexlab{b}}, \apjs, 263, 2

\bibitem[{{Ivezic} {et~al.}(2008){Ivezic}, {Axelrod}, {Brandt}, {Burke}, {Claver}, {Connolly}, {Cook}, {Gee}, {Gilmore}, {Jacoby}, {Jones}, {Kahn}, {Kantor}, {Krabbendam}, {Lupton}, {Monet}, {Pinto}, {Saha}, {Schalk}, {Schneider}, {Strauss}, {Stubbs}, {Sweeney}, {Szalay}, {Thaler}, {Tyson}, \& {LSST Collaboration}}]{ivezic2008}
{Ivezic}, Z., {Axelrod}, T., {Brandt}, W.~N., {et~al.} 2008, Serbian Astronomical Journal, 176, 1

\bibitem[{{Ivezi{\'c}} {et~al.}(2019){Ivezi{\'c}}, {Kahn}, {Tyson}, {Abel}, {Acosta}, {Allsman}, {Alonso}, {AlSayyad}, {Anderson}, {Andrew}, {Angel}, {Angeli}, {Ansari}, {Antilogus}, {Araujo}, {Armstrong}, {Arndt}, {Astier}, {Aubourg}, {Auza}, {Axelrod}, {Bard}, {Barr}, {Barrau}, {Bartlett}, {Bauer}, {Bauman}, {Baumont}, {Bechtol}, {Bechtol}, {Becker}, {Becla}, {Beldica}, {Bellavia}, {Bianco}, {Biswas}, {Blanc}, {Blazek}, {Blandford}, {Bloom}, {Bogart}, {Bond}, {Booth}, {Borgland}, {Borne}, {Bosch}, {Boutigny}, {Brackett}, {Bradshaw}, {Brandt}, {Brown}, {Bullock}, {Burchat}, {Burke}, {Cagnoli}, {Calabrese}, {Callahan}, {Callen}, {Carlin}, {Carlson}, {Chandrasekharan}, {Charles-Emerson}, {Chesley}, {Cheu}, {Chiang}, {Chiang}, {Chirino}, {Chow}, {Ciardi}, {Claver}, {Cohen-Tanugi}, {Cockrum}, {Coles}, {Connolly}, {Cook}, {Cooray}, {Covey}, {Cribbs}, {Cui}, {Cutri}, {Daly}, {Daniel}, {Daruich}, {Daubard}, {Daues}, {Dawson}, {Delgado}, {Dellapenna}, {de Peyster}, {de Val-Borro}, {Digel}, {Doherty}, {Dubois},
  {Dubois-Felsmann}, {Durech}, {Economou}, {Eifler}, {Eracleous}, {Emmons}, {Fausti Neto}, {Ferguson}, {Figueroa}, {Fisher-Levine}, {Focke}, {Foss}, {Frank}, {Freemon}, {Gangler}, {Gawiser}, {Geary}, {Gee}, {Geha}, {Gessner}, {Gibson}, {Gilmore}, {Glanzman}, {Glick}, {Goldina}, {Goldstein}, {Goodenow}, {Graham}, {Gressler}, {Gris}, {Guy}, {Guyonnet}, {Haller}, {Harris}, {Hascall}, {Haupt}, {Hernandez}, {Herrmann}, {Hileman}, {Hoblitt}, {Hodgson}, {Hogan}, {Howard}, {Huang}, {Huffer}, {Ingraham}, {Innes}, {Jacoby}, {Jain}, {Jammes}, {Jee}, {Jenness}, {Jernigan}, {Jevremovi{\'c}}, {Johns}, {Johnson}, {Johnson}, {Jones}, {Juramy-Gilles}, {Juri{\'c}}, {Kalirai}, {Kallivayalil}, {Kalmbach}, {Kantor}, {Karst}, {Kasliwal}, {Kelly}, {Kessler}, {Kinnison}, {Kirkby}, {Knox}, {Kotov}, {Krabbendam}, {Krughoff}, {Kub{\'a}nek}, {Kuczewski}, {Kulkarni}, {Ku}, {Kurita}, {Lage}, {Lambert}, {Lange}, {Langton}, {Le Guillou}, {Levine}, {Liang}, {Lim}, {Lintott}, {Long}, {Lopez}, {Lotz}, {Lupton}, {Lust}, {MacArthur}, {Mahabal},
  {Mandelbaum}, {Markiewicz}, {Marsh}, {Marshall}, {Marshall}, {May}, {McKercher}, {McQueen}, {Meyers}, {Migliore}, {Miller}, \& {Mills}}]{ivezic2019}
{Ivezi{\'c}}, {\v{Z}}., {Kahn}, S.~M., {Tyson}, J.~A., {et~al.} 2019, \apj, 873, 111

\bibitem[{{Just} {et~al.}(2015){Just}, {Bauswein}, {Ardevol Pulpillo}, {Goriely}, \& {Janka}}]{Just2015}
{Just}, O., {Bauswein}, A., {Ardevol Pulpillo}, R., {Goriely}, S., \& {Janka}, H.~T. 2015, \mnras, 448, 541

\bibitem[{{Kalogera} {et~al.}(2021){Kalogera}, {Sathyaprakash}, {Bailes}, {Bizouard}, {Buonanno}, {Burrows}, {Colpi}, {Evans}, {Fairhurst}, {Hild}, {Kasliwal}, {Lehner}, {Mandel}, {Mandic}, {Nissanke}, {Alessandra Papa}, {Reddy}, {Rosswog}, {Van Den Broeck}, {Ajith}, {Anand}, {Andreoni}, {Arun}, {Barausse}, {Baryakhtar}, {Belgacem}, {Berry}, {Bertacca}, {Brito}, {Caprini}, {Chatziioannou}, {Coughlin}, {Cusin}, {Dietrich}, {Dirian}, {East}, {Fan}, {Figueroa}, {Foffa}, {Ghosh}, {Hall}, {Harms}, {Harry}, {Hinderer}, {Janka}, {Justham}, {Kasen}, {Kotake}, {Lovelace}, {Maggiore}, {Mangiagli}, {Mapelli}, {Maselli}, {Matas}, {McIver}, {Messer}, {Mezzacappa}, {Mills}, {Mueller}, {M{\"u}ller}, {P{\"u}rrer}, {Pani}, {Pratten}, {Regimbau}, {Sakellariadou}, {Schneider}, {Sesana}, {Shao}, {Sotiriou}, {Tamanini}, {Tauris}, {Thrane}, {Valiante}, {van de Meent}, {Varma}, {Vines}, {Vitale}, {Yang}, {Yunes}, {Zumalacarregui}, {Punturo}, {Reitze}, {Couvares}, {Katsanevas}, {Kajita}, {Lueck}, {McClelland}, {Rowan}, {Sanders},
  {Shoemaker}, \& {van den Brand}}]{Kalogera2021}
{Kalogera}, V., {Sathyaprakash}, B.~S., {Bailes}, M., {et~al.} 2021, arXiv e-prints, arXiv:2111.06990

\bibitem[{{Kawaguchi} {et~al.}(2015){Kawaguchi}, {Kyutoku}, {Nakano}, {Okawa}, {Shibata}, \& {Taniguchi}}]{kawagichi2015}
{Kawaguchi}, K., {Kyutoku}, K., {Nakano}, H., {et~al.} 2015, \prd, 92, 024014

\bibitem[{{Kawaguchi} {et~al.}(2016){Kawaguchi}, {Kyutoku}, {Shibata}, \& {Tanaka}}]{kawaguchi2016}
{Kawaguchi}, K., {Kyutoku}, K., {Shibata}, M., \& {Tanaka}, M. 2016, \apj, 825, 52

\bibitem[{{Kiziltan} {et~al.}(2013){Kiziltan}, {Kottas}, {De Yoreo}, \& {Thorsett}}]{Kiziltan2013}
{Kiziltan}, B., {Kottas}, A., {De Yoreo}, M., \& {Thorsett}, S.~E. 2013, \apj, 778, 66

\bibitem[{{Kr{\"u}ger} \& {Foucart}(2020)}]{Kruger2020}
{Kr{\"u}ger}, C.~J. \& {Foucart}, F. 2020, \prd, 101, 103002

\bibitem[{{Lattimer} \& {Schramm}(1974)}]{Lattimer1974}
{Lattimer}, J.~M. \& {Schramm}, D.~N. 1974, \apjl, 192, L145

\bibitem[{{Lazzati} \& {Perna}(2019)}]{Lazzati2019}
{Lazzati}, D. \& {Perna}, R. 2019, \apj, 881, 89

\bibitem[{{Lazzati} {et~al.}(2018){Lazzati}, {Perna}, {Morsony}, {Lopez-Camara}, {Cantiello}, {Ciolfi}, {Giacomazzo}, \& {Workman}}]{Lazzati2018}
{Lazzati}, D., {Perna}, R., {Morsony}, B.~J., {et~al.} 2018, \prl, 120, 241103

\bibitem[{{Levan} {et~al.}(2024){Levan}, {Gompertz}, {Salafia}, {Bulla}, {Burns}, {Hotokezaka}, {Izzo}, {Lamb}, {Malesani}, {Oates}, {Ravasio}, {Rouco Escorial}, {Schneider}, {Sarin}, {Schulze}, {Tanvir}, {Ackley}, {Anderson}, {Brammer}, {Christensen}, {Dhillon}, {Evans}, {Fausnaugh}, {Fong}, {Fruchter}, {Fryer}, {Fynbo}, {Gaspari}, {Heintz}, {Hjorth}, {Kennea}, {Kennedy}, {Laskar}, {Leloudas}, {Mandel}, {Martin-Carrillo}, {Metzger}, {Nicholl}, {Nugent}, {Palmerio}, {Pugliese}, {Rastinejad}, {Rhodes}, {Rossi}, {Saccardi}, {Smartt}, {Stevance}, {Tohuvavohu}, {van der Horst}, {Vergani}, {Watson}, {Barclay}, {Bhirombhakdi}, {Breedt}, {Breeveld}, {Brown}, {Campana}, {Chrimes}, {D'Avanzo}, {D'Elia}, {De Pasquale}, {Dyer}, {Galloway}, {Garbutt}, {Green}, {Hartmann}, {Jakobsson}, {Kerry}, {Kouveliotou}, {Langeroodi}, {Le Floc'h}, {Leung}, {Littlefair}, {Munday}, {O'Brien}, {Parsons}, {Pelisoli}, {Sahman}, {Salvaterra}, {Sbarufatti}, {Steeghs}, {Tagliaferri}, {Th{\"o}ne}, {de Ugarte Postigo}, \& {Kann}}]{levan2024}
{Levan}, A.~J., {Gompertz}, B.~P., {Salafia}, O.~S., {et~al.} 2024, \nat, 626, 737

\bibitem[{{Li} \& {Paczy{\'n}ski}(1998)}]{li1998}
{Li}, L.-X. \& {Paczy{\'n}ski}, B. 1998, \apjl, 507, L59

\bibitem[{{Loffredo} {et~al.}(2024){Loffredo}, {Hazra}, {Dupletsa}, {Branchesi}, {Ronchini}, {Santoliquido}, {Perego}, {Banerjee}, {Bisero}, {Ricigliano}, {Vergani}, {Andreoni}, {Cantiello}, {Harms}, {Mapelli}, \& {Oganesyan}}]{loffredo2024}
{Loffredo}, E., {Hazra}, N., {Dupletsa}, U., {et~al.} 2024, arXiv e-prints, arXiv:2411.02342

\bibitem[{{Madau} \& {Dickinson}(2014)}]{Madau2014}
{Madau}, P. \& {Dickinson}, M. 2014, Annual Review of Astronomy and Astrophysics, 52, 415

\bibitem[{Maggiore(2007)}]{Maggiore:2007ulw}
Maggiore, M. 2007, {Gravitational Waves. Vol. 1: Theory and Experiments} (Oxford University Press)

\bibitem[{{Maggiore} {et~al.}(2020){Maggiore}, {Van Den Broeck}, {Bartolo}, {Belgacem}, {Bertacca}, {Bizouard}, {Branchesi}, {Clesse}, {Foffa}, {Garc{\'\i}a-Bellido}, {Grimm}, {Harms}, {Hinderer}, {Matarrese}, {Palomba}, {Peloso}, {Ricciardone}, \& {Sakellariadou}}]{maggiore2020}
{Maggiore}, M., {Van Den Broeck}, C., {Bartolo}, N., {et~al.} 2020, \jcap, 2020, 050

\bibitem[{{Mainieri} {et~al.}(2024){Mainieri}, {Anderson}, {Brinchmann}, {Cimatti}, {Ellis}, {Hill}, {Kneib}, {McLeod}, {Opitom}, {Roth}, {Sanchez-Saez}, {Smiljanic}, {Tolstoy}, {Bacon}, {Randich}, {Adamo}, {Annibali}, {Arevalo}, {Audard}, {Barsanti}, {Battaglia}, {Bayo Aran}, {Belfiore}, {Bellazzini}, {Bellini}, {Beltran}, {Berni}, {Bianchi}, {Biazzo}, {Bisero}, {Bisogni}, {Bland-Hawthorn}, {Blondin}, {Bodensteiner}, {Boffin}, {Bonito}, {Bono}, {Bouche}, {Bowman}, {Braga}, {Bragaglia}, {Branchesi}, {Brucalassi}, {Bryant}, {Bryson}, {Busa}, {Camera}, {Carbone}, {Casali}, {Casali}, {Casasola}, {Castro}, {Catelan}, {Cavallo}, {Chiappini}, {Cioni}, {Colless}, {Colzi}, {Contarini}, {Couch}, {D'Ammando}, {d'Assignies D.}, {D'Orazi}, {da Silva}, {Dainotti}, {Damiani}, {Danielski}, {De Cia}, {de Jong}, {Dhawan}, {Dierickx}, {Driver}, {Dupletsa}, {Escoffier}, {Escorza}, {Fabrizio}, {Fiorentino}, {Fontana}, {Fontani}, {Forero Sanchez}, {Franois}, {Galindo-Guil}, {Gallazzi}, {Galli}, {Garcia}, {Garcia-Rojas},
  {Garilli}, {Grand}, {Guarcello}, {Hazra}, {Helmi}, {Herrero}, {Iglesias}, {Ilic}, {Irsic}, {Ivanov}, {Izzo}, {Jablonka}, {Joachimi}, {Kakkad}, {Kamann}, {Koposov}, {Kordopatis}, {Kovacevic}, {Kraljic}, {Kuncarayakti}, {Kwon}, {La Forgia}, {Lahav}, {Laigle}, {Lazzarin}, {Leaman}, {Leclercq}, {Lee}, {Lee}, {Lehnert}, {Lira}, {Loffredo}, {Lucatello}, {Magrini}, {Maguire}, {Mahler}, {Zahra Majidi}, {Malavasi}, {Mannucci}, {Marconi}, {Martin}, {Marulli}, {Massari}, {Matsuno}, {Mattheee}, {McGee}, {Merc}, {Merle}, {Miglio}, {Migliorini}, {Minchev}, {Minniti}, {Miret-Roig}, {Monreal Ibero}, {Montano}, {Montet}, {Moresco}, {Moretti}, {Moscardini}, {Moya}, {Mueller}, {Nanayakkara}, {Nicholl}, {Nordlander}, {Onori}, {Padovani}, {Pala}, {Panda}, {Pandey-Pommier}, {Pasquini}, {Pawlak}, {Pessi}, {Pisani}, {Popovic}, {Prisinzano}, {Raddi}, {Rainer}, {Rebassa-Mansergas}, {Richard}, {Rigault}, {Rocher}, {Romano}, {Rosati}, {Sacco}, {Sanchez-Janssen}, {Sander}, {Sanders}, {Sargent}, {Sarpa}, {Schimd}, {Schipani},
  {Sefusatti}, {Smith}, {Spina}, {Steinmetz}, {Tacchella}, {Tautvaisiene}, {Theissen}, {Thomas}, {Ting}, {Travouillon}, {Tresse}, {Trivedi}, {Tsantaki}, {Tsedrik}, {Urrutia}, {Valenti}, {Van der Swaelmen}, {Van Eck}, {Verdiani}, {Verdier}, {Vergani}, {Verhamme}, \& {Vernet}}]{mainieri2024}
{Mainieri}, V., {Anderson}, R.~I., {Brinchmann}, J., {et~al.} 2024, arXiv e-prints, arXiv:2403.05398

\bibitem[{{Makhathini} {et~al.}(2021){Makhathini}, {Mooley}, {Brightman}, {Hotokezaka}, {Nayana}, {Intema}, {Dobie}, {Lenc}, {Perley}, {Fremling}, {Mold{\`o}n}, {Lazzati}, {Kaplan}, {Balasubramanian}, {Brown}, {Carbone}, {Chandra}, {Corsi}, {Camilo}, {Deller}, {Frail}, {Murphy}, {Murphy}, {Nakar}, {Smirnov}, {Beswick}, {Fender}, {Hallinan}, {Heywood}, {Kasliwal}, {Lee}, {Lu}, {Rana}, {Perkins}, {White}, {J{\'o}zsa}, {Hugo}, \& {Kamphuis}}]{Makhathini2021}
{Makhathini}, S., {Mooley}, K.~P., {Brightman}, M., {et~al.} 2021, \apj, 922, 154

\bibitem[{Mapelli \& Giacobbo(2018)}]{Mapelli2018}
Mapelli, M. \& Giacobbo, N. 2018, Monthly Notices of the Royal Astronomical Society, 479, 4391–4398

\bibitem[{{Marconi} {et~al.}(2022){Marconi}, {Abreu}, {Adibekyan}, {Alberti}, {Albrecht}, {Alcaniz}, {Aliverti}, {Allende Prieto}, {Alvarado G{\'o}mez}, {Amado}, {Amate}, {Andersen}, {Artigau}, {Baker}, {Baldini}, {Balestra}, {Barnes}, {Baron}, {Barros}, {Bauer}, {Beaulieu}, {Bellido-Tirado}, {Benneke}, {Bensby}, {Bergin}, {Biazzo}, {Bik}, {Birkby}, {Blind}, {Boisse}, {Bolmont}, {Bonaglia}, {Bonfils}, {Borsa}, {Brandeker}, {Brandner}, {Broeg}, {Brogi}, {Brousseau}, {Brucalassi}, {Brynnel}, {Buchhave}, {Buscher}, {Cabral}, {Calderone}, {Calvo-Ortega}, {Canto Martins}, {Cantalloube}, {Carbonaro}, {Chauvin}, {Chazelas}, {Cheffot}, {Cheng}, {Chiavassa}, {Christensen}, {Cirami}, {Cook}, {Cooke}, {Coretti}, {Covino}, {Cowan}, {Cresci}, {Cristiani}, {Cunha Parro}, {Cupani}, {D'Odorico}, {de Castro Le{\~a}o}, {De Cia}, {De Medeiros}, {Debras}, {Debus}, {Demangeon}, {Dessauges-Zavadsky}, {Di Marcantonio}, {Dionies}, {Doyon}, {Dunn}, {Ehrenreich}, {Faria}, {Feruglio}, {Fisher}, {Fontana}, {Fumagalli}, {Fusco}, {Fynbo},
  {Gabella}, {Gaessler}, {Gallo}, {Gao}, {Genolet}, {Genoni}, {Giacobbe}, {Giro}, {Gon{\c{c}}alves}, {Gonzalez}, {Gonz{\'a}lez Hern{\'a}ndez}, {Gracia T{\'e}mich}, {Haehnelt}, {Haniff}, {Hatzes}, {Helled}, {Hoeijmakers}, {Huke}, {J{\"a}rvinen}, {J{\"a}rvinen}, {Kaminski}, {Korn}, {Kouach}, {Kowzan}, {Kreidberg}, {Landoni}, {Lanotte}, {Lavail}, {Li}, {Liske}, {Lovis}, {Lucatello}, {Lunney}, {MacIntosh}, {Madhusudhan}, {Magrini}, {Maiolino}, {Malo}, {Man}, {Marquart}, {Marques}, {Martins}, {Martins}, {Maslowski}, {Mason}, {Mason}, {McCracken}, {Mergo}, {Micela}, {Mitchell}, {Molli{\`e}re}, {Monteiro}, {Montgomery}, {Mordasini}, {Morin}, {Mucciarelli}, {Murphy}, {N'Diaye}, {Neichel}, {Niedzielski}, {Niemczura}, {Nortmann}, {Noterdaeme}, {Nunes}, {Oggioni}, {Oliva}, {{\"O}nel}, {Origlia}, {{\"O}stlin}, {Palle}, {Papaderos}, {Pariani}, {Pe{\~n}ate Castro}, {Pepe}, {Perreault Levasseur}, {Petit}, {Pino}, {Piqueras}, {Pollo}, {Poppenhaeger}, {Quirrenbach}, {Rauscher}, {Rebolo}, {Redaelli}, {Reffert}, {Reid},
  {Reiners}, {Richter}, {Riva}, {Rivoire}, {Rodr{\'\i}guez-L{\'o}pez}, {Roederer}, {Romano}, {Rousseau}, {Rowe}, {Salvadori}, {Santos}, {Santos Diaz}, {Sanz-Forcada}, {Sarajlic}, {Sauvage}, {Sch{\"a}fer}, {Schiavon}, {Schmidt}, {Selmi}, {Sivanandam}, {Sordet}, {Sordo}, {Sortino}, {Sosnowska}, {Sousa}, {Stempels}, {Strassmeier}, {Su{\'a}rez Mascare{\~n}o}, \& {Sulich}}]{marconi2022}
{Marconi}, A., {Abreu}, M., {Adibekyan}, V., {et~al.} 2022, in Society of Photo-Optical Instrumentation Engineers (SPIE) Conference Series, Vol. 12184, Ground-based and Airborne Instrumentation for Astronomy IX, ed. C.~J. {Evans}, J.~J. {Bryant}, \& K.~{Motohara}, 1218424

\bibitem[{{Mei} {et~al.}(2022){Mei}, {Banerjee}, {Oganesyan}, {Salafia}, {Giarratana}, {Branchesi}, {D'Avanzo}, {Campana}, {Ghirlanda}, {Ronchini}, {Shukla}, \& {Tiwari}}]{Mei2022}
{Mei}, A., {Banerjee}, B., {Oganesyan}, G., {et~al.} 2022, \nat, 612, 236

\bibitem[{{Metzger}(2019)}]{metzger2019_kn}
{Metzger}, B.~D. 2019, Living Reviews in Relativity, 23, 1

\bibitem[{{Miceli} \& {Nava}(2022)}]{Miceli2022}
{Miceli}, D. \& {Nava}, L. 2022, Galaxies, 10, 66

\bibitem[{{Mochkovitch} {et~al.}(1993){Mochkovitch}, {Hernanz}, {Isern}, \& {Martin}}]{Mochkovitch1993}
{Mochkovitch}, R., {Hernanz}, M., {Isern}, J., \& {Martin}, X. 1993, \nat, 361, 236

\bibitem[{{Mooley} {et~al.}(2018){Mooley}, {Frail}, {Dobie}, {Lenc}, {Corsi}, {De}, {Nayana}, {Makhathini}, {Heywood}, {Murphy}, {Kaplan}, {Chandra}, {Smirnov}, {Nakar}, {Hallinan}, {Camilo}, {Fender}, {Goedhart}, {Groot}, {Kasliwal}, {Kulkarni}, \& {Woudt}}]{Mooley2018}
{Mooley}, K.~P., {Frail}, D.~A., {Dobie}, D., {et~al.} 2018, \apjl, 868, L11

\bibitem[{{Nandra} {et~al.}(2013){Nandra}, {Barret}, {Barcons}, {Fabian}, {den Herder}, {Piro}, {Watson}, {Adami}, {Aird}, {Afonso}, {Alexander}, {Argiroffi}, {Amati}, {Arnaud}, {Atteia}, {Audard}, {Badenes}, {Ballet}, {Ballo}, {Bamba}, {Bhardwaj}, {Stefano Battistelli}, {Becker}, {De Becker}, {Behar}, {Bianchi}, {Biffi}, {B{\^\i}rzan}, {Bocchino}, {Bogdanov}, {Boirin}, {Boller}, {Borgani}, {Borm}, {Bouch{\'e}}, {Bourdin}, {Bower}, {Braito}, {Branchini}, {Branduardi-Raymont}, {Bregman}, {Brenneman}, {Brightman}, {Br{\"u}ggen}, {Buchner}, {Bulbul}, {Brusa}, {Bursa}, {Caccianiga}, {Cackett}, {Campana}, {Cappelluti}, {Cappi}, {Carrera}, {Ceballos}, {Christensen}, {Chu}, {Churazov}, {Clerc}, {Corbel}, {Corral}, {Comastri}, {Costantini}, {Croston}, {Dadina}, {D'Ai}, {Decourchelle}, {Della Ceca}, {Dennerl}, {Dolag}, {Done}, {Dovciak}, {Drake}, {Eckert}, {Edge}, {Ettori}, {Ezoe}, {Feigelson}, {Fender}, {Feruglio}, {Finoguenov}, {Fiore}, {Galeazzi}, {Gallagher}, {Gandhi}, {Gaspari}, {Gastaldello}, {Georgakakis},
  {Georgantopoulos}, {Gilfanov}, {Gitti}, {Gladstone}, {Goosmann}, {Gosset}, {Grosso}, {Guedel}, {Guerrero}, {Haberl}, {Hardcastle}, {Heinz}, {Alonso Herrero}, {Herv{\'e}}, {Holmstrom}, {Iwasawa}, {Jonker}, {Kaastra}, {Kara}, {Karas}, {Kastner}, {King}, {Kosenko}, {Koutroumpa}, {Kraft}, {Kreykenbohm}, {Lallement}, {Lanzuisi}, {Lee}, {Lemoine-Goumard}, {Lobban}, {Lodato}, {Lovisari}, {Lotti}, {McCharthy}, {McNamara}, {Maggio}, {Maiolino}, {De Marco}, {de Martino}, {Mateos}, {Matt}, {Maughan}, {Mazzotta}, {Mendez}, {Merloni}, {Micela}, {Miceli}, {Mignani}, {Miller}, {Miniutti}, {Molendi}, {Montez}, {Moretti}, {Motch}, {Naz{\'e}}, {Nevalainen}, {Nicastro}, {Nulsen}, {Ohashi}, {O'Brien}, {Osborne}, {Oskinova}, {Pacaud}, {Paerels}, {Page}, {Papadakis}, {Pareschi}, {Petre}, {Petrucci}, {Piconcelli}, {Pillitteri}, {Pinto}, {de Plaa}, {Pointecouteau}, {Ponman}, {Ponti}, {Porquet}, {Pounds}, {Pratt}, {Predehl}, {Proga}, {Psaltis}, {Rafferty}, {Ramos-Ceja}, {Ranalli}, {Rasia}, {Rau}, {Rauw}, {Rea}, {Read}, {Reeves},
  {Reiprich}, {Renaud}, {Reynolds}, {Risaliti}, {Rodriguez}, {Rodriguez Hidalgo}, {Roncarelli}, {Rosario}, {Rossetti}, {Rozanska}, {Rovilos}, {Salvaterra}, {Salvato}, {Di Salvo}, {Sanders}, {Sanz-Forcada}, {Schawinski}, {Schaye}, {Schwope}, \& {Sciortino}}]{nandra2013}
{Nandra}, K., {Barret}, D., {Barcons}, X., {et~al.} 2013, arXiv e-prints, arXiv:1306.2307

\bibitem[{{Neijssel} {et~al.}(2019){Neijssel}, {Vigna-G{\'o}mez}, {Stevenson}, {Barrett}, {Gaebel}, {Broekgaarden}, {de Mink}, {Sz{\'e}csi}, {Vinciguerra}, \& {Mandel}}]{Neijssel2019}
{Neijssel}, C.~J., {Vigna-G{\'o}mez}, A., {Stevenson}, S., {et~al.} 2019, \mnras, 490, 3740

\bibitem[{{{\"O}zel} \& {Freire}(2016)}]{ozel2016}
{{\"O}zel}, F. \& {Freire}, P. 2016, \araa, 54, 401

\bibitem[{{{\"O}zel} {et~al.}(2010){{\"O}zel}, {Psaltis}, {Narayan}, \& {McClintock}}]{ozel2010}
{{\"O}zel}, F., {Psaltis}, D., {Narayan}, R., \& {McClintock}, J.~E. 2010, \apj, 725, 1918

\bibitem[{{{\"O}zel} {et~al.}(2012){{\"O}zel}, {Psaltis}, {Narayan}, \& {Santos Villarreal}}]{ozel2012}
{{\"O}zel}, F., {Psaltis}, D., {Narayan}, R., \& {Santos Villarreal}, A. 2012, \apj, 757, 55

\bibitem[{{Pannarale} {et~al.}(2015){Pannarale}, {Berti}, {Kyutoku}, {Lackey}, \& {Shibata}}]{pannarale2015}
{Pannarale}, F., {Berti}, E., {Kyutoku}, K., {Lackey}, B.~D., \& {Shibata}, M. 2015, \prd, 92, 084050

\bibitem[{{Pian} {et~al.}(2017){Pian}, {D'Avanzo}, {Benetti}, {Branchesi}, {Brocato}, {Campana}, {Cappellaro}, {Covino}, {D'Elia}, {Fynbo}, {Getman}, {Ghirlanda}, {Ghisellini}, {Grado}, {Greco}, {Hjorth}, {Kouveliotou}, {Levan}, {Limatola}, {Malesani}, {Mazzali}, {Melandri}, {M{\o}ller}, {Nicastro}, {Palazzi}, {Piranomonte}, {Rossi}, {Salafia}, {Selsing}, {Stratta}, {Tanaka}, {Tanvir}, {Tomasella}, {Watson}, {Yang}, {Amati}, {Antonelli}, {Ascenzi}, {Bernardini}, {Bo{\"e}r}, {Bufano}, {Bulgarelli}, {Capaccioli}, {Casella}, {Castro-Tirado}, {Chassande-Mottin}, {Ciolfi}, {Copperwheat}, {Dadina}, {De Cesare}, {di Paola}, {Fan}, {Gendre}, {Giuffrida}, {Giunta}, {Hunt}, {Israel}, {Jin}, {Kasliwal}, {Klose}, {Lisi}, {Longo}, {Maiorano}, {Mapelli}, {Masetti}, {Nava}, {Patricelli}, {Perley}, {Pescalli}, {Piran}, {Possenti}, {Pulone}, {Razzano}, {Salvaterra}, {Schipani}, {Spera}, {Stamerra}, {Stella}, {Tagliaferri}, {Testa}, {Troja}, {Turatto}, {Vergani}, \& {Vergani}}]{Pian2017}
{Pian}, E., {D'Avanzo}, P., {Benetti}, S., {et~al.} 2017, \nat, 551, 67

\bibitem[{{Planck Collaboration} {et~al.}(2020){Planck Collaboration}, {Aghanim}, {Akrami}, {Ashdown}, {Aumont}, {Baccigalupi}, {Ballardini}, {Banday}, {Barreiro}, {Bartolo}, \& et~al.}]{Planck2020}
{Planck Collaboration}, {Aghanim}, N., {Akrami}, Y., {et~al.} 2020, \aap, 641, A6

\bibitem[{{Punturo} {et~al.}(2010){Punturo}, {Abernathy}, {Acernese}, {Allen}, {Andersson}, {Arun}, {Barone}, {Barr}, {Barsuglia}, {Beker}, {Beveridge}, {Birindelli}, {Bose}, {Bosi}, {Braccini}, {Bradaschia}, {Bulik}, {Calloni}, {Cella}, {Chassande Mottin}, {Chelkowski}, {Chincarini}, {Clark}, {Coccia}, {Colacino}, {Colas}, {Cumming}, {Cunningham}, {Cuoco}, {Danilishin}, {Danzmann}, {De Luca}, {De Salvo}, {Dent}, {De Rosa}, {Di Fiore}, {Di Virgilio}, {Doets}, {Fafone}, {Falferi}, {Flaminio}, {Franc}, {Frasconi}, {Freise}, {Fulda}, {Gair}, {Gemme}, {Gennai}, {Giazotto}, {Glampedakis}, {Granata}, {Grote}, {Guidi}, {Hammond}, {Hannam}, {Harms}, {Heinert}, {Hendry}, {Heng}, {Hennes}, {Hild}, {Hough}, {Husa}, {Huttner}, {Jones}, {Khalili}, {Kokeyama}, {Kokkotas}, {Krishnan}, {Lorenzini}, {L{\"u}ck}, {Majorana}, {Mandel}, {Mandic}, {Martin}, {Michel}, {Minenkov}, {Morgado}, {Mosca}, {Mours}, {M{\"u}ller{\textendash}Ebhardt}, {Murray}, {Nawrodt}, {Nelson}, {Oshaughnessy}, {Ott}, {Palomba}, {Paoli}, {Parguez},
  {Pasqualetti}, {Passaquieti}, {Passuello}, {Pinard}, {Poggiani}, {Popolizio}, {Prato}, {Puppo}, {Rabeling}, {Rapagnani}, {Read}, {Regimbau}, {Rehbein}, {Reid}, {Rezzolla}, {Ricci}, {Richard}, {Rocchi}, {Rowan}, {R{\"u}diger}, {Sassolas}, {Sathyaprakash}, {Schnabel}, {Schwarz}, {Seidel}, {Sintes}, {Somiya}, {Speirits}, {Strain}, {Strigin}, {Sutton}, {Tarabrin}, {Th{\"u}ring}, {van den Brand}, {van Leewen}, {van Veggel}, {van den Broeck}, {Vecchio}, {Veitch}, {Vetrano}, {Vicere}, {Vyatchanin}, {Willke}, {Woan}, {Wolfango}, \& {Yamamoto}}]{punturo2010}
{Punturo}, M., {Abernathy}, M., {Acernese}, F., {et~al.} 2010, Classical and Quantum Gravity, 27, 194002

\bibitem[{{Qin} {et~al.}(2018){Qin}, {Fragos}, {Meynet}, {Andrews}, {S{\o}rensen}, \& {Song}}]{Qin2018}
{Qin}, Y., {Fragos}, T., {Meynet}, G., {et~al.} 2018, \aap, 616, A28

\bibitem[{{Rastinejad} {et~al.}(2022){Rastinejad}, {Gompertz}, {Levan}, {Fong}, {Nicholl}, {Lamb}, {Malesani}, {Nugent}, {Oates}, {Tanvir}, {de Ugarte Postigo}, {Kilpatrick}, {Moore}, {Metzger}, {Ravasio}, {Rossi}, {Schroeder}, {Jencson}, {Sand}, {Smith}, {Ag{\"u}{\'\i} Fern{\'a}ndez}, {Berger}, {Blanchard}, {Chornock}, {Cobb}, {De Pasquale}, {Fynbo}, {Izzo}, {Kann}, {Laskar}, {Marini}, {Paterson}, {Escorial}, {Sears}, \& {Th{\"o}ne}}]{rastinejad2022}
{Rastinejad}, J.~C., {Gompertz}, B.~P., {Levan}, A.~J., {et~al.} 2022, \nat, 612, 223

\bibitem[{{Reitze} {et~al.}(2019){Reitze}, {Adhikari}, {Ballmer}, {Barish}, {Barsotti}, {Billingsley}, {Brown}, {Chen}, {Coyne}, {Eisenstein}, {Evans}, {Fritschel}, {Hall}, {Lazzarini}, {Lovelace}, {Read}, {Sathyaprakash}, {Shoemaker}, {Smith}, {Torrie}, {Vitale}, {Weiss}, {Wipf}, \& {Zucker}}]{Reitze2019}
{Reitze}, D., {Adhikari}, R.~X., {Ballmer}, S., {et~al.} 2019, in Bulletin of the American Astronomical Society, Vol.~51, 35

\bibitem[{{Rom{\'a}n-Garza} {et~al.}(2021{\natexlab{a}}){Rom{\'a}n-Garza}, {Bavera}, {Fragos}, {Zapartas}, {Misra}, {Andrews}, {Coughlin}, {Dotter}, {Kovlakas}, {Serra}, {Qin}, {Rocha}, \& {Tran}}]{RomanGarza2021}
{Rom{\'a}n-Garza}, J., {Bavera}, S.~S., {Fragos}, T., {et~al.} 2021{\natexlab{a}}, \apjl, 912, L23

\bibitem[{{Rom{\'a}n-Garza} {et~al.}(2021{\natexlab{b}}){Rom{\'a}n-Garza}, {Bavera}, {Fragos}, {Zapartas}, {Misra}, {Andrews}, {Coughlin}, {Dotter}, {Kovlakas}, {Serra}, {Qin}, {Rocha}, \& {Tran}}]{RomanGarza:2021ApJ...912L..23R}
{Rom{\'a}n-Garza}, J., {Bavera}, S.~S., {Fragos}, T., {et~al.} 2021{\natexlab{b}}, \apjl, 912, L23

\bibitem[{{Ronchini} {et~al.}(2022){Ronchini}, {Branchesi}, {Oganesyan}, {Banerjee}, {Dupletsa}, {Ghirlanda}, {Harms}, {Mapelli}, \& {Santoliquido}}]{ronchini2022}
{Ronchini}, S., {Branchesi}, M., {Oganesyan}, G., {et~al.} 2022, \aap, 665, A97

\bibitem[{Safarzadeh \& Berger(2019)}]{Safarzadeh2019}
Safarzadeh, M. \& Berger, E. 2019, The Astrophysical Journal Letters, 878, L12

\bibitem[{{Salafia} {et~al.}(2020){Salafia}, {Barbieri}, {Ascenzi}, \& {Toffano}}]{Salafia2020}
{Salafia}, O.~S., {Barbieri}, C., {Ascenzi}, S., \& {Toffano}, M. 2020, \aap, 636, A105

\bibitem[{{Salafia} \& {Ghirlanda}(2022)}]{Salafia2022}
{Salafia}, O.~S. \& {Ghirlanda}, G. 2022, Galaxies, 10, 93

\bibitem[{{Salafia} {et~al.}(2019){Salafia}, {Ghirlanda}, {Ascenzi}, \& {Ghisellini}}]{salafia2019}
{Salafia}, O.~S., {Ghirlanda}, G., {Ascenzi}, S., \& {Ghisellini}, G. 2019, \aap, 628, A18

\bibitem[{{Salafia} \& {Giacomazzo}(2021)}]{Salafia2021}
{Salafia}, O.~S. \& {Giacomazzo}, B. 2021, \aap, 645, A93

\bibitem[{{Salafia} {et~al.}(2023){Salafia}, {Ravasio}, {Ghirlanda}, \& {Mandel}}]{Salafia2023}
{Salafia}, O.~S., {Ravasio}, M.~E., {Ghirlanda}, G., \& {Mandel}, I. 2023, arXiv e-prints, arXiv:2306.15488

\bibitem[{{Salafia} {et~al.}(2022){Salafia}, {Ravasio}, {Yang}, {An}, {Orienti}, {Ghirlanda}, {Nava}, {Giroletti}, {Mohan}, {Spinelli}, {Zhang}, {Marcote}, {Cim{\`o}}, {Wu}, \& {Li}}]{Salafia2022b}
{Salafia}, O.~S., {Ravasio}, M.~E., {Yang}, J., {et~al.} 2022, \apjl, 931, L19

\bibitem[{{Schutz}(2011)}]{schutz2011}
{Schutz}, B.~F. 2011, Classical and Quantum Gravity, 28, 125023

\bibitem[{{Schwab} {et~al.}(2010){Schwab}, {Podsiadlowski}, \& {Rappaport}}]{schwab2010}
{Schwab}, J., {Podsiadlowski}, P., \& {Rappaport}, S. 2010, \apj, 719, 722

\bibitem[{{Steiner} {et~al.}(2013){Steiner}, {Lattimer}, \& {Brown}}]{Steiner2013}
{Steiner}, A.~W., {Lattimer}, J.~M., \& {Brown}, E.~F. 2013, \apjl, 765, L5

\bibitem[{{Troja} {et~al.}(2022){Troja}, {Fryer}, {O'Connor}, {Ryan}, {Dichiara}, {Kumar}, {Ito}, {Gupta}, {Wollaeger}, {Norris}, {Kawai}, {Butler}, {Aryan}, {Misra}, {Hosokawa}, {Murata}, {Niwano}, {Pandey}, {Kutyrev}, {van Eerten}, {Chase}, {Hu}, {Caballero-Garcia}, \& {Castro-Tirado}}]{troja2022}
{Troja}, E., {Fryer}, C.~L., {O'Connor}, B., {et~al.} 2022, \nat, 612, 228

\bibitem[{{Urrutia} {et~al.}(2021){Urrutia}, {De Colle}, {Murguia-Berthier}, \& {Ramirez-Ruiz}}]{Urrutia2021}
{Urrutia}, G., {De Colle}, F., {Murguia-Berthier}, A., \& {Ramirez-Ruiz}, E. 2021, \mnras, 503, 4363

\bibitem[{{Valentim} {et~al.}(2011){Valentim}, {Rangel}, \& {Horvath}}]{valentim2011}
{Valentim}, R., {Rangel}, E., \& {Horvath}, J.~E. 2011, \mnras, 414, 1427

\bibitem[{Vallisneri(2008)}]{Vallisneri:2007ev}
Vallisneri, M. 2008, Phys. Rev. D, 77, 042001

\bibitem[{Veitch {et~al.}(2015)Veitch, Raymond, Farr, Farr, Graff, Vitale, Aylott, Blackburn, Christensen, Coughlin, Del~Pozzo, Feroz, Gair, Haster, Kalogera, Littenberg, Mandel, O'Shaughnessy, Pitkin, Rodriguez, R\"over, Sidery, Smith, Van Der~Sluys, Vecchio, Vousden, \& Wade}]{Veitch2015}
Veitch, J., Raymond, V., Farr, B., {et~al.} 2015, Phys. Rev. D, 91, 042003

\bibitem[{{Villar} {et~al.}(2017){Villar}, {Guillochon}, {Berger}, {Metzger}, {Cowperthwaite}, {Nicholl}, {Alexander}, {Blanchard}, {Chornock}, {Eftekhari}, {Fong}, {Margutti}, \& {Williams}}]{Villar2017}
{Villar}, V.~A., {Guillochon}, J., {Berger}, E., {et~al.} 2017, \apjl, 851, L21

\bibitem[{{Xing} {et~al.}(2024{\natexlab{a}}){Xing}, {Bavera}, {Fragos}, {Kruckow}, {Rom{\'a}n-Garza}, {Andrews}, {Dotter}, {Kovlakas}, {Misra}, {Srivastava}, {Rocha}, {Sun}, \& {Zapartas}}]{Xing:2024A&A...683A.144X}
{Xing}, Z., {Bavera}, S.~S., {Fragos}, T., {et~al.} 2024{\natexlab{a}}, \aap, 683, A144

\bibitem[{{Xing} {et~al.}(2024{\natexlab{b}}){Xing}, {Kalogera}, {Fragos}, {Andrews}, {Bavera}, {Briel}, {Gossage}, {Kovlakas}, {Kruckow}, {Rocha}, {Sun}, {Srivastava}, \& {Zapartas}}]{xing2024}
{Xing}, Z., {Kalogera}, V., {Fragos}, T., {et~al.} 2024{\natexlab{b}}, arXiv e-prints, arXiv:2410.20415

\bibitem[{{Yang} {et~al.}(2024){Yang}, {Troja}, {O'Connor}, {Fryer}, {Im}, {Durbak}, {Paek}, {Ricci}, {Bom}, {Gillanders}, {Castro-Tirado}, {Peng}, {Dichiara}, {Ryan}, {van Eerten}, {Dai}, {Chang}, {Choi}, {De}, {Hu}, {Kilpatrick}, {Kutyrev}, {Jeong}, {Lee}, {Makler}, {Navarete}, \& {P{\'e}rez-Garc{\'\i}a}}]{yang2024}
{Yang}, Y.-H., {Troja}, E., {O'Connor}, B., {et~al.} 2024, \nat, 626, 742

\bibitem[{{You} {et~al.}(2024){You}, {Zhu}, {Liu}, {M{\"u}ller}, {Heger}, {Stevenson}, {Thrane}, {Chen}, {Sun}, {Lasky}, {Galloway}, {Bailes}, {Hobbs}, {Manchester}, {Gao}, \& {Zhu}}]{you2024}
{You}, Z.-Q., {Zhu}, X., {Liu}, X., {et~al.} 2024, arXiv e-prints, arXiv:2412.05524

\bibitem[{{Zevin} {et~al.}(2020){Zevin}, {Spera}, {Berry}, \& {Kalogera}}]{zevin2020}
{Zevin}, M., {Spera}, M., {Berry}, C. P.~L., \& {Kalogera}, V. 2020, \apjl, 899, L1

\bibitem[{{Zhu} {et~al.}(2021){Zhu}, {Wu}, {Yang}, {Zhang}, {Gao}, {Yu}, {Li}, {Cao}, {Liu}, {Huang}, \& {Zhang}}]{zhu2021}
{Zhu}, J.-P., {Wu}, S., {Yang}, Y.-P., {et~al.} 2021, \apj, 917, 24

\end{thebibliography}
\normalsize

\begin{appendix}

\section{Population model}
\begin{figure*}
    \centering
    \includegraphics[width=.9\textwidth]{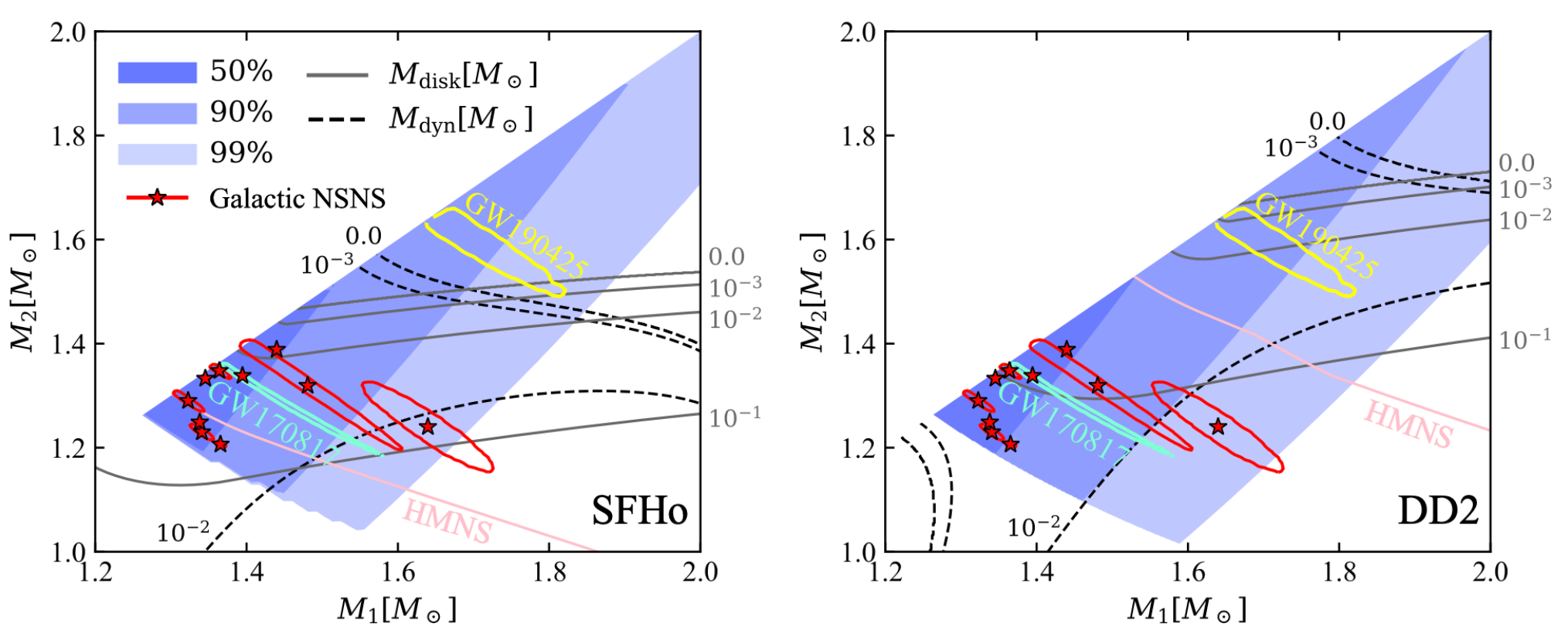}
    \caption{$M_1, M_2$ plane showing the mass distribution for our NSNS population. The filled blue colored regions contain $50\%$, $90\%$ and $99\%$ of the binaries. The black dashed lines and the grey lines represent respectively the contours for the predicted dynamical ejecta and disk mass, assuming the SFHo EoS (left panel) and the DD2 EoS (right panel). The pink line indicates the condition for a HMNS remnant ($M_\mathrm{rem}>1.2M_\mathrm{TOV}$). Red stars and contours show the best fit and $90\%$ credible regions for the known Galactic NSNS \citep{ozel2016,farrow2019} systems that merge within a Hubble time. Yellow and aquamarine lines represent the $50\%$ confidence regions for the component masses in GW190425 \citep{Abbott2019_GW190425} and GW170817 \citep{Abbott2019_GW170817_properties}, both constructed using the publicly available low-spin-prior posterior samples.}
    \label{fig:nsns_mass}
\end{figure*}

\begin{figure*}
    \centering
    \includegraphics[width=.9\textwidth]{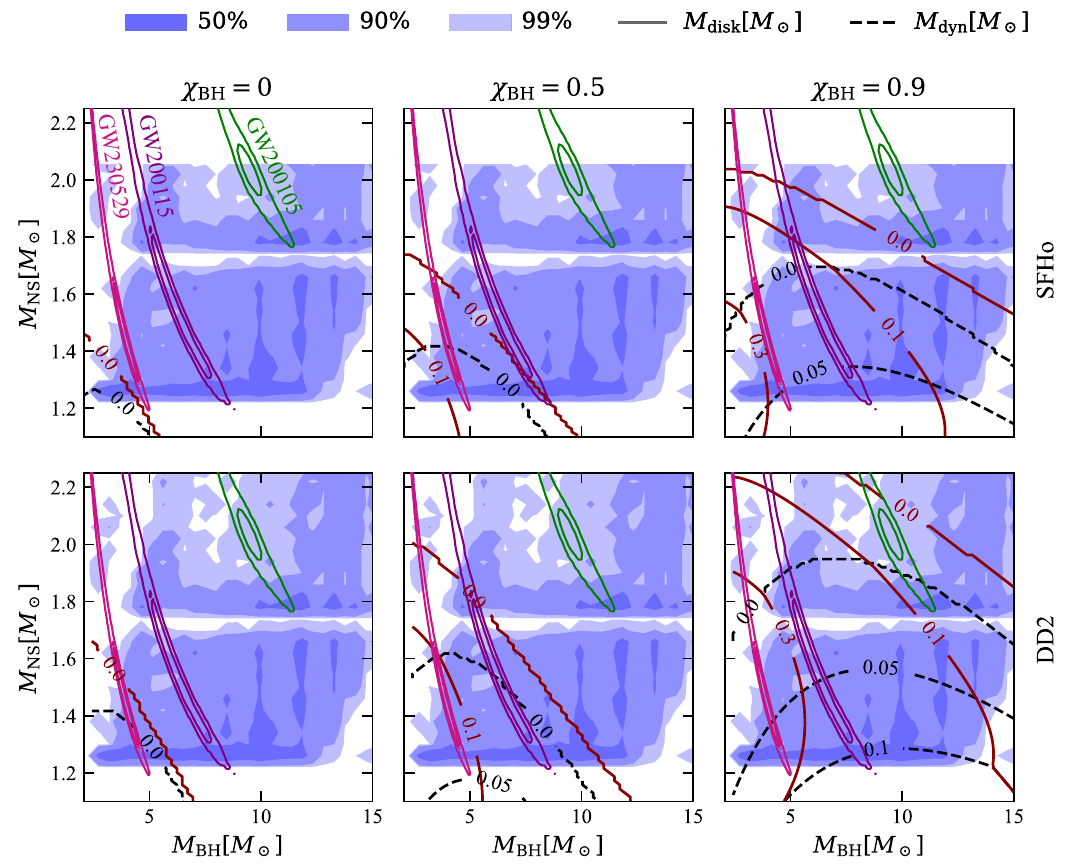}
    \caption{$M_\mathrm{NS}, M_\mathrm{BH}$ plane showing the mass distribution for our BHNS population at redshift $z=0$ (fiducial model in \citep{Broekgaarden2021}). The filled blue colored regions contain $50\%$, $90\%$ and $99\%$ of the binaries. The black dashed lines and the dark red lines represent respectively the contours for the predicted dynamical ejecta and disk mass, assuming the SFHo EoS (upper panel) and the DD2 EoS (lower panel).  Violet and green lines represent the $50\%$ and $90\%$ confidence regions for the component masses in GW200115, GW200105 \citep{Abbott2021} and GW230529, both constructed using the publicly available low-spin-prior posterior samples.}
    \label{fig:bhns_mass}
\end{figure*}

Figure~\ref{fig:nsns_mass} illustrates the mass distribution from \cite{colombo2022} compared against observational data on the $(M_1,M_2)$ plane. The plot also includes iso-contours of ejecta and accretion disk masses derived using the adopted fitting formulae \citep{Kruger2020,barbieri2021} and the equations of state (EoS) SFHo (left panel) and DD2 (right panel). These contours provide a visual representation of the lack of EM counterparts for events located in the upper right region of the plane and highlight the general trends of ejecta and disk mass distributions within the population. Additionally, the pink line indicates the condition for forming a HMNS remnant ($M_\mathrm{rem}>1.2M_\mathrm{TOV}$), which is also the threshold for launching a relativistic jet, together with the requirement of a non-negligible accretion disk ($m_\mathrm{disk}>0$).

In Figure~\ref{fig:bhns_mass}, we display analogous information to that shown in Figure~\ref{fig:nsns_mass}, but for the BHNS population. The figure illustrates our fiducial mass distribution for BHNS binaries at redshift $z = 0$ \citep{Broekgaarden2021} on the $(M_\mathrm{NS},M_\mathrm{BH})$ plane. The shaded blue regions represent the areas containing 50\%, 90\%, and 99\% of the binary systems. Iso-contours of ejecta and accretion disk masses are also included, computed using the fitting formulae from \citet{Kruger2020} and \citet{kawaguchi2016} for two equations of state (EoS), SFHo (upper panel) and DD2 (right panel), under three different black hole spin configurations: $\chi_\mathrm{BH}=0$, $\chi_\mathrm{BH}=0.5$, and $\chi_\mathrm{BH}=0.9$. The figure highlights the general trends in the distribution of ejecta and disk masses within the population and emphasizes the strong dependence of ejecta production on the black hole spin.

\section{Detection rates as a function of the assumed EM detection threshold}\label{App:det_lim}

\begin{figure*}
    \centering
    \includegraphics[width=\textwidth]{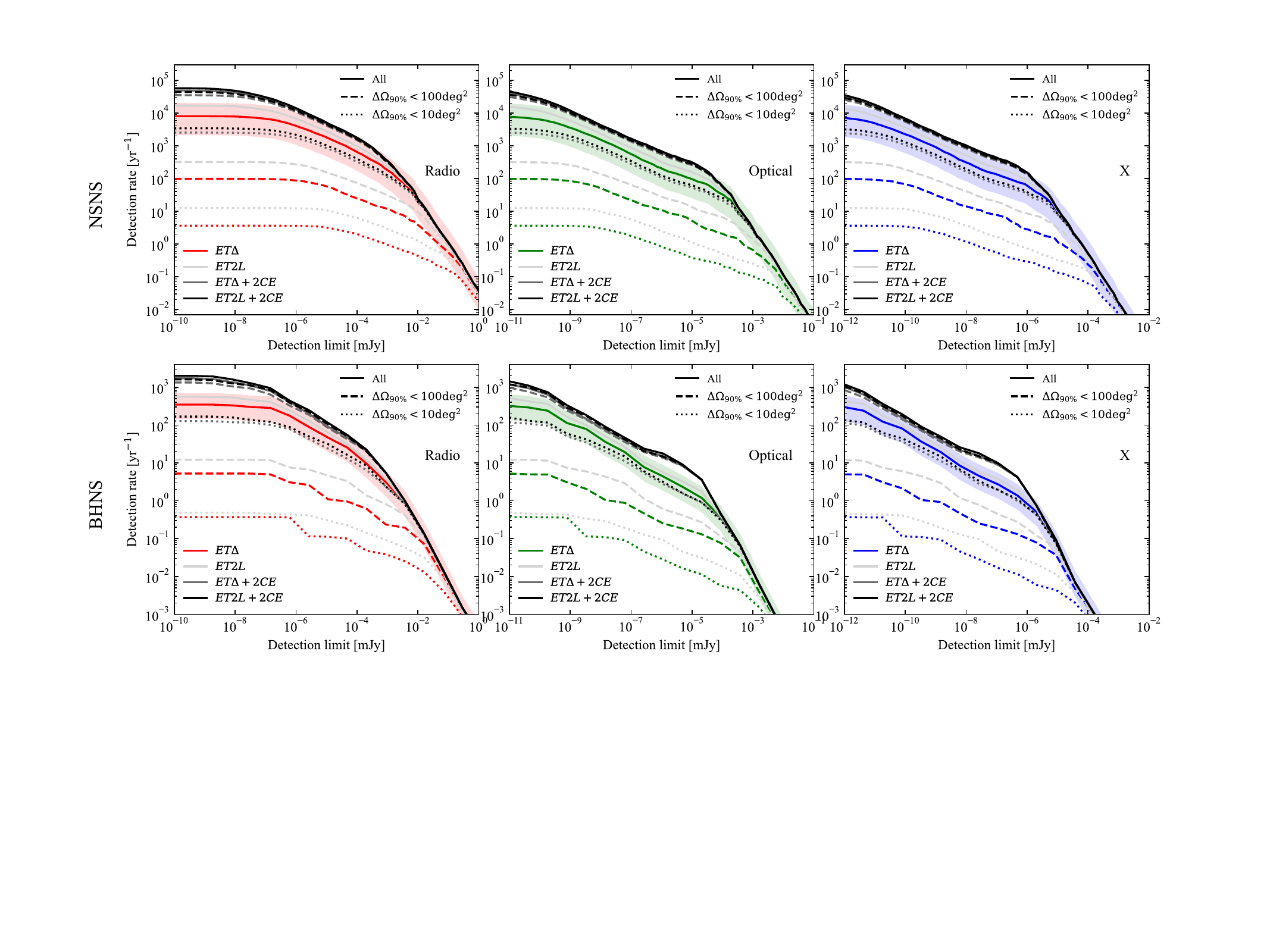}
    \caption{GRB afterglow detection rate as a function of the EM detection limit threshold for our fiducial NSNS  (upper panels) and BHNS (lower panels) populations. The red, green and blue colors indicate the radio, optical and X bands, respectively, assuming the ET$\Delta$ configuration. In each panel we also report in gray and black the ET2L and ET$\Delta$+2CE configurations. The solid line indicates all the detectable binaries, the dashed and dotted lines the detectable binaries with $\Delta\Omega_{{\rm 90}\%}<100\mathrm{deg}^2$ and the ones with $\Delta\Omega_{{\rm 90}\%}<10\mathrm{deg}^2$, respectively.}
    \label{fig:after_rate_lim}
\end{figure*}

\begin{figure*}
    \centering
    \includegraphics[width=.9\textwidth]{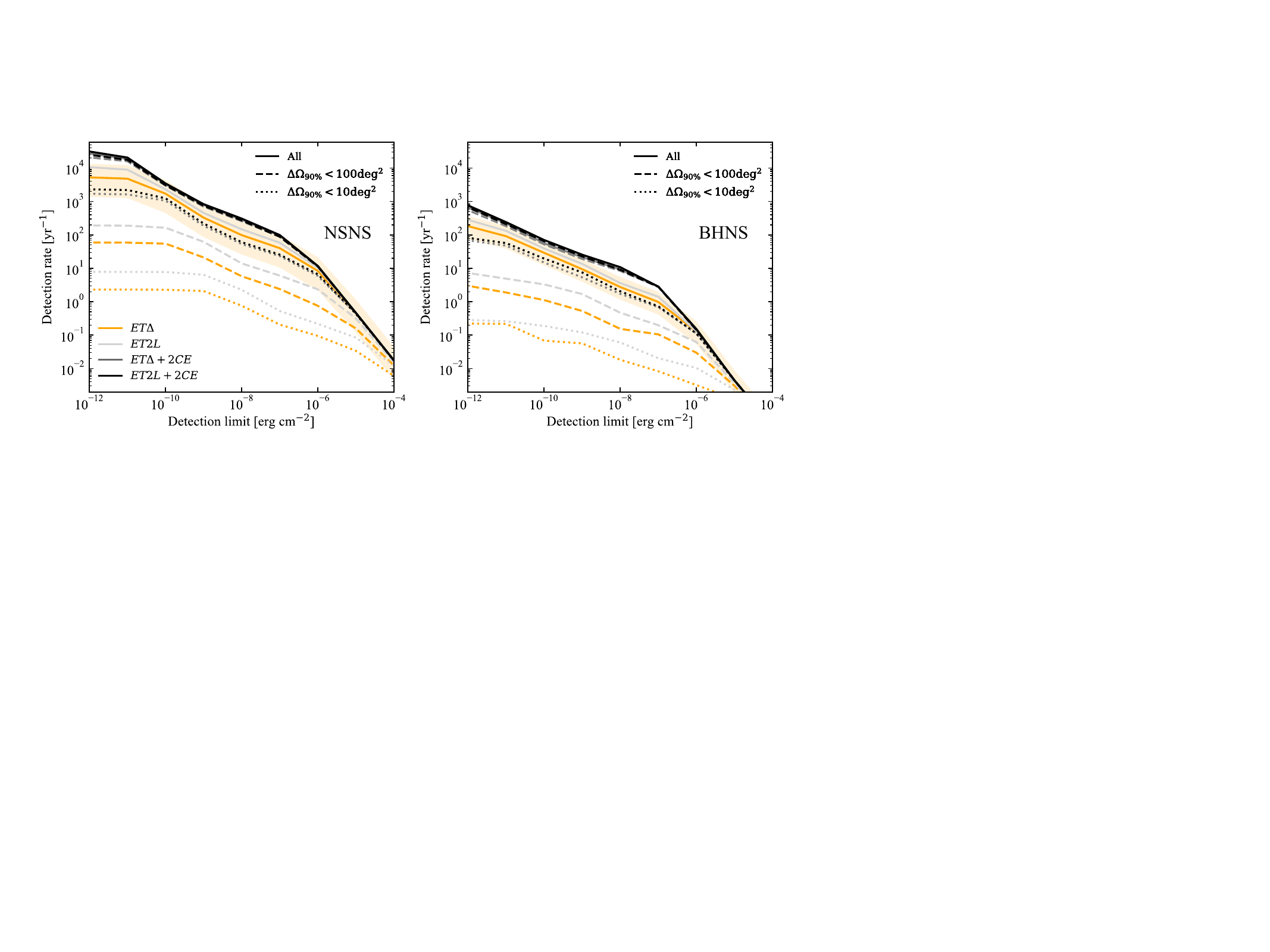}
    \caption{GRB prompt detection rate as a function of the EM detection limit threshold for our fiducial NSNS  (left panel) and BHNS (right panel) populations. We account for the \textit{Fermi}/GBM duty cycle. In each panel we report in orange, gray and black the ET$\Delta$, ET2L and ET$\Delta$+2CE configurations, respectively. The solid line indicates all the detectable binaries, the dashed and dotted lines the detectable binaries with $\Delta\Omega_{{\rm 90}\%}<100\mathrm{deg}^2$ and the ones with $\Delta\Omega_{{\rm 90}\%}<10\mathrm{deg}^2$, respectively.}
    \label{fig:prompt_rate_lim}
\end{figure*}

Figures \ref{fig:after_rate_lim} and \ref{fig:prompt_rate_lim} present the same information as in Figure \ref{fig:kn_rate_lim}, but for GRB afterglow+GW events and for GRB prompt+GW events, respectively. In Figure \ref{fig:after_rate_lim}, the upper panel refers to the NSNS population, while the lower panel refers to the BHNS population. The red, green, and blue colors indicate the radio, optical, and X bands. The solid, dashed, and dotted lines correspond to all GW events, and to those with $\Delta\Omega_{{\rm 90}\%}<100\,\mathrm{deg}^2$ and $\Delta\Omega_{{\rm 90}\%}<10\,\mathrm{deg}^2$, respectively. The configurations ET$\Delta$, ET2L, ET$\Delta$+2CE, and ET2L+2CE are represented by the colored, light gray, dark gray, and black lines, respectively.

In Figure \ref{fig:prompt_rate_lim}, all lines refer to the bolometric fluence. The left panel refers to the NSNS population and the right panel to the BHNS population.

\end{appendix}
\end{document}